\documentclass[12pt]{article}
\usepackage[english]{babel}
\usepackage[utf8]{inputenc}
\usepackage[T1]{fontenc} 

\usepackage{fullpage}
\usepackage{setspace}

\usepackage[margin=1.25in, letterpaper]{geometry}

\usepackage[indentafter]{titlesec}
\usepackage{sectsty}
\usepackage{microtype}
\usepackage{enumerate}
\usepackage[shortlabels]{enumitem}
\usepackage{bm}
\usepackage[ruled,vlined,linesnumbered]{algorithm2e}
\usepackage{bbm}

\usepackage{amsmath, amssymb, amsthm, amsfonts, latexsym, amscd, euscript, mathrsfs, mathtools}
\usepackage[colorlinks=true, linkcolor=darkred, citecolor=darkblue]{hyperref}
\usepackage{graphicx, xcolor}
\definecolor{darkred}{RGB}{144,0,0}
\definecolor{darkblue}{RGB}{0,0,144}
\definecolor{darkgreen}{RGB}{0,100,0}
\definecolor{lightyellow}{RGB}{255,255,224}
\usepackage[colorinlistoftodos]{todonotes}

\usepackage[retainorgcmds]{IEEEtrantools}
\usepackage{tabularx}
\usepackage{array}
\usepackage{caption}
\usepackage{xcolor}
\usepackage{subfig}
\usepackage{float}
\usepackage{tcolorbox}
\usepackage{tabularx}
\usepackage{threeparttable}
\usepackage{enumitem}
\usepackage{multirow}
\usepackage{lscape, pdflscape}
\usepackage[authoryear]{natbib}
\usepackage{bibunits}
\defaultbibliographystyle{chicago}
\defaultbibliography{EASI_SGD}
\makeatletter
\renewcommand{\@bibunitname}{\jobname-bu\the\@bibunitauxcnt}
\newcommand{\slimbibscope}[1]{%
  \def\@extra@b@citeb{.#1}%
  \def\@extra@binfo{.#1}%
}
\makeatother
\usepackage{fancyhdr}

\usepackage{booktabs}
\usepackage{array}
\newcolumntype{L}[1]{>{\raggedright\let\newline\\\arraybackslash\hspace{0pt}}m{#1}}
\newcolumntype{C}[1]{>{\centering\let\newline\\\arraybackslash\hspace{0pt}}m{#1}}
\newcolumntype{R}[1]{>{\raggedleft\let\newline\\\arraybackslash\hspace{0pt}}m{#1}}

\def\oP{o_{\rm P}}
\def\wto{\rightsquigarrow}
\def\dto{\overset{d}{\to}}
\def\pto{\overset{\mathbb{P}}{\to}}
\def\ppto{\overset{\mathbb{P}}{\to}}
\def\p{\mathbb{P}}
\def\psto{\overset{\mathbb{P}^*}{\to}}
\def\asto{\overset{\mathrm{a.s.}}{\to}}

\newcommand{\stt}{T}
\newcommand{\tmt}{\tau}
\newcommand{\loclevel}{\mathcal M}

\renewcommand{\P}{P}
\newcommand{\Ns}{{N^{\star}}}
\newcommand{\N}{{T}}
\newcommand{\gs}[1][t]{{\gamma_{#1}^{\star}}}
\newcommand{\tto}[2][{}]{{#1 \theta}_{{#2}}^{\circ}}
\renewcommand{\H}{\mathcal{H}}
\newcommand{\fo}{\mbox{first-order}}
\newcommand{\ts}{\mbox{second-order}}
\newcommand{\wpa}{\mbox{w.p.a.$1$}}

\newcommand{\argmin}{\operatornamewithlimits{arg\hspace{0.1em} min}}

\newcommand{\oo}{o}
\newcommand{\dtheta}{d}
\newcommand{\dg}{d_g}

\newcommand{\I}[1]{\mathbbm{1}\left\{{#1}\right\}}

\def\re{\operatornamewithlimits{Re}}

\usepackage{accents}
\makeatletter
\renewcommand{\underbar}[1]{\underaccent{\bar}{#1}}
\makeatother

\sectionfont{\Large}

\theoremstyle{definition} 
\theoremstyle{remark} \newtheorem{remark}{Remark}
\theoremstyle{plain} \newtheorem{theorem}{Theorem}
\theoremstyle{plain} \newtheorem{proposition}{Proposition}
\theoremstyle{plain} \newtheorem{lemma}{Lemma}
\theoremstyle{plain} \newtheorem{corollary}{Corollary}
\theoremstyle{plain} 
\theoremstyle{plain} \newtheorem{assumption}{Assumption}
\theoremstyle{plain} \newtheorem*{assumption*}{\assumptionletter}
\providecommand{\assumptionletter}{}

\usepackage{tikz}
\usetikzlibrary{shapes.geometric, arrows.meta, positioning, calc}

\tikzstyle{startstop} = [rectangle, rounded corners, minimum width=4cm, minimum height=1cm, text centered, draw=black, fill=gray!20]
\tikzstyle{process} = [rectangle, minimum width=4.6cm, minimum height=1cm, text centered, draw=black, fill=blue!10]
\tikzstyle{decision} = [ellipse, minimum height=1cm, minimum width=3cm, align=center,draw=black, fill=orange!20]
\tikzstyle{note} = [rectangle, text width=3.2cm, text centered, draw=black, fill=yellow!20]
\tikzstyle{arrow} = [thick,->,>=stealth]

\title{SLIM: Stochastic Learning and Inference \\in Nonlinear GMM Models\thanks{This is a revised version of arXiv:2510.20996v1. We thank participants at numerous conferences and workshops for helpful comments. We gratefully acknowledge access to Revelio Labs data through Yale University’s institutional academic license. Zhong's research is supported by the National Natural Science Foundation of China (project 72603281).}}

\author{Xiaohong Chen\footnote{Department of Economics, Yale University and Cowles Foundation for Research in Economics. Email: xiaohong.chen@yale.edu} 
\and
Min Seong Kim\footnote{Department of Economics, University of Connecticut. Email: min\_seong.kim@uconn.edu}
\and
Sokbae Lee\footnote{Department of Economics, Columbia University and Centre for Microdata Methods and Practice. Email: sl3841@columbia.edu}
\and
Myung Hwan Seo\footnote{Department of Economics, Hong Kong University of Science and Technology. Email: myunghseo@ust.hk}
\and
Myunghyun Song\footnote{Department of Economics, Columbia University. Email: ms6347@columbia.edu}
\and
Ling Zhong\footnote{Research Institute of Economics and Management, Southwestern University of Finance and Economics. Email: zhongling@swufe.edu.cn}
}
\date{October 2026}

\begin{document}
\begin{bibunit}
\slimbibscope{main}

% Declare new goemetry for the title page only.
%\newgeometry{margin=0.8in}

\maketitle
%\tableofcontents
% \thispagestyle{fancy}
% \phantom{}

%\newpage
%

\vspace{-6ex}

\begin{abstract}
We propose SLIM (Stochastic Learning and Inference in overidentified Models), a new scalable stochastic approximation algorithm for nonlinear GMMs with many nonseparable moments and parameters. SLIM is shown to be consistent and asymptotically normal without a consistent initial estimator or a globally convex GMM criterion. Random-scaling and plug-in based confidence set constructions remain valid when computational uncertainty can be non-negligible relative to sampling uncertainty. An optional second-order refinement achieves full-sample GMM efficiency. We develop $J$-tests for SLIM. The theoretical results are extended to models with clustered data of unequal cluster sizes. Monte Carlo experiments for a nonlinear demand system with 576 moments and 380 parameters demonstrate computational gains and scalability to one million observations. A LinkedIn application with 4.8 million college graduates illustrates clustered estimation and inference for first employment destinations.

\smallskip

\noindent \textbf{Keywords}: U-statistics, generalized method of moments, debiased overidentification test, random scaling, cluster-robust inference
\end{abstract}

\smallskip

% Ends the declared geometry for the title page
%\restoregeometry

%\newpage

%\doublespacing
\onehalfspacing

\section{Introduction}

Building on stochastic approximation introduced by \citet{Robbins-Monro-1951}, stochastic gradient descent (SGD) and its variants use inexpensive updates to handle large datasets, many parameters, and complex nonlinear objectives. These methods are central computational ingredients in modern deep learning, including large language models \citep{bottou2018optimization,dubey2024llama3}.

SGD has been widely applied in the context of M-estimation, where the parameter of interest \(\theta_\oo\) is defined as the minimizer of a population loss function \(\mathbb{E}[\ell(z_i, \theta)]\), with \(\ell(z_i, \theta)\) a known real-valued function of the observation \(z_i\) and parameter \(\theta\). The standard SGD for M-estimation takes the form
\begin{align}\label{SGD-m-est}
\theta_t = \theta_{t-1} - \gamma_t \frac{\partial}{\partial \theta} \ell(z_t, \theta_{t-1}),
\end{align}
where $t$ indexes the iteration count, and $\gamma_t \downarrow 0$ denotes the learning rate.

The application of SGD in economics and econometrics remains limited, however. A central challenge is that many parameters of interest in economics are defined through moment restrictions:
\[
\mathbb{E}[g(z_i, \theta_\oo)] = 0,
\]
where \(g(z_i, \theta)\) is a vector-valued function with \(\dim(g) \geq \dim(\theta)\). The model is termed overidentified when \(\dim(g) > \dim(\theta)\), and the unknown parameter $\theta$ is typically estimated using the generalized method of moments (GMM) introduced by \citet{GMM}. Adapting SGD to estimate such overidentified moment models poses significant challenges because the standard notion of a stochastic gradient does not directly apply.

To illustrate these differences, consider the standard GMM setup. Let \(\theta_\oo \in \Theta \subset \mathbb{R}^d\) denote the true parameter vector, where \(\Theta\) is the parameter space. Let \(z_{1:n} := (z_i)_{i=1}^n\) be a sample of size \(n\), and let \(g(z, \theta)\) be a vector of moment functions such that \(\theta \mapsto g(z, \theta)\) is differentiable for each \(z\). Define the sample averages of the moment function and its Jacobian as:
\begin{align}\label{def:full-sample:g_and_G}
\bar{g}_n(\theta) = \frac{1}{n} \sum_{i=1}^n g(z_i, \theta),
\qquad
\bar{G}_n(\theta) = \frac{1}{n} \sum_{i=1}^n \frac{\partial g(z_i, \theta)}{\partial \theta^\prime}.
\end{align}
The standard GMM estimator is then given by
\[
\hat{\theta}_{n,W_n} := \argmin_{\theta \in \Theta} \bar{g}_n(\theta)' W_n \bar{g}_n(\theta),
\]
where \(W_n\) is a possibly data-dependent weighting matrix. When \(g(z, \theta)\) is nonlinear in \(\theta\), computing \(\hat{\theta}_{n,W_n}\) typically requires iterative numerical methods. Conventional inference often treats the difference between the computed estimate and the fully optimized estimator as negligible relative to sampling error. Achieving that precision can be demanding with large datasets and many parameters.

Conventional full-sample iterative methods for computing $\hat{\theta}_{n,W_n}$ evaluate the moment function and its derivative over the entire dataset at each update, which can become computationally burdensome or infeasible when $n$ is large. Moreover, the usual form of SGD in~\eqref{SGD-m-est} does not directly apply, since the gradient of the GMM objective function involves a product of two sample averages: \( 2 \bar{G}_n(\theta)' W_n \bar{g}_n(\theta) \). Using the same observation or mini-batch to estimate both factors generally produces a biased estimate of this gradient.

To address these computational and inferential challenges, we develop SLIM (Stochastic Learning and Inference in overidentified Models), a scalable stochastic approximation framework for estimation and inference in nonlinear overidentified GMM. Our joint limit theory allows valid inference when computational uncertainty is non-negligible relative to sampling uncertainty.

Establishing convergence guarantees for algorithms that compute nonlinear GMM estimators has remained a longstanding challenge, particularly in settings where the objective function is nonconvex or a consistent initial estimate is difficult to obtain without exhaustive search. \citet{robinson1988} analyzes the use of grid search to construct a consistent initial estimator. Given such an estimator \( \hat{\theta}_0 \), applying \( t \) iterations of a Newton–Raphson (NR) procedure yields a refined estimator \( \hat{\theta}_t \) that is consistent and asymptotically normal, provided the initial estimator is sufficiently accurate. In Robinson’s framework, this initial estimator is obtained through a global search over a regular grid, which becomes computationally infeasible in large-scale problems. Building on this idea, \citet{andrews1997} proposes a stopping rule to obtain a consistent first-step estimator, followed by a fixed number of NR updates. While this method avoids pre-specifying grid resolution, it can still require a global search, which limits its scalability. More recently, \citet{Forneron:Zhong:23} study optimization methods for nonlinear GMM without assuming convexity or requiring a consistent initial estimator, focusing on iterative procedures such as gradient descent and Gauss–Newton. When their global-convergence conditions fail, these methods can be supplemented with a global search over candidate parameter values, whose computational burden can grow with the dimension of $\theta$ and, through full-sample evaluations at each step, with $n$. SLIM reduces the computational cost of each update through mini-batches and, under suitable regularity conditions, yields consistent estimators without requiring a consistent initial estimator or global convexity.

Two related papers develop stochastic and online approaches to overidentified moment conditions. \citet{SGMM:2023} extend stochastic approximation to GMM, referred to as SGMM, and develop methods for online inference in the context of linear instrumental variable (IV) regression, where $g(z_i, \theta)$ is linear in $\theta$. SGMM employs a second-order optimization method, in the terminology of \citet[Section 6]{bottou2018optimization}, and achieves asymptotic efficiency comparable to two-stage least squares (2SLS) and GMM. Their nonlinear extension initializes with a preliminary GMM estimate and leaves its asymptotic properties for future research. Its second-order nature limits scalability relative to first-order methods. As a related approach, \citet{OGMM:2025} develop an online generalized method of moments (OGMM), designed for stationary and ergodic time series data that arrive in batches. They assume that the initial batch size grows to infinity and that a consistent GMM estimator can be constructed from this batch. Moment restrictions are subsequently updated using a first-order Taylor approximation.\footnote{The literature on stochastic approximation for overidentified moment restrictions is sparse. Beyond \citet{SGMM:2023} and \citet{OGMM:2025}, there are a few papers on online linear IV regressions in computer science and engineering communities: see, for example, \citet{venkatraman2016online}, \citet{Chen:Roy:Hu:Balasubramanian:2024}, \citet{della2023online} and the references therein. These works do not develop asymptotic distribution theory for their online linear IV estimators, however.}

SLIM combines a first-order algorithm with an optional second-order refinement that can attain full-sample GMM efficiency. The updates use independent mini-batches of the moment function and its Jacobian, so that, as in a U-statistic, the update directions are conditionally unbiased for the corresponding full-sample criterion direction. It is the direction minus its conditional mean, given the data and past algorithmic draws, that forms a martingale difference sequence. For the refinement, we decompose the asymptotic variance into sampling uncertainty, moment-batch noise, and Jacobian-batch noise, clarifying how iteration counts and batch sizes affect precision.

For inference we offer two routes. Random scaling \citep{lee2021fast,lee2025fast} studentizes with a term updated inside either algorithm and is preferred when loading the full dataset is costly; plug-in inference uses full-sample averages and is preferred when full-sample access is relatively cheap. Debiased $J$-tests remove the first-order contribution of estimation error. They permit specification testing using full-sample evaluations at a suitably accurate approximate estimate, without requiring iterative full-sample optimization. We also extend SLIM to clustered data: drawing clusters rather than observations, with appropriately weighted cluster contributions, leaves the full-sample GMM estimand unchanged but makes the mini-batch noise inherit the cluster-level second moment of the moments, so random-scaling and plug-in inference and the debiased $J$-test become cluster-robust with no change to the form of the procedures.

Furthermore, we address the challenge of instability in single-pass stochastic approximation, where each observation is accessed only once, by formally establishing asymptotic theory for multi-pass stochastic approximation with mini-batches. This allows additional computation to reduce approximation error by repeatedly sampling from the observed data. While \citet{SGMM:2023} demonstrate the empirical benefits of multi-pass strategies in linear IV regression, they do not provide formal theoretical guarantees for these multi-pass strategies, which we develop in this paper.

We illustrate SLIM with the Exact Affine Stone Index (EASI) demand system of \citet{EASI}, in which budget shares depend nonlinearly on an implicit utility that is itself endogenous through the Stone index. With $n = 100{,}000$, 576 moment conditions, and 380 parameters,
on a powerful laptop, the Stata GMM routine requires more than 18 hours, exceeding our computational budget, whereas SLIM completes estimation and inference in under 1.4 hours. At smaller sample sizes, SLIM achieves precision comparable to full-sample GMM. All three SLIM $J$-tests have null rejection frequencies close to the nominal level in these experiments. SLIM remains feasible at $n = 10^6$.

We also apply SLIM to work choices in LinkedIn profiles for 4.8 million college graduates. The nonlinear choice model contains 52 parameters and 53 moments, with state $\times$ graduation year $\times$ undergraduate major clusters. The application relates first reported employment destinations to group-specific destination propensities and demographic characteristics. We report plug-in and random-scaling intervals that account for sampling and computational uncertainty. For the destination-propensity coefficient, the variance decomposition attributes about half of the estimated plug-in variance to computational uncertainty, predominantly through Jacobian-batch noise.

The paper is organized as follows. Section~\ref{sec:SGD} introduces the setting and the SLIM algorithm, gives a heuristic account of why stochastic approximation works for GMM, and discusses related methods. Section~\ref{sec:asymp:theory} develops asymptotic theory under random sampling, and Section~\ref{sec:inference} proposes random-scaling inference. Section~\ref{sec:extensions} adds the second-order refinement and plug-in inference, compares the two inference approaches, and extends the Sargan--Hansen $J$-test to plug-in, debiased plug-in, and recursive mini-batch versions. Section~\ref{sec:MC} reports the Monte Carlo experiments, Section~\ref{sec:linkedin} presents the LinkedIn application, and Section~\ref{sec:conclusion} concludes.
%The proof of consistency, the paper's central result, follows the conclusion.
Appendix~\ref{sec:setting:fixed} develops the fixed-sample theory; Appendix~\ref{sec:cluster} extends SLIM to clustered data, with asymptotic theory, random-scaling and plug-in inference, and a debiased $J$-test; Appendix~\ref{sec:algo:alt} introduces an optional warm-start algorithm; Appendix~\ref{appendix:asymp:regimes} complements Section~\ref{sec:inference}. Appendix~\ref{subsec:Selection_gamma0} gives learning-rate and mini-batch guidance for both applications. Appendix~\ref{app:linkedin} provides data construction, additional discussion of the economic interpretation of the discrete choice model, and implementation details for the LinkedIn application. Appendix~\ref{appx:proofs-thms} proves the main-text theorems; Appendix~\ref{appx:lemmas-proofs} supplies the technical setup, supporting lemmas, the proofs of Corollaries~\ref{cor:inference:random} and~\ref{cor:plugin:eff}, and proofs for Appendices~\ref{sec:setting:fixed} and~\ref{sec:cluster}; Appendix~\ref{appx:proofs-lems} proves the lemmas. Appendix~\ref{app:easi:details} gives the EASI calibration and Monte Carlo details. Appendix~\ref{app:linkedin:results} reports the complete LinkedIn coefficient results. Finally, Appendix~\ref{appx:multi-local} studies convergence when the sample GMM criterion has multiple stationary points.

\section{SLIM: Stochastic Learning for Nonlinear GMM}\label{sec:SGD}

\begin{figure}[!htbp]

\centering

\resizebox{.83\textwidth}{!}{%
\begin{tikzpicture}

\node (start) [startstop] {SLIM: Stochastic learning for nonlinear GMM};

% Main explanation before warm start
\node (consistency) [note, below=0.5cm of start, text width=5cm, align=center] {An initial consistent estimator is \textbf{not} required};

\node (warmstart) [process, below=0.5cm of consistency] {Preliminary step (optional): warm start};

\node (firstorder) [process, below=0.5cm of warmstart] {Step 1 (required): first-order U-statistics approach};

\node (secondorder) [process, below=0.5cm of firstorder] {Step 2 (optional but recommended): second-order update for efficiency};

%\node (finalest) [startstop, below=0.8cm of secondorder] {Obtain final estimates and/or random scaling statistics};

\node (inference) [decision, below=.5cm of secondorder, text width=5cm] {Is full data accessible efficiently?};

% Shorter inference branches
\node (randomscale) [process, below left=1.5cm and .1cm of inference] {Random scaling inference};
\node (plugin) [process, below right=1.5cm and .1cm of inference] {Plug-in inference};

% Arrows
\draw [dashed] (start) -- (consistency);
\draw [arrow] (warmstart) -- (firstorder);
\draw [arrow] (firstorder) -- (secondorder);
%\draw [arrow] (secondorder) -- (finalest);
%\draw [arrow] (finalest) -- (inference);

% Inference split
\coordinate (midpoint) at ($(inference.south) + (0,-0.3)$);
\draw [arrow] (inference.south) -- (randomscale.north east);
\draw [arrow] (inference.south) -- (plugin.north west);

% Option labels
\node at ($(midpoint)!0.5!(randomscale.north east) + (-0.3,0.2)$) {\scriptsize No};
\node at ($(midpoint)!0.5!(plugin.north west) + (0.3,0.2)$) {\scriptsize Yes};

\end{tikzpicture}
}

\caption{Outline for SLIM}\label{fig:sgd-flowchart}

\end{figure}

We begin by providing a roadmap for stochastic approximation in nonlinear GMM. Figure~\ref{fig:sgd-flowchart} outlines the recommended procedure. The roadmap emphasizes that an initial consistent estimator is not required, regardless of whether a warm start (described in Appendix~\ref{sec:algo:alt}) is used. The first main step, which is required, applies a first-order approach based on U-statistics (see the current section for details). The second step, which is optional but recommended, refines this estimator using a second-order update to improve efficiency (see Section~\ref{sec:extensions}). The procedure branches into two alternative inference methods: the random scaling approach, described in Section~\ref{sec:inference}, preferred when loading the full dataset is computationally expensive; and the plug-in approach, described in Section~\ref{sec:extensions}, preferred when full-sample access is feasible.

\subsection{Setting}

In our setting, computational efficiency is as important as statistical efficiency. 
Thus, for the moment, we assume that the weighting matrix is fixed and positive-definite but not necessarily optimal. To reduce computational costs, we use stochastic approximation, where $\theta$ is iteratively updated using randomly selected subsamples of the data. 
Specifically, at each iteration, two independent mini-batches of observations are used to compute unbiased estimates of $\bar{g}_n(\theta)$ and $\bar{G}_n(\theta)$, conditional on the observed data and all information available before that iteration.
We show that this approach significantly reduces the computational burden while remaining consistent and asymptotically normal under suitable regularity conditions, even though it may sacrifice statistical efficiency compared to the fully efficient GMM estimator. We return to the issue of efficiency and optimal weighting in Section~\ref{sec:extensions}.

Since the weighting matrix $W_n$ remains fixed throughout the iterations, we further simplify the optimization objective by setting $W_n$ to the identity matrix, without loss of generality for the algorithm.\footnote{It is implicitly assumed that it is computationally straightforward to compute $W_n^{1/2}$ and pre-multiply it to $g(z_i, \theta)$, so that we can re-define $g(z_i, \theta)$ to be $W_n^{1/2}g(z_i, \theta)$.} This leads to the following simplified formulation of the estimator:
\begin{align}\label{def:full-sample-gmm}
    \hat{\theta}_n := \argmin_{\theta \in \Theta} \bar{g}_n(\theta)' \bar{g}_n(\theta).
\end{align}
The following subsection  details our stochastic approximation.

\subsection{Algorithm}\label{sec:algo}

We begin by specifying the input for our stochastic approximation algorithm. 
Let $\mathcal{F}_n:=\sigma(z_{1:n})$, and let $\mathcal{F}_{n,0}^*$ denote the initial information, including the observed data and any preliminary draws used to construct the starting value. Let $\theta_0^* \in \Theta$ denote the initial value of the parameter vector, which is $\mathcal{F}_{n,0}^*$-measurable and satisfies $\|\theta_0^*\|\le M_0$ almost surely for some nonrandom $M_0<\infty$ independent of $n$. For example, $\theta_0^*$ can be a vector of predetermined constants (e.g., zeros) or a computationally feasible estimator computed over a fixed bounded subset of $\Theta$ using a random subsample of $z_{1:n}$ when $n$ is very large. A key advantage of our approach is that consistency and asymptotic normality of the final estimator do not depend on the initial estimator being consistent, although the quality of the starting value can have a non-negligible impact on finite-sample performance. A useful starting value may be obtained from the warm-start algorithm in Appendix~\ref{sec:algo:alt} or another preliminary optimization method. For example, the strategic Monte Carlo optimization method of \citet{chen2026smco} could be applied to a GMM criterion constructed from a manageable subsample over a bounded rectangular parameter region. A preliminary estimate used as $\theta_0^*$ must satisfy the boundedness condition above, for example by restriction to a fixed bounded subset of $\Theta$.

In addition, the algorithm takes as input
the total number of iterations $N$, 
the learning rate schedule $\{\gamma_t : t=1, \ldots, N\}$, 
and the mini-batch sizes $B_G$ and $B_g$, where $B = B_G + B_g$ denotes the total mini-batch size. 
These quantities are predetermined before the algorithm begins. 
Conditional on $\mathcal{F}_{n,0}^*$, let $\{\tilde{z}_j : j=1, \ldots, B N\}$ be independent draws from the empirical distribution of $z_{1:n}$.
That is, for $j = 1$ to $B N$, sample $i(j)$ independently and uniformly, with replacement, from $\{1, \ldots, n\}$ and let $\tilde{z}_j = z_{i(j)}$.
This sequence provides the mini-batches of data used for the stochastic updates. 
When $n$ is relatively large, we set $B N = n$; otherwise, we allow $B N \geq n$, which is often referred to as \emph{multi-pass} algorithms. When $n$ is relatively small, multi-pass algorithms can improve the finite-sample performance as they allow the algorithm to iterate over the data multiple times. 

Write $G(z,\theta):=\partial g(z,\theta)/\partial\theta^{\prime}$ for the observation-level Jacobian, and let $\mathcal{F}_{n,t-1}^*$ contain the initial information and the mini-batch draws through iteration $t-1$. At each iteration $t \geq 1$, the parameter vector is updated using a mini-batch gradient descent step, which can be expressed as:
\begin{align}\label{def:sgd}
\theta_t^* = \theta_{t-1}^* - \gamma_t \tilde{G}_{t} (\theta_{t-1}^*)' \tilde{g}_{t} (\theta_{t-1}^*),    
\end{align}
where
\begin{align}\label{def:SGD:g_and_G}
\tilde{G}_{t} (\theta) := \frac{1}{B_G} \sum_{i=1}^{B_G} G(\tilde{z}_{(t-1)B + i}, \theta) 
\ \ \text{ and } \ \
\tilde{g}_{t} (\theta) := \frac{1}{B_g} \sum_{i=1}^{B_g} g(\tilde{z}_{(t-1)B + B_G + i}, \theta).
\end{align}
In this update, conditional on $\mathcal{F}_{n,t-1}^*$, the derivative matrix $\tilde{G}_{t} (\theta_{t-1}^*)$ serves as an unbiased estimate of the full-sample derivative matrix $\bar{G}_n(\theta_{t-1}^*)$, while the moment vector $\tilde{g}_{t} (\theta_{t-1}^*)$ is an unbiased estimate of the full-sample moment vector $\bar{g}_n(\theta_{t-1}^*)$.
By construction, $\tilde{G}_{t} (\theta_{t-1}^*)$ and $\tilde{g}_{t} (\theta_{t-1}^*)$ are independent of each other conditional on $\mathcal{F}_{n,t-1}^*$.
On one hand, the use of independent mini-batches for these two components ensures that the stochastic update directions remain unbiased for $\bar G_n(\theta_{t-1}^*)^{\prime}\bar g_n(\theta_{t-1}^*)$ conditional on $\mathcal{F}_{n,t-1}^*$, while significantly reducing the computational burden relative to full-sample evaluations.
On the other hand, mini-batches also provide variance reduction compared to stochastic approximation with $B_G = B_g = 1$. This variance reduction is crucial in finite samples, as each iteration involves a product of two noisy elements, despite both being unbiased and independent of each other under this conditioning.

At each iteration, the \citet{Polyak1990}-\citet{ruppert1988efficient} average is computed \citep[see also][]{polyak1992acceleration}:
\[
\bar{\theta}^*_t = \frac{1}{t} \theta^*_t + \frac{t-1}{t} \bar{\theta}^*_{t-1}
\]
and our proposed estimator is $\bar{\theta}^*_{N}$.
The averaging step helps mitigate noise in the stochastic updates.
Our proposed estimator is consistent and asymptotically normal under suitable regularity conditions as both $n$ and $N$ tend to infinity.
Furthermore, we show how to conduct inference using the stochastic path $\{\theta_t^*: t=1,\ldots, N \}$. 
In summary, the proposed first-order method is outlined in Algorithm~\ref{alg:mini-batch}.

\begin{algorithm}[ht!]
%\caption{Mini-Batch Stochastic Approximation Algorithm for GMM Estimation}
\caption{First-Order U-statistic Approach}
\label{alg:mini-batch}
\KwIn{
Sample $z_{1:n} = (z_i)_{i=1}^n$; 
mini-batch sizes $B_G$ and $B_g$; 
number of iterations $N$; 
schedule of learning rates $(\gamma_t)_{t=1}^N$; 
initial value $\theta_0^*$.
}

Set $t \leftarrow 1$ and $B \leftarrow B_G + B_g$.

Initialize $\theta^* \leftarrow \theta_0^*$.

Initialize $\bar{\theta}^* \leftarrow \theta_0^*$.

\While{$t \le N$}{ 

Sample $i(1), \ldots, i(B)$ independently and uniformly, with replacement, from $\{1, \ldots, n\}$, independently of the initial information $\mathcal{F}_{n,0}^*$. % and all previous draws.

Compute $\tilde{G} = \frac{1}{B_G}\sum_{j=1}^{B_G} G(z_{i(j)}, \theta^*)$ and $\tilde{g} = \frac{1}{B_g}\sum_{j=B_G+1}^B g(z_{i(j)}, \theta^*)$.

Update $\theta^* \leftarrow \theta^* - \gamma_t \tilde{G}' \tilde{g}$.

Update $\bar{\theta}^* \leftarrow \frac{t-1}{t} \bar{\theta}^* + \frac{1}{t} \theta^*$.

Set $t \leftarrow t + 1$.

}

Set $\theta_N^* \leftarrow \theta^*$ and $\bar{\theta}_N^* \leftarrow \bar{\theta}^*$.

\KwOut{
Final stochastic update $\theta_N^*$; 
final estimator $\bar{\theta}_N^*$.
}
\end{algorithm}

\section{Asymptotic Theory}\label{sec:asymp:theory}

We begin by stating the basic assumptions that underpin our analysis. The asymptotic analysis holds the parameter and moment dimensions fixed.

\begin{assumption}[Uniform Sampling with Replacement]\label{assm:uniform}  
Conditional on the initial information $\mathcal{F}_{n,0}^*$, including the observed data, $\{\tilde{z}_j\}_{j=1}^{BN}$ are independent draws from the empirical distribution assigning probability $1/n$ to each observed data index in $\{1, \ldots, n\}$.
\end{assumption}  
 
\begin{assumption}[Learning Rates]\label{assm:lr}
The sequence of learning rates satisfies $\gamma_t = \gamma_0 t^{-a}$ for $t \ge 1$, where $\gamma_0 > 0$ and $a \in (1/2, 1)$.  
\end{assumption}

\begin{assumption}[Parameter Space and Smoothness of the Moment Function]\label{assm:diff}
The domain $\Theta$ is convex and contains the minimizer and all iterates of the algorithm under consideration, almost surely. There exists a set $\mathcal{Z}$ such that $z_{1:n} \subseteq \mathcal{Z}$ and, for all $z \in \mathcal{Z}$, $g(z, \theta)$ is twice continuously differentiable in $\theta \in \Theta$.
\end{assumption}

The sequence $\{ \tilde{z}_j \}_{j=1}^{BN}$ is generated by uniform sampling with replacement from $z_{1:n}$, and is partitioned into $N$ mini-batches of size $B$. When $BN = c \cdot n$ for some $c > 1$, our algorithm, as described in Section~\ref{sec:algo}, samples each data point $c$ times on average. This framework is commonly referred to as a \emph{multi-pass algorithm}. Appendix~\ref{sec:algo:alt} describes an optional warm start that instead reshuffles the data in each epoch; it only produces a starting value and is not covered by Assumption~\ref{assm:uniform}.

The schedule $\gamma_t=\gamma_0t^{-a}$ in Assumption~\ref{assm:lr}, with $a\in(1/2,1)$, satisfies $\sum_{t=1}^\infty \gamma_t=\infty$ and $\sum_{t=1}^\infty \gamma_t^2<\infty$. These standard learning-rate conditions are used together with the remaining regularity assumptions to establish convergence.
We also assume the moment function $g(z,\theta)$ is twice differentiable in $\theta$, as is customary in the nonlinear GMM setting.

We develop asymptotic theory under the i.i.d. sampling. 
Extensions of this framework to settings with time series dependence are left for future research, while cluster dependence is accommodated in Appendix~\ref{sec:cluster} by resampling at the cluster level.\footnote{In Appendix~\ref{sec:theory:fixed}, we consider an alternative asymptotic framework in which the data are treated as fixed, and the objective is to approximate the full-sample GMM estimator $\hat{\theta}_n$, defined in~\eqref{def:full-sample-gmm}. This setting is particularly relevant when evaluating $\hat{\theta}_n$ is computationally intensive, such as when the sample size $n$ is very large.}
To state the further regularity conditions, we introduce additional notation.
Define the sample Lyapunov function
\begin{align}
\bar{Q}_n(\theta) &:= \frac{1}{2}\bar{g}_n(\theta)'\bar{g}_n(\theta) - \frac{1}{2}\bar{g}_n(\hat{\theta}_{n})'\bar{g}_n(\hat{\theta}_{n}),\label{def:sample:Lyapunov}
\end{align}
where $\bar{g}_n(\theta)$ and $\hat{\theta}_{n}$ are given in \eqref{def:full-sample:g_and_G} and \eqref{def:full-sample-gmm}, respectively.
 %{\color{red}[N.B. Use $Q_n$ instead of $V_n$]}
The population moment function and its Jacobian are also denoted by
\begin{equation*}
g(\theta) = \mathbb{E}[g(z_i, \theta)] \quad \text{and} \quad G(\theta) = \mathbb{E}[G(z_i, \theta)].    
\end{equation*}
Let $I_{\dtheta}$ denote the $\dtheta$-dimensional identity matrix. For square matrices $A$ and $B$, we write $A \le B$, $A\ge 0$, and $A >0$ if $B - A$ is positive semidefinite, $A$ is positive semidefinite, and positive definite, respectively. 
We use $\|\cdot\|$ to denote the Euclidean norm on vectors and the spectral norm on matrices.

\begin{assumption}[Random Sampling and Regularity Conditions]\label{asm:regularity}
The following conditions hold for some constants $c > 0$, $\delta > 0$, $M < \infty$, and $p \ge 1$.
\begin{enumerate}[({A\ref{asm:regularity}}.1)]
\item \label{asm:item:iid}
The observations $(z_i)_{i=1}^n \subseteq \mathcal{Z}$ are i.i.d.\ draws from the population distribution $\mathbb{P}$.

\item \label{asm:item:interior}

The true parameter $\theta_\oo\in\operatorname{int}(\Theta)$ satisfies $g(\theta)=0$ if and only if $\theta=\theta_\oo$.

\item \label{asm:item:pd}
$G_\oo' G_\oo \ge c I_{\dtheta}$ and $\Omega_\oo \ge c I_{\dg}$, where
$G_\oo := G(\theta_\oo)$ and $\Omega_\oo := \mathbb{E}[g(z_i, \theta_\oo) g(z_i, \theta_\oo)']$.

\item \label{asm:item:integrability} $\mathbb{E}[\|g(z_i, \theta_\oo)\|^{2p}] < \infty$, and there exists a measurable function $H : \mathcal{Z} \to \mathbb{R}$ such that $\|G(z, \theta)\| \le H(z)$ for all $z \in \mathcal{Z}$ and $\theta \in \Theta$, and $\mathbb{E}[H(z_i)^{2p}] < \infty$.

\item \label{asm:item:Lipschitz-G}
The map $\theta \mapsto G(z, \theta)$ is Lipschitz for each $z \in \mathcal{Z}$. That is,  $
\|G(z, \theta) - G(z, \tilde{\theta})\| \le L(z) \|\theta - \tilde{\theta}\| \quad \text{for all $\theta$ and $\tilde{\theta} \in \Theta$},
$
where $L : \mathcal{Z} \to \mathbb{R}$ satisfies $\mathbb{E}[L(z_i)^{2p}] < \infty$.

\item \label{asm:item:local-convex-Lojasiewicz}
For all $\|\theta - \hat{\theta}_n\| \le \delta$, $\frac{\partial^2 \bar{Q}_n(\theta)}{\partial \theta \partial \theta'} \ge c I_{\dtheta}$ with probability approaching one (\wpa).

\item \label{asm:item:id}
For all $\|\theta - \hat{\theta}_n\| \ge \delta$, $\bar{Q}_n(\theta) \ge c$ and $\|\bar{G}_n(\theta)' \bar{g}_n(\theta)\|^2 \ge c$ \wpa.

\item \label{asm:item:stability}
For all $\theta \in \Theta$, $
\frac{\partial^2 \bar{Q}_n(\theta)}{\partial \theta \partial \theta'} \le M I_{\dtheta}$ and $\frac{1}{n} \sum_{i=1}^n \|g(z_i, \theta)\|^{2p} \le M (\bar{Q}_n(\theta)^p + 1)$ \wpa.
\end{enumerate}
\end{assumption}

%Here and throughout, a condition stated to hold for all $\theta$ in a given set \wpa\ is required to hold simultaneously for all such $\theta$; that is, the probability of the event on which it holds for every such $\theta$ tends to one.
The bulk of Assumption~\ref{asm:regularity} reflects standard regularity requirements under a random-sampling framework. Condition~\ref{asm:item:iid} assumes that the data are i.i.d. Condition~\ref{asm:item:interior} imposes global point identification of the true parameter $\theta_\oo$, while~\ref{asm:item:pd} provides local identification through a rank condition on the Jacobian $G(\theta_\oo)$; its condition on $\Omega_\oo$ ensures a nondegenerate sampling distribution and an invertible efficient weighting matrix. Condition~\ref{asm:item:integrability} imposes integrability on $g(\cdot,\theta_\oo)$ and the envelope function for the Jacobian $G(\cdot, \theta)$.
% \footnote{Strictly speaking, $F_K$ is a \emph{local} envelope for the map $\theta \mapsto g(z,\theta)$ over a compact subset $K$ of the parameter space.} 
Condition~\ref{asm:item:Lipschitz-G} requires Lipschitz continuity of $\theta \mapsto G(z, \theta)$, with an integrable Lipschitz constant $L(z)$. Condition~\ref{asm:item:stability} assumes the existence of a constant $M$ such that the difference between $M I_{\dtheta}$ and the Hessian of the sample Lyapunov function $\bar{Q}_n(\theta)$ is uniformly positive semidefinite, and that the ratio of $n^{-1} \sum_{i=1}^n \|g(z_i, \theta)\|^{2p}$ to $\bar{Q}_n(\theta)^p + 1$ is uniformly bounded by $M$, both with probability approaching one. This condition is used to control the variability of the stochastic errors.

It remains to discuss Conditions~\ref{asm:item:local-convex-Lojasiewicz} and~\ref{asm:item:id}, which impose requirements on the behavior of the sample Lyapunov function in the regions where $\|\theta - \hat{\theta}_n\| \le \delta$ and $\|\theta - \hat{\theta}_n\| \ge \delta$, respectively. The local lower bound in~\ref{asm:item:local-convex-Lojasiewicz} ensures that $\bar{Q}_n(\theta)$ is strongly convex in a neighborhood of $\hat{\theta}_n$. In turn,~\ref{asm:item:id} requires that both $\bar{Q}_n(\theta)$ and the squared norm of the gradient of the sample Lyapunov function, $\|\bar{G}_n(\theta)' \bar{g}_n(\theta)\|^2$, are bounded away from zero for all $\theta$ such that $\|\theta - \hat{\theta}_n\| \ge \delta$. Notably, we do not impose global convexity of the criterion or compactness of the parameter space; instead, convergence is ensured by the two-region structure encoded in~\ref{asm:item:local-convex-Lojasiewicz} and~\ref{asm:item:id}. These two conditions jointly serve a role analogous to global strong convexity in ensuring convergence of stochastic approximation algorithms. Condition~\ref{asm:item:local-convex-Lojasiewicz} provides local curvature near the solution, ensuring that stochastic iterates are attracted to the unique minimizer $\hat{\theta}_n$ eventually. Meanwhile, Condition~\ref{asm:item:id} prevents the algorithm from stagnating far from the solution by requiring that the gradient magnitude remains bounded away from zero outside the neighborhood. This two-region structure, consisting of local strong convexity near the optimum and a sharp separation condition away from it, allows us to establish stability and convergence without relying on global convexity or compactness. Such an approach is particularly well suited to overidentified nonlinear GMM settings.

\begin{remark}[Additional Assumptions Beyond Standard GMM Theory]
The standard asymptotic analysis of GMM estimators does not use
Conditions~\ref{asm:item:local-convex-Lojasiewicz}--\ref{asm:item:stability}.
Classical econometric theory implicitly assumes that GMM estimators can be computed exactly 
or that any computational errors are asymptotically negligible relative to sampling variability. 
In our setting, we explicitly account for both statistical and computational errors, 
which necessitates these additional regularity conditions. 
These conditions are formulated in terms of the sample Lyapunov function to better separate 
computational errors from sampling variability. 
For i.i.d.\ data, analogous conditions can also be formulated in terms of population quantities 
under suitable uniform convergence arguments via the uniform law of large numbers, 
though we omit these for brevity.
\end{remark}

\begin{remark}[Comparison with \citet{Forneron:Zhong:23}]
Our assumptions also differ from those in \citet{Forneron:Zhong:23}. 
Although their analysis does not require convexity of the GMM criterion function, it relies on global conditions involving
\(G(\theta)\), the weighting matrix, and certain integral transforms of \(G(\theta)\),
along with other regularity requirements.
More precisely, under correct specification and their accompanying regularity conditions, Propositions 1--2 establish that their Assumption 2(b) is equivalent to requiring that the
population GMM objective function \(Q(\theta)\) satisfy a global quadratic lower bound around its minimizer and the
Polyak–Łojasiewicz inequality:
\[
\Big\| \frac{\partial Q(\theta)}{\partial \theta} \Big\|^2 \geq \mu\,\big(Q(\theta) - Q(\theta_\oo)\big),
\]
for all \(\theta \in \Theta\) and some constant \(\mu > 0\).
%Our two-region conditions also imply a truncated gradient-dominance inequality for the sample criterion; see~\eqref{eq:eq8}.
\end{remark}

To analyze the stochastic properties of our algorithm, it is useful to decompose 
$(\bar{\theta}^*_{N} - \theta_\oo)$ into two parts: 
\begin{align*}
\bar{\theta}^*_{N} - \theta_\oo
&=
\underbrace{\big(\bar{\theta}^*_{N} - \hat{\theta}_{n}\big)}_{\text{computational error}}
+
\underbrace{\big(\hat{\theta}_{n} - \theta_\oo\big)}_{\text{statistical error}},    
\end{align*}
where the first term captures the discrepancy arising from stochastic approximation relative to the full-sample GMM estimator, and the second reflects the estimation error of the full-sample GMM estimator relative to the true parameter vector.  
In econometrics, the conventional analysis focuses only on the statistical error, implicitly assuming that the computational error is asymptotically negligible.  
Intuitively, this decomposition separates the error from using a stochastic algorithm (computational) from the inherent sampling uncertainty in estimation (statistical).  

We now introduce some notation. Let $\mathbb{P}^*_n$ denote the conditional distribution induced by the subsequent mini-batch draws $(\tilde{z}_j)_{j \ge 1}$ given $\mathcal{F}_{n,0}^*$, and let $\mathbb{E}^*_n$ denote the corresponding expectation. With this in place, we are ready to state our first theoretical result.

\begin{theorem}[Consistency]
\label{thm:consistency}
Let Assumptions \ref{assm:uniform}, \ref{assm:lr}, \ref{assm:diff}, and \ref{asm:regularity} hold.
Then, $\theta_N^*$ and $\bar{\theta}_N^*$ are consistent for $\theta_\oo$ as $n \to \infty$ and $N \to \infty$. Furthermore, there exists a sequence of events $E_n \in \mathcal{F}_n$ such that $\mathbb{P}(E_n) \to 1$, and for every $\varepsilon > 0$,
\begin{equation*}
\sup_{n \in \mathbb{N}} \left\{\mathbb{P}_n^*\left( \|\theta_N^* - \hat{\theta}_n \| \ge \varepsilon \right) \mathbbm{1}_{E_n}\right\} \to 0 \quad \text{as } N \to \infty.
\end{equation*}
\end{theorem}

%The proof is given in Appendix~\ref{appx:proof-consistency}.

%To prove Theorem~\ref{thm:consistency}, we establish that 
The theorem controls the computational error uniformly over sample sizes on events whose probability approaches one.
The localized Lyapunov bound in Lemma~\ref{lem:sgd-conv-rate}, together with the stopping-time bounds in Lemma~\ref{lem:stopping-time-and-tail}, yields $\theta_N^*-\hat{\theta}_n=O_p(N^{-a/2})$ as $n,N\to\infty$.

Under the conditions of Theorem~\ref{thm:asymp:normal}, averaging over the stochastic path yields computational error $\bar\theta_N^*-\hat\theta_n=O_p(N^{-1/2})$, where $\bar\theta_N^*=N^{-1}\sum_{t=1}^N\theta_t^*$. The total estimation error is $\bar\theta_N^*-\theta_\oo=O_p(n^{-1/2}+N^{-1/2})$.
This estimator also satisfies the (martingale) central limit theorem (CLT), as established in Theorem~\ref{thm:asymp:normal}.
The limiting covariance of $\sqrt N(\bar\theta_N^*-\hat\theta_n)$ equals $1/B_g$ times the limiting covariance of $\sqrt n(\hat\theta_n-\theta_\oo)$, plus a contribution from the mini-batch approximation of the Jacobian.
To describe it precisely, let us introduce the following matrix:
\begin{equation}
\label{eq:sgd-score-variance}
\Sigma_\oo = \frac{1}{B_g} \left( G_\oo' \Omega_\oo G_\oo + \frac{1}{B_G}\mathbb{E}\Big[ (G(z_i,\theta_\oo)-G(\theta_\oo))' \Omega_\oo (G(z_i,\theta_\oo)-G(\theta_\oo))\Big]\right),
\end{equation}
which represents $\operatornamewithlimits{plim}_{n \to \infty} \bar \Sigma_n$, the asymptotic variance of the stochastic error, where $\bar \Sigma_n$ is a sample analog of $\Sigma_\oo$ and formally defined in the appendix (see \eqref{def:bar-Sigma-n}). 
The second term reflects the variability in the mini-batch approximation of $G(\theta)$, which decays at a rate inversely proportional to the corresponding mini-batch size.
The next theorem presents the CLT applied to the averaged estimator.

\begin{theorem}
\label{thm:asymp:normal}
Let Assumptions \ref{assm:uniform}, \ref{assm:lr}, \ref{assm:diff}, and \ref{asm:regularity} hold with $p > 1$.
Then, as $n \to \infty$, $N \to \infty$, and $N^{1-a}/n \to 0$, it holds that
\begin{align*}
     \left( \begin{matrix}
            \sqrt{n}(\hat{\theta}_n - \theta_\oo) \\    
    \sqrt{N} (\bar \theta^*_{N} - \hat{\theta}_n)    
\end{matrix} \right) \dto \mathcal{N}\left( 0, \left( \begin{matrix}
   (G_\oo'G_\oo)^{-1} G_\oo'\Omega_\oo G_\oo (G_\oo' G_\oo)^{-1} & 0\\    
   0 & (G_\oo'G_\oo)^{-1} \Sigma_\oo (G_\oo'G_\oo)^{-1} 
\end{matrix} \right)\right).
\end{align*}
\end{theorem}

Theorem~\ref{thm:asymp:normal} establishes that \(\sqrt{n}(\hat{\theta}_n - \theta_\oo)\) and \(\sqrt{N} (\bar{\theta}_N^* - \hat{\theta}_n)\) are jointly asymptotically normal and asymptotically independent. 
While the convergence result for the full-sample estimator \(\hat{\theta}_n\) is standard, the result for the averaged stochastic estimator \(\bar{\theta}_N^*\) is new. 
It shows that \(\bar{\theta}_N^*\), centered at \(\hat{\theta}_n\), is asymptotically normal, achieves the \(O_p(N^{-1/2})\) convergence rate, and has an asymptotic covariance matrix whose ``meat'' component includes an additional term
\begin{align}\label{def:add:var:ineff}
B_G^{-1} \, \mathbb{E}\big[(G(z_i, \theta_\oo) - G(\theta_\oo))' \Omega_\oo (G(z_i, \theta_\oo) - G(\theta_\oo))\big],    
\end{align}
which captures the noise introduced by approximating the Jacobian with finite mini-batches. 
This variance inflation diminishes as the mini-batch size \(B_G\) increases.

By leveraging global information from $(\bar \theta_t^*)_{t=1}^N$, it is possible to extract additional information about the asymptotic variance of the stochastic approximation estimator.
This idea is formalized in the following theorem, which establishes the functional central limit theorem (FCLT) for the partial sum process of stochastic updates, generalizing Theorem~\ref{thm:asymp:normal} under a slightly stronger moment condition.

Throughout, let $Z$ denote a standard normal vector and $W(r)$ a standard multivariate Wiener process on $[0,1]$, independent of $Z$. The Brownian Bridge is denoted by $\bar{W}(r) = W(r) -r W(1)$. When necessary to evade confusion, we specify their dimensions by adding a subscript, e.g., $Z_d$. Let $\wto$ denote weak convergence in $\ell^\infty([0,1])$. When a random vector and a process appear together, their convergence is joint.

\begin{theorem}
\label{thm:FCLT}
Let Assumptions \ref{assm:uniform}, \ref{assm:lr}, \ref{assm:diff}, and \ref{asm:regularity} hold with $p > (1-a)^{-1}$.
Then, as $n \to \infty$, $N \to \infty$, and $N^{1-a}/n \to 0$, it holds
\begin{equation*}
    \left( \begin{matrix}
   \sqrt{n} (\hat{\theta}_n - \theta_\oo) \\    
   \frac{1}{\sqrt{N}}  \sum_{t=1}^{\lfloor N r\rfloor}(\theta^*_t - \hat \theta_n)    
\end{matrix} \right)    \wto \left( \begin{matrix}
    (G_\oo' G_\oo)^{-1}( G_\oo'    \Omega_\oo G_\oo)^{1/2} Z \\
    (G_\oo'G_\oo)^{-1} \Sigma_\oo^{1/2}  W(r) 
\end{matrix} \right).
\end{equation*}
% $Z \sim \mathcal{N}(0, I_{\dtheta})$ is a standard normal vector, and $W(r)$ denotes a standard $\dtheta$-dimensional Wiener process independent of $Z$.
\end{theorem}

Theorem~\ref{thm:FCLT} strengthens Theorem~\ref{thm:asymp:normal} by establishing weak convergence of the normalized partial-sum process of the centered stochastic iterates rather than only its terminal average. The limiting process is a multidimensional Wiener process, scaled by the same asymptotic variance as in the CLT.
This result enables the development of inference procedures that utilize the full stochastic path without recomputing averages of the sample moment function or Jacobian over the entire dataset. 
As a result, inference can be conducted efficiently even in large-scale settings where loading the full dataset is computationally infeasible. 
In summary, Theorem~\ref{thm:FCLT} provides the theoretical foundation for the random scaling inference methods introduced in the next section.

We conclude this section by commenting on the restrictions imposed on \( p \), \( a \), and \( N \) by our asymptotic framework. To illustrate, consider the case \( a = 0.501 \), which is the value used in our Monte Carlo experiments. The moment condition \( p > (1 - a)^{-1} \) then requires \(p>1/0.499\approx2.004008\), which is relatively mild. This reflects the fact that a slower decay of the learning rate (i.e., smaller \( a \)) permits weaker moment conditions. At the same time, the requirement \( N^{1 - a} / n \to 0 \) imposes a restriction on how fast the number of stochastic updates \( N \) can grow relative to the sample size \( n \). For \( a = 0.501 \), the restriction is $N=o(n^{1/0.499})$, with $1/0.499\approx2.004008$. This example highlights the trade-off in choosing \( a \): slower decay relaxes the required moment conditions but tightens the allowable growth rate of the number of iterations.

%\footnote{See, for example, \citet{lee2021fast} and \citet{lee2025fast} for applications of random scaling methods to mean and quantile regression models, respectively, within the context of stochastic approximation.}

\section{Random Scaling Inference}\label{sec:inference}

In this section, we develop methods for conducting inference on the true parameter vector $\theta_\oo$. 
We focus on testing $\ell \leq d$ linear restrictions of the form
\[
H_0 : R \theta_\oo = c,
\]
where $R$ is a known $(\ell \times d)$ matrix of rank $\ell$, and $c$ is a known vector in $\mathbb{R}^\ell$.

We consider two approaches to inference: one based on random scaling and the other on plug-in estimation of the asymptotic variance matrix. 
%We begin with the random scaling approach.
Random scaling inference has been used in two largely disconnected branches of the literature.
In time series econometrics, it has long been used to conduct inference without consistently estimating the long-run variance, following \citet{kiefer2000simple}.
More recently, \citet{lee2021fast} demonstrated its usefulness for inference with SGD, after which it has been adopted in stochastic approximation methods across various estimation problems.
For instance, \citet{lee2025fast} applied it to quantile regression models, while \citet{SGMM:2023} used it for linear GMM estimation.
Additional applications of random scaling can be found in \citet{li2021statistical,ChenLai2021,Du:Zhu:Wu:Na:2025}, among others.
The plug-in method is deferred to Section~\ref{sec:extensions}, where it is naturally integrated with the efficient estimation procedure developed there.

\subsection{Inference Based on Random Scaling}

Define the $(\ell \times \ell)$ random scaling matrix by
\begin{equation}
    V_t(R) := \frac{1}{t} \sum_{s=1}^t \left\{ \frac{1}{\sqrt{t}} \sum_{j=1}^s (R \theta^*_j - R \bar{\theta}^*_t) \right\} \left\{ \frac{1}{\sqrt{t}} \sum_{j=1}^s (R \theta^*_j - R \bar{\theta}^*_t) \right\}'. \label{eq:V_tR}
\end{equation}
This matrix admits the following iterative updating rule for each $t \geq 1$:
\begin{align}
    A_t(R) &= A_{t-1}(R) + t^2 R \bar{\theta}^*_t {\bar{\theta}^*_t}' R', \label{eq:A_t} \\
    b_t(R) &= b_{t-1}(R) + t^2 R \bar{\theta}^*_t, \label{eq:b_t} \\
    V_t(R) &= t^{-2} \left( A_t(R) - R \bar{\theta}^*_t b_t(R)' - b_t(R) {\bar{\theta}^*_t}' R' + R \bar{\theta}^*_t {\bar{\theta}^*_t}' R' \cdot \frac{t(t+1)(2t+1)}{6} \right), \label{eq:V_t}
\end{align}
with initial values $A_0 = 0_{\ell \times \ell}$ and $b_0 = 0_\ell$. %The FCLT in Theorem~\ref{thm:FCLT} supports random-scaling inference using a pivotal reference distribution. Inference for the population parameter can be conservative when Jacobian batches remain finite, as described below.

\medskip

\noindent\textbf{Computational advantage:} The iterative structure in \eqref{eq:A_t}--\eqref{eq:V_t} enables inference based on random scaling to be implemented efficiently. After forming $R\bar{\theta}_t^*$, the recursion updates and stores only $\ell$-dimensional vectors and $\ell\times\ell$ matrices, avoiding full $d\times d$ matrix updates. This leads to substantial computational gains, particularly when the number of restrictions $\ell$ is much smaller than the parameter dimension $d$. For instance, in many applications we are interested in a single linear combination of the parameter vector, so that $\ell = 1$, making the gain especially pronounced.

\subsection{Asymptotic Regimes}

Since we can decompose $(\bar \theta^*_{N} - \theta_\oo)$ as the sum of 
$(\bar \theta^*_{N} - \hat{\theta}_n)$ and $(\hat{\theta}_n - \theta_\oo)$,
the relative rates at which $n$ and $N$ diverge affect asymptotics.
We consider four regimes: two intermediate cases and two polar extremes.
The leading case is discussed below, and three other cases are in Appendix~\ref{appendix:asymp:regimes}.

\subsubsection{Intermediate Case I. $N \asymp n$ and $B_G < \infty$}\label{subsec:asymp:regimes}

We begin with the scenario where $N$ and $n$ tend to infinity at the same rate.
Define the Wald statistic as
\begin{align}\label{def:wald:stat}
T_{n,N} &:= \underbrace{(1 + n/(NB_g))^{-1}}_{\text{deflating factor}} \cdot \underbrace{n (R \bar\theta^*_N - c)' (B_g V_N(R))^{-1} (R \bar\theta^*_N - c)}_{\text{Wald stat. w. random scaling}} \\
&=(n^{-1} + (NB_g)^{-1})^{-1} (R \bar\theta^*_N - c)' (B_g V_N(R))^{-1} (R \bar\theta^*_N - c).
\end{align}
Here, the scaling factor $(n^{-1} + (NB_g)^{-1})^{-1}$ accounts for the combined uncertainty from both stochastic approximation and random sampling. 
From the decomposition
\[
R \bar\theta^*_N - c = R(\bar\theta^*_N - \hat \theta_n) + R(\hat \theta_n - \theta_\oo),
\]
under $H_0 : R \theta_\oo = c$, and Theorem~\ref{thm:asymp:normal}, the asymptotic variance of $R \bar\theta^*_N - c$ is given by
\begin{align*}
& N^{-1} R (G_\oo'G_\oo)^{-1} \Sigma_\oo (G_\oo'G_\oo)^{-1} R' + n^{-1} R(G_\oo' G_\oo)^{-1} G_\oo' \Omega_\oo G_\oo     (G_\oo' G_\oo)^{-1}R' \\
& \le    (n^{-1} + (NB_g)^{-1}) B_g R (G_\oo'G_\oo)^{-1} \Sigma_\oo (G_\oo'G_\oo)^{-1} R'
\end{align*}
because $B_g \Sigma_\oo \ge G_\oo' \Omega_\oo G_\oo$.
Since $V_N(R)$ is random, we combine the variance bound with a stochastic-dominance argument to obtain the following corollary from Theorem~\ref{thm:FCLT}.

\begin{corollary}\label{cor:inference:random}
Let the conditions of Theorem~\ref{thm:FCLT} hold. For $\alpha\in(0,1)$, let
$\mathrm{cv}_{\ell,1-\alpha}$ denote the $(1-\alpha)$ quantile of
\[
Z' \left( \int_0^1 \bar{W}(r)\bar{W}(r)' dr \right)^{-1} Z,
\]
where $Z\sim\mathcal N(0,I_\ell)$ is independent of the standard $\ell$-dimensional Brownian motion $W$, and $\bar W(r)=W(r)-rW(1)$. Under $H_0 : R\theta_\oo=c$,
\[
\limsup_{\substack{n,N\to\infty\\N\asymp n}}\mathbb P\big(T_{n,N}>\mathrm{cv}_{\ell,1-\alpha}\big)\le\alpha.
\]
\end{corollary}

In Corollary~\ref{cor:inference:random}, we have the following distributional equivalence:
\begin{align}\label{iden-dist-results}
 Z'\left( \int_0^1 \bar{W}(r)\bar{W}(r)' dr \right)^{-1} Z \ \overset{d}{=} \  W(1)' \left( \int_0^1 \bar{W}(r)\bar{W}(r)' dr \right)^{-1} W(1),
\end{align}
which follows from the fact that \( V_N(R) \) is asymptotically a functional of the process $\bar{W}(r)=  W(r) - r W(1)$, up to an unknown variance matrix. Since \( W(1) \) is projected out, \( \operatorname{Cov}(\bar{W}(r) , W(1)) = 0 \) and thus the process $\bar{W}(r)$ is independent of \( W(1) \) by joint Gaussianity.
%As a result, the critical values remain unchanged regardless of whether the left-hand or right-hand expression in~\eqref{iden-dist-results} is used to simulate them.

\subsubsection{Universality of Asymptotic Validity in Inference via Random Scaling} \label{subsubsec:RS}

Recall that the test statistic $T_{n,N}$ defined in~\eqref{def:wald:stat} takes the form
\begin{align*}
T_{n,N} = \left(n^{-1} + (N B_g)^{-1} \right)^{-1} (R \bar{\theta}^*_N - c)' (B_g V_N(R))^{-1} (R \bar{\theta}^*_N - c).
\end{align*}

Under the assumptions and growth conditions of Theorem~\ref{thm:FCLT}, including $N^{1-a}/n\to0$, the regimes in Section~\ref{subsec:asymp:regimes} and Appendix~\ref{appendix:asymp:regimes} motivate the following decision rule:

\begin{align*}
\text{Reject } H_0: R\theta_\oo = c \text{ if } T_{n,N} > \mathrm{cv}_{\ell,1-\alpha},
\end{align*}
where $\mathrm{cv}_{\ell,1-\alpha}$ depends only on the rank of $R$ and the nominal significance level.
This general validity stems from the scaling factor
\[
\tau_n := \left(n^{-1} + (N B_g)^{-1} \right)^{-1},
\]
which adapts automatically to different asymptotic regimes. When $N/n \to 0$, we have $\tau_n / (N B_g) \to 1$; when $n/N \to 0$, we have $\tau_n / n \to 1$. As a result, the procedure accommodates both sources of uncertainty—stochastic approximation and random sampling—without requiring explicit knowledge of their relative magnitudes.
If inference precision is a priority, it is advisable to increase both $N B_g$ and $B_G$ as much as the available computational budget allows.

\subsection{Summary and Practical Implementation}\label{sec:alg:random-scaling} 

We conclude this section with a summary of the random scaling inference procedure and practical guidance for its implementation. The test statistic $T_{n,N}$ enables valid inference across a wide range of asymptotic regimes by adapting to the relative growth of $n$ and $N$. Its implementation is computationally efficient: after forming $R\bar{\theta}_t^*$, the recursive formulas for $V_N(R)$ update only $\ell$-dimensional vectors and $\ell\times\ell$ matrices, avoiding full $d\times d$ matrix updates. The method also bypasses the need to estimate the asymptotic variance explicitly, making it particularly attractive for large-scale problems.

For practical use, we recommend choosing mini-batch sizes $B_G$ and $B_g$ to balance statistical precision and computational cost. Increasing $B_G$ improves efficiency but raises computational burden. Run Algorithm~\ref{alg:mini-batch} for a sufficiently large number of iterations $N$, chosen jointly with the mini-batch sizes and the available computational budget. Since each iteration uses $B=B_G+B_g$ draws, the ratio $BN/n$ measures the expected number of passes through the sample. Inference can be conducted by adding
the random scaling inference component to 
%Algorithm~\ref{alg:random-scaling} 
%in parallel 
Algorithm~\ref{alg:mini-batch}.
This procedure provides a unified and robust method for inference across different asymptotic regimes. The reference distribution in~\eqref{iden-dist-results} is pivotal and depends only on the number of restrictions $\ell$. With finite Jacobian batches, inference can be conservative. When $\ell = 1$, critical values $\mathrm{cv}_{1 - \alpha/2}$ are given in \citet[Table I]{Abadir:Paruolo:97}. For instance, $\mathrm{cv}_{1 - \alpha/2} = 6.747$ when $\alpha = 0.05$, and $\mathrm{cv}_{1 - \alpha/2} = 5.323$ when $\alpha = 0.10$. For a scalar restriction, reject when $T_{n,N}>\mathrm{cv}_{1-\alpha/2}^{2}$, or equivalently when $\sqrt{T_{n,N}}>\mathrm{cv}_{1-\alpha/2}$. Thus, the quadratic critical values are $6.747^2=45.522009$ at the 5\% level and $5.323^2=28.334329$ at the 10\% level.
%The unsquared value remains the confidence-interval multiplier.
The corresponding $(1 - \alpha)$ confidence interval for $R \theta_\oo$ is given by
\[
\left[ R \bar{\theta}_N^* \pm \mathrm{cv}_{1 - \alpha/2} \cdot \sqrt{n^{-1} + (N B_g)^{-1}} \cdot (B_g V_N(R))^{1/2} \right].
\]
For general $\ell > 1$, compare $T_{n,N}/\ell$ with the corresponding F-type critical value in \citet[Table II]{kiefer2000simple}.

\begin{remark}[Cluster sampling and automatic cluster robustness]\label{rem:cluster:main}
When the data are clustered, it suffices to replace the sampling unit in Algorithm~\ref{alg:mini-batch} by the cluster: at each iteration, $B = B_G + B_g$ cluster indices $k \in \{1, \ldots, K\}$, where $K$ denotes the number of clusters, are drawn uniformly with replacement, and the moment and Jacobian updates are computed from the weighted within-cluster sums $\hat{g}_k(\theta) = (K/n) \sum_{i=1}^{n_k} g(z_{ik}, \theta)$.
Appendix~\ref{sec:cluster} establishes the corresponding convergence results using the exact covariance induced by cluster resampling.
For inference, the random-scaling statistic keeps the same form as in this section, with the cluster count replacing the observation count and batches measured in clusters. Under its stated assumptions, Corollary~\ref{cor:clpop:RS} in Appendix~\ref{sec:cluster} establishes asymptotic validity with the same critical values.
\end{remark}

\section{Efficient Estimation}\label{sec:extensions}

Although the stochastic approximation steps in \eqref{def:sgd} yield asymptotically normal estimators, they are generally not efficient. The potential gains from improving efficiency can be substantial, especially when the sample size \(n\) is not exceedingly large. To address this limitation, we propose optional (but recommended) refinement steps that incorporate second-order information.

Let \(\bar{\theta}^*_N\) denote the averaged estimator after the \(N\)-th iteration. Compute:
\begin{align}\label{def:Phi:W}
\Phi_n &:= \frac{1}{n} \sum_{i=1}^{n} G\bigl(z_i, \bar{\theta}^*_N\bigr),
\quad \text{and} \quad
W_{\mathrm{MB}} := 
\left( 
\frac{B_g}{M_{\mathrm{MB}}} \sum_{\ell=1}^{M_{\mathrm{MB}}} \tilde{g}_\ell \bigl( \bar \theta_N^*\bigr)\,\tilde{g}_\ell \bigl( \bar \theta_N^*\bigr)' 
\right)^{\dagger},
\end{align}
where $W_{\mathrm{MB}}$ is based on $M_{\mathrm{MB}}$ randomly sampled mini-batches, \({\dagger}\) denotes the generalized inverse, and $M_{\mathrm{MB}}$ is a tuning parameter that diverges as \(n \to \infty\). The mini-batches in the definition of $W_{\mathrm{MB}}$ are drawn before refinement and are separate from the subsequent refinement batches.

For refinement, set \(B_{G,N+s}=b_s\), \(s\ge1\), where \((b_s)\) is a fixed nondecreasing sequence of positive integers independent of \(n,N,T\); for example, \(B_{G,t} = B_{G,0} + \lfloor\log(t-N)\rfloor\) for $t>N$, with $B_{G,t}=B_{G,0}$ for $t\le N$, where $B_{G,0}$ is the mini-batch size used in the first step. Define \(B_t := \sum_{s=1}^{t} (B_{G,s} + B_g)\) as the total number of mini-batch draws used for parameter updates through iteration $t$. Choose a predetermined integer $T>N$, and sample \(\{\tilde{z}_j\}_{j=B_N+1}^{B_T}\) independently and uniformly, with replacement, from the original dataset $\{z_1,\ldots,z_n\}$, conditional on the observed data and all information available before refinement.
For \(t = N+1\) to \(t = T\), sequentially update:
\begin{align}
\theta_t^* &= \theta_{t-1}^* 
- \gamma_t 
\left(\Phi_n' W_{\mathrm{MB}} \Phi_n \right)^{\dagger} 
\,\tilde{G}_{t} (\theta_{t-1}^*)' 
\,W_{\mathrm{MB}} 
\,\tilde{g}_{t} (\theta_{t-1}^*), \label{def:SGD:eff} \\
\bar \theta_t^* &= \frac{t-N-1}{t-N} \bar \theta_{t-1}^* + \frac{1}{t-N} \theta_t^*, \label{def:average:eff}
\end{align}
where \(\tilde{G}_{t} (\theta)\) and \(\tilde{g}_{t} (\theta)\) are re-defined as
\begin{align}\label{def:SGD:g_and_G_t}
\tilde{G}_{t} (\theta) := \frac{1}{B_{G,t}} \sum_{i=1}^{B_{G,t}} G(\tilde{z}_{B_{t-1} + i}, \theta),
\quad \text{and} \quad
\tilde{g}_{t} (\theta) := \frac{1}{B_g} \sum_{i=1}^{B_g} g(\tilde{z}_{B_{t-1} + B_{G,t} + i}, \theta),
\end{align}
replacing the definitions in \eqref{def:SGD:g_and_G}. The refinement replaces lines 7 and 8 in Algorithm~\ref{alg:mini-batch} with \eqref{def:SGD:eff} and \eqref{def:average:eff}, respectively. It starts at $\theta_N^*$ or $\bar\theta_N^*$ and continues the learning-rate schedule $\gamma_t=\gamma_0t^{-a}$ without restarting. Theorem~\ref{thm:asymp:normal:second:order} covers both initialization choices.

It is important to emphasize that \(\Phi_n\) and \(W_{\mathrm{MB}}\) are computed only once and reused throughout the refinement step. Since \(\bar{\theta}_N^*\) is already a consistent estimator, there is no need to recompute \(\left(\Phi_n' W_{\mathrm{MB}} \Phi_n\right)^{\dagger}\) or \(W_{\mathrm{MB}}\). However, updating \(\tilde{G}_{t}(\theta)\) and \(\tilde{g}_{t}(\theta)\) at each iteration remains essential, as it leverages the Lyapunov-function-based convergence guarantees of the algorithm.

\paragraph{Choice of \(\Phi_n\) and \(W_{\mathrm{MB}}\)}
Alternative choices for \(\Phi_n\) and \(W_{\mathrm{MB}}\) are possible, as long as they satisfy the consistency conditions:
\[
(\Phi_n' W_{\mathrm{MB}} \Phi_n)^{-1} \ppto (G_\oo' \Omega_\oo^{-1} G_\oo)^{-1},
\quad \text{and} \quad
W_{\mathrm{MB}} \ppto \Omega_\oo^{-1}.
\]
Conditional on the data and the preliminary estimate, independent empirical draws satisfy, at $\theta=\bar\theta_N^*$,
\[
\mathbb{E}_n^*\!\left[B_g\tilde g_\ell(\theta)\tilde g_\ell(\theta)'\mid\bar\theta_N^*\right]
=C_n(\theta)+B_g\bar g_n(\theta)\bar g_n(\theta)',
\]
where $C_n(\theta)=n^{-1}\sum_{i=1}^n g(z_i,\theta)g(z_i,\theta)'-\bar g_n(\theta)\bar g_n(\theta)'$ is the centered empirical moment covariance. The mean term is amplified by $B_g$; it vanishes under the maintained consistency conditions with fixed $B_g$ but can distort finite-sample weighting. A centered alternative is
\begin{align}\label{def:centered W}
    W_{\mathrm{MB}}^c :=
\left(
\frac{B_g}{M_{\mathrm{MB}}} \sum_{\ell=1}^{M_{\mathrm{MB}}} \left(\tilde{g}_\ell \bigl( \bar \theta_N^*\bigr)-\bar{g}_n\bigl( \bar \theta_N^*\bigr)\right)\,\left(\tilde{g}_\ell \bigl( \bar \theta_N^*\bigr)-\bar{g}_n\bigl( \bar \theta_N^*\bigr)\right)'
\right)^{\dagger}.
\end{align}
The refinement updates can use either $W_{\mathrm{MB}}$ or $W_{\mathrm{MB}}^c$. In what follows, $W_{\mathrm{MB}}$ in the refinement denotes whichever of the two is used, and the results below hold for either choice.

When the sample size \(n\) is relatively small, a natural alternative to $W_{\mathrm{MB}}$ is the full-sample analog:
\[
W_n = 
\left( 
 \frac{1}{n} \sum_{i=1}^{n} 
  g(z_i, \bar \theta_N^*)
  g(z_i, \bar \theta_N^*)'
\right)^{\dagger}.
\]
By contrast, \(\Phi_n\) is based on the full sample by default, since it does not involve matrix inversion and its mini-batch counterpart tends to yield similar results due to averaging. 
% However, when \(n\) is too large for full-sample computation, one may replace \(\Phi_n\) with a random subsample version or a mini-batch approximation:
% \[
% \Phi_{\mathrm{MB}} := \frac{1}{M_{\mathrm{MB}}} \sum_{\ell=1}^{M_{\mathrm{MB}}} \tilde{G}_\ell \bigl(\bar{\theta}^*_N\bigr).
% \]
However, when \(n\) is too large for full-sample computation, one can approximate \(\Phi_n\) using either of the following alternatives: (i) a simple average over a randomly selected subsample of the data, or (ii) an average over multiple randomly drawn mini-batches.
% The latter takes the form
% \[
% \Phi_{\mathrm{MB}} := \frac{1}{M_{\mathrm{MB}}} \sum_{\ell=1}^{M_{\mathrm{MB}}} \tilde{G}_\ell \bigl(\bar{\theta}^*_N\bigr),
% \]
% where each \(\tilde{G}_\ell(\bar{\theta}_N^*)\) denotes the average of \(G(z_i, \bar{\theta}_N^*)\) over a randomly drawn mini-batch. 
Both approaches offer computational savings and yield similar results due to averaging. For the centered weighting matrix, $\bar{g}_n\bigl( \bar \theta_N^*\bigr)$ can be approximated in the same way.

Define the modified Lyapunov function, 
\begin{equation*}
\bar{Q}_{n,W}(\theta) := \frac{1}{2}\left[\bar{g}_n(\theta)' W_{\mathrm{MB}} \bar{g}_n(\theta) - \bar{g}_n(\hat{\theta}_{n,W})' W_{\mathrm{MB}} \bar{g}_n(\hat{\theta}_{n,W})\right]
\end{equation*}
and the corresponding GMM estimator,
\begin{equation*}
\hat{\theta}_{n, W} = \argmin_{\theta \in \Theta} \bar{g}_n(\theta)' W_{\mathrm{MB}} \bar{g}_n(\theta).
\end{equation*}
We are now ready to state the additional assumptions required for this section.

\begin{assumption}
\label{assm:configuration-iteration-number-and-mini-batch-size}
The following holds for some $c > 0$, $\delta > 0$, $M < \infty$, and $p \ge 1$.
\begin{enumerate}[({A\ref{assm:configuration-iteration-number-and-mini-batch-size}}.1)]
\item \label{asm:item:local-convex-Lojasiewicz:WMB}
$\frac{\partial^2 \bar{Q}_{n,W}(\theta)}{\partial \theta \partial \theta'}  \ge c I_{\dtheta}$ for all $\|\theta - \hat{\theta}_{n,W}\|\le \delta$ \wpa.

\item \label{asm:item:id:WMB}
$\bar{Q}_{n,W}(\theta) \ge c$ and $\|\tfrac{\partial}{\partial  \theta} \bar{Q}_{n,W}(\theta)\|^2 \ge c$ for all $\|\theta - \hat{\theta}_{n,W} \| \ge \delta$ \wpa.

\item \label{asm:item:stability:WMB}
$\frac{\partial^2 \bar{Q}_{n,W}(\theta)}{\partial \theta \partial \theta'}  \le M I_{\dtheta}$ and $\frac{1}{n} \sum_{i=1}^n \|g(z_i, \theta)\|^{2p} \le M (\bar{Q}_{n,W}(\theta)^p + 1)$ for all $\theta \in \Theta$ \wpa.

\item \label{asm:item:eff:growing-B-G}
As \(s \to \infty\), \(b_s \to B_G \equiv B_{G,\infty}\), where either \(B_G \in \mathbb{N}\) or \(B_G = \infty \).

\item \label{asm:item:eff:growing-N-T}
The iteration numbers \((N, T)\) satisfy \(N = O(T - N)\), with \((T - N)^{1 - a}/n \to 0\) and \(N^a/n = O(1)\).
\end{enumerate}
\end{assumption}

Conditions~\ref{asm:item:local-convex-Lojasiewicz:WMB}–\ref{asm:item:stability:WMB} mirror Conditions~\ref{asm:item:local-convex-Lojasiewicz}–\ref{asm:item:stability}, but are formulated for the weighted objective \(\bar{Q}_{n,W}(\theta)\) in place of \(\bar{Q}_n(\theta)\). Condition~\ref{asm:item:eff:growing-B-G} allows for the local sequence \((b_s)\) to either converge to a finite constant or diverge. Condition~\ref{asm:item:eff:growing-N-T} governs the joint growth of \(N\) and \(T\) relative to the sample size \(n\). To interpret this condition, suppose \(a = 0.501\), as used in our Monte Carlo experiments. The requirement \((T - N)^{1 - a}/n \to 0\) implies that \(T - N\) must grow slower than \(n^{1/(1 - a)}\), or approximately \(n^{2.004}\). Likewise, the condition \(N^a/n = O(1)\) is satisfied when \(N = O(n^{1/a})\), or approximately \(O(n^{1.996})\). It also implies \(N^{1 - a}/n \to 0\), because \(N^{1-a}/n=(N^a/n)N^{1-2a}\) and \(a>1/2\). Thus, for polynomial growth $N\asymp n^b$ and $T-N\asymp n^c$, these iteration-growth conditions require $0<b\le 1/a$ and $b\le c<1/(1-a)$.

Denote the additional variance component by
\[
V_{{G}} = (G_\oo' \Omega_\oo^{-1} G_\oo)^{-1}  \mathbb{E}\left[ \left(G(z_i, \theta_\oo) - G(\theta_\oo)\right)' \Omega_\oo^{-1} \left(G(z_i, \theta_\oo) - G(\theta_\oo)\right) \right] (G_\oo' \Omega_\oo^{-1} G_\oo)^{-1},
\]
which reflects the variance of the mini-batch estimator $\tilde{G}_t(\theta_{t-1}^*)$ as $\theta_{t-1}^* \to \theta_\oo$.\footnote{Note that \(V_G/B_G\) differs from \eqref{def:add:var:ineff} because we now employ the optimal weighting matrix.}

\begin{theorem}[Second-Order Algorithm: CLT and FCLT]
\label{thm:asymp:normal:second:order}
\begin{enumerate}

\item [(i)]
Let Assumptions \ref{assm:uniform}, \ref{assm:lr}, \ref{assm:diff}, \ref{asm:regularity} and
\ref{assm:configuration-iteration-number-and-mini-batch-size} hold with $p > 1$. 
Then, as $n$, $N$, $M_{\mathrm{MB}}$, and $T-N$ all tend to infinity, it holds
{\setlength{\arraycolsep}{2pt}%
\begin{align*}
\hspace{-1.5em}\left( \begin{matrix}
            \sqrt{n}(\hat{\theta}_{n,W} - \theta_\oo) \\    
    \sqrt{T-N} (\bar \theta^*_{T} - \hat{\theta}_{n,W})    
\end{matrix} \right) & \dto \mathcal{N}\left( 0, \left( \begin{matrix}
   (G_\oo' \Omega_\oo^{-1} G_\oo)^{-1} & 0\\    
   0 & \frac{1}{B_g} \left( (G_\oo' \Omega_\oo^{-1} G_\oo)^{-1} + \frac{1}{B_G} V_{{G}}\right)
\end{matrix} \right)\right).
\end{align*}}

\item [(ii)]
Let Assumptions \ref{assm:uniform}, \ref{assm:lr}, \ref{assm:diff}, \ref{asm:regularity} and
\ref{assm:configuration-iteration-number-and-mini-batch-size} hold with $p > (1-a)^{-1}$.
Then, as $n$, $N$, $M_{\mathrm{MB}}$, and $T-N$ all tend to infinity, it holds
\begin{align*}
 \hspace{-1.75em}\left( \begin{matrix}
   \sqrt{n} (\hat{\theta}_{n,W} - \theta_\oo) \\    
   \frac{1}{\sqrt{T-N}}   \sum_{t=N+1}^{N+\lfloor (T-N) r\rfloor}(\theta^*_{t} - \hat \theta_{n,W})  
\end{matrix} \right) \wto \left( \begin{matrix}
    (G_\oo' \Omega_\oo^{-1} G_\oo)^{-1/2}  Z \\
    \frac{1}{B_g^{1/2}} \left( (G_\oo' \Omega_\oo^{-1} G_\oo)^{-1} + \frac{1}{B_G} V_{{G}}\right)^{1/2} W(r)
\end{matrix} \right).
\end{align*}
\end{enumerate}
\end{theorem}

Theorem~\ref{thm:asymp:normal:second:order} establishes the CLT and FCLT for the refinement, extending Theorems~\ref{thm:asymp:normal} and \ref{thm:FCLT}. The asymptotic variance of $\bar\theta_T^*-\theta_\oo$ separates into
\begin{equation}\label{eq:refinement:variance-decomposition}
\underbrace{\frac{(G_\oo'\Omega_\oo^{-1}G_\oo)^{-1}}{n}}_{\text{sampling}}
+
\underbrace{\frac{(G_\oo'\Omega_\oo^{-1}G_\oo)^{-1}}{(T-N)B_g}}_{\text{moment-batch noise}}
+
\underbrace{\frac{V_G}{(T-N)B_gB_G}}_{\text{Jacobian-batch noise}},
\end{equation}
with $1/B_G=0$ when $B_G=\infty$. More refinement updates reduce both computational components, while larger Jacobian batches specifically reduce the third component, holding the other quantities fixed. When $B_G=\infty$ and $n/((T-N)B_g)\to\tau<\infty$, the asymptotic covariance of $\sqrt n(\bar\theta_T^*-\theta_\oo)$ is $(1+\tau)(G_\oo'\Omega_\oo^{-1}G_\oo)^{-1}$. Full-sample GMM efficiency is attained when computational uncertainty is negligible relative to sampling uncertainty ($\tau=0$), including with finite $B_G$.

\paragraph{Random Scaling Inference}

Just as Theorem~\ref{thm:FCLT} provides a foundation for random scaling inference, Theorem~\ref{thm:asymp:normal:second:order}(ii) implies that the algorithm described in Section~\ref{sec:inference}
%Algorithm~\ref{alg:random-scaling} 
remains applicable with minor modifications. Reset the random-scaling accumulators at the start of refinement and apply \eqref{eq:A_t}--\eqref{eq:V_t} using the refinement-stage index $s=t-N$ and the refinement averages $\bar\theta_{N+s}^*$. In particular, all time-dependent factors in these recursions use $s$ rather than $t$. Write $V_T(R)$ for the resulting refinement-stage matrix evaluated at $s=T-N$. The test statistic becomes
\[
(n^{-1} + ((T - N) B_g)^{-1})^{-1} (R \bar{\theta}_T^* - c)' (B_g V_T(R))^{-1} (R \bar{\theta}_T^* - c),
\]
while continuing to use the same critical values as before. 

When \(R \theta_\oo\) is scalar (i.e., \(\ell = 1\)), the \((1 - \alpha)\)-level random scaling confidence interval for \(R \theta_\oo\) becomes
\begin{align}\label{CI:eff:RS}
\left[
R \bar{\theta}_T^* \pm \mathrm{cv}_{1 - \alpha/2} \sqrt{n^{-1} + ((T - N) B_g)^{-1}} \left( B_g V_T(R) \right)^{1/2}
\right],
\end{align}
where \(\mathrm{cv}_{1 - \alpha/2}\) is the critical value defined in Section~\ref{sec:inference}.
%defined in Algorithm~\ref{alg:random-scaling}.

\paragraph{Plug-In Inference}

We now turn to an important special case in which \(B_{G,t} \to B_G = \infty\). Let \(\chi^2_\ell\) denote a chi-squared variable with \(\ell\) degrees of freedom. In this setting, the Jacobian-noise contribution vanishes, yielding the following plug-in inference result.

\begin{corollary}[Second-Order Algorithm: Plug-in Inference]\label{cor:plugin:eff}
Assume the conditions in Theorem~\ref{thm:asymp:normal:second:order}(i) hold with \(B_G = \infty\). Then, under \(H_0 : R \theta_\oo = c\),
\begin{equation}\label{def:Wald:eff:in-feasible}
\frac{(R\bar{\theta}_T^* - c)' \left(R (G_\oo' \Omega_\oo^{-1} G_\oo)^{-1} R' \right)^{-1} (R \bar{\theta}_T^* - c)}{n^{-1} + ((T - N) B_g)^{-1}} \dto \chi^2_{\ell}.
\end{equation}
%where \(R \in \mathbb{R}^{\ell \times \dtheta}\) is of full row rank.
%, and \(\chi^2_\ell\) denotes the chi-squared distribution with \(\ell\) degrees of freedom.
\end{corollary}

The corollary provides a Wald-type test statistic based on \(\bar{\theta}_T^*\), using a plug-in estimate of the asymptotic variance implied by Theorem~\ref{thm:asymp:normal:second:order}(i). Since plug-in inference is most appropriate when the sample size \(n\) is relatively modest, we estimate \(G_\oo\) and \(\Omega_\oo^{-1}\) using full-sample quantities evaluated at \(\bar{\theta}_T^*\). Specifically, define
\[
\Phi_n(\bar \theta_T^*) := \frac{1}{n} \sum_{i=1}^{n} G\bigl(z_i, \bar \theta_T^* \bigr), \quad 
W_n(\bar \theta_T^*) := \left( 
 \frac{1}{n} \sum_{i=1}^{n} 
  g(z_i, \bar \theta_T^*)
  g(z_i, \bar \theta_T^*)'
\right)^{\dagger}.
\]
Then, a feasible version of the Wald statistic in \eqref{def:Wald:eff:in-feasible} is
\begin{equation}\label{def:Wald:eff:feasible}
\frac{(R\bar{\theta}_T^* - c)' \left(R \left(\Phi_n(\bar \theta_T^*)' W_n(\bar \theta_T^*) \Phi_n(\bar \theta_T^*)\right)^{-1} R' \right)^{-1} (R \bar{\theta}_T^* - c)}{n^{-1} + ((T - N) B_g)^{-1}},
\end{equation}
and the null \(H_0 : R \theta_\oo = c\) is rejected if this statistic exceeds the \(\chi^2_\ell\) critical value.

As a counterpart to \eqref{CI:eff:RS}, a \((1 - \alpha)\) plug-in confidence interval for the scalar case of \(R \theta_\oo\) is given by
\begin{align}\label{CI:eff:plug-in}
\left[ R \bar{\theta}_T^* \pm z_{1 - \alpha/2} \sqrt{n^{-1} + ((T - N) B_g)^{-1}} 
\left( R \left( \Phi_n(\bar \theta_T^*)' W_n(\bar \theta_T^*) \Phi_n(\bar \theta_T^*) \right)^{-1} R' \right)^{1/2} \right],
\end{align}
where \(z_{1 - \alpha/2}\) denotes the \(1 - \alpha/2\) quantile of the standard normal distribution.

\subsection{Comparison of Estimation and Inference Methods}  

Both random scaling and plug-in approaches provide asymptotically valid inference for the refined GMM estimator $\bar{\theta}_T^*$, but each offers distinct practical advantages. Random scaling uses the stochastic path for inference without requiring an additional full-sample evaluation of the moment covariance or Jacobian. By contrast, the plug-in method relies on full-sample estimates of the Jacobian and the optimal GMM weighting matrix, which can deliver more accurate and tighter confidence intervals when $n$ is moderate and full-sample access is feasible. The choice between the two methods should depend on computational constraints and the precision required.

To sharpen the discussion, it is useful to distinguish three scenarios. In the first, a \emph{strictly online} setting possibly involving real-time decision-making, each data batch is observed only once and cannot be revisited. This also covers cases in which privacy concerns prevent storage of the full dataset. 
With fresh i.i.d.\ observations, SLIM can be implemented as a single-pass stochastic algorithm using independent incoming batches for the moment and Jacobian updates. For inference, random scaling remains viable without revisiting past data.

\begin{remark}[SLIM with fresh i.i.d.\ observations]\label{rem:iid-online}
In Algorithm~\ref{alg:mini-batch}, replace empirical resampling by fresh i.i.d.\ observations in disjoint batches of fixed sizes $B_G,B_g$, independent of the past $\mathcal F_{t-1}$. Maintain Assumptions~\ref{assm:lr}, \ref{assm:diff}, and the population version of Assumption~\ref{asm:regularity}: replace $\bar Q_n,\hat\theta_n,\bar G_n$ by $Q(\theta)=\tfrac12g(\theta)'g(\theta),\theta_\oo,G$, replace empirical averages by expectations, and impose the bounds deterministically, with $p>(1-a)^{-1}$. Let $\Theta$ be convex and contain the iterates, starting from a fixed $\theta_0^*\in\Theta$.

Batch independence gives
$\mathbb E[\widetilde G_t(\theta)'\widetilde g_t(\theta)\mid\mathcal F_{t-1}]=G(\theta)'g(\theta)$
for each past-measurable $\theta$. The localization and linearization arguments for Theorems~\ref{thm:consistency} and \ref{thm:FCLT} therefore apply. Since $g(\theta_\oo)=0$, the update-noise covariance at $\theta_\oo$ is $\Sigma_\oo$ in \eqref{eq:sgd-score-variance}. Consequently, $\theta_N^*\to\theta_\oo$ almost surely and
\[
\frac1{\sqrt N}\sum_{t=1}^{\lfloor Nr\rfloor}(\theta_t^*-\theta_\oo)
\wto \Lambda_\oo^{1/2}W(r),\qquad
\Lambda_\oo=(G_\oo'G_\oo)^{-1}\Sigma_\oo(G_\oo'G_\oo)^{-1}.
\]
For $R$ of full row rank $\ell$, under $H_0:R\theta_\oo=c$, we have 
\begin{equation}\label{eq:iid-online:RS}
T_N^{\mathrm{on}}:=N(R\bar\theta_N^*-c)'V_N(R)^{-1}(R\bar\theta_N^*-c)
\dto W_\ell(1)'\left[\int_0^1\bar W_\ell(r)\bar W_\ell(r)'\,dr\right]^{-1}W_\ell(1),
\end{equation}
where $\bar W_\ell(r)=W_\ell(r)-rW_\ell(1)$. Thus the critical values are unchanged.
\end{remark}

The second scenario involves \emph{sequential streaming} of a stored dataset that is too large to fit in memory. Data are processed in parts and can be revisited for multiple passes. Stochastic algorithms are well suited to these memory constraints. For inference, both random scaling and plug-in methods can be applied, although the plug-in approach requires an additional pass through the data to estimate the Jacobian and covariance matrix.

% \begin{table}[htbp]
% \centering
% \caption{Estimation and inference methods across scenarios}
% \label{tab:scenarios}
% \begin{tabular}{p{0.28\linewidth} p{0.28\linewidth} p{0.32\linewidth}}
% \hline\hline
% Scenario & Estimation & Inference \\
% \hline
% Online (real-time, no storage)
% & \leavevmode{}Single-pass stochastic; i.i.d.\ draws
% & \leavevmode{}Random scaling \\
% \hline
% Streaming (too large for memory)
% & Stochastic; multi-pass possible
% & Both feasible; \leavevmode{}plug-in requires an additional data pass \\
% \hline
% In-memory but intensive
% & Stochastic if overall cost is lower
% & Both feasible; differences minor \\
% \hline\hline
% \end{tabular}
% \end{table}

In the third scenario, the dataset fits into memory but batch algorithms that process the entire dataset in each iteration are \emph{computationally burdensome}. The choice between stochastic and full-sample optimization depends on their total computational cost at the required precision. Both random scaling and plug-in inference methods are feasible, and the computational difference between them may be negligible since the plug-in method involves only sample averages and matrix inversions rather than repeated iterations over the data. Our numerical experiments in Section~\ref{sec:MC} correspond to this third setting.

In summary, random scaling is well suited to strictly online environments, advantageous for sequential streaming, and also feasible when the data fit in memory, while the plug-in approach is attractive when full-sample access is feasible and tighter confidence intervals are desired.
%Table~\ref{tab:scenarios} provides a compact overview.

\subsection{Sargan--Hansen Overidentification Tests}\label{subsec:J-test}

We propose Sargan--Hansen specification tests based on the refined GMM estimator $\bar{\theta}_T^*$. The first is the plug-in $J$-statistic,
\[
J := n \, \bar{g}_n(\bar{\theta}_T^*)' W_{\mathrm{MB}}^c \bar{g}_n(\bar{\theta}_T^*).
\]
 We use the centered weight in~\eqref{def:centered W} to improve finite-sample performance.

Because $\bar{\theta}_T^*$ need not satisfy the first-order condition $\Phi_{n}(\theta)' W_{n} \bar{g}_{n}(\theta)=0$ of the full-sample GMM criterion $\bar{g}_{n}(\theta)' W_{n} \bar{g}_{n}(\theta)$ with $W_n$ held fixed, we debias the plug-in moment condition before constructing the test. Let $\bar{\Phi}=\Phi_{n}(\bar{\theta}_T^*)$, $\bar{W}=W_{n}(\bar{\theta}_T^*)$, and $\bar{g}=\bar{g}_{n}(\bar{\theta}_T^*)$. Define
\[
\bar{g}_{n}^{D}=\Big(I-\bar{W}^{1/2}\bar{\Phi}
(\bar{\Phi}' \bar{W} \bar{\Phi})^{-1}\bar{\Phi}'\bar{W}^{1/2}\Big)\bar{W}^{1/2}\bar{g},
\]
which reduces to $\bar{W}^{1/2}\bar{g}$ if and only if $\bar\Phi'\bar W\bar g=0$, provided $\bar W$ is positive definite and $\bar\Phi$ has full column rank.
For example, this condition holds at an interior full-sample GMM solution computed with the weighting matrix held fixed at $\bar W$.
The associated debiased $J$-statistic is
\[
J_{\mathrm{D}} := n \, \bar{g}_{n}^{D}{}' \bar{g}_{n}^{D}
= n \, \bar{g}' \!\left(\bar{W}-\bar{W}\bar{\Phi}(\bar{\Phi}' \bar{W}\bar{\Phi})^{-1}\bar{\Phi}'\bar{W}\right)\! \bar{g}.
\]

Finally, we consider a recursive mini-batch $J$-test that averages moment evaluations along the refinement path during resampling of the observed sample. Starting at iteration $t=N+1$, update the averaged moment function recursively as
\[
\bar{g}_t^* = \frac{t-N-1}{t-N}\,\bar{g}_{t-1}^* + \frac{1}{t-N}\,\tilde{g}_t(\theta_{t-1}^*), \qquad 
t=N+1,\ldots,T,
\]
with $\bar{g}_N^*=0$. The corresponding mini-batch $J$-statistic is
\[
J^* := \Big(n^{-1} + ((T-N)B_g)^{-1}\Big)^{-1} \, \bar{g}_T^{*'} W_{\mathrm{MB}}^c \bar{g}_T^*.
\]

The next theorem establishes the asymptotic distributions of $J$, $J_{\mathrm{D}}$, and $J^*$.  

\begin{theorem}[Plug-in, Debiased Plug-in, and Recursive Mini-Batch $J$-Tests]
\label{thm:Plug-in-and-online-J-tests}
Let $\dg > \dtheta$, where $\dg$ denotes the dimension of the moment function, and $\chi^2_{\dg - \dtheta}$ and $\chi^2_{\dtheta}$ be independent chi-squares.
Suppose further that $\frac{n}{(T - N) B_g} \to \tau < \infty$ as $n \to \infty$.
Then, under the same conditions as in Theorem~\ref{thm:asymp:normal:second:order}(i), with $p>(1-a)^{-1}$, a fixed moment batch size $B_g$, and $B_{G,t}\to\infty$,
$$J  \dto \chi^2_{\dg - \dtheta} + \tau \chi^2_{\dtheta},\quad J_D \dto\chi^2_{\dg - \dtheta}, \quad \text{and} \quad 
J^*  \dto \chi^2_{\dg - \dtheta}.$$
\end{theorem}

Under the null and the conditions of Theorem~\ref{thm:Plug-in-and-online-J-tests}, the plug-in statistic $J$ is biased upward when $\tau>0$, even asymptotically. Its limit contains
the additional term $\tau\chi^2_{\dtheta}$, which raises the mean by $\tau\dtheta$,
because the computational error $\bar\theta_T^*-\hat\theta_{n,W}$ is non-negligible relative to sampling error. Centering the weight does not eliminate this additional asymptotic term. The same
$\tau$ governs efficiency: under optimal weighting and $B_{G,t}\to\infty$, the asymptotic covariance of
$\sqrt n(\bar\theta_T^*-\theta_\oo)$ is $(1+\tau)(G_\oo'\Omega_\oo^{-1}G_\oo)^{-1}$
(Theorem~\ref{thm:asymp:normal:second:order}). Hence $J$ has the classical
$\chi^2_{\dg-\dtheta}$ limit only when $\tau=0$, the case in which SLIM attains
full-sample GMM efficiency.

The other two statistics have the classical limit for any $\tau<\infty$.
$J_{\mathrm D}$ is the orthogonalized form of the overidentification statistic
\citep{GMM,newey1985}; the projection removes the first-order effect of the
estimation error. Full-sample moment, Jacobian, and weighting-matrix evaluations at the computed estimate do not require iterative full-sample optimization. By Lemma~\ref{lem:debiased-J:general} with $K=n$ and under its conditions,
$J_{\mathrm D}\dto\chi^2_{\dg-\dtheta}$ at any estimate $\tilde\theta$ with
$\tilde\theta-\theta_\oo=o_p(n^{-1/4})$, without efficient refinement or a diverging
Jacobian batch. $J^*$ instead corrects the scale of the averaged moment:
$n\,\bar g_T^{*\prime}W_{\mathrm{MB}}^c\bar g_T^*\dto(1+\tau)\chi^2_{\dg-\dtheta}$, and its
normalization removes the factor $1+\tau$. The condition $B_{G,t}\to\infty$ in
Theorem~\ref{thm:Plug-in-and-online-J-tests} removes the Jacobian-noise covariance in
Theorem~\ref{thm:asymp:normal:second:order}; $J_{\mathrm D}$ does not require it.

In practice, we recommend $J_{\mathrm D}$ when full-sample moment evaluations are
feasible and $J^*$ when they are costly. When $\tau$ is small, that is,
$n\ll(T-N)B_g$, the null distributions of all three statistics are well approximated by $\chi^2_{\dg-\dtheta}$.

\section{Monte Carlo Experiments}\label{sec:MC}

We evaluate SLIM in Monte Carlo experiments calibrated to the Exact Affine Stone Index (EASI) demand system of \citet{EASI}, with 576 moment conditions and 380 parameters. Sample sizes range from 20,000 to one million observations. We examine estimation accuracy, inference, and computational cost, including the performance of the $J$-tests.

\subsection{EASI Model and Calibration}\label{subsec:EASI GMM}

The implicit Marshallian budget shares $\mathbf{w}\in\mathbb R^J$ for $J$ goods are given by
\begin{align}\label{EASI:eq}
\mathbf{w}=\sum_{r=0}^5 \mathbf{b}_r y^r+\mathbf{C}\bm{z}+\mathbf{D}\bm{z} y+\sum_{l=0}^L z_l \mathbf{A}_l \mathbf{p}+\mathbf{B p} y+\bm{\varepsilon},
\end{align}
where $y$ is the implicit utility defined as
\begin{align}\label{y:eq}
y=\frac{x-\mathbf{p}^{\prime} \mathbf{w}+\frac{1}{2} \sum_{l=0}^L z_l \mathbf{p}^{\prime} \mathbf{A}_l \mathbf{p}}{1-\frac{1}{2} \mathbf{p}^{\prime} \mathbf{B} \mathbf{p}},
\end{align}
where $x$ is log nominal total expenditure, $\mathbf p$ is the vector of log prices, $\bm z=(z_1,\ldots,z_L)'$ contains demographic characteristics, $z_0=1$, and $\bm\varepsilon$ denotes unobserved preference shocks. The vectors $\mathbf b_r$ and matrices $\mathbf C$, $\mathbf D$, $\mathbf A_l$, and $\mathbf B$ are unknown coefficients. The system is nonlinear in its parameters because $y$ depends on $\mathbf A_l$ and $\mathbf B$. Moreover, $y$ is endogenous through the budget shares in the Stone index, $\mathbf p'\mathbf w$.

The empirical specification has $J=9$ expenditure categories and $L=5$ demographic characteristics. We omit the last share equation and express prices relative to the omitted category. Imposing Slutsky symmetry on $\mathbf A_l$ and $\mathbf B$ leaves 380 parameters in the eight estimated equations. We use the instruments in \citet{Pendakur:2015},
\begin{align}\label{GMM:inst:simple}
\bm{q}=[1, x, \ldots, x^5, \bm{p}', \bm{z}', \bm{z}'x, \bm{p}'x, \bm{p}'z_1, \ldots, \bm{p}'z_L]',
\end{align}
where $\bm p$ now denotes the eight relative log prices. Log nominal expenditure $x$ is taken to be exogenous but correlated with $y$. Multiplying the residual from each share equation by these 72 instruments gives 576 moment conditions. The model is therefore overidentified under the symmetry restrictions.

We calibrate the experiments using full-sample GMM estimates from the Canadian expenditure sample of 4,847 single-person renter households. The estimated coefficients and residual variances determine the simulation parameters. Appendix~\ref{app:easi:calibration} describes the sample and reports the empirical estimates, including the estimated Engel curves.

We consider two estimation targets. At $\mathbf p=\bm0$ and $\bm z=\bm0$, with preference shocks set to zero, the reference Engel curves are $\mathbf w(x)=\sum_{r=0}^5\mathbf b_r x^r$. We also consider the compensated own-price semi-elasticity of the rent budget share for the reference individual at $x=0$. This is the third diagonal element of $\mathbf A_0$, whose calibrated value is $\mathbf A_{0,33}=0.0533374$. The two targets allow us to examine both function estimation and inference about a scalar parameter.

\subsection{Simulation Design and Implementation}\label{subsec:easi:design}

Because budget shares and utility are jointly determined through higher-order utility terms, we simplify the data-generating process. We resample expenditure, prices, and normalized demographics from the calibration sample and construct utility by replacing the budget shares in~\eqref{y:eq} with their sample averages. We generate shares from~\eqref{EASI:eq} using the calibrated coefficients and independent normal shocks with the estimated residual variances. Utility is exogenous in this data-generating process, but estimation remains nonlinear because utility depends on candidate parameters. Generated shares need not lie in $[0,1]$, a restriction not imposed by the moment conditions. Appendix~\ref{app:easi:dgp} gives details.

We use 500 Monte Carlo replications at $n=20{,}000$, $50{,}000$, $100{,}000$, and $1{,}000{,}000$. SLIM is implemented in MATLAB; the full-sample benchmark uses the Stata GMM code of \citet{Pendakur:2015}, with the same pregenerated datasets. The reported timings are specific to these implementations and their stated tuning choices. Full-sample access is feasible, but repeated optimization passes are costly, as in the third setting of Section~\ref{sec:extensions}. Simulations run on a high-performance computing (HPC) cluster; timings use five separate replications on an Apple M1 Pro laptop. Appendix~\ref{app:easi:implementation} gives implementation details.

For $n\le100{,}000$, second-order SLIM uses a warm start from $\bm0_{380}$ followed by first-order updates with the initial 2SLS weight. Refinement uses the centered weight $W_{\mathrm{MB}}^c$ in~\eqref{def:centered W}, with the weight and curvature adjustment evaluated at the first-order average. Averaging and random scaling restart at refinement. We set $a=0.501$, and $B_G=B_g=512$ in the warm-start and first-order stages and $B_{G,0}=B_g=512$ with logarithmically increasing Jacobian batches during refinement. The first-order stage ends at $N=20{,}000$; $T=40{,}000$ or $60{,}000$ gives $T-N$ refinement updates. We use learning rates calibrated as in Appendix~\ref{subsec:Selection_gamma0}. Random-scaling (RS) and plug-in (PI) inference use~\eqref{CI:eff:RS} and~\eqref{CI:eff:plug-in}.

%The million-observation exercise also includes the first-order variants described below.

\subsection{Estimation, Inference, and Computational Performance}\label{subsec:easi:results}

We first examine Engel-curve estimation. Our precision measure is the root integrated mean squared error (RIMSE),
\[
\sqrt{\sum_{j=1}^{J-1}\int_{\underline x}^{\overline x}
\mathbb E\left[\left\{\sum_{r=0}^5(\widehat b_{j,r}-b_{j,r})x^r\right\}^2\right]\,dx},
\]
where $b_{j,r}$ is the $j$th element of $\mathbf b_r$ and $[\underline x,\overline x]=[-0.7,0.9]$.

\begin{table}[!htb]
\caption{Simulation Results for Engel Curves}
\begin{center}
\begin{tabular}{cccccc}
\hline\hline
  & Full-sample &  \multicolumn{4}{c}{SLIM} \\
  & GMM &  \multicolumn{4}{c}{} \\
 & (1) & (2) & (3) & (4) & (5) \\ \hline
Sample Size   & &  \multicolumn{4}{c}{($N,T $)} \\
$(n)$ & & \multicolumn{2}{c}{(20,000, \ 40,000)} & \multicolumn{2}{c}{(20,000, \ 60,000)} \\
 \hline
 & & \multicolumn{4}{c}{($s_0,E_{\mathrm{ws}}$)} \\
 \ 20,000 &       & $(5, 25)$ & $(8, 25)$& $(5, 25)$& $(8, 25)$\\
 \ 50,000 &       & $(3, \ 4)$ & $(8,\ 4)$ & $(3,\ 4)$ & $(8,\ 4)$\\
100,000 &       &   $(3,\ 1)$ & $(8,\ 1)$ & $(3,\ 1)$ & $(8,\ 1)$\\
\hline
\multicolumn{6}{c}{Panel A. Precision (Root Integrated Mean Squared Error)} \\
\ 20,000 & 0.024  & 0.023 & 0.023 & 0.024 & 0.023  \\
\ 50,000 & 0.015 &  0.015  & 0.015  & 0.015 & 0.015   \\
100,000 & NA & 0.011  & 0.012 & 0.011  &   0.011 \\
\hline
\multicolumn{6}{c}{Panel B. Computational Cost (Hours)} \\
\ 20,000 & \ 1.93  & 1.09  & 1.06 & 1.31 & 1.30 \\
 \ 50,000 & \ 7.84 & 1.05  & 1.06 & 1.33  &  1.33 \\
100,000 & $18.93^\ast$ & 1.06 & 1.07  & 1.35  &   1.35 \\
\hline \hline
\multicolumn{6}{l}{\footnotesize $\ast$ This exceeds our computational budget.}
\end{tabular}
\par\smallskip\begin{minipage}{\textwidth}\footnotesize
Notes: $N$ ends the first-order stage and $T$ is the terminal iteration, so refinement uses $T-N$ updates. $s_0$ calibrates the learning rate and $E_{\mathrm{ws}}$ is the number of warm-start epochs. Results use 500 Monte Carlo replications; hours are measured separately using five laptop replications. Settings are detailed in Appendix~\ref{app:easi:implementation}.
\end{minipage}
\end{center}
\label{tab:sim1}
\end{table}

Table~\ref{tab:sim1} shows comparable precision to full-sample GMM at smaller sample sizes and increasing computational gains with $n$. At $n=50{,}000$, GMM takes 7.84 hours versus 1.05--1.33 for SLIM, with the same RIMSE of 0.015. At $n=100{,}000$, SLIM takes 1.06--1.35 hours, whereas the full-sample benchmark exceeds 18 hours and our budget, so its Monte Carlo precision results are unavailable. SLIM RIMSEs decline with $n$ at both iteration budgets.

\begin{table}[!htb]
\caption{Simulation Results for Price Effect on Rent Budget Share}
\centering
\setlength{\tabcolsep}{4.5pt}
\begin{tabular}{cccccccccc}
\hline\hline
  & Full-Sample &  \multicolumn{8}{c}{SLIM} \\
  & GMM &  \multicolumn{8}{c}{} \\
 & (1) & (2) & (3) & (4) & (5)  & (6)  & (7)  & (8)  & (9) \\
 \hline
 Sample Size   & &  \multicolumn{8}{c}{($N,T$)} \\
$(n)$ & & \multicolumn{4}{c}{(20,000, \ 40,000)} & \multicolumn{4}{c}{(20,000, \ 60,000)} \\
 \hline
 & & \multicolumn{8}{c}{($s_0,E_{\mathrm{ws}}$)} \\
 \ 20,000 &       & \multicolumn{2}{c}{$(5, 25)$} & \multicolumn{2}{c}{$(8, 25)$}& \multicolumn{2}{c}{$(5, 25)$}& \multicolumn{2}{c}{$(8, 25)$}\\
 \ 50,000 &       & \multicolumn{2}{c}{$(3,\ 4)$} & \multicolumn{2}{c}{$(8,\ 4)$} & \multicolumn{2}{c}{$(3,\ 4)$} & \multicolumn{2}{c}{$(8, \ 4)$}\\
100,000 &       &   \multicolumn{2}{c}{$(3,\ 1)$} & \multicolumn{2}{c}{$(8,\ 1)$} & \multicolumn{2}{c}{$(3, \ 1)$} & \multicolumn{2}{c}{$(8, \ 1)$}\\
\hline
\multicolumn{10}{c}{Panel A. RMSE} \\
\ 20,000 & 0.019  &  \multicolumn{2}{c}{ 0.019 } & \multicolumn{2}{c}{ 0.018 } & \multicolumn{2}{c}{ 0.019 } & \multicolumn{2}{c}{ 0.019 }  \\
\ 50,000 &  0.013 &  \multicolumn{2}{c}{0.013 } & \multicolumn{2}{c}{0.012 } & \multicolumn{2}{c}{ 0.013}  & \multicolumn{2}{c}{0.013 }  \\
100,000 &  NA   &  \multicolumn{2}{c}{ 0.009 } & \multicolumn{2}{c}{ 0.009} & \multicolumn{2}{c}{0.009 }  & \multicolumn{2}{c}{0.009 }  \\
\hline
\multicolumn{10}{c}{Panel B. Coverage Probability (Nominal Level: 0.95)} \\
        &       & \underline{RS} &  \underline{PI} &  \underline{RS} &  \underline{PI} &  \underline{RS} &  \underline{PI} &  \underline{RS} &  \underline{PI} \\
\ 20,000  & 0.962 &  0.974   & 0.966    &   0.972  & 0.972  & 0.978 & 0.962 & 0.978 & 0.968 \\
\ 50,000  & 0.948 & 0.944 & 0.946 & 0.924 & 0.950 & 0.970 & 0.948  & 0.954 & 0.950 \\
100,000 & NA    & 0.954 & 0.952  & 0.940 & 0.960 &  0.960     &   0.954    &  0.954  & 0.956   \\
\hline
\multicolumn{10}{c}{Panel C. CI Length (Nominal Level: 0.95)} \\
        &       & \underline{RS} &  \underline{PI} &  \underline{RS} &  \underline{PI} &  \underline{RS} &  \underline{PI} &  \underline{RS} &  \underline{PI} \\
\ 20,000 & 0.077  & 0.202  & 0.077  & 0.237   & 0.077  &0.179 & 0.077 & 0.220 & 0.077\\
\ 50,000 & 0.049  & 0.079 & 0.049 & 0.085 & 0.049 & 0.077 & 0.049  & 0.085 & 0.049\\
100,000 & NA    & 0.054 & 0.035 & 0.057 & 0.035 & 0.056      &   0.035     & 0.059 & 0.035      \\
\hline
\multicolumn{10}{c}{Panel D. Computational Cost (Hours)} \\
        &       & \underline{RS} &  \underline{PI} &  \underline{RS} &  \underline{PI} &  \underline{RS} &  \underline{PI} &  \underline{RS} &  \underline{PI} \\
\ 20,000   & \ 1.93       &  1.05  &  1.04  &  1.03  &  1.05  &  1.33  &  1.29  &   1.36 &  1.30  \\
\ 50,000  & \ 7.84       &  1.06  &  1.06  &  1.07  & 1.06   &  1.34  &  1.36  & 1.34   &  1.34  \\
100,000    & 18.93$^\ast$ & 1.07   & 1.09   & 1.07   & 1.08   &  1.38  &  1.38  &   1.36 &  1.37   \\
\hline \hline
\multicolumn{10}{l}{\footnotesize $\ast$ This exceeds our computational budget.}
\end{tabular}
\par\smallskip\begin{minipage}{\textwidth}\footnotesize
Notes: RS and PI denote random-scaling and plug-in inference. $N$, $T$, $s_0$, and $E_{\mathrm{ws}}$ are defined in Table~\ref{tab:sim1}. Coverage and average interval length concern nominal 95\% intervals. Hours are measured separately using five laptop replications. Table~\ref{tab:easi:rent-full} also reports bias and standard deviation.
\end{minipage}
\label{tab:sim2}
\end{table}

Table~\ref{tab:sim2} reports rent-effect results; Table~\ref{tab:easi:rent-full} adds bias and standard deviation. SLIM has comparable RMSE to full-sample GMM at $n=20{,}000$ and $50{,}000$. SLIM estimation and inference take under 1.4 hours in these settings.

Coverage is generally close to the nominal 0.95 level at the larger sample sizes, but there are finite-sample differences between procedures and tuning choices. %At $n=20{,}000$, PI coverage ranges from 0.966 to 0.972.
For example, at $n=50{,}000$, RS coverage is 0.924 with $s_0=8$ and $T=40{,}000$, compared with 0.970 with $s_0=3$ and $T=60{,}000$. PI intervals have lengths similar to the full-sample GMM intervals, whereas RS intervals are longer. The similar RS and PI computation times are consistent with the availability of full-sample access in this experiment.

\begin{figure}[!htb]
	\begin{center}
		\includegraphics[width=\textwidth]{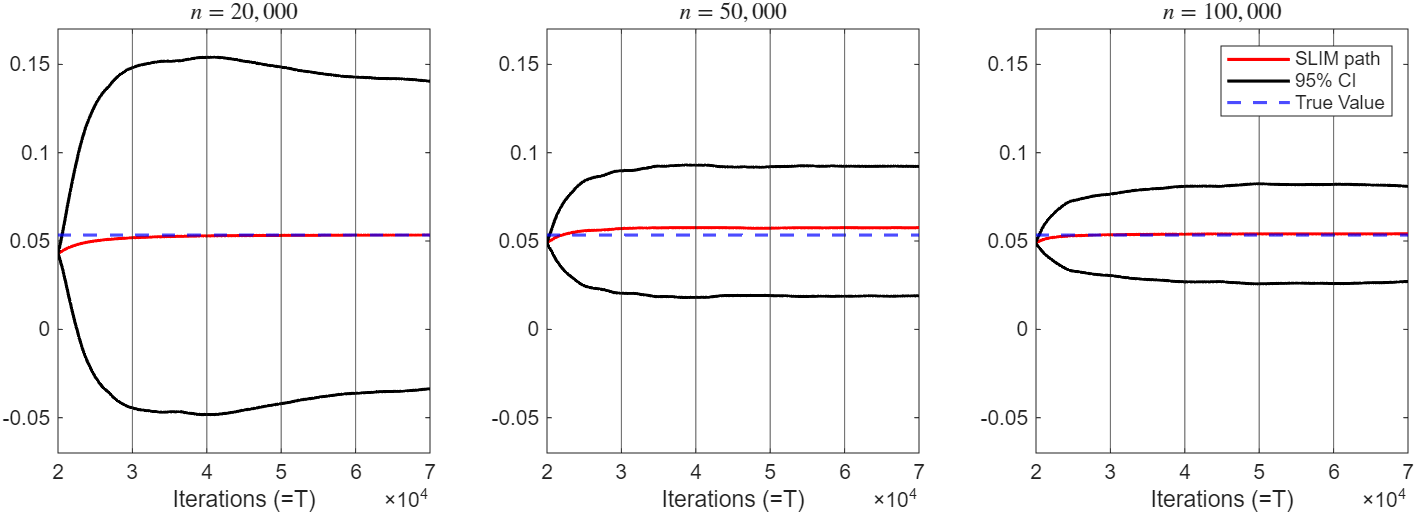}
	\end{center}
        \caption{SLIM Path and Random Scaling Confidence Intervals over Iterations for the Price Effect on Rent Budget Share}
	\label{figure:CIs}
\end{figure}

Figure~\ref{figure:CIs} shows how the SLIM estimates and RS intervals evolve with the terminal iteration $T$, averaged across replications. We use $(s_0,E_{\mathrm{ws}})=(5,25)$, $(3,4)$, and $(3,1)$ for $n=20{,}000$, $50{,}000$, and $100{,}000$, respectively. The intervals begin to stabilize around $T=40{,}000$ in these designs.

We then increase the sample size to one million to compare the first-order and refined procedures of Section~\ref{sec:extensions}. In addition to second-order SLIM, we consider first-order SLIM with the original 2SLS weight and first-order SLIM with the centered weighting matrix $W_{\mathrm{MB}}^c$, which omits the curvature factor $(\Phi_n'W_{\mathrm{MB}}\Phi_n)^\dagger$ from~\eqref{def:SGD:eff}. First-order SLIM with the 2SLS weight approximates one-step GMM, which may be inefficient. We use random subsamples for the warm start, initial weight construction, and computation of $\Phi_n$ and $\bar{g}_n\bigl( \bar \theta_N^*\bigr)$ for $W_{\mathrm{MB}}^c$ in the second-order method. Inference uses RS. A full-sample GMM comparison at this scale is beyond our computational budget.

\begin{table}[!htb]
\caption{Simulation Results when $n=$ 1,000,000}
\begin{center}
\begin{tabular}{ccccccc}
\hline\hline
 & \multicolumn{2}{c}{First-order SLIM}&  \multicolumn{2}{c}{First-order SLIM}&  \multicolumn{2}{c}{Second-order SLIM} \\
 & \multicolumn{2}{c}{(2SLS Weight)}& \multicolumn{2}{c}{(Refined Weight)}&   \\
 & (1) & (2) & (3)  & (4) & (5) &(6) \\ \hline
\hline
 $N$ &  300,000  &  500,000  & 20,000  & 20,000 & 20,000 & 20,000 \\
 $T$ &           &           & 120,000 & 140,000 & 40,000 & 60,000 \\
\hline
\multicolumn{7}{c}{Panel A. Engel Curves } \\
RIMSE & 0.020 & 0.020 & 0.030 & 0.030 & 0.006 & 0.005  \\
\hline
\multicolumn{7}{c}{Panel B. Price Effect on Rent Budget Share } \\
RMSE & 0.005 & 0.005 & 0.003 & 0.003 & 0.003 & 0.003  \\
Coverage Probability & 0.866 & 0.924 & 0.974& 0.968& 0.944 & 0.968 \\
 CI Length &  0.024 & 0.035 & 0.030 & 0.031 &  0.018 & 0.019 \\
\hline
\multicolumn{7}{c}{Panel C. Computational Cost (Hours)} \\
 & 4.98 & 7.74  & 2.09 & 2.38 &  1.11 & 1.38 \\
\hline \hline
\end{tabular}
\par\smallskip\begin{minipage}{\textwidth}\footnotesize
Notes: Coverage and average interval length use nominal 95\% random-scaling intervals. $E_{\mathrm{ws}}=1$ for all methods. $N$ ends the first-order stage; where reported, $T-N$ is the number of refinement updates. Table~\ref{tab:easi:million-full} reports bias, standard deviation, and the warm-start, pilot, and tuning settings. Hours are measured separately using five laptop replications.
\end{minipage}
\end{center}
\label{tab:sim4}
\end{table}

Table~\ref{tab:sim4} shows substantial gains from second-order refinement: Engel RIMSE falls to 0.005--0.006, rent-effect RMSE to 0.003, and computation takes 1.11--1.38 hours. First-order SLIM with the 2SLS weight takes 4.98--7.74 hours, with RS coverage improving from 0.866 to 0.924 as its iteration budget increases. Refining the weight alone reduces rent-effect RMSE and improves coverage, but yields larger Engel RIMSEs. Second-order SLIM further improves precision and shortens intervals and runtime, with coverage similar to that of the refined-weight first-order method. Table~\ref{tab:easi:million-full} reports the complete results and settings.

\subsection{J-Tests}\label{subsec:easi:jtests}

Finally, we evaluate the SLIM $J$-tests developed in Section~\ref{subsec:J-test}. %Full-sample moment evaluations are feasible in this simulation, so we focus on the debiased and plug-in $J$-tests.
We construct instruments using $x-c\,\mathbf p'\mathbf w$ in place of $x$ in~\eqref{GMM:inst:simple}, where $c\in\{0,1,2\}$ controls the strength of endogeneity. At $c=0$ the instruments are exogenous; larger values induce stronger endogeneity. We compare the debiased, plug-in, and recursive mini-batch $J$-tests with the standard full-sample GMM test.

\begin{table}[!htb]
\caption{Empirical Rejection Probabilities of Sargan--Hansen Overidentification Tests}
\begin{center}
\begin{tabular}{ccccc}
\hline\hline
Sample Size   & Full-Sample &  \multicolumn{3}{c}{SLIM} \\
$(n)$  & GMM &  Debiased & Plug-in & Recursive
 \\ \hline
\multicolumn{5}{c}{$c=0$} \\
\ 20,000 & 0.044  & 0.044 & 0.038 & 0.036 \\
 \ 50,000 & 0.048 &  0.046  & 0.040 & 0.038 \\
100,000 & NA$^\ast$ & 0.052  & 0.060 & 0.060 \\
\hline
\multicolumn{5}{c}{ $c=1$} \\
\ 20,000 & 0.090  & 0.090  & 0.104 & 0.104 \\
 \ 50,000 & 0.198 & 0.196  & 0.208 & 0.220 \\
100,000 & NA$^\ast$ & 0.488 & 0.510 & 0.514 \\
\hline
\multicolumn{5}{c}{ $c=2$} \\
\ 20,000 & 0.304  & 0.302  & 0.332 & 0.340 \\
 \ 50,000 & 0.906 & 0.904  & 0.910 & 0.918 \\
100,000 & NA$^\ast$ & 1.000 & 1.000 & 1.000 \\
\hline \hline
\multicolumn{5}{l}{\((s_0, E_{\mathrm{ws}}) = (5, 25), (3, 4)\), and \((3, 1)\) for \(n = 20{,}000\), \(50{,}000\), }\\
\multicolumn{5}{l}{and \(100{,}000\), respectively, and $(N,T)=(20,000,\ 40,000)$}\\
\multicolumn{5}{l}{$\ast$ This exceeds our computational budget.}
\end{tabular}
\end{center}
\label{tab:sim3}
\end{table}

Table~\ref{tab:sim3} reports rejection frequencies at the 0.05 nominal level. All three SLIM $J$-tests have null rejection frequencies close to 0.05, and their power increases with both endogeneity and sample size. Their performance is similar to that of the full-sample test where the comparison is available. The plug-in and recursive mini-batch tests tend to have slightly higher power in our experiments. As in the refinement stage, we employ the centered weighting matrix in constructing the plug-in and recursive mini-batch $J$-statistics. Simulation results not reported here show that, without centering, these tests exhibit substantial underrejection. Full-sample rejection frequencies at $n=100{,}000$ are unavailable because of the computational budget.

% The debiased test has null rejection frequencies close to 0.05, and its power increases with both endogeneity and sample size. Its performance is similar to the full-sample test where that comparison is available. The plug-in test underrejects under the null and has little power at $n=20{,}000$ and $50{,}000$; its power improves at $n=100{,}000$ with $c=2$. Full-sample rejection frequencies at $n=100{,}000$ are unavailable because of the computational budget.

%We next apply SLIM to observed employment destinations for 4.8 million individuals, using the clustered extension.

\section{College-to-Work Destination Choices: A LinkedIn Application}\label{sec:linkedin}

\subsection{Economic Question and Data}

Our application studies how college graduates sort into their first reported industry--occupation destinations. Fields of study can be linked to particular destinations through curricula, professional networks, employer recruiting, and information about employment opportunities. We summarize the resulting destination patterns with a destination-propensity measure for each group and allow destination preferences to vary with individual demographic characteristics. The resulting nonlinear choice model provides an application of SLIM to millions of observations with clustered moment restrictions.

The data are structured LinkedIn profiles provided by Revelio Labs. We use undergraduate education, graduation year, geographic assignment, the first reported post-college job, and predicted gender and race/ethnicity probabilities. The analysis focuses on bachelor degree holders who enter employment without first entering graduate school and have sufficient information to construct the choice set and model covariates. The final sample contains $n=4{,}812{,}451$ individuals in $K=32{,}430$ state $\times$ graduation year $\times$ undergraduate major groups.

The outcome is a graduate's first reported labor-market destination after college, defined by the job's industry and occupation. We combine six broad industry groups and 22 major occupation groups into 132 potential destinations, indexed by $a=(h,o)$. Each individual's estimation choice set contains at least two of these destinations and need not include all 132. Across groups, there are 513,950 group--destination combinations. Expanding individuals over their choice sets would yield 250,172,143 person--destination rows, so the implementation stores individual characteristics and group-specific destination attributes separately. Appendix~\ref{app:linkedin} describes the sample and construction rules.

%These profiles are not a probability sample of U.S. college graduates. The outcome is the first post-college job reported in the profile, which need not be the first job actually held. The empirical interpretation therefore concerns sorting within the observed sample, subject to the limitations of self-reported career histories and predicted demographic characteristics.

\subsection{Model and Moment Restrictions}

We model each graduate's first reported labor market destination after college as a choice among industry--occupation pairs in the individual-specific choice set $\mathcal A_i$. The specification is a static Roy-style model of sorting across labor market destinations \citep{roy1951thoughts}. We parameterize systematic utility using three components. The first component captures group-specific differences in destination choices through a destination-share index, $\widetilde C_{ia}$, with coefficient $\delta$. The second consists of additive industry and occupation components of mean utility, denoted by $\alpha_h$ and $\eta_o$. The third allows for demographic differences in sorting across industries. Let $x_i$ collect the five predicted demographic probabilities and let $\lambda_h$ denote the corresponding industry-specific coefficients.

Let $h(a)$ and $o(a)$ denote the industry and occupation of destination $a$. Collect the model coefficients in
\[
\theta
=
\left(
\delta,
\{\alpha_h\}_{h\ne h_0},
\{\eta_o\}_{o\ne o_0},
\{\lambda_{hr}\}_{h\ne h_0,\ r\in\mathcal D}
\right).
\]
The systematic utility from destination $a$ is
\begin{equation}\label{eq:linkedin:utility}
v_{ia}(\theta)
=
\delta\widetilde C_{ia}
+\alpha_{h(a)}
+\eta_{o(a)}
+x_i'\lambda_{h(a)}.
\end{equation}
We normalize $\alpha_{h_0}=0$, $\eta_{o_0}=0$, and $\lambda_{h_0}=0$. The reference industry $h_0$ is information, finance, and business, and the reference occupation $o_0$ is management.

Field-specific skills and access to employment opportunities can shape
graduates' sorting across labor market destinations
\citep{Traiberman2019-tg}. \citet{Johnson2013-xw} provide a related precedent, using a lagged aggregate occupation share in utility. We summarize these destination patterns using $\widetilde C_{ia}$, the standardized negative log odds of the share choosing destination $a$ in individual $i$'s state $\times$ graduation year $\times$ major group. Higher values indicate less commonly chosen destinations. A negative $\delta$ therefore assigns higher systematic utility to destinations with larger group shares, holding other covariates fixed. The index is an empirical proxy for sorting patterns. We treat $\widetilde C_{ia}$ as endogenous because it is constructed from contemporaneous choices. Our instruments are the first two graduation-cohort lags of the same index constructed from pooled outside-state counts. Persistence in major-specific destination shares motivates relevance. The identifying assumption is that these instruments are uncorrelated with the differences between observed choices and model-implied choice probabilities. Appendix~\ref{app:linkedin:economic-interpretation-destination-choice} discusses these economic connections and the interpretation of the index.

Let $\mathcal D$ index the five demographic probabilities. The demographic covariates $x_{ir}$, $r\in\mathcal D$, are the predicted probabilities of being female, Asian/Pacific Islander, Hispanic, Black, and other race/ethnicity. White is the omitted race/ethnicity category; other combines the Native American and multiple-race probabilities. We use these probabilities without rescaling. Individual demographic main effects are constant across destinations and therefore cancel from the choice probabilities.

Individual $i$'s utility from destination $a$ is
\[
U_{ia}=v_{ia}(\theta)+\varepsilon_{ia}.
\]
Motivated by the standard random-utility model with exogenous covariates and i.i.d. Type I extreme-value shocks, we specify model-implied choice probabilities in the conditional-logit form \citep{mcfadden1974conditional},
\begin{equation}
p_{ia}(\theta)
=
\frac{\exp\{v_{ia}(\theta)\}}
{\sum_{b\in\mathcal A_i}\exp\{v_{ib}(\theta)\}}.
\end{equation}

There are 52 parameters: one destination-propensity coefficient ($\delta$), five industry intercepts ($\alpha_h$), 21 occupation intercepts ($\eta_o$), and 25 industry--demographic interaction coefficients ($\lambda_{hr}$).

Let $f_{ia}$ be the 51-vector of industry indicators, occupation indicators, and industry--demographic interactions in~\eqref{eq:linkedin:utility}, and let $\widetilde Z^{\mathrm{out},1}_{ia}$ and $\widetilde Z^{\mathrm{out},2}_{ia}$ be the two standardized excluded instruments. Let $d_{ia}$ indicate whether individual $i$ chooses destination $a$. With the destination-share index treated as endogenous, we estimate $\theta$ by GMM under the following maintained unconditional moment restrictions:
\begin{equation}\label{eq:linkedin:moments}
g(z_i,\theta)=\sum_{a\in\mathcal A_i}
\begin{pmatrix} f_{ia}\\ \widetilde Z^{\mathrm{out},1}_{ia}\\
\widetilde Z^{\mathrm{out},2}_{ia}\end{pmatrix}
\{d_{ia}-p_{ia}(\theta)\},
\qquad \mathbb E[g(z_i,\theta_\oo)]=0.
\end{equation}
Thus, we have 53 moments and one overidentifying restriction.
% %The exogenous utility terms serve as their own instruments; the current destination-propensity measure does not.

% %Persistence in destination shares motivates instrument relevance, but exclusion remains a maintained assumption. Persistent national destination shocks, omitted wages, or labor-demand conditions could affect both lagged outside-state shares and current utility. The selected two-instrument specification follows earlier exploratory comparisons of own-state and leave-state lag combinations; it is an illustrative specification rather than a prespecified confirmatory design. Its overidentification test is not evidence that these threats are absent.

% Because larger values of the destination measure indicate less commonly chosen destinations, we expect $\delta$ to be negative. The coefficient may reflect adjustment frictions or access to opportunities \citep{Johnson2013-xw,Traiberman2019-tg}, including mechanisms associated with switching costs \citep{Artuc2010-jm}. Appendix \ref{app:linkedin:economic-interpretation-destination-choice} provides a detailed discussion of the economic interpretation of the destination-choice model.

\subsection{Clustered Estimation and Inference}

We treat state $\times$ graduation year $\times$ major groups as clusters, allowing dependence among individuals sharing the group-level destination measures. For cluster $k$, containing $n_k$ individuals, its contribution is
\begin{equation}\label{eq:linkedin:cluster}
\hat g_k(\theta)=\frac{K}{n}\sum_{i\in k}g(z_i,\theta),
\qquad \frac{1}{K}\sum_{k=1}^K\hat g_k(\theta)=\bar g_n(\theta).
\end{equation}
Thus, uniform cluster draws preserve the person-average GMM criterion despite unequal cluster sizes. Moment and Jacobian batches are drawn independently, uniformly, and with replacement. Population inference relies on the sampling and regularity conditions in Appendix~\ref{sec:cluster}.
%In particular, clustering does not by itself account for every consequence of constructing regressors and instruments: the shared outside-state counts and any first-order contribution from their estimation must satisfy the maintained sampling argument. The reported inference is conditional on those requirements.

The procedure uses 4,096 first-order updates followed by 8,192 updates in the second-order refinement stage. Counting both moment and Jacobian evaluations, these correspond to approximately 8.1 and 18.0 expected passes through the sample, respectively. Additional computations for calibration and inference are described in Appendix~\ref{app:linkedin:implementation}.
%The first-order stage starts from a previously computed tuning average; the refinement starts from the first-order average. The refinement weight and preconditioner are constructed at that average and held fixed, with fresh draws and a new averaging sequence during refinement. Moment batches contain 32 clusters. Jacobian batches contain 32 clusters in the first-order stage and increase from 32 to 41 during refinement according to the logarithmic schedule in
We report both plug-in and random-scaling confidence intervals for the averaged second-order SLIM estimator.
For the destination-propensity coefficient, the clustered counterpart of the decomposition in~\eqref{eq:refinement:variance-decomposition} attributes approximately 48.6\% of the estimated plug-in variance to sampling uncertainty, 6.0\% to moment-batch noise, and 45.4\% to Jacobian-batch noise. Computational uncertainty therefore accounts for 51.4\% of the total, predominantly through the finite Jacobian batches.

%Plug-in inference uses the actual optimization weight and finite Jacobian-batch contribution to account separately for sampling and computational variation. Random scaling uses the refinement path and can be conservative under the finite-Jacobian-batch regime described in Appendix~\ref{sec:cluster}. We retain both methods rather than selecting an interval according to its realized width. Computational uncertainty remains material in this run: it accounts for approximately 51.4\% of the destination-propensity coefficient's plug-in variance.

\subsection{Results and Computational Performance}

Table~\ref{tab:linkedin:main} reports the destination-propensity coefficient and the five industry interactions with predicted female probability. The destination coefficient is $-1.2957$, with a plug-in 95\% interval of $[-1.3001,-1.2912]$ and an RS interval of $[-1.3043,-1.2870]$. Holding the other index terms fixed, a one-standard-deviation increase in the negative-log-odds measure lowers the destination's systematic utility by about 1.30. %Thus destinations that are less common for a graduate's group receive lower model-implied utility. %This magnitude is an index association, not a structural cost or a change in choice probability.

The demographic interactions describe sorting relative to the information, finance, and business industry group. A higher predicted female probability is associated with higher relative utility for education, health, and public services and for consumer and other services, and lower relative utility for natural resources, utilities, and construction and for manufacturing. The trade and transportation interaction is positive and smaller. For example, a 0.1 increase in predicted female probability raises relative systematic utility for education, health, and public services by about 0.0296 and lowers it for natural resources, utilities, and construction by about 0.0466.
%These comparisons hold the other covariates fixed and concern predicted probabilities, not causal effects of gender or unconditional employment shares.
Tables~\ref{tab:linkedin:full-levels}--\ref{tab:linkedin:full-demographics} in Appendix~\ref{app:linkedin:results} report all 52 coefficient estimates with plug-in and random-scaling confidence intervals. Most occupation intercepts have relatively narrow confidence intervals, with wider intervals for farming, fishing, and forestry and for building and grounds maintenance. Plug-in intervals are generally shorter than random-scaling intervals, except for the trade-and-transportation interaction with predicted female probability. Across all 52 coefficients, the median RS-to-plug-in interval-width ratio is approximately 3.8. Finite Jacobian batches can make RS inference conservative, as described in Appendix~\ref{subsec:cluster:population-inference}; the observed width differences also reflect the random scaling matrix.

% Generated by Code/linkedin/build_paper_application_outputs.R; do not hand-edit numbers.
\begin{table}[!htbp]
\centering
\caption{Selected destination-choice estimates and confidence intervals}\label{tab:linkedin:main}
\begingroup
\small
\setstretch{1}\setlength{\tabcolsep}{3pt}\renewcommand{\arraystretch}{1.18}
\begin{tabularx}{\textwidth}{@{}Xrrr@{}}\toprule
Coefficient & Estimate & Plug-in 95\% CI & RS 95\% CI \\
\midrule
Destination propensity & $-1.2957$ & $[-1.3001, -1.2912]$ & $[-1.3043, -1.2870]$ \\
\addlinespace\multicolumn{4}{@{}l}{Industry $\times$ predicted female probability}\\
Natural resources, utilities, construction & $-0.4660$ & $[-0.5072, -0.4247]$ & $[-0.7496, -0.1824]$ \\
Manufacturing & $-0.1135$ & $[-0.1246, -0.1025]$ & $[-0.1731, -0.0539]$ \\
Trade and transportation & $0.0281$ & $[0.0133, 0.0430]$ & $[0.0175, 0.0388]$ \\
Education, health, public services & $0.2959$ & $[0.2777, 0.3140]$ & $[0.1841, 0.4076]$ \\
Consumer and other services & $0.2393$ & $[0.2176, 0.2611]$ & $[0.1436, 0.3350]$ \\
\bottomrule\end{tabularx}
\par\smallskip
\begin{minipage}{\textwidth}\footnotesize Notes: All estimates use the averaged second-order SLIM estimator. The destination measure is standardized negative log odds. Demographic probabilities enter in their original 0--1 units. The reference industry is information, finance, and business. Inference uses state $\times$ graduation year $\times$ major clusters. Plug-in intervals include sampling and computational uncertainty; RS denotes pointwise random scaling. Table~\ref{tab:linkedin:full-levels} and Table~\ref{tab:linkedin:full-demographics} report the complete model.\end{minipage}
\endgroup
\end{table}

\begin{figure}[!htbp]
\centering
\includegraphics[width=\textwidth]{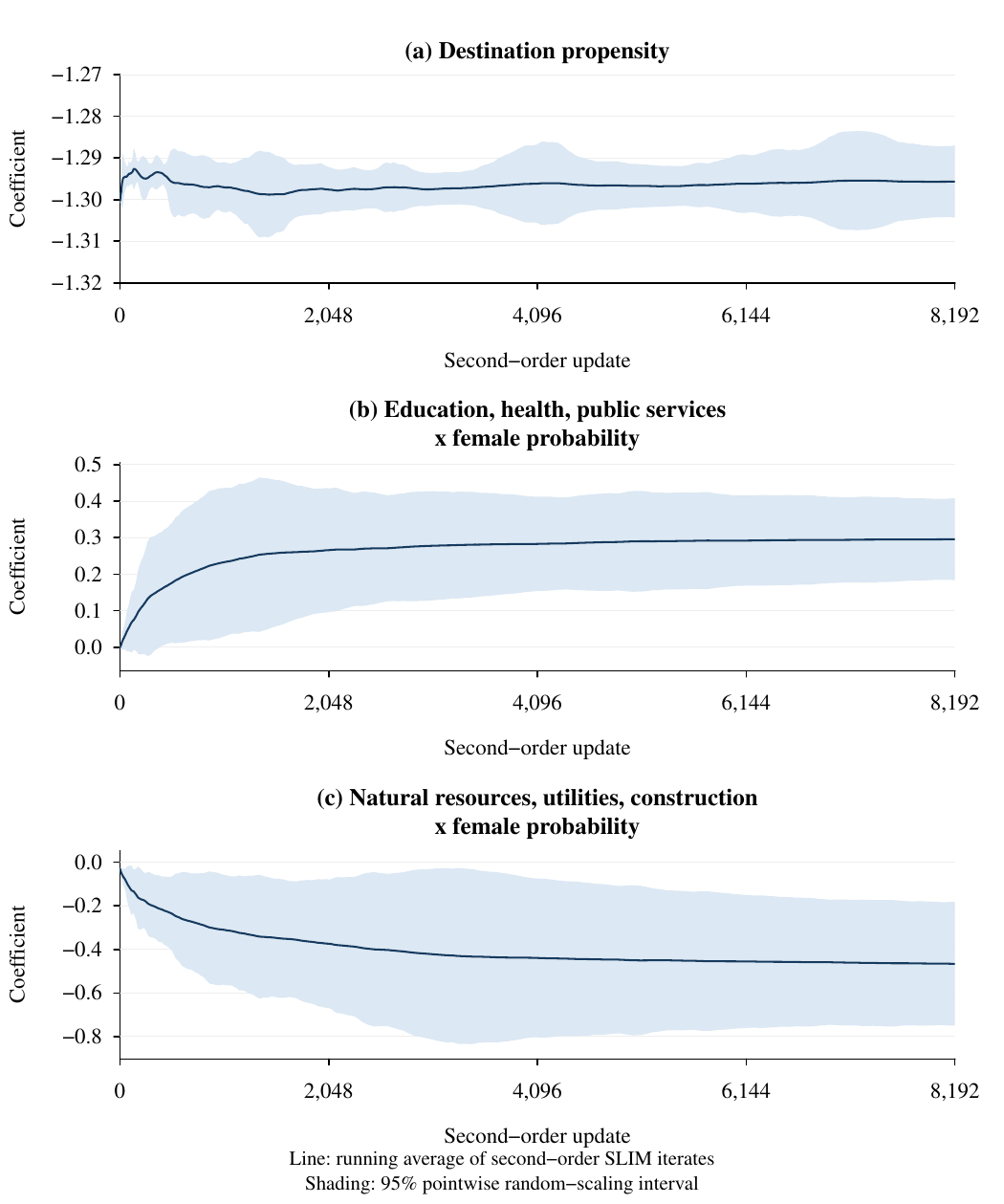}
\caption{Running averages of the second-order SLIM iterates and pointwise random-scaling intervals. The line is the stage-specific running average and the shading is its 95\% pointwise RS interval. The running averages include every second-order update. The first average is shown, but the undefined first-step interval is omitted. Coefficient scales differ across panels.}\label{fig:linkedin:paths}
\end{figure}

Figure~\ref{fig:linkedin:paths} follows the running averages of the second-order SLIM iterates and their pointwise RS intervals (shaded regions) for the destination coefficient and two industry--female-probability interactions.
The cumulative averages exhibit an early adjustment, with some continued movement over the reported refinement run. The pointwise RS intervals differ substantially across coefficients and reflect the realized stochastic paths. The cluster-robust debiased $J$ statistic is 3.04385 with one degree of freedom, giving $p=0.08104$.

For this application, SLIM is implemented in Julia. The production run took less than three hours on an HPC cluster, using an NVIDIA RTX A6000 GPU for mini-batch evaluations and CPU passes for final full-sample quantities. %This elapsed time includes calibration, both estimation stages, inference, and export; it excludes upstream data preparation and earlier tuning and development.
 Appendix~\ref{app:linkedin:implementation} records the settings and timing components.
 %The application demonstrates a feasible estimation and inference workflow at this sample size.

\section{Conclusion}\label{sec:conclusion}

This paper develops a new algorithm, SLIM, for scalable estimation and inference in overidentified nonlinear GMM. Under suitable conditions, it yields consistent estimators without a consistent initial estimator, while second-order refinement achieves full-sample GMM efficiency. Our simple inference methods, plug-in and random-scaling, account for both sampling and computational uncertainty. The Monte Carlo experiments based on a nonlinear demand model demonstrate computational gains in the reported comparisons and scalability to one million observations. The LinkedIn application illustrates clustered estimation and inference in a nonlinear model of industry--occupation choice with 4.8 million individuals. This application also demonstrates that for some parameters, the computational uncertainty is non-negligible relative to sampling uncertainty.

Future research could extend SLIM to nonsmooth moment functions, following \citet{Chen:Liao:2015}, and to inference under model misspecification, building on \citet{hansen2021inference}. Other directions include principled choices of mini-batch sizes, stopping rules, and learning rates, as well as extensions to settings in which the number of parameters grows with the sample size.

%%%%%%%%%%%%%%%%%%%%

{\singlespacing
\putbib
}
\end{bibunit}

\clearpage

\appendix

\renewcommand{\thepage}{S-\arabic{page}}
\setcounter{page}{1}

\onehalfspacing
\begin{bibunit}
\slimbibscope{appendix}

\section*{Supplemental Appendix to ``SLIM: Stochastic Learning and Inference in Nonlinear GMM Models''}

\section{Approximation to Full-Sample GMM Estimate}\label{sec:setting:fixed}

We study first-order SLIM conditional on the observed sample, targeting the full-sample GMM estimator
\begin{equation*}
\hat{\theta}_{n} = \argmin_{\theta \in \Theta} \bar{g}_n(\theta)' \bar{g}_n(\theta),
\end{equation*}
with uncertainty arising only from the mini-batch draws. No sampling model for the observed data is needed.

\subsection{Asymptotic Theory When Data are Treated as Fixed}\label{sec:theory:fixed}

Regularity conditions are imposed directly on the sample moments. Define
\begin{equation*}
 \bar{G}_n := \frac{1}{n} \sum_{i=1}^n G(z_i, \hat \theta_n)\quad \text{and} \quad    \bar \Omega_n := \frac{1}{n} \sum_{i=1}^n g(z_i,\hat \theta_n) g(z_i,\hat \theta_n)'.
\end{equation*}
Define the following quantities:
\begin{align*}
\bar{Q}_n(\theta) &:= \frac{1}{2}\bar{g}_n(\theta)'\bar{g}_n(\theta) - \frac{1}{2}\bar{g}_n(\hat\theta_n)'\bar{g}_n(\hat\theta_n), \\
\bar{\H}_n &:= \frac{\partial^2 \bar{Q}_n(\hat \theta_n)}{\partial \theta \partial \theta'} = \bar{G}_n'\bar{G}_n + \sum_{j=1}^{\dg} \bar{g}_{nj}(\hat \theta_n) \frac{\partial^2}{\partial \theta \partial \theta'} \bar{g}_{nj}(\hat \theta_n),
\end{align*}
where $\bar{g}_{nj}(\theta)$ denotes the $j$th element of $\bar{g}_{n}(\theta)$ for $j=1,\ldots, \dg$.
{
Write $q_n=\bar g_n(\hat\theta_n)$, $C_n=\bar\Omega_n-q_nq_n'$, and $\Delta_i=G(z_i,\hat\theta_n)-\bar G_n$. For independent moment and Jacobian batches of fixed sizes, the exact conditional gradient covariance is
\begin{align}\label{def:bar-Sigma-n}
\bar\Sigma_n
&:=\frac{1}{B_g}\bar G_n'C_n\bar G_n
 +\frac{1}{nB_G}\sum_{i=1}^n
 \Delta_i'\left(\frac{C_n}{B_g}+q_nq_n'\right)\Delta_i.
\end{align}
This follows by decomposing the stochastic gradient into
$\bar G_n'(\tilde g-q_n)+(\tilde G-\bar G_n)'\tilde g$;
the two terms are uncorrelated because the batches are independent.
Although $\bar G_n'q_n=0$ at an interior minimum, $\Delta_i'q_n$ need not vanish, so the conditional covariance retains the $q_n$ terms. The linearization uses the full objective Hessian $\bar{\H}_n$, including its moment-weighted second derivatives.
}
Here $\bar Q_n$ is the Lyapunov function, and the matrices above are evaluated at $\hat\theta_n$.

\begin{assumption}[Data are Fixed]\label{assm:regularity:data:fixed:1}
The data $z_{1:n} = (z_i)_{i=1}^n$ are fixed. Moreover, the following holds for some finite constants $c > 0$, $\delta > 0$, and $M < \infty$:
\begin{enumerate}[({A\ref{assm:regularity:data:fixed:1}}.1)]
% \item 
% \label{assm:fixed:item:pd}
% % $c I_{\dtheta} \le \bar{G}_n'\bar{G}_n$,
% % $\bar \Omega_n \ge c I_{\dtheta}$
% % and
% $\bar{\H}_n \ge c I_{\dtheta};$

\item 
\label{assm:fixed:item:G-bounded}
$\|\bar{G}_n(\theta)\|  \le M$ for all $\theta \in \Theta;$

\item 
\label{assm:fixed:item:G-Lip}
$\|\bar{G}_n(\theta) - \bar{G}_n(\tilde{\theta})\| \le M \|\theta - \tilde{\theta}\|$ for all $\theta$ and $\tilde{\theta} \in \Theta;$

\item 
\label{assm:fixed:item:g-Hessian-Lip}
For each $j=1,\ldots, \dg$, $\|\frac{\partial^2 \bar{g}_{nj}}{\partial \theta \partial \theta'}(\theta) - \frac{\partial^2 \bar{g}_{nj}}{\partial \theta \partial \theta'}(\tilde{\theta})\| \le M \|\theta - \tilde{\theta}\|$ for all $\theta$ and $\tilde{\theta}$ in $\{\theta \in \Theta : \|\theta - \hat{\theta}_{n}\| \le \delta\}$;

\item 
\label{assm:fixed:item:V-local-convexity}
$\frac{\partial^2 \bar{Q}_n(\theta)}{\partial \theta \partial \theta'} \ge c I_{\dtheta}$  for all $\|\theta - \hat{\theta}_{n}\|\le \delta;$

\item 
\label{assm:fixed:item:V-iden}
$\bar{Q}_n(\theta) \ge c$ and $\|\bar{G}_n(\theta)' \bar{g}_n(\theta)\|^2 \ge c$ for all $\|\theta - \hat{\theta}_{n}\|\ge \delta;$

\item 
\label{assm:fixed:item:V-bounded-curvature}
$\frac{\partial^2 \bar{Q}_n(\theta)}{\partial \theta \partial \theta'} \le M I_{\dtheta}$ for all $\theta \in \Theta;$

\end{enumerate}
\end{assumption}

Assumption~\ref{assm:regularity:data:fixed:1} is the deterministic counterpart of Assumption~\ref{asm:regularity}, imposing smoothness, local strong convexity, separation, and bounded curvature. These conditions identify $\hat\theta_n$ as the unique global minimizer.
The additional local Lipschitz condition on the moment Hessians controls terms multiplied by $\bar g_n(\hat\theta_n)$, which need not vanish when the sample is fixed.

Let $\mathbb P_n^*$ and $\mathbb E_n^*$ denote probability and expectation over the mini-batch draws conditional on the observed sample, and let $\rightsquigarrow^*$ denote weak convergence under $\mathbb P_n^*$. For each $\theta\in\Theta$, define the stochastic error
\begin{align}\label{def:xi}
\xi_t(\theta) = \tilde{G}_{t} (\theta)' \tilde{g}_{t} (\theta) - \bar{G}_n(\theta)'\bar{g}_n(\theta),
\end{align}
where $\tilde G_t$ and $\tilde g_t$ are defined in~\eqref{def:SGD:g_and_G}. Independent sampling gives the martingale difference property
$\mathbb{E}_n^*[\xi_{t}({\theta}) | \mathcal F_{n,t-1}^* ] = 0$ for all $t\ge1$ and $\theta\in\Theta$.

\begin{assumption}[Stochastic Errors]\label{assm:regularity:data:fixed:2}
The following hold for some finite constants $M < \infty$ and $p > (1-a)^{-1}$, where $a \in (1/2,1)$ denotes the learning rate exponent defined in Assumption~\ref{assm:lr}$:$
\begin{enumerate}[({A\ref{assm:regularity:data:fixed:2}}.1)]
\item 
\label{assm:fixed:item:5}
$\mathbb{E}_n^*[\|\xi_t(\theta) - \xi_t(\hat{\theta}_{n})\|^2] \le M \|\theta - \hat{\theta}_{n}\|^2$ for all $\theta \in \Theta;$

\item 
\label{assm:fixed:item:6}
$\mathbb{E}_n^*[\|\xi_t(\theta)\|^{2p}]  \le M (\bar{Q}_n(\theta)^p + 1)$ for all $\theta\in \Theta.$
\end{enumerate}
\end{assumption}

Condition~\ref{assm:fixed:item:5} ensures convergence of the conditional noise covariance to $\bar\Sigma_n$ as the iterates approach $\hat\theta_n$, while Condition~\ref{assm:fixed:item:6} controls noise growth and supports stability.

% \todo[inline]{Myunghyun, please again check whether this assumption is correct. I could have made missed or have duplicated conditions.
% In addition, please add some comments on each of the conditions above regarding what sense they are needed.}

\begin{theorem}\label{thm:data:fixed}
Let Assumptions \ref{assm:uniform}, \ref{assm:lr}, \ref{assm:diff}, \ref{assm:regularity:data:fixed:1}, and \ref{assm:regularity:data:fixed:2} hold.
Then, both $\theta^*_N$ and $\bar{\theta}^*_N$ converge $\mathbb{P}_n^*$-almost surely to $\hat \theta_n$ as $N \to \infty$. Moreover, as $N \to \infty$,
\begin{equation*}
\frac 1 {\sqrt {N}}  \sum_{t=1}^{\lfloor N r\rfloor}(\theta^*_t - \hat \theta_n)
    \rightsquigarrow^* (\bar{\H}_n^{-1} \bar \Sigma_n \bar{\H}_n^{-1})^{1/2} W(r).
\end{equation*}

\end{theorem}

Setting $r=1$ in Theorem~\ref{thm:data:fixed} yields asymptotic normality of the averaged estimator; the functional convergence supports random-scaling inference in Section~\ref{subsec:inference-fixed-data}.

\subsection{Inference with Fixed Data}
\label{subsec:inference-fixed-data}

For fixed-data inference, consider $\ell\leq d$ linear restrictions
\[
H^*_{0}: R \hat{\theta}_n = c,
\]
where $R$ is a known $\ell\times d$ matrix of rank $\ell$ and $c$ is a known $\ell$-vector. Using $V_t(R)$ and its recursive updates in~\eqref{eq:V_tR}--\eqref{eq:V_t}, Theorem~\ref{thm:data:fixed} yields the following pivotal limit.

{
\begin{corollary}\label{cor:Wald:fixed}
Let the conditions assumed in Theorem \ref{thm:data:fixed} hold.
Suppose that $H^*_{0}: R \hat{\theta}_n = c$ holds with 
$\mathrm{rank}(R)=\ell$ and
$R\bar{\H}_n^{-1}\bar\Sigma_n\bar{\H}_n^{-1}R'$ is positive definite. Then, as $N \to \infty$,
\begin{align*}
N B_g \left(R \bar\theta^*_N-c\right)'\left(B_g V_N(R)\right)^{-1}\left(R \bar\theta^*_N-c\right)
\rightsquigarrow^*
W_\ell\left(1\right)'\left(\int_{0}^{1}\bar{W}_\ell(r)\bar{W}_\ell(r)'dr\right)^{-1}W_\ell\left(1\right).
\end{align*}
\end{corollary}
}

Corollary~\ref{cor:Wald:fixed} quantifies computational uncertainty around the full-sample estimate, conditional on the observed data.

\section{SLIM with Clustered Data}\label{sec:cluster}

{
This appendix develops conditional convergence and population inference for cluster resampling. Proofs are in Section~\ref{appx:proofs-cluster}; supporting lemmas are stated in Appendix~\ref{appx:lemmas-proofs} and proved in Appendix~\ref{appx:proofs-lems}.
}

\subsection{Cluster Sampling Scheme}\label{sec:cluster:scheme}

{
Cluster $k=1,\ldots,K$ contains $n_k$ observations $z_{ik}$, with $n=\sum_{k=1}^K n_k$ and arbitrary dependence within clusters. For the pooled sample, use the full-sample definitions in~\eqref{def:full-sample:g_and_G}, \eqref{def:full-sample-gmm}, and Appendix~\ref{sec:setting:fixed}.
}

Define the cluster weights
\begin{equation}\label{eq:cluster:weights}
w_k := \frac{K n_k}{n}, \qquad k = 1, \ldots, K,
\end{equation}
and the weighted cluster-level moment and Jacobian
\begin{equation}\label{eq:cluster:moments}
\hat{g}_k(\theta) := \frac{w_k}{n_k} \sum_{i=1}^{n_k} g(z_{ik}, \theta) = \frac{K}{n} \sum_{i=1}^{n_k} g(z_{ik}, \theta),
\qquad
\hat{G}_k(\theta) := \frac{K}{n} \sum_{i=1}^{n_k} G(z_{ik}, \theta).
\end{equation}
These weights preserve the person-average moments despite unequal cluster sizes:
\begin{equation}\label{eq:cluster:identity}
\frac{1}{K} \sum_{k=1}^{K} \hat{g}_k(\theta) = \bar{g}_n(\theta)
\qquad \text{and} \qquad
\frac{1}{K} \sum_{k=1}^{K} \hat{G}_k(\theta) = \bar{G}_n(\theta)
\qquad \text{for all } \theta \in \Theta.
\end{equation}
Uniform cluster draws therefore target the same full-sample GMM estimate $\hat\theta_n$.

At iteration $t\ge1$, draw $B=B_G+B_g$ cluster indices $k_t(1),\ldots,k_t(B)$ independently and uniformly with replacement from $\{1,\ldots,K\}$, using fresh draws each iteration. The first $B_G$ draws form the Jacobian batch and the remaining $B_g$ the moment batch:
\begin{equation}\label{eq:cluster:tilde}
\tilde{G}_t(\theta) := \frac{1}{B_G} \sum_{j=1}^{B_G} \hat{G}_{k_t(j)}(\theta)
\qquad \text{and} \qquad
\tilde{g}_t(\theta) := \frac{1}{B_g} \sum_{j=B_G+1}^{B} \hat{g}_{k_t(j)}(\theta),
\end{equation}
with the parameter update in~\eqref{def:sgd},
\begin{equation}\label{eq:cluster:update}
\theta_t^* = \theta_{t-1}^* - \gamma_t \, \tilde{G}_t(\theta_{t-1}^*)' \tilde{g}_t(\theta_{t-1}^*),
\end{equation}
followed by averaging $\bar{\theta}_t^* = t^{-1}\theta_t^* + (1 - t^{-1})\bar{\theta}_{t-1}^*$. Let $\mathbb P_n^*$ and $\mathbb E_n^*$ denote probability and expectation over cluster draws conditional on the data. Uniform sampling and~\eqref{eq:cluster:identity} give
\begin{equation}\label{eq:cluster:unbiased}
\mathbb{E}_n^*\big[\tilde{g}_t(\theta)\big] = \bar{g}_n(\theta)
\qquad \text{and} \qquad
\mathbb{E}_n^*\big[\tilde{G}_t(\theta)\big] = \bar{G}_n(\theta)
\qquad \text{for all } \theta \in \Theta,
\end{equation}
and batch independence gives $\mathbb{E}_n^*[\tilde{G}_t(\theta)'\tilde{g}_t(\theta)] = \bar{G}_n(\theta)'\bar{g}_n(\theta)$, preserving the conditionally unbiased gradient.

Define the cluster second-moment matrix
\begin{equation}\label{eq:cluster:meat}
\bar{\Omega}^{\mathrm{cl}}_K := \frac{1}{K} \sum_{k=1}^{K} \hat{g}_k(\hat{\theta}_n)\, \hat{g}_k(\hat{\theta}_n)',
\end{equation}
{
whose centered version enters the conditional gradient covariance below. Section~\ref{subsec:cluster:population} gives its sampling-covariance normalization.
}

\subsection{Conditional Convergence and Random Scaling}\label{sec:cluster:assumptions}
\label{subsec:cluster:fclt}\label{subsec:cluster:RS}

{For conditional convergence, impose Assumptions~\ref{assm:uniform} (with uniform cluster draws), \ref{assm:lr}, \ref{assm:diff}, and \ref{assm:regularity:data:fixed:1}, together with the following bounds on the weighted cluster moments and Jacobians.}

\begin{assumption}[Cluster Sampling Regularity]\label{assm:cluster:reg}
The following hold for some finite constants $\bar{C} < \infty$, $M < \infty$, and $p > (1-a)^{-1}$, where $a \in (1/2,1)$ is the learning rate exponent of Assumption~\ref{assm:lr}$:$
\begin{enumerate}[({A\ref{assm:cluster:reg}}.1)]
\item \label{assm:cluster:item:weights}
$\max_{1 \le k \le K} w_k \le \bar{C}$;

\item \label{assm:cluster:item:stability}
$\displaystyle \frac{1}{K} \sum_{k=1}^{K} \|\hat{g}_k(\theta)\|^{2p} \le M\big(\bar{Q}_n(\theta)^p + 1\big)$ for all $\theta \in \Theta$;

\item \label{assm:cluster:item:envelope}
there exist constants $\hat{H}_k$ such that $\sup_{\theta \in \Theta} \|\hat{G}_k(\theta)\| \le \hat{H}_k$ for each $k \le K$ and $\displaystyle \frac{1}{K} \sum_{k=1}^{K} \hat{H}_k^{2p} \le M$;

\item \label{assm:cluster:item:Lipschitz}
$\|\hat{G}_k(\theta) - \hat{G}_k(\tilde{\theta})\| \le L_k \|\theta - \tilde{\theta}\|$ for all $\theta, \tilde{\theta} \in \Theta$ and each $k \le K$, where $\displaystyle \frac{1}{K}\sum_{k=1}^{K} L_k^2 \le M$.
\end{enumerate}
\end{assumption}

Condition~\ref{assm:cluster:item:weights} bounds relative cluster sizes $n_k/(n/K)$ from above: unequal sizes are allowed, but no cluster may exceed the average size by more than a fixed factor.
It implies $\max_k n_k^2/n\to0$ when $n=o(K^2)$, a cluster-size restriction used in \citet{hansen2019clustered} and \citet{djogbenou2019asymptotic}. The remaining conditions impose empirical moment, envelope, and Lipschitz bounds for conditional resampling.
With bounded weights, suitable empirical observation-level bounds imply the corresponding cluster bounds by Jensen's inequality.
Population inference additionally requires the sampling and covariance conditions in Assumption~\ref{assm:clpop} below.

\paragraph{Conditional convergence.}

Let $\mathcal F_{n,t-1}^*$ contain the initial information and the cluster draws through iteration $t-1$. For $\theta\in\Theta$, define the stochastic error
\begin{align}\label{eq:cluster:xi}
\xi_t(\theta) := \tilde{G}_t(\theta)' \tilde{g}_t(\theta) - \bar{G}_n(\theta)' \bar{g}_n(\theta),
\end{align}
the cluster analogue of~\eqref{def:xi}. Fresh draws and independent batches give
\begin{equation*}
\mathbb{E}_n^*[\xi_t(\theta) \mid \mathcal{F}_{n,t-1}^*]
= \mathbb{E}_n^*[\tilde{G}_t(\theta)]' \mathbb{E}_n^*[\tilde{g}_t(\theta)] - \bar{G}_n(\theta)'\bar{g}_n(\theta) = 0
\end{equation*}
for all $t \ge 1$ and $\theta \in \Theta$ by \eqref{eq:cluster:unbiased}.
{Thus $(\xi_t(\theta))_{t\ge1}$ is a martingale difference sequence.}

{
Define $q_n=\bar g_n(\hat\theta_n)$,
$C_K^{\mathrm{cl}}=\bar\Omega_K^{\mathrm{cl}}-q_nq_n'$, and
$\Delta_k=\hat G_k(\hat\theta_n)-\bar G_n$. The exact cluster-sampling variance is
\begin{align}\label{eq:cluster:Sigma}
\bar\Sigma_K^{\mathrm{cl}}
&:=\frac{1}{B_g}\bar G_n'C_K^{\mathrm{cl}}\bar G_n
 +\frac{1}{KB_G}\sum_{k=1}^K
 \Delta_k'\left(\frac{C_K^{\mathrm{cl}}}{B_g}+q_nq_n'\right)\Delta_k.
\end{align}
Here $B_g,B_G$ are fixed and all quantities are evaluated at $\hat\theta_n$. This is the weighted-cluster version of~\eqref{def:bar-Sigma-n}; residual moments $q_n$ need not vanish, and their contribution persists as $N\to\infty$.
}

{
Applying Appendix~\ref{sec:setting:fixed} gives the following result.
}

\begin{theorem}[{Consistency and FCLT under Cluster Sampling}]
\label{thm:cluster:FCLT}
{
Let Assumptions~\ref{assm:uniform} (with uniform cluster draws), \ref{assm:lr}, \ref{assm:diff}, \ref{assm:regularity:data:fixed:1}, and \ref{assm:cluster:reg} hold.
% [G7 patch] Removed the repeated convexity hypothesis and the unnecessary unit residual bound: convexity is part of Assumption 3, and Assumption 8.2 gives a finite residual bound by Jensen's inequality. The proof does not require that bound to be one (see the proof of Lemma 16, steps (c)--(d)).
}
Then both $\theta_N^*$ and $\bar{\theta}_N^*$ converge $\mathbb{P}_n^*$-almost surely to $\hat{\theta}_n$ as $N \to \infty$. Moreover, as $N \to \infty$,
\begin{equation*}
\frac{1}{\sqrt{N}} \sum_{t=1}^{\lfloor N r \rfloor} (\theta_t^* - \hat{\theta}_n)
\rightsquigarrow^* \big( \bar{\H}_n^{-1} \bar{\Sigma}^{\mathrm{cl}}_K \bar{\H}_n^{-1} \big)^{1/2} W(r),
\end{equation*}
where $\bar{\H}_n$ is defined in Appendix~\ref{sec:setting:fixed} and $W$ is a standard $\dtheta$-dimensional Wiener process.
\end{theorem}

Setting $r=1$ yields asymptotic normality of $\bar\theta_N^*$; the FCLT supports conditional random-scaling inference as in Section~\ref{subsec:inference-fixed-data}.

{
\begin{remark}[Sampling and Computational Uncertainty]
\label{rem:cluster:two-layers}
Theorem~\ref{thm:cluster:FCLT} holds with the observed data and the number of clusters fixed as $N\to\infty$. Section~\ref{subsec:cluster:joint-population} accounts for sampling uncertainty as the number of clusters grows, under additional sampling assumptions and uniform bounds on the stochastic approximation error.
\end{remark}
}

\paragraph{Conditional random scaling.}

For conditional inference, consider $\ell\le\dtheta$ linear restrictions
\[
H^*_{0}: R \hat{\theta}_n = c,
\]
where $R$ is a known $\ell\times\dtheta$ matrix of rank $\ell$ and $c$ a known $\ell$-vector. Define the path-based matrix
\begin{equation}\label{eq:cluster:V_NR}
V_N(R) := \frac{1}{N^2} \sum_{t=1}^{N}
\left\{ \sum_{s=1}^{t} R \left(\theta_s^* - \bar{\theta}_N^*\right) \right\}
\left\{ \sum_{s=1}^{t} R \left(\theta_s^* - \bar{\theta}_N^*\right) \right\}',
\end{equation}
with recursive updates~\eqref{eq:A_t}--\eqref{eq:V_t}, requiring no path storage.

The FCLT gives the following conditional pivotal limit.

{
\begin{corollary}[Conditional Random Scaling under Cluster Resampling]\label{cor:cluster:RS}
Let the conditions assumed in Theorem~\ref{thm:cluster:FCLT} hold.
Suppose that $H^*_{0}: R \hat{\theta}_n = c$ holds with $\mathrm{rank}(R)=\ell$, and suppose
$R\bar{\H}_n^{-1}\bar\Sigma_K^{\mathrm{cl}}\bar{\H}_n^{-1}R'$ is positive definite. Then, as $N \to \infty$,
\begin{align*}
N \left(R \bar{\theta}_N^* - c\right)' V_N(R)^{-1} \left(R \bar{\theta}_N^* - c\right)
\rightsquigarrow^*
W_\ell(1)' \left( \int_{0}^{1} \bar{W}_\ell(r) \bar{W}_\ell(r)' dr \right)^{-1} W_\ell(1).
\end{align*}
\end{corollary}
}

{
Population inference adds sampling uncertainty to this conditional result.

\subsection{Asymptotic Theory}
\label{subsec:cluster:joint-population}

For growing $K$, write $u_k(\theta)=\hat g_k(\theta)$ and $D_k(\theta)=\hat G_k(\theta)$. We use a nondegenerate $\sqrt K$ sampling limit; other cluster rates require a corresponding normalization \citep{hansen2019clustered}.

For first-order estimation or refinement, consider $L$ updates
\begin{equation}\label{eq:clpop:update}
\vartheta_t=\vartheta_{t-1}-\gamma_0(s+t)^{-a}
 P_K\widetilde D_t(\vartheta_{t-1})'W_K\widetilde u_t(\vartheta_{t-1}),
\qquad \bar\vartheta_L=L^{-1}\sum_{t=1}^L\vartheta_t.
\end{equation}
Here $s$ is a deterministic learning-rate offset, $P_K,W_K$ are symmetric positive definite and fixed during the stage, and $\widetilde D_t$ and $\widetilde u_t$ are the averages of $D_k$ and $u_k$ over independent uniform Jacobian and moment batches of sizes $b_t$ and $B_g$, with $B_g$ fixed. Draws are conditionally independent across iterations given initial information $\mathcal F_{K,0}$, which includes the data, weights, preconditioner, starting value, and preliminary draws. Let $\hat\theta_{n,W}$ minimize the full-sample criterion with weight $W_K$, and define
\[
Q_K(\theta)=\tfrac12\{\bar g_n(\theta)'W_K\bar g_n(\theta)
 -\bar g_n(\hat\theta_{n,W})'W_K\bar g_n(\hat\theta_{n,W})\}.
\]

\begin{assumption}[Population Cluster Regularity]\label{assm:clpop}
The dimensions are fixed, $K\to\infty$, and the following conditions hold.
\begin{enumerate}[({A\ref{assm:clpop}}.1)]
\item \label{assm:clpop:item:sampling} The sampled clusters are independent. For an interior parameter $\theta_\oo$, deterministic matrices $\Omega>0$ and $D_0$ of full column rank satisfy
\begin{align*}
\sqrt K\,\bar g_n(\theta_\oo)&\dto\mathcal N(0,\Omega),&
\bar G_n(\theta_\oo)&\pto D_0,\\
\frac1K\sum_{k=1}^K u_k(\theta_\oo)u_k(\theta_\oo)'&\pto\Omega.
\end{align*}
\item \label{assm:clpop:item:weights} $W_K\pto W>0$ and $P_K\pto P>0$. For $\Delta_{k,0}=D_k(\theta_\oo)-\bar G_n(\theta_\oo)$, there is a finite deterministic matrix $\Xi_W$ such that
\[
\frac1K\sum_{k=1}^K\Delta_{k,0}'W\Omega W\Delta_{k,0}\pto \Xi_W.
\]
\item \label{assm:clpop:item:regularity} For some $p>(1-a)^{-1}$ and every $\eta>0$, there are constants $c,\delta>0$ and $M<\infty$, depending on $\eta$ but not on $K$, such that the following hold on $\mathcal F_{K,0}$-measurable events of probability at least $1-\eta$ for all large $K$:
\begin{enumerate}[(a)]
\item Conditions~\ref{assm:cluster:item:stability}--\ref{assm:cluster:item:Lipschitz} hold with $Q_K$ in place of $\bar Q_n$, and the Lipschitz constants in Condition~\ref{assm:cluster:item:Lipschitz} satisfy $K^{-1}\sum_{k=1}^KL_k^{2p}\le M$;
\item $\partial^2Q_K(\theta)/\partial\theta\partial\theta'\ge cI_{\dtheta}$ if $\|\theta-\hat\theta_{n,W}\|\le\delta$, and $\partial^2Q_K(\theta)/\partial\theta\partial\theta'\le MI_{\dtheta}$ for all $\theta\in\Theta$;
\item $Q_K(\theta)\ge c$ and $\|\partial Q_K(\theta)/\partial\theta\|^2\ge c$ if $\|\theta-\hat\theta_{n,W}\|\ge\delta$.
\end{enumerate}
Moreover, Assumption~\ref{assm:diff} holds for the cluster update, with $\hat\theta_{n,W}$ as the minimizer.
\item \label{assm:clpop:item:initialization} The initialization satisfies $\|\vartheta_0\|=O_p(1)$ and $Q_K(\vartheta_0)=O_p((s+1)^{-a})$. The learning rate has $1/2<a<1$, $s/L=O(1)$, and $L^{1-a}/K\to0$.
\item \label{assm:clpop:item:batches} The batch schedule is deterministic and, for some $\beta\in[0,1]$,
\[
\sup_{0\le r\le1}\left|\frac1L\sum_{t=1}^{\lfloor Lr\rfloor}\frac1{b_t}-\beta r\right|\longrightarrow0.
\]
\end{enumerate}
\end{assumption}

Condition~\ref{assm:clpop:item:regularity} requires the moment, envelope, and Lipschitz bounds of Assumption~\ref{assm:cluster:reg} uniformly in $K$, and its parts~(b) and~(c) mirror Conditions~\ref{asm:item:local-convex-Lojasiewicz:WMB}--\ref{asm:item:stability:WMB} for the weighted criterion $Q_K$. Independence within clusters is not required. Independent mean-zero cluster sums with suitable moment, Lindeberg, and law-of-large-numbers conditions suffice for the sampling and covariance limits, for example with nonrandom cluster sizes. With random sizes, $u_k=(K/n)\sum_{i=1}^{n_k}g(z_{ik},\cdot)$ depends on all sizes through $n$, so one may instead require independent pairs $\big(\sum_{i=1}^{n_k}g(z_{ik},\theta_\oo),n_k\big)$ and $n/K\pto\mu\in(0,\infty)$. Averaging heterogeneous cluster means alone does not ensure matching covariance limits.

Condition~\ref{assm:clpop:item:batches} covers fixed Jacobian batches ($\beta=1/B_G$) and diverging schedules whose average inverse size vanishes ($\beta=0$), including triangular stages. Joint convergence also requires the uniform algorithmic and initialization bounds in Conditions~\ref{assm:clpop:item:regularity}--\ref{assm:clpop:item:initialization}.

Define
\begin{align}\label{eq:clpop:covariances}
J_W&=D_0'WD_0,&
A_W&=J_W^{-1}D_0'W\Omega WD_0J_W^{-1},\notag\\
\Sigma_W&=B_g^{-1}(D_0'W\Omega WD_0+\beta \Xi_W),&
C_W&=J_W^{-1}\Sigma_WJ_W^{-1}.
\end{align}
In particular,
\begin{equation}\label{eq:clpop:dominance}
B_g C_W=A_W+\beta J_W^{-1}\Xi_WJ_W^{-1}\succeq A_W.
\end{equation}

\begin{theorem}[Joint Sampling and Computational Limit]\label{thm:clpop:joint}
Under Assumption~\ref{assm:clpop}, as $K,L\to\infty$,
\begin{equation}\label{eq:clpop:joint}
\begin{pmatrix}
\sqrt K(\hat\theta_{n,W}-\theta_\oo)\\
L^{-1/2}\displaystyle\sum_{t=1}^{\lfloor Lr\rfloor}(\vartheta_t-\hat\theta_{n,W})
\end{pmatrix}
\wto
\begin{pmatrix}A_W^{1/2}Z_d\\ C_W^{1/2}\mathbb B_d(r)\end{pmatrix},
\end{equation}
where $Z_d$ is standard normal and $\mathbb B_d$ is an independent standard Wiener process. Thus the asymptotic variance of $\bar\vartheta_L-\theta_\oo$ is $K^{-1}A_W+L^{-1}C_W$. The preconditioner $P$ does not enter these covariances.
\end{theorem}

The asymptotic variance can be decomposed as
\begin{equation}\label{eq:clpop:variance-decomposition}
\frac{A_W}{K}+\frac{C_W}{L}
=
\underbrace{\frac{A_W}{K}}_{\text{sampling}}
+
\underbrace{\frac{A_W}{LB_g}}_{\text{moment-batch noise}}
+
\underbrace{\frac{\beta}{LB_g}J_W^{-1}\Xi_WJ_W^{-1}}_{\text{Jacobian-batch noise}}.
\end{equation}
This decomposition applies to any admissible positive definite optimization weight $W$. Its empirical counterpart uses the realized average inverse Jacobian batch size $\beta_L=L^{-1}\sum_{t=1}^L b_t^{-1}$.

\subsection{Inference}\label{subsec:cluster:population-inference}
\label{subsec:cluster:population}

\paragraph{Covariance Estimation and Normalization.}
For a preliminary estimate $\tilde\theta$, let
$S_k(\theta)=\sum_{i=1}^{n_k}g(z_{ik},\theta)$,
$\hat g=\bar g_n(\tilde\theta)$, and
$\hat D=\bar G_n(\tilde\theta)$, the full-sample Jacobian. Define
\begin{equation}\label{eq:cluster:terminal-cov}
\hat V=\frac{1}{n^2}\sum_{k=1}^K S_k(\tilde\theta)S_k(\tilde\theta)',
\qquad \hat\Omega_K=K\hat V,
\qquad \hat\Omega_n=n\hat V.
\end{equation}
Here $\hat V$ is in sample-mean covariance units; $\hat\Omega_K$ and $\hat\Omega_n$ correspond to $\sqrt K$ and $\sqrt n$ normalization. At $\tilde\theta=\hat\theta_n$, $\hat\Omega_K=\bar\Omega_K^{\mathrm{cl}}$ from~\eqref{eq:cluster:meat}, and $\hat\Omega_n=(n/K)\bar\Omega_K^{\mathrm{cl}}$.

Consistency requires appropriate sampling and covariance conditions, such as mean-zero independent cluster contributions and suitable laws of large numbers; Assumption~\ref{assm:clpop} provides a sufficient $\sqrt K$ framework. Averaging heterogeneous cluster means alone does not justify this uncentered covariance. Constructed regressors and instruments must be accounted for when they affect first-order moment variation.

\paragraph{Inference Based on Random Scaling.}
Construct $V_L(R)$ from the stage path $(\vartheta_t)_{t=1}^L$ by \eqref{eq:V_tR}, resetting its accumulators at the start of the stage. For $H_0:R\theta_\oo=c$, define
\begin{equation}\label{eq:clpop:RS}
T_{K,L}=\tau_{K,L}(R\bar\vartheta_L-c)'(B_gV_L(R))^{-1}(R\bar\vartheta_L-c),
\qquad \tau_{K,L}=\left(K^{-1}+(LB_g)^{-1}\right)^{-1}.
\end{equation}

\begin{corollary}[Random Scaling Inference]\label{cor:clpop:RS}
Let Assumption~\ref{assm:clpop} hold and let $R$ have fixed full row rank $\ell$. If $c_{\ell,1-\alpha}$ is the $(1-\alpha)$ quantile of
\[
\mathcal T_\ell=Z_\ell'\left(\int_0^1\overline{\mathbb B}_\ell(r)
\overline{\mathbb B}_\ell(r)'dr\right)^{-1}Z_\ell,
\]
where $Z_\ell$ and the Brownian bridge $\overline{\mathbb B}_\ell$ are independent, then under $H_0$,
\[
\limsup_{K,L\to\infty}\mathbb P(T_{K,L}>c_{\ell,1-\alpha})\le\alpha.
\]
Equality holds if $\beta=0$, or if $L/K\to0$. The same exact limit also holds for contrasts satisfying $R J_W^{-1}\Xi_WJ_W^{-1}R'=0$. The growth conditions of Assumption~\ref{assm:clpop} apply in each case.
\end{corollary}

Use the path updates and critical values of Section~\ref{sec:inference}, with $K$ clusters and batches counted in clusters. Finite Jacobian batches generally yield conservative inference; the conditional statistic in Corollary~\ref{cor:cluster:RS} omits sampling uncertainty for $R\theta_\oo$. A scalar $(1-\alpha)$ interval is
\[
R\bar\vartheta_L\ \pm\ \sqrt{c_{1,1-\alpha}\left(K^{-1}+(LB_g)^{-1}\right)B_gV_L(R)}.
\]

\paragraph{Weighted refinement and plug-in inference.}
First-order SLIM uses $s=0$, $L=N$, and $P_K=W_K=I$; refinement uses $s=N$, $L=T-N$, and the cluster update~\eqref{def:SGD:eff} with a consistent weight and preconditioner fixed during the stage. Condition~\ref{assm:clpop:item:initialization} requires accuracy relative to the new weighted criterion. A pilot with $\|\vartheta_0-\theta_\oo\|=O_p(K^{-1/2}+N^{-a/2})$ satisfies it if $N^a/K=O(1)$ and $N/L=O(1)$: the Hessian bound gives $Q_K(\vartheta_0)=O_p(K^{-1}+N^{-a})$. This covers a final pilot iterate or an averaged pilot with rate $O_p(K^{-1/2}+N^{-1/2})$. Refinement requires fresh draws and restarted averaging.

\begin{corollary}[Plug-in Inference]\label{cor:clpop:plugin}
Let Assumption~\ref{assm:clpop} hold, let $R$ have fixed full row rank $\ell$, and let $\widehat A_W\pto A_W$ and $\widehat C_W\pto C_W$ be consistent symmetric plug-in estimates. Under $H_0:R\theta_\oo=c$, as $K,L\to\infty$,
\begin{equation}\label{eq:clpop:plugin}
(R\bar\vartheta_L-c)'\left\{R\left(\frac{\widehat A_W}{K}
 +\frac{\widehat C_W}{L}\right)R'\right\}^{-1}(R\bar\vartheta_L-c)
\dto\chi^2_\ell.
\end{equation}
\end{corollary}

Evaluate empirical counterparts of~\eqref{eq:clpop:covariances} at a consistent estimate, using the cluster moment covariance and centered Jacobian deviations, with $L^{-1}\sum_t b_t^{-1}$ replacing $\beta$. The matching limits and local bounds in Assumption~\ref{assm:clpop} give consistency. The random matrix $V_L(R)$ is not a consistent estimate of $RC_WR'$ and cannot replace the plug-in covariance.

For $W=\Omega^{-1}$ and $\beta=0$, $A_W=A_{\mathrm{eff}}=(D_0'\Omega^{-1}D_0)^{-1}$ and $C_W=A_{\mathrm{eff}}/B_g$. Consistent estimates $\widehat\Omega_K$ of $\Omega$ and $\widehat D$ of $D_0$, evaluated at the final parameter estimate, give
\[
\widehat A_{\mathrm{eff}}=(\widehat D'\widehat\Omega_K^{-1}\widehat D)^{-1},
\qquad
\widehat{\operatorname{Var}}(\bar\vartheta_L)
=\left(K^{-1}+(LB_g)^{-1}\right)\widehat A_{\mathrm{eff}}.
\]
For other optimization weights or a nonzero Jacobian-noise contribution, use the general covariance $\widehat A_W/K+\widehat C_W/L$. Singleton clusters recover the observation-sampling formulas under matching assumptions.

\paragraph{Accuracy for the debiased $J$-test.}
Theorem~\ref{thm:clpop:joint} implies
$\bar\vartheta_L-\theta_\oo=O_p(K^{-1/2}+L^{-1/2})$. Therefore
\begin{equation}\label{eq:clpop:J-accuracy}
L/\sqrt K\longrightarrow\infty
\quad\Longrightarrow\quad
\bar\vartheta_L-\theta_\oo=o_p(K^{-1/4}).
\end{equation}
With the covariance and residual-rank conditions in Lemma~\ref{lem:debiased-J:general}, this rate validates the general debiased test. Efficient refinement below meets it when $K/(LB_g)$ has a finite limit. The curvature and separation conditions of Assumption~\ref{assm:clpop} remain required.

\subsection{Cluster-Robust Debiased $J$-Test}\label{subsec:cluster:debiased-J}
For the refinement in \eqref{eq:clpop:update}, set $s=N$ and $L=T-N$. At the refined average $\bar\vartheta_L$, define
\[
\bar g=\bar g_n(\bar\vartheta_L),\qquad
\bar\Phi=\bar G_n(\bar\vartheta_L),\qquad
\bar W_K=\hat\Omega_K(\bar\vartheta_L)^{-1},
\]
where
\[
\hat\Omega_K(\theta)=\frac{K}{n^2}\sum_{k=1}^K S_k(\theta)S_k(\theta)'.
\]
Evaluate the testing weight $\bar W_K$ at the final estimate, holding the optimization weight $W_K$ fixed during refinement. As in Section~\ref{subsec:J-test}, project onto the orthogonal complement of the weighted Jacobian:
\begin{equation}\label{eq:cluster:projected-weight}
\bar g_K^D=\Big[I-\bar W_K^{1/2}\bar\Phi
 (\bar\Phi'\bar W_K\bar\Phi)^{-1}\bar\Phi'\bar W_K^{1/2}\Big]
 \bar W_K^{1/2}\bar g.
\end{equation}
The clustered debiased statistic is
\begin{equation}\label{eq:cluster:debiased-J}
J_D^{\mathrm{cl}}=K\,\bar g_K^{D\prime}\bar g_K^D
=K\,\bar g'\Big[\bar W_K-\bar W_K\bar\Phi
 (\bar\Phi'\bar W_K\bar\Phi)^{-1}\bar\Phi'\bar W_K\Big]\bar g.
\end{equation}

\begin{theorem}[Cluster-Robust Debiased $J$-Test]\label{thm:cluster:debiased-J}
Let $\dg>\dtheta$ and let Assumption~\ref{assm:clpop} hold for the refinement stage, with any weight limit $W>0$ and any $\beta\in[0,1]$ (the efficient choice $W=\Omega^{-1}$ and $\beta=0$ is the leading case). Suppose, as $K,L\to\infty$,
\[
\frac{K}{LB_g}\longrightarrow\tau<\infty,
\qquad \hat\Omega_K(\bar\vartheta_L)\pto\Omega.
\]
Then, under the maintained moment restrictions,
\[
J_D^{\mathrm{cl}}\dto\chi^2_{\dg-\dtheta}.
\]
\end{theorem}

Here $\beta=0$ means a vanishing average inverse Jacobian batch size. Lemma~\ref{lem:debiased-J:general} also validates the statistic under $\tilde\theta-\theta_\oo=o_p(K^{-1/4})$ and its sampling, covariance, and rank conditions, without efficient refinement.

Projection removes first-order estimation error, so no computational variance adjustment is needed, even when $\tau>0$. Using $\hat\Omega_n=(n/K)\hat\Omega_K$ and an outer factor $n$ gives the same statistic; singleton clusters recover the statistic in Section~\ref{subsec:J-test}.

One full-sample pass at $\bar\vartheta_L$ supplies the moments, Jacobian, and cluster sums; positive-definite factorizations solve the covariance and Jacobian cross-product systems.
}

\section{Warm-Start Algorithm}\label{sec:algo:alt}

Algorithm~\ref{alg:sequential-reshuffle} provides an optional warm start for the method in Section~\ref{sec:algo}. A useful starting value can improve finite-sample performance, although consistency of the initial estimator is not required. The resulting starting value must satisfy the boundedness condition in Section~\ref{sec:algo}. Each epoch reshuffles the observations and cycles through all ordered pairs of distinct batches for the Jacobian and moment vector. The Jacobian batch indices stay fixed within each outer-loop cycle, but both the Jacobian and moment vector are recomputed at the current parameter at every step. Parameter updates and averaging occur at every step, while the learning rate changes by epoch.

\begin{algorithm}[!htbp]
\caption{Warm-Start Algorithm for SLIM}
\label{alg:sequential-reshuffle}
\KwIn{
Sample $z_{1:n} = (z_i)_{i=1}^n$; 
mini-batch size $B_{\mathrm{ws}}$; 
number of epochs $E_{\mathrm{ws}}$; 
learning rate schedule $(\gamma_e)_{e=1}^{E_{\mathrm{ws}}}$; 
initial value $\theta_{0,\mathrm{ws}}^*$ for a warm start.
}

Set $t \leftarrow 1$.

Initialize $\theta^* \leftarrow \theta_{0,\mathrm{ws}}^*$ and $\bar \theta^* \leftarrow \theta_{0,\mathrm{ws}}^*$.

\For{epoch $e = 1,\ldots, E_{\mathrm{ws}}$}{

Shuffle $\{1, \ldots, n\}$ into a permutation $\{m(1), \ldots, m(n)\}$ such that $m(i) \ne m(i')$ for $i \ne i'$.

Set $\tilde{z}_{1:n} \leftarrow (z_{m(i)})_{i=1}^n$.

\For{$j = 1,\ldots, \lfloor n / B_{\mathrm{ws}} \rfloor$}{

\For{$k = 1,\ldots, \lfloor n / B_{\mathrm{ws}} \rfloor$, $k \ne j$}{

Compute $\tilde{G}_j \leftarrow \frac{1}{B_{\mathrm{ws}}} \sum_{i = (j-1)B_{\mathrm{ws}} + 1}^{jB_{\mathrm{ws}}} G(\tilde{z}_i, \theta^*)$.

Compute $\tilde{g}_k \leftarrow \frac{1}{B_{\mathrm{ws}}} \sum_{i = (k-1)B_{\mathrm{ws}} + 1}^{kB_{\mathrm{ws}}} g(\tilde{z}_i, \theta^*)$.

Update $\theta^* \leftarrow \theta^* - \gamma_e \tilde{G}_j' \tilde{g}_k$.

Update $\bar \theta^* \leftarrow \frac{t-1}{t} \bar \theta^* + \frac{1}{t} \theta^*$.

Increment $t \leftarrow t + 1$.

}

}

}

\KwOut{
Final estimator $\bar \theta^*$, to be used as the initial value $\theta_0^*$ in Algorithm~\ref{alg:mini-batch}
}
\end{algorithm}

\section{Alternative Asymptotic Regimes}\label{appendix:asymp:regimes}

Section~\ref{subsec:asymp:regimes} considers $N\asymp n$ with finite $B_G$. Here we consider three alternatives under the relevant conditions of Theorems~\ref{thm:asymp:normal} and~\ref{thm:FCLT}, including $N^{1-a}/n\to0$.

\subsection{Intermediate Case II. $n \asymp N$ and $B_G \rightarrow \infty$}

As $B_G \rightarrow \infty$, $B_g \Sigma_\oo \rightarrow G_\oo' \Omega_\oo G_\oo$, reducing Jacobian noise and the conservativeness of random-scaling inference. Theorems~\ref{thm:asymp:normal} and~\ref{thm:FCLT} are stated for a fixed $B_G$; diverging Jacobian batches $B_{G,t}\to\infty$ are covered by the framework of Appendix~\ref{appx:lemmas-proofs} (Condition~\ref{asm:appx:batch-size} with $B_G=\infty$), in which case $B_g\Sigma_\oo=G_\oo'\Omega_\oo G_\oo$ and random-scaling inference is asymptotically exact. Larger Jacobian batches increase computational cost. Second-order SLIM in Section~\ref{sec:extensions} additionally uses an efficient weight and curvature adjustment.

\subsection{Polar Case I. Very large $n$ such that $N/n \rightarrow 0$}

When $N/n \rightarrow 0$, Theorem~\ref{thm:asymp:normal} gives
\begin{align*}
    \sqrt{N} (\bar \theta^*_{N} - \theta_\oo) 
    &= \sqrt{N} (\bar \theta^*_{N} - \hat{\theta}_n) + \left( \frac{N}{n} \right)^{1/2} \sqrt{n}(\hat{\theta}_n - \theta_\oo) \\
    &= \sqrt{N} (\bar \theta^*_{N} - \hat{\theta}_n) + o_p(1),
\end{align*}
where the remainder is $o_p(1)$ under the joint distribution of $(\hat \theta_n, \bar \theta^*_N)$. Computational error dominates sampling variability, so centering at $\hat\theta_n$ or $\theta_\oo$ gives the same limiting distribution. Inference can therefore proceed as if the data were fixed. In particular, $\tau_n/(NB_g)\to1$, so $T_{n,N}$ is asymptotically equivalent to $N(R\bar\theta_N^*-c)'V_N(R)^{-1}(R\bar\theta_N^*-c)$, and random-scaling inference is asymptotically exact in this regime even with finite $B_G$. Related observations arise in random sketching for linear models \citep[see, e.g.,][]{Lee:Ng:2020,Lee:Ng:2022}.

\subsection{Polar Case II. Very large $N$ such that $N/n \rightarrow \infty$}

When $N/n \rightarrow \infty$, Theorem~\ref{thm:asymp:normal} gives
\begin{align*}
    \sqrt{n} (\bar \theta^*_{N} - \theta_\oo) 
    &= \left( \frac{n}{N} \right)^{1/2} \sqrt{N} (\bar \theta^*_{N} - \hat{\theta}_n) +  \sqrt{n}(\hat{\theta}_n - \theta_\oo) \\
    &= \sqrt{n}(\hat{\theta}_n - \theta_\oo) + o_p(1).
\end{align*}
Computational error is negligible relative to sampling variation, so conventional GMM inference applies when the sampling covariance is consistently estimated. The random-scaling statistic in~\eqref{def:wald:stat} reduces asymptotically to
\[
n (R \bar\theta^*_N - c)' (B_g V_N(R))^{-1} (R \bar\theta^*_N - c).
\]
With finite $B_G$, random-scaling inference can remain conservative because $B_g V_N(R)$ still reflects Jacobian noise.

\section{Learning Rates and Mini-Batch Sizes}\label{subsec:Selection_gamma0}

The main-stage learning-rate calibration below is used for the EASI experiments and, with cluster batches and its own settings, the LinkedIn implementation in Section~\ref{app:linkedin:implementation}. We describe hyperparameter choices for both the warm-start and main stages. In the EASI warm-start stage, limited prior information makes learning rate selection difficult. For Algorithm~\ref{alg:sequential-reshuffle}, we set
$
\gamma_e = \gamma_{0,\mathrm{ws}} e^{-a},
$
where $e$ indexes epochs and \(a = 0.501\). We explore several candidate pairs \((\gamma_{0,\mathrm{ws}}, B_{\mathrm{ws}})\), adopting a linear scaling rule where \(\gamma_{0,\mathrm{ws}}\) increases linearly with \(B_{\mathrm{ws}}\).

For the main stage, we recommend \(\gamma_t = \gamma_0 t^{-a}\) with the same exponent. To determine \(\gamma_0\), we compute \(\tilde{G}_{t}(\bar{\theta}^*_{\mathrm{ws}})\) using mini-batches of size \(B_{\mathrm{ws}}\) and the warm-start estimate \(\bar{\theta}^*_{\mathrm{ws}}\). If \(n\) is moderate, we use all \(\lfloor n / B_{\mathrm{ws}} \rfloor\) batches; otherwise, we sample a large random subset. Let \(\mathbb{T}\) denote the selected batch indices.

Next, in the first-order stage, suppose we set the mini-batch size to \(B_{\mathrm{main}} = B_g = B_G\), which may differ from \(B_{\mathrm{ws}}\). To specify \(\gamma_0\), define
\[
\Psi_0 \;:=\; \text{median}\Bigl\{
    \bigl\| \tilde{G}_{t}(\bar{\theta}^*_{\mathrm{ws}})' \tilde{G}_{t}(\bar{\theta}^*_{\mathrm{ws}})\bigr\|_2
    : t \in \mathbb{T}\Bigr\},
\]
where \(\|\cdot\|_2\) denotes the spectral norm.
We then propose the following for \(\gamma_0\):
\begin{equation} \label{selection:gamma_0}
\gamma_0 = \frac{1}{s_0 \Psi_0} \cdot \frac{B_{\mathrm{main}}}{B_{\mathrm{ws}}},
\end{equation}
where \(s_0 > 0\) is a tuning constant. This calibration is intended to limit early-stage instability, and the factor $B_{\mathrm{main}}/B_{\mathrm{ws}}$ is a heuristic that scales $\gamma_0$ linearly with the batch size. Since \(\gamma_0\) is one of the most critical hyperparameters, we strongly recommend tuning \(s_0\) based on performance, as shown in our Monte Carlo results.

In the LinkedIn application, the averaged estimate from preliminary tuning replaces $\bar{\theta}^*_{\mathrm{ws}}$, and batch sizes count clusters; Section~\ref{app:linkedin:implementation} gives its calibration settings.

\section{LinkedIn Application Details}\label{app:linkedin}

This appendix provides data construction, economic interpretation, and implementation details for the LinkedIn application in Section~\ref{sec:linkedin}. Complete coefficient results are reported in Appendix~\ref{app:linkedin:results}.

\subsection{Sample and Alternative Construction}\label{app:linkedin:data}

The data are structured LinkedIn profiles provided by Revelio Labs; \citet{dorn2025linkedin} describe the underlying database. We use education records on degree type, field of study, attendance and completion dates, and institution identifiers linked to school locations; employment records on the start date, location, industry, and occupation of the first reported full-time post-college job; and individual identifiers and predicted gender and race/ethnicity probabilities.

The sample consists of four-year bachelor's degree holders from U.S. institutions with a recorded graduation year and undergraduate major, an identifiable state, a reported job with industry and occupation codes covered by our classification, and the required demographic probabilities. For multiple qualifying bachelor's degrees, we use the earliest. Individuals who work before entering graduate school remain in the sample; those entering graduate school immediately after college are excluded.

We assign the undergraduate institution's state, recovering missing locations from its identifier where possible. If that state remains unavailable, we use the state of the first reported post-college job. This assignment defines the individual's choice-set type and links to the destination counts. Codes are standardized to the 50 states and the District of Columbia.

Table~\ref{tab:linkedin:industry-groups} reports the six industry groups and their two-digit NAICS codes. Occupations comprise the 22 major groups with codes 11, 13, 15, 17, 19, 21, 23, 25, 27, 29, 31, 33, 35, 37, 39, 41, 43, 45, 47, 49, 51, and 53. Management (11) is the reference occupation; the remaining labels appear in Table~\ref{tab:linkedin:full-levels}.

\begin{table}[!htbp]
\centering
\caption{Industry classification for the LinkedIn application}\label{tab:linkedin:industry-groups}
\begingroup\small\setstretch{1}\renewcommand{\arraystretch}{1.15}
\begin{tabularx}{\textwidth}{@{}Xl@{}}\toprule
Industry group & Two-digit NAICS codes\\\midrule
Natural resources, utilities, and construction & 11, 21, 22, 23\\
Manufacturing & 31--33\\
Trade and transportation & 42, 44--45, 48--49\\
Information, finance, and business & 51--56\\
Education, health, and public services & 61, 62, 92\\
Consumer and other services & 71, 72, 81\\
\bottomrule
\end{tabularx}
\par\smallskip
\begin{minipage}{\textwidth}\footnotesize Notes: Information, finance, and business is the reference industry for both intercepts and demographic interactions. Unclassified industry code 99 and codes outside the listed groups are excluded.\end{minipage}
\endgroup
\end{table}

The demographic covariates are continuous predicted female, Asian/Pacific Islander, Hispanic, Black, and other race/ethnicity probabilities; other is the sum of Native American and multiple-race probabilities. They enter in their original units, without assignment to the most likely category or a minimum probability threshold. White is omitted; the included race/ethnicity probabilities are not rescaled to sum to one, and industry interactions are not separately standardized.

%To use the same sample across the candidate instrument specifications, we require a valid current own-state destination measure and both lags of the own-state and outside-state measures for every included group--alternative combination. Although the reported specification uses only the two outside-state lags as excluded instruments, it uses this same sample.
Each person's observed choice must belong to a choice set containing at least two alternatives. %Wage variables play no role in this sample restriction or utility specification.
The sample covers graduation years 1961 and 1963--2021; no individuals from 1962 qualify. The three 1961 graduates are included under the same rules. The final sample contains 4,812,451 individuals, 32,430 groups, and 513,950 group--alternative combinations. %The implied 250,172,143 expanded person--alternative rows are an accounting quantity; they are not stored as a separate expanded dataset.

\subsection{Destination Measures and Instruments}\label{app:linkedin:measures}

Let $c_{s,y,m,a}$ denote the destination count in the source transition panel and $N_{s,y,m}$ its group total, where $s$, $y$, and $m$ index state, graduation cohort, and undergraduate major. Detailed industry counts are aggregated into the six groups before constructing the measures; source group denominators are counted once, not summed across destination rows. Shares use counts preceding the final choice-set and individual-covariate restrictions and are neither recomputed from the final sample nor renormalized over its choice sets.

For valid counts, define
\begin{align}\label{eq:linkedin:measure}
p^{\mathrm{own}}_{s,y,m,a}
&=\min\!\left\{1-10^{-6},\max\!\left\{10^{-6},
\frac{c_{s,y,m,a}+0.5}{N_{s,y,m}+1}\right\}\right\},\notag\\
C^{\mathrm{own}}_{s,y,m,a}
&=-\log\!\left(\frac{p^{\mathrm{own}}_{s,y,m,a}}
{1-p^{\mathrm{own}}_{s,y,m,a}}\right).
\end{align}
Count validation requires nonmissing, consistent counts with positive denominators and admissible numerators. Smoothed probabilities must lie strictly between $10^{-12}$ and $1-10^{-12}$ before the numerical bounds in~\eqref{eq:linkedin:measure} are imposed. Missing source cells are not generally interpreted as observed zero shares.

Outside-state counts are constructed by subtraction,
\begin{equation}\label{eq:linkedin:outside-counts}
c_{-s,y,m,a}=\sum_{s'}c_{s',y,m,a}-c_{s,y,m,a},
\qquad
N_{-s,y,m}=\sum_{s'}N_{s',y,m}-N_{s,y,m}.
\end{equation}
The same smoothing, validity checks, and probability bounds define $C^{\mathrm{out}}_{s,y,m,a}$. National denominators count each state $\times$ graduation year $\times$ major total once. Outside-state construction requires valid national and own-state denominators and a positive difference; an absent destination numerator can contribute zero when the corresponding totals are available and valid. The four candidate instruments are $C^{\mathrm{own}}_{s,y-\ell,m,a}$ and $C^{\mathrm{out}}_{s,y-\ell,m,a}$ for $\ell=1,2$, where lags index graduation cohorts. The reported 53-moment specification uses only the two outside-state lags, selected after exploratory comparisons of own-state and outside-state lag combinations.

Sample construction requires the current own-state measure and both lags of the own-state and outside-state measures for every included group--alternative combination.

The current measure and four candidate instruments are standardized separately over the same $M=513{,}950$ included group--alternative combinations, weighting combinations equally. For any such variable $V_j$,
\begin{equation}\label{eq:linkedin:standardization}
\widetilde V_j=\frac{V_j-\bar V}{s_V},\qquad
\bar V=\frac{1}{M}\sum_{j=1}^M V_j,\qquad
s_V^2=\frac{1}{M-1}\sum_{j=1}^M(V_j-\bar V)^2.
\end{equation}
Standardization uses no person weights or additional clipping. The reported specification uses the same sample and scales; demographic probabilities remain in their original units.

\subsection{Economic Interpretation of the Discrete Choice Model}
\label{app:linkedin:economic-interpretation-destination-choice}

We model the first reported post-college destination as a choice among industry--occupation pairs in the individual's choice set. This follows the Roy tradition of reward-maximizing self-selection across occupations with heterogeneous potential outcomes \citep{roy1951thoughts}. The standard random-utility model with exogenous covariates and i.i.d. Type-I extreme-value alternative-specific shocks motivates our conditional-logit specification \citep{mcfadden1974conditional}. Related multinomial structures appear in the occupational-choice models of \citet{Johnson2013-xw} and \citet{Traiberman2019-tg}. Our specification provides a parsimonious description of conditional sorting without their full dynamic structure.

The endogenous destination-share index summarizes contemporaneous sorting within state--cohort--major groups: larger smoothed negative log odds indicate less commonly chosen destinations, as constructed in Section~\ref{app:linkedin:measures}. The closest precedent is \citet{Johnson2013-xw}, who include a lagged aggregate occupation share in utility to capture persistence in occupational choices and discuss habit persistence, labor adjustment costs, and easier access to offers in larger sectors as possible explanations. In our setting, persistent variation may similarly summarize durable links between a field of study and particular labor market destinations---including curricula, professional networks, employer recruiting, information, and adjustment frictions---but the model does not separately identify these channels. The index is therefore an empirical proxy, rather than a structural switching cost recovered from a dynamic labor-adjustment model \citep{Artuc2010-jm}.

Under the moment restrictions in~\eqref{eq:linkedin:moments}, $\delta$ measures the association between the destination-share index and systematic utility, conditional on industry and occupation intercepts and demographic interactions. The negative estimate assigns higher systematic utility to destinations with larger group shares.

\subsection{Implementation and Inference Details}\label{app:linkedin:implementation}

The first-order stage uses identity weighting without preconditioning, starts from the preliminary tuning average with fresh mini-batch draws, and restarts averaging. A pass through all clusters at the first-order average computes the second-moment matrix of the cluster contributions $\widehat\Omega_0$ and Jacobian $\widehat D_0$. Refinement starts from that average, fixes $W=\widehat\Omega_0^{-1}$ and $P=(\widehat D_0'W\widehat D_0)^{-1}$, and restarts averaging and random scaling. The recorded covariance and curvature matrices have full rank, so no singular-matrix fallback is used. The learning-rate clock continues across stages: $\gamma_t=\gamma_0t^{-0.501}$ for first-order updates and $\gamma_{N+\ell}=\gamma_0(N+\ell)^{-0.501}$ for second-order update $\ell$.

Calibration evaluates the largest eigenvalue of each batch Jacobian cross-product at the averaged estimate from preliminary tuning, takes the median over 128 batches of 16 clusters, and applies~\eqref{selection:gamma_0} with $s_0=8$ and main batch size 32. This gives $\gamma_0\approx0.4122$.
%The saved initialization and earlier tuning are prerequisites to the measured production job, not evidence that the procedure began from an arbitrary point in this application.

\begin{table}[!htb]
\centering
\caption{LinkedIn implementation settings and recorded time}\label{tab:linkedin:implementation}
\begingroup\small\setstretch{1}\renewcommand{\arraystretch}{1.15}
\begin{tabularx}{\textwidth}{@{}Xr@{}}\toprule
Quantity & Setting or recorded value\\\midrule
Individuals; clusters & 4,812,451; 32,430\\
Parameters; moments & 52; 53\\
First-order updates $N$; second-order updates $L$ & 4,096; 8,192\\
Moment batch size, both stages & 32 clusters\\
First-order Jacobian batch size & 32 clusters\\
Second-order Jacobian batch at local step $\ell$ & $32+\lfloor\log\ell\rfloor$\\
Realized second-order Jacobian batch range & 32--41 clusters\\
Average inverse Jacobian batch size $\beta_L$ & 0.025374\\
Learning-rate exponent; calibration multiplier & 0.501; $s_0=8$\\
Calibration batches; calibration batch size & 128; 16 clusters\\
GPU; Julia version & NVIDIA RTX A6000; 1.10.2\\
CUDA.jl; CUDA runtime & 6.2.0; 12.9.0\\\midrule
Calibration & 0.71 minutes\\
First-order stage & 30.66 minutes\\
Refinement weight and preconditioner & 12.42 minutes\\
Second-order stage & 95.38 minutes\\
Final cluster pass & 13.13 minutes\\
Jacobian-noise calculation & 13.51 minutes\\
Matrix inference & 6.25 seconds\\
Complete job, including overhead and export & 2:51:50\\\bottomrule
\end{tabularx}
\par\smallskip\begin{minipage}{\textwidth}\footnotesize
Notes: Batch sizes count clusters. Mini-batch evaluations use the GPU; final full-sample passes use the CPU. Stage timings exclude some startup, monitoring, archiving, and export overhead and therefore do not sum to the complete-job time. Data preparation and earlier tuning/development runs are excluded.\end{minipage}
\endgroup
\end{table}

Final passes through all clusters at the averaged second-order SLIM estimator compute the moment vector, Jacobian, cluster-contribution second moments, and Jacobian-noise contribution. Plug-in covariance uses the general sampling-plus-computational expression in Appendix~\ref{subsec:cluster:population-inference}, with the fixed optimization weight and realized $\beta_L=L^{-1}\sum_{\ell=1}^L(32+\lfloor\log\ell\rfloor)^{-1}$. For the decomposition in~\eqref{eq:clpop:variance-decomposition}, we evaluate the empirical covariance quantities at the same averaged estimate, retaining the fixed optimization weight and the realized $\beta_L$. The percentages for a coefficient are the diagonal entries of the three covariance components divided by the corresponding diagonal entry of the total plug-in covariance. The moment-batch component equals $K/(LB_g)$ times the sampling component; the Jacobian-batch component is the remaining computational component.
%We do not substitute the efficient-limit covariance for this finite-computation covariance.
%Scalar normal intervals use multiplier 1.959964; scalar RS intervals use multiplier 6.747 and the population normalization in Appendix~\ref{sec:cluster}.
Random-scaling intervals follow Appendix~\ref{subsec:cluster:population-inference}. The debiased $J$ test uses a testing weight evaluated at this estimator, separately from the fixed optimization weight, and is scaled by the number of clusters, $K$.

%All reported intervals concern the same averaged estimator. Their validity depends on the paper's sampling, identification, initialization, and rate conditions; a completed run or full numerical rank does not verify those conditions. The observed logarithmic Jacobian schedule and fresh draws implement the stated procedure, while the general plug-in calculation preserves the finite Jacobian-noise contribution. Table~\ref{tab:linkedin:implementation} documents computation rather than a matched comparison with another estimator.

\section{Proofs of Main Results}\label{appx:proofs-thms}

This appendix proves Theorems~\ref{thm:consistency}--\ref{thm:Plug-in-and-online-J-tests}, using the setup in Section~\ref{appx:proof-setup} and the supporting lemmas in Section~\ref{appx:key-lemmas}. Proofs of the lemmas are collected in Appendix~\ref{appx:proofs-lems}.

Theorems~{\ref{thm:consistency}}–\ref{thm:FCLT} concern the stochastic approximation estimator produced by our first-order SLIM algorithm, in which $\Ns = 0$ and both $\P_n$ and $W_n$ are chosen to be identity matrices.
In the {proofs} of Theorems~{\ref{thm:consistency}}–\ref{thm:FCLT}, we use $N$ to denote the number of iterations.
We suppress $W$ in the subscripts of $\hat{\theta}_{n,W}$ and $\bar{Q}_{n,W}(\cdot)$ and write $\hat{\theta}_{n}$ and $\bar{Q}_{n}(\theta)$ instead since $W_n = I_{\dg}$.
Let $Q_t^*  = \bar{Q}_n(\theta_t^*)$.
We also use the following definitions:
\begin{align*}
\stt_{\loclevel}^*  = \inf\left\{t \ge 0 : Q_t^* \ge \loclevel \right\}, \quad \text{and}\quad \tmt_{\varepsilon}^*  = \inf\left\{t \ge 0 : \sup_{s \ge t} Q_s^* < \varepsilon \right\}.
\end{align*}
In contrast, the second-order SLIM algorithm, as used in Theorems~\ref{thm:asymp:normal:second:order} and \ref{thm:Plug-in-and-online-J-tests}, continues from $\Ns = N$ iterations in the preliminary first-order step.

\subsection{Proof of Theorem~\ref{thm:consistency}}\label{appx:proof-consistency}

\begin{proof}
We write $\theta_N^* - \theta_\oo =     (\theta_N^*  - \hat{\theta}_{n}) + (\hat{\theta}_{n}- \theta_\oo).$
The latter term satisfies $\hat{\theta}_{n}- \theta_\oo {\pto} 0$ as $n \to \infty$ by Lemma~\ref{lem:consistency-and-normality}. Thus, it suffices to verify that $\theta_N^*  - \hat{\theta}_{n} \ppto 0$ as $n \to \infty$ and $N \to \infty$. The same applies to $\bar{\theta}_N^*$.
{
We apply the appendix lemmas with $W_n=I_{\dg}$, $\P_n=I_{\dtheta}$, and $\Ns=0$.
On the event $E_n$ defined after Lemma~\ref{lem:sample-regularity}, the upper Hessian bound and the uniform bounds on $\theta_0^*$ and $\hat\theta_n$ give
\[
\bar Q_n(\theta_0^*)\le \frac{M}{2}\|\theta_0^*-\hat\theta_n\|^2\le C.
\]
Since $\gamma_0>0$ is fixed, Condition~\ref{asm:appx:initial} holds on $E_n$ after enlarging its constant.
}
By Lemma~\ref{lem:stopping-time-and-tail}, we can find an increasing deterministic sequence $(\tilde{T}_k)_{k\ge 1}$ such that, for all $k \in \mathbb{N}$,
\begin{equation*}
\sup_{n \in \mathbb{N}} \mathbb{P}_n^{*}(\tmt_{k^{-2} \underbar{c}}^* > \tilde{T}_k) \mathbbm{1}_{E_n} < \frac{1}{k},
\end{equation*}
where $\underbar{c}>0$ is given in Lemma~\ref{lem:sample-regularity} and $\tmt_{k^{-2}\underbar{c}}^*  = \inf\{t \ge 0 : \sup_{s \ge t} \bar{Q}_n({\theta_s^*}) < k^{-2} \underbar{c}\}$.
Define the event $\mathcal{E}_{k,n} := \{\tmt_{k^{-2} \underbar{c}}^* \le \tilde{T}_k \} \cap E_n$, where {$E_n$ is the event defined after Lemma~\ref{lem:sample-regularity}, specialized to the first-order algorithm, and $\mathbb{P}(E_n)\to1$.} Because $W_n$ and $\P_n$ are nonrandom, $E_n\in\mathcal F_n$.
By construction of $\mathcal{E}_{k,n}$, it holds that
\begin{equation*}
\forall t \ge \tilde{T}_k: \ \ \ \underbar{c} (\| \theta_t^* - \hat{\theta}_n\|^2 \wedge \delta^2) \le  \bar{Q}_n(\theta_t^*) \le k^{-2} \underbar{c} \quad \Longrightarrow \quad \| \theta_t^* - \hat{\theta}_n\| \le \frac{1}{k}
\end{equation*}
on the event $\mathcal{E}_{k,n}$ for $k > 1/\delta$, where $\delta>0$ is given in Assumption~\ref{asm:regularity}.
Since $\p(\mathcal{E}_{k,n}^c) = \mathbb{E} [\mathbb{P}_n^*(\tmt_{k^{-2} \underbar{c}}^* > \tilde{T}_k) \mathbbm{1}_{E_n} ] + \mathbb{P}(E_n^c) \le \frac{1}{k} + \mathbb{P}(E_n^c)$, we find that $\limsup_{n\to\infty}\p(\mathcal{E}_{k,n}^c) \le 1/k$ for all $k > 1/\delta$.

We first show $\|\theta_N^* - \hat{\theta}_n\| \ppto 0$.
Observe that, for all $\varepsilon >0$ and $k \in \mathbb{N}$ such that $1/k < (\varepsilon \wedge \delta)$, it holds
\begin{align*}
\p\left( \|\theta_N^* - \hat{\theta}_n\| > \varepsilon \right) \le \p\left( \|\theta_N^* - \hat{\theta}_n\| > 1/k, \mathcal{E}_{k,n} \right) + \p(\mathcal{E}_{k,n}^c) \le \mathbbm{1}\{N < \tilde{T}_k\} + \p(\mathcal{E}_{k,n}^c).
\end{align*}
As $n \to \infty$ and $N \to \infty$, this implies that $\limsup_{n, N \to \infty} \p( \|\theta_N^* - \hat{\theta}_n\| > \varepsilon ) \le 1/k$.
Since $1/k$ can be chosen arbitrarily close to $0$ for any given $\varepsilon>0$, this shows $\lim_{n, N \to \infty} \p( \|\theta_N^* - \hat{\theta}_n\| > \varepsilon ) = 0$, establishing $\theta_N^* - \theta_\oo \ppto 0$.

{
The same tail bound also yields the uniform conditional convergence in the theorem. For any $\varepsilon>0$, choose $k$ such that $1/k<\varepsilon\wedge\delta$. For $N\ge\tilde T_k$,
\[
\sup_{n\in\mathbb N}\left\{
\mathbb P_n^*\bigl(\|\theta_N^*-\hat\theta_n\|\ge\varepsilon\bigr)\mathbbm1_{E_n}
\right\}
\le \sup_{n\in\mathbb N}\left\{
\mathbb P_n^*(\tmt_{k^{-2}\underbar c}^*>\tilde T_k)\mathbbm1_{E_n}
\right\}<\frac1k.
\]
Letting $N\to\infty$ and then $k\to\infty$ proves the stated convergence.
}

Next, we show $\|\bar{\theta}_N^* - \hat{\theta}_n\| \ppto 0$.
Let $\varepsilon > 0 $ be given, and pick $k \in \mathbb{N}$ such that $1/k < (\varepsilon \wedge \delta)$.
{For $N\ge\tilde T_k$,} we write $\bar{\theta}_N^* - \hat{\theta}_n = \frac{1}{N} \sum_{t=1}^{\tilde{T}_k} (\theta_t^* - \hat{\theta}_n ) + \frac{1}{N} \sum_{t= \tilde{T}_k+1}^{N}(\theta_t^* - \hat{\theta}_n )$.
Since $\|\theta_t^*-\hat{\theta}_n\|{\le}1/k<\varepsilon$ for all $t\ge \tilde{T}_k$ on $\mathcal{E}_{k,n}$, we have, by the triangle inequality,
\begin{align}
\label{eq:eq1-thm1}
\p \left( \|\bar{\theta}_N^* - \hat{\theta}_n\| \ge 2 \varepsilon \right)
\le &\  \p \left( \{\|\bar{\theta}_N^* - \hat{\theta}_n\| \ge 2 \varepsilon \} \cap \mathcal{E}_{k,n}\right) + \p(\mathcal{E}_{k,n}^c) \nonumber\\
\le &\ \p \Bigg( \frac{1}{N} \sum_{t=1}^{\tilde{T}_k} \|\theta_t^* - \hat{\theta}_n\| \ge \varepsilon  \Bigg) + \p(\mathcal{E}_{k,n}^c).
\end{align}
We now show that $\lim_{n, N \to \infty} \p ( N^{-1} \sum_{t=1}^{\tilde{T}_k} \|\theta_t^* - \hat{\theta}_n\| \ge \varepsilon) = 0$ for all $k$.
Since $\tilde{T}_k$ is a fixed integer, it suffices to establish that $N^{-1} \|\theta_t^* - \hat{\theta}_n\| \ppto 0$ for all $t \ge 0$.
We proceed by induction on $t\ge 0$.
Observe that the statement is true for $ t = 0$, because $\theta_0^* = O_{\p}(1)$ by assumption, and $\hat{\theta}_{n} \pto \theta_\oo$ by Lemma~\ref{lem:consistency-and-normality}.
Therefore, for the induction step, it suffices to show $N^{-1}\|\theta_t^* - \theta_{t-1}^*\| = N^{-1} \gamma_t \| \tilde{G}_t(\theta_{t-1}^*)'\tilde{g}_t(\theta_{t-1}^*)  \| \ppto 0$ for all $t \ge 1$.

%Again, let $\varepsilon > 0$ be given.
By Jensen's inequality and the moment bounds included in $E_n$, we have, on the event $E_n$,
\begin{equation*}
    \mathbb{E}_n^*[\| \tilde{G}_t(\theta)\|^2] \le C,\quad \mathbb{E}_n^*[\| \tilde{g}_t(\theta)\|^2] \le C(\bar{Q}_{n}(\theta) + 1), \quad \forall \theta \in \Theta.
\end{equation*}
Let $\eta >0$ and pick $\loclevel > 0$ such that $\sup_{n \in\mathbb{N}} \mathbb{P}_n^*(\stt_{\loclevel}^* < \infty) \mathbbm{1}_{E_n} < \eta$ by Lemma~\ref{lem:stopping-time-and-tail}, where $\stt_{\loclevel}^*  = \inf\{t \ge 0 : \bar{Q}_n(\theta_t^*) \ge \loclevel \}$.
Let $\tilde{\mathcal{E}}_{t,n} := \{\stt_{\loclevel}^* \ge t\} \cap E_n$.
Then, $\p(\tilde{\mathcal{E}}_{t,n}^c) \le \mathbb{E}[\mathbb{P}_n^*(\stt_{\loclevel}^* < \infty) \mathbbm{1}_{E_n}]+\mathbb{P}(E_n^c)$, and therefore
\begin{align*}
& \p\left( \| \tilde{G}_t(\theta_{t-1}^*)'\tilde{g}_t(\theta_{t-1}^*)  \| \ge N \varepsilon \right) \\
\le &\ \p\left( \| \tilde{G}_t(\theta_{t-1}^*)'\tilde{g}_t(\theta_{t-1}^*)  \| \ge N \varepsilon, \tilde{\mathcal{E}}_{t,n} \right) + \mathbb{P}(\tilde{\mathcal{E}}_{t,n}^c) \\
\le &\ \mathbb{E}\left[ \mathbb{P}_n^*\left( \| \tilde{G}_t(\theta_{t-1}^*)'\tilde{g}_t(\theta_{t-1}^*)  \| \ge N \varepsilon | \mathcal{F}_{n,t-1}^*\right) \mathbbm{1}_{\tilde{\mathcal{E}}_{t,n}}\right] + \mathbb{P}(E_n^c) + \mathbb{E}[\mathbb{P}_n^*(\stt_{\loclevel}^* < \infty) \mathbbm{1}_{E_n}] \\
\le &\ \frac{C}{N^2 \varepsilon^2}\mathbb{E}\left[ (Q_{t-1}^* +1)\I{\stt_{\loclevel}^* \ge t}\mathbbm{1}_{E_n}\right] + \mathbb{P}(E_n^c)  + \eta \le\frac{C (\loclevel+1)}{N^2 \varepsilon^2}+ \mathbb{P}(E_n^c)  + \eta .
\end{align*}
Since $\eta > 0$ is arbitrary, we obtain $\lim_{n, N \to \infty} \p( N^{-1}\| \tilde{G}_t(\theta_{t-1}^*)'\tilde{g}_t(\theta_{t-1}^*)  \| \ge  \varepsilon ) = 0$ for all $t \ge 1$ and $\varepsilon>0$.
This establishes that $N^{-1} \|\theta_t^* - \hat{\theta}_n\| \ppto 0$ for all $t\ge 0$, and hence, $\lim_{n, N \to \infty} \p ( N^{-1} \sum_{t=1}^{\tilde{T}_k} \|\theta_t^* - \hat{\theta}_n\| \ge \varepsilon  ) = 0$ for all $k \ge 1$.
This, in turn, implies that
\begin{equation*}
\limsup_{n, N \to \infty} \p \left( \|\bar{\theta}_N^* - \hat{\theta}_n\| \ge 2 \varepsilon \right) \le  \frac{1}{k}
\end{equation*}
for all $1/k < (\varepsilon \wedge \delta)$ by \eqref{eq:eq1-thm1}.
Letting $k \to\infty$ completes the proof.
\end{proof}

%%%%%%%%%%%%%%%%%%%%

\subsection{Proof of Theorem~\ref{thm:asymp:normal}}

We begin by writing
\begin{equation*}
\sqrt{N}( \bar \theta_N^* - \hat \theta_n)  =  \sqrt{N}( \tto[\bar]{N}- \hat \theta_n) + \sqrt{N}( \bar \theta_N^* - \tto[\bar]{N}) =: J_1 + J_2.
\end{equation*}
For the first term, apply Lemma~\ref{lem:CLT-extension} to get
$J_1 \dto ((G_\oo' G_\oo)^{-1}  \Sigma_\oo  (G_\oo' G_\oo)^{-1})^{1/2}Z\ \quad \text{as}\ \ n\to\infty, N \to \infty.$
The second term converges in probability to zero by Lemma~\ref{lem:control-approx-error} since
\begin{equation*}
\|J_2\| \le \frac{1}{\sqrt{N}}\sup_{1\le t\le N} \left\|\sum_{s=1}^t (\theta_s^* - \tto{s})\right\| \ppto 0
\end{equation*}
under $n \to \infty$, $N \to \infty$, and $N^{1-a}/n \to 0$.
Putting these together, we have
\begin{equation*}
\sqrt{N}( \bar \theta_N^* - \hat \theta_n) \dto ((G_\oo' G_\oo)^{-1}  \Sigma_\oo  (G_\oo' G_\oo)^{-1})^{1/2}Z
\end{equation*}
as $n\to\infty$, $N \to \infty$, and $N^{1-a}/n \to 0$.
Finally, note that $\sqrt{n} (\hat{\theta}_{n} - \theta_\oo) \dto \mathcal{N}(0, (G_\oo' G_\oo)^{-1}  G_\oo' \Omega_\oo G_\oo (G_\oo' G_\oo)^{-1})$ by Lemma~\ref{lem:consistency-and-normality}.
Since $\sqrt{n} (\hat{\theta}_{n} - \theta_\oo)$ is $\mathcal F_n$-measurable, and hence $\mathcal F_{n,0}^\star$-measurable, the stable convergence in Lemma~\ref{lem:CLT-extension}, applied with $X_n=\sqrt{n} (\hat{\theta}_{n} - \theta_\oo)$, gives the joint convergence of $(\sqrt{n} (\hat{\theta}_{n} - \theta_\oo),J_1)$ with independent limits. Since $J_2=o_p(1)$, the proof is complete.

%%%%%%%%%%%%%%%%%%%%
\subsection{Proof of Theorem~\ref{thm:FCLT}}

We begin by writing
\begin{align*}
I(r) = \frac{1}{\sqrt{N}}\sum_{t=1}^{\lfloor N r \rfloor} (\theta_t^{*} - \hat{\theta}_{n}) & = \frac{1}{\sqrt{N}}\sum_{t=1}^{\lfloor N r \rfloor} (\tto{t} - \hat{\theta}_{n}) + \frac{1}{\sqrt{N}}\sum_{t=1}^{\lfloor Nr \rfloor} (\theta_t^{*} - \tto{t}) \\
& =: J_{1}(r) + J_{2}(r), \quad  r \in [0,1].
\end{align*}
Lemma~\ref{lem:FCLT-extension} then shows
$J_{1}(\cdot) \wto \left( (G_\oo' G_\oo)^{-1}  \Sigma_\oo  (G_\oo' G_\oo)^{-1}\right)^{1/2} W(\cdot)
$
as $n\to\infty$ and $N \to \infty$.
To address the remainder term, we show that $\|J_2\|_\infty := \sup_{0 \le r \le 1} \|J_{2}(r)\|$ converges in probability to zero.
For every $\varepsilon > 0$, $n \in \mathbb{N}$, and $N \in \mathbb{N}$, note that
\begin{align*}
{\p\left( \|J_{2}\|_\infty > \varepsilon \mid\mathcal F_{n,0}^\star \right)}
& \le \mathbbm 1_{E_n^c} + \mathbb{P}_n^*\left( \|J_{2}\|_\infty > \varepsilon \right)\mathbbm 1_{E_n}.
\end{align*}
Letting $n\to\infty$ and $N \to \infty$ at the rate $N^{1-a}/n \to 0$, Lemma~\ref{lem:control-approx-error} implies
\begin{equation*}
\mathbbm 1_{E_n^c} + \mathbb{P}_n^*\left( \|J_{2}\|_\infty > \varepsilon \right)\mathbbm 1_{E_n}\pto 0,
\end{equation*}
thereby establishing $\|J_{2}\|_\infty \overset{\p}{\to} 0$.
Putting these pieces together, we obtain
\begin{equation*}
I(\cdot) \wto \left( (G_\oo' G_\oo)^{-1}  \Sigma_\oo  (G_\oo' G_\oo)^{-1}\right)^{1/2} W(\cdot).
\end{equation*}
Lemma~\ref{lem:consistency-and-normality} states that $\sqrt{n} (\hat{\theta}_{n} - \theta_\oo) \dto \mathcal{N}(0, (G_\oo' G_\oo)^{-1}  G_\oo' \Omega_\oo G_\oo (G_\oo' G_\oo)^{-1})$.
Since $\sqrt{n} (\hat{\theta}_{n} - \theta_\oo)$ is $\mathcal F_{n,0}^\star$-measurable, the linear representation and stable convergence in Lemma~\ref{lem:FCLT-extension}, applied with $X_n=\sqrt{n} (\hat{\theta}_{n} - \theta_\oo)$, together with $\|J_2\|_\infty=o_p(1)$, give the joint weak convergence with independent limits:
\begin{equation*}
\left( \begin{matrix}
   \sqrt{n} (\hat{\theta}_{n} - \theta_\oo) \\
   I(r)
\end{matrix} \right)
\wto
\left( \begin{matrix}
\left( (G_\oo' G_\oo)^{-1}  G_\oo' \Omega_\oo G_\oo (G_\oo' G_\oo)^{-1} \right)^{1/2} Z \\
\left( (G_\oo' G_\oo)^{-1}  \Sigma_\oo  (G_\oo' G_\oo)^{-1}\right)^{1/2} W(r)
\end{matrix} \right).
\end{equation*}

\subsection{Proof of Theorem~\ref{thm:asymp:normal:second:order}}

Lemmas~\ref{lem:refinement-weight} and~\ref{lem:refinement-initial} verify Conditions~\ref{asm:appx:conditioning-and-weighting} and~\ref{asm:appx:initial} for the refinement stage under the assumptions in Theorem~\ref{thm:asymp:normal:second:order}, which also impose the remaining conditions of Assumption~\ref{asm:appx:regularity}. Hence Lemmas~\ref{lem:consistency-and-normality}--\ref{lem:CLT-extension} apply to the refinement stage, with $E_n$ replaced by $E_n^{(2)}$ and $A_{n,\eta}$ given by Lemma~\ref{lem:refinement-initial}.
\bigskip

\noindent\textit{Proof of Part~(i):}
Let us write
\begin{align*}
\sqrt{T-N} (\bar \theta_{T}^{*} - \hat{\theta}_{n,W}) & =  J_1 + J_2,
\end{align*}
where
\begin{align*}
J_1 & = \sqrt{T-N} (\tto[\bar]{T} - \hat{\theta}_{n,W}) = \frac{1}{\sqrt{T-N}} \sum_{t=N+1}^T (\tto{t} - \hat{\theta}_{n,W}), \\
J_2 & = \sqrt{T-N} (\bar \theta_{T}^{*} - \tto[\bar]{T}) = \frac{1}{\sqrt{T-N}} \sum_{t=N+1}^T (\theta_t^* - \tto{t}).
\end{align*}
Here, $(\tto{t})_{t=N}^T$ is the approximating sequence of Lemma~\ref{lem:control-approx-error} for the refinement stage, which starts at $t = N$ with $\tto{N}$ equal to the initial value of the refinement ($\theta_N^*$ or $\bar\theta_N^*$).
Since $N/(T-N) = O(1)$ under the given assumptions, Lemma~\ref{lem:CLT-extension} applies to $J_1$, yielding
\begin{equation*}
J_1 \dto \mathcal{N}(0, (G_\oo' \Omega_\oo^{-1} G_\oo)^{-1}  \Sigma_\oo  (G_\oo' \Omega_\oo^{-1} G_\oo)^{-1}),
\end{equation*}
where the limiting normal random variable is independent of $(z_i)_{i\ge 1}$, and
\begin{equation*}
\Sigma_\oo = \frac{1}{B_g} \left( G_\oo' \Omega_\oo^{-1} G_\oo + \frac{1}{B_G} \mathbb{E}\left[ (G(z_i, \theta_\oo)-G_\oo)' \Omega_\oo^{-1}(G(z_i, \theta_\oo)-G_\oo)\right]\right).
\end{equation*}
The same argument as in the proof of Theorem~\ref{thm:asymp:normal} implies $J_2 \ppto 0$ under $(T-N)^{1-a}/n \to 0$.
Putting $J_1$ and $J_2$ together, we obtain
\begin{equation*}
\sqrt{T-N} (\bar \theta_{T}^{*} - \hat{\theta}_{n,W}) \dto     \mathcal{N}(0, (G_\oo' \Omega_\oo^{-1} G_\oo)^{-1}  \Sigma_\oo  (G_\oo' \Omega_\oo^{-1} G_\oo)^{-1}).
\end{equation*}
{
Put $A=G_\oo'\Omega_\oo^{-1}G_\oo$. Lemma~\ref{lem:consistency-and-normality} gives the sampling covariance $A^{-1}$, since $W_n=W_{\mathrm{MB}}\pto\Omega_\oo^{-1}$. The estimator $\hat\theta_{n,W}$ is measurable with respect to $\mathcal F_{n,0}^\star$. The stable convergence in Lemma~\ref{lem:CLT-extension}, applied with $X_n=\sqrt n(\hat\theta_{n,W}-\theta_\oo)$, therefore yields
\[
\begin{pmatrix}
\sqrt n(\hat\theta_{n,W}-\theta_\oo)\\
\sqrt{T-N}(\bar\theta_T^*-\hat\theta_{n,W})
\end{pmatrix}
\dto
\begin{pmatrix}
A^{-1/2}Z\\
(A^{-1}\Sigma_\oo A^{-1})^{1/2}W
\end{pmatrix},
\]
where $Z$ and $W$ are independent standard normal vectors. Preliminary weight randomness is negligible in the sampling expansion of Lemma~\ref{lem:consistency-and-normality} and is included in the conditioning for subsequent iterations.
}

\bigskip

\noindent\textit{Part~(ii):}
Start with the decomposition
\begin{align*}
I(r) &= \frac{1}{\sqrt{T-N}}\sum_{t=N + 1}^{N + \lfloor (T-N) r \rfloor} (\theta_t^{*} - \hat{\theta}_{n,W}) \\
& = \frac{1}{\sqrt{T-N}}\sum_{t=1}^{\lfloor (T-N) r \rfloor} (\tto{t+N} - \hat{\theta}_{n,W}) + \frac{1}{\sqrt{T-N}}\sum_{t=1}^{\lfloor (T-N) r \rfloor} (\theta_{t+N}^{*} - \tto{t+N}) \\
& =: J_{1}(r) + J_{2}(r), \quad r \in [0,1].
\end{align*}
By Lemma~\ref{lem:FCLT-extension}, under $p > (1-a)^{-1}$, we have
\begin{equation*}
J_1(\cdot) \wto \left( (G_\oo' \Omega_\oo^{-1} G_\oo)^{-1}  \Sigma_{\oo} (G_\oo' \Omega_\oo^{-1} G_\oo)^{-1}\right)^{1/2} W(\cdot),
\end{equation*}
By the same argument as in the proof of Theorem~\ref{thm:FCLT},
$\|J_2\|_\infty \ppto 0.
$
This establishes $I(\cdot) \wto \left( (G_\oo' \Omega_\oo^{-1} G_\oo)^{-1}  \Sigma_{\oo} (G_\oo' \Omega_\oo^{-1} G_\oo)^{-1}\right)^{1/2} W(\cdot)$.
{
Since $\sqrt n(\hat\theta_{n,W}-\theta_\oo)$ is $\mathcal F_{n,0}^\star$-measurable, the stable convergence in Lemma~\ref{lem:FCLT-extension} and the sampling limit in Lemma~\ref{lem:consistency-and-normality} give the claimed joint convergence with independent limits.
}

\subsection{Proof of Theorem~\ref{thm:Plug-in-and-online-J-tests}}
{
Put $M=T-N$, $A=G_\oo'\Omega_\oo^{-1}G_\oo$, and
$\Pi=\Omega_\oo^{-1/2}G_\oo A^{-1}G_\oo'\Omega_\oo^{-1/2}$.
The matrix $\Pi$ is an orthogonal projector of rank $\dtheta$.
Throughout this proof $B_g$ is fixed and $B_{G,t}\to\infty$.

\paragraph{Joint moment and iterate approximation.}
Let $\epsilon_t=\tilde g_t(\theta_{t-1}^*)-\bar g_n(\theta_{t-1}^*)$ and
\[
U_M=\sqrt{\frac{B_g}{M}}\,\Omega_\oo^{-1/2}
       \sum_{t=N+1}^T\epsilon_t,\qquad
Z_n=\sqrt n\,\Omega_\oo^{-1/2}\bar g_n(\hat\theta_{n,W}).
\]
By Lemma~\ref{lem:consistency-and-normality}, which applies to the refinement stage with $W=\Omega_\oo^{-1}$ by Lemma~\ref{lem:refinement-weight}, $Z_n=(I_m-\Pi)\Omega_\oo^{-1/2}\sqrt n\,\bar g_n(\theta_\oo)+o_p(1)$, where $m=\dg$; hence
$Z_n\dto Z\sim\mathcal N(0,I_m-\Pi)$.
Lemma~\ref{lem:stable-mclt}, applied at $r=1$ with $M$ in place of $\N$ to the martingale-difference array $(\sqrt{B_g}\,\Omega_\oo^{-1/2}\epsilon_t)_{N<t\le T}$ and with $X_n=Z_n$, gives $U_M\dto U\sim\mathcal N(0,I_m)$ jointly with the $\mathcal F_{n,0}^\star$-measurable $Z_n$, with independent limits $Z$ and $U$.
To verify the conditions of that lemma, the predictable covariance of $\epsilon_t$ is
$B_g^{-1}\tilde\Omega_n(\theta_{t-1}^*)$, with $\tilde\Omega_n$ denoting the centered moment covariance.
Its average converges to $B_g^{-1}\Omega_\oo$ by consistency and the moment bounds. The stronger moment condition in the theorem supplies the conditional Lindeberg bound as in Lemma~\ref{lem:FCLT-extension}.

For the weighted gradient error used in refinement, decompose
\[
\xi_t(\theta)=\bar G_n(\theta)'W_{\mathrm{MB}}
       \{\tilde g_t(\theta)-\bar g_n(\theta)\}
       +\eta_t(\theta),
\qquad
\eta_t(\theta)=(\tilde G_t(\theta)-\bar G_n(\theta))'
                    W_{\mathrm{MB}}\tilde g_t(\theta).
\]
By independence of the batches, the second term is a martingale difference. After localization to the bounded-objective events used in Lemma~\ref{lem:sgd-conv-rate}, its conditional second moment is bounded by a constant times $1/B_{G,t}$. Consequently,
\[
\sqrt{\frac{B_g}{M}}\sum_{t=N+1}^T\eta_t(\theta_{t-1}^*)=o_p(1),
\]
because $M^{-1}\sum_{t=N+1}^T B_{G,t}^{-1}=M^{-1}\sum_{s=1}^M b_s^{-1}\to0$ by averaging the fixed local sequence $b_s^{-1}\to0$. This is the step that requires vanishing Jacobian noise. The same localization and consistency allow $\bar G_n(\theta_{t-1}^*)'W_{\mathrm{MB}}$ to be replaced by $G_\oo'\Omega_\oo^{-1}$ in the normalized martingale sum.

The approximation in Lemma~\ref{lem:control-approx-error} and the linear representation in Lemma~\ref{lem:CLT-extension}, together with $\bar{\H}_n\pto A$ and the refinement initialization in Lemma~\ref{lem:refinement-initial}, therefore give
\begin{equation}\label{eq:J:joint-linearization}
\sqrt{\frac{B_g}{M}}\sum_{t=N+1}^T
 (\theta_{t-1}^*-\hat\theta_{n,W})
 =-A^{-1}G_\oo'\Omega_\oo^{-1/2}U_M+o_p(1).
\end{equation}
The preconditioner cancels in the averaged linear recursion, and the objective Hessian converges to $A$. Shifting the sum from $t-1$ to $t$ changes it by
$\sqrt{B_g/M}(\theta_T^*-\theta_N^*)=o_p(1)$.
Moreover, the localized rate bound of Lemma~\ref{lem:sgd-conv-rate} and local identification imply
\begin{equation}\label{eq:J:path-remainder}
\frac{1}{\sqrt M}\sum_{t=N+1}^T
       \|\theta_{t-1}^*-\hat\theta_{n,W}\|^2=o_p(1).
\end{equation}
Indeed the bound is proportional to
$M^{-1/2}\sum_{s=1}^M(N+s)^{-a}=O(M^{1/2-a})$ after localization; the initial term is negligible by Lemma~\ref{lem:refinement-initial}. The localization error can be made arbitrarily small. The maintained restrictions $N=O(M)$ and $M^{1-a}/n\to0$ also give the approximation conditions in those lemmas.

\paragraph{Plug-in tests.}
Equation~\eqref{eq:J:joint-linearization} implies
$\bar\theta_T^*-\hat\theta_{n,W}=O_p(M^{-1/2})$.
A Taylor expansion at $\hat\theta_{n,W}$ then yields
\[
z_n:=\sqrt n\,\Omega_\oo^{-1/2}\bar g_n(\bar\theta_T^*)
   =Z_n-\sqrt{\frac{n}{MB_g}}\,\Pi U_M+o_p(1).
\]
The second-order remainder is $O_p(\sqrt n/M)=o_p(1)$ because $n/(MB_g)$ is bounded and $B_g$ is fixed.
For any consistent weight $\hat W\pto\Omega_\oo^{-1}$, the exact whitening identity is
\[
n\bar g_n(\bar\theta_T^*)'\hat W\bar g_n(\bar\theta_T^*)
=z_n'\Omega_\oo^{1/2}\hat W\Omega_\oo^{1/2}z_n.
\]
Hence $J\dto\|Z-\sqrt\tau\,\Pi U\|^2$, which has distribution
$\chi^2_{m-\dtheta}+\tau\chi^2_{\dtheta}$ with independent terms.
For the debiased statistic, define
$\mathcal W_n=\bar W-\bar W\bar\Phi(\bar\Phi'\bar W\bar\Phi)^{-1}\bar\Phi'\bar W$.
Then $J_D=z_n'\Omega_\oo^{1/2}\mathcal W_n\Omega_\oo^{1/2}z_n$ and
$\Omega_\oo^{1/2}\mathcal W_n\Omega_\oo^{1/2}\pto I_m-\Pi$.
The projection removes $\Pi U$, so $J_D\dto\chi^2_{m-\dtheta}$.

\paragraph{Recursive mini-batch test.}
Taylor expansion of each $\bar g_n(\theta_{t-1}^*)$ around $\hat\theta_{n,W}$, with \eqref{eq:J:path-remainder}, gives
\[
\sqrt n\,\Omega_\oo^{-1/2}\bar g_T^*
 =Z_n+\sqrt{\frac{n}{MB_g}}(I_m-\Pi)U_M+o_p(1).
\]
Here the moment-batch noise and the iterate error share the same $U_M$; their tangent components cancel by \eqref{eq:J:joint-linearization}.
The limit is $\mathcal N(0,(1+\tau)(I_m-\Pi))$.
Since $(n^{-1}+(MB_g)^{-1})^{-1}=n/(1+n/(MB_g))$ and
$W_{\mathrm{MB}}^c\pto\Omega_\oo^{-1}$ (Lemma~\ref{lem:refinement-weight}), it follows that
$J^*\dto\chi^2_{m-\dtheta}$.
}

\section{Technical Setup, Supporting Lemmas, and Additional Proofs}\label{appx:lemmas-proofs}

{
This appendix provides the common setup and supporting lemmas used in Appendix~\ref{appx:proofs-thms}, followed by proofs of Corollaries~\ref{cor:inference:random} and~\ref{cor:plugin:eff} and of the results in Appendices~\ref{sec:setting:fixed} and~\ref{sec:cluster}.
Proofs of the supporting lemmas are collected in Appendix~\ref{appx:proofs-lems}.
}
{The setup below uses i.i.d. observations; the fixed-data and cluster results use the corresponding assumptions in Appendices~\ref{sec:setting:fixed} and~\ref{sec:cluster}. The assumptions for Lemma~\ref{lem:debiased-J:general} are stated separately.}

\subsection{Setup}\label{appx:proof-setup}

\paragraph{Notation.}
We adopt the following notation.
Generic absolute positive constants are denoted by $\underbar{c}$ and $C$, which may differ from line to line.
Absolute constants are allowed to depend on (fixed) quantities appearing in Assumption~\ref{asm:appx:regularity}, but not on the sample size $n$, the time index $t$, or the iteration numbers, $\Ns$ and $\N$.
We write
% $\p$ for the joint probability distribution of the sample and the mini-batch draws, and
$\mathbb{P}$ for the joint distribution of the sample $(z_{i})_{i \ge 1}$ and the algorithmic draws, with the corresponding expectation denoted by $\mathbb{E}$.
{
Throughout Appendices~\ref{appx:proofs-thms}--\ref{appx:proofs-lems}, $\mathbb P_n^\star$ and $\mathbb E_n^\star$ denote the law and expectation of the subsequent mini-batch draws conditional on the initial sigma-algebra $\mathcal F_{n,0}^\star$ defined below. This includes the data and any randomness already used to construct the starting value, weight, and preconditioner. Unconditional probability integrates over both the sample and this initial randomness.
}
We write $a \vee b$ (resp. $a \wedge b$) for $\max\{a,b\}$ (resp. $\min\{a,b\}$).
Throughout the paper, \wpa\ abbreviates \emph{with probability approaching one}.
A condition required to hold for all $\theta$ in a set $\Theta_n\subseteq\Theta$ \wpa,
where $\Theta_n$ may depend on the data (for example,
$\Theta_n=\{\theta\in\Theta:\|\theta-\hat\theta_{n,W}\|\le\delta\}$), means
\[
\lim_{n\to\infty}\mathbb{P}\big(\text{the condition holds for every }\theta\in \Theta_n\big)=1;
\]
for conditions involving $W_{\mathrm{MB}}$, the limit is taken as $n$, $\Ns$, and
$M_{\mathrm{MB}}\to\infty$. Thus the condition must hold simultaneously for all
$\theta\in \Theta_n$ on a single event whose probability tends to one, which is stronger
than requiring it to hold \wpa\ for each fixed $\theta$. For example,
Condition~\ref{asm:appx:item:local-convex-Lojasiewicz} means
\[
\mathbb{P}\Big(\frac{\partial^2\bar Q_{n,W}(\theta)}{\partial\theta\,\partial\theta'}
\ge c\,I_{\dtheta}\ \text{ for all }\theta\in\Theta\text{ with }
\|\theta-\hat\theta_{n,W}\|\le\delta\Big)\to1.
\]

\paragraph{Algorithm.}

We first set out a unified framework that encompasses both the \fo\ and \ts\ algorithms in the paper.
Let $z_{1:n} = (z_i)_{i=1}^n$ denote the sample of size $n$.
Let $(B_{G,t})_{t \ge 1}$ denote a sequence of possibly time-varying mini-batch sizes used to approximate the Jacobian of the moment function.
Here $t$ indexes updates within the stage, and the batch-size sequence is fixed independently of $n$, the learning-rate offset $\Ns$, and the stage length $\N$.
Let $B_g \ge 1$ be a fixed mini-batch size used for the moment function.
Let $B_t = \sum_{s = 1}^t (B_{G,s} + B_g)$ denote the total number of mini-batch draws up to iteration $t$.
Let $(\tilde{z}_i)_{i = 1}^{B_{\N}}$ denote mini-batch draws up to iteration $\N$, each of which is sampled uniformly at random from $z_{1:n}$.
Define the mini-batch approximations to the Jacobian and the moment function as follows:
\begin{equation*}
\tilde{G}_t(\theta) = \frac{1}{B_{G,t}}\sum_{j=1}^{B_{G,t}} G(\tilde{z}_{j+B_{t-1}}, \theta),\quad \tilde{g}_t(\theta) = \frac{1}{B_g}\sum_{j=1}^{B_g} g(\tilde{z}_{j+B_{t-1} + B_{G,t}}, \theta).
\end{equation*}
We write $\P_n$ for a pre-conditioning matrix and $W_n$ for a weighting matrix, both of which are allowed to be random and vary with $n$; both are symmetric and, with probability approaching one, positive definite (in the refinement, they are generalized inverses).

We use the following iterative algorithm in general form. Let $\theta_0^{\star} \in \Theta$ denote an initial value. The stochastic iterates $(\theta_t^{\star})_{t=0}^T$ evolve according to the updating rule,
\begin{equation}
\label{eq:appx:MB-GD-updating-rule}
\theta_t^{\star} = \theta_{t-1}^{\star} - \gs \P_n \tilde{G}_t(\theta_{t-1}^{\star})' W_n \tilde{g}_t(\theta_{t-1}^{\star})\quad \text{for}\ t \ge 1,
\end{equation}
where $(\gs)_{t\ge 1}$ denotes the learning rate (step size) schedule as specified below.
Let $\gamma_0 > 0$ and $a \in (1/2,1)$ be the initial learning rate and the learning rate exponent, respectively.
The learning rate is now defined by $\gs = \gamma_0 ((t+\Ns)\vee 1)^{-a},\quad t \ge 0$, where $\Ns \ge 0$ is an integer representing the number of first-order iterations in the SLIM algorithm.
When $\Ns = 0$, $\gs$ reduces to the standard learning rate form $\gamma_0 t^{-a}$ for $t\ge1$, which corresponds to no previous learning steps.
In this case, the choice of $\P_n = I_{\dtheta}$ and $W_n = I_{d_g}$ yields the first-order algorithm.
Otherwise, $\Ns > 0$ indicates that the algorithm has already advanced $\Ns$ iterations.
The second-order refinement step in Section~\ref{sec:extensions} then takes the form \eqref{eq:appx:MB-GD-updating-rule}, where $\Ns$ denotes the iteration number of the first-order algorithm, and the resulting stochastic iterate $\theta_{\Ns}^*$ (or the average $\bar\theta_{\Ns}^*$; see Lemma~\ref{lem:refinement-initial}) is relabeled as $\theta_0^{\star}$.
% \footnote{For instance, our second-order refinement step in Section~\ref{sec:extensions} takes the form of \eqref{eq:appx:MB-GD-updating-rule}, with $\Ns$ the iteration number of the first-order algorithm, and we relabel the resulting stochastic iterate, $\theta_{\Ns}^*$, as $\theta_0^{\star}$.}

% \phantom{Other defs.}

After $\N$ iterations, where $\N$ is chosen by the user, the algorithm yields the averaged stochastic approximation estimator $\bar{\theta}_{\N}^{\star} = \frac{1}{\N}\sum_{t=1}^{\N} \theta_t^{\star},$
which can also be updated recursively via Algorithm~\ref{alg:mini-batch}.
The approximand is the full-sample GMM estimator $\hat{\theta}_{n,W}$ under the weighting matrix $W_n$:
\begin{equation*}
\hat{\theta}_{n,W} = \argmin_{\theta \in \Theta} \bar{g}_n(\theta)' W_n \bar{g}_n(\theta).
\end{equation*}
The sample Lyapunov function is accordingly defined as
\begin{equation*}
\bar{Q}_{n,W}(\theta) = \frac{1}{2}\left( \bar{g}_n(\theta)' W_n \bar{g}_n(\theta) - \bar{g}_n(\hat{\theta}_{n,W})' W_n \bar{g}_n(\hat{\theta}_{n,W})\right) \ge 0.
\end{equation*}
Additionally, we define the following moment quantities:
{
\begin{align*}
\bar\Omega_n(\theta)&=\frac1n\sum_{i=1}^n g(z_i,\theta)g(z_i,\theta)',
\qquad \tilde\Omega_n(\theta)=\bar\Omega_n(\theta)-\bar g_n(\theta)\bar g_n(\theta)',\\
\bar\Omega_n&=\bar\Omega_n(\hat\theta_{n,W}),\qquad
\tilde\Omega_n=\tilde\Omega_n(\hat\theta_{n,W}),\qquad
\bar G_n=\bar G_n(\hat\theta_{n,W}),\\
\bar{\H}_n&=\bar G_n'W_n\bar G_n+
 \sum_{j=1}^{\dg}[W_n\bar g_n(\hat\theta_{n,W})]_j
 \frac{\partial^2\bar g_{nj}(\hat\theta_{n,W})}{\partial\theta\partial\theta'}.
\end{align*}
The last expression is the Hessian of $\bar Q_{n,W}$ for the fixed symmetric weight $W_n$; each moment Hessian is multiplied by the corresponding component of $W_n\bar g_n$.
}
Let $\xi_t(\theta)$ denote the stochastic approximation error at iteration $t$,
\begin{equation*}
\xi_t(\theta) = \tilde{G}_t(\theta)' W_n \tilde{g}_t(\theta) - \bar{G}_n(\theta)' W_n \bar{g}_n(\theta), \quad \theta \in \Theta,
\end{equation*}
so that \eqref{eq:appx:MB-GD-updating-rule} can be rewritten as
\begin{equation*}
\theta_t^{\star} = \theta_{t-1}^{\star} - \gs \P_n \bar{G}_n(\theta_{t-1}^{\star})' W_n \bar{g}_n(\theta_{t-1}^{\star}) - \gs \P_n \xi_t(\theta_{t-1}^{\star}).
\end{equation*}
{
For the weighted gradient error, write $q(\theta)=\bar g_n(\theta)$,
$C_n(\theta)=\tilde\Omega_n(\theta)$, the centered moment covariance defined above, and
$\Delta_i(\theta)=G(z_i,\theta)-\bar G_n(\theta)$.
Independence of the two batches gives the exact covariance
\begin{align*}
\Sigma_{n,t}(\theta)
&=\frac{1}{B_g}\bar G_n(\theta)'W_n C_n(\theta)W_n\bar G_n(\theta)\\
&\quad+\frac{1}{nB_{G,t}}\sum_{i=1}^n
 \Delta_i(\theta)'W_n\left(\frac{C_n(\theta)}{B_g}
                      +q(\theta)q(\theta)'\right)W_n\Delta_i(\theta).
\end{align*}
Write $\Sigma_{n,t}=\Sigma_{n,t}(\hat\theta_{n,W})$.
Under the maintained population assumptions, $q(\hat\theta_{n,W})=o_p(1)$ and the empirical derivative moments and weights are bounded in probability. Thus the terms involving $q$ disappear in the growing-sample limit below, but remain in the fixed-data covariance.
}

The asymptotic covariance matrix is then given by
\begin{align*}
\Sigma_t=\operatornamewithlimits{plim}_{n\to\infty}\Sigma_{n,t}
&=\frac1{B_g}G_\oo'W\Omega_\oo WG_\oo\\
&\quad+\frac1{B_gB_{G,t}}\mathbb E\big[
 (G(z_i,\theta_\oo)-G_\oo)'W\Omega_\oo W
 (G(z_i,\theta_\oo)-G_\oo)\big].
\end{align*}
{
Define the time-stationary covariance by
$\Sigma_n=\lim_{t\to\infty}\Sigma_{n,t}(\hat\theta_{n,W})$,
using the exact expression above with $B_{G,t}$ replaced by $B_G$ and $1/B_G=0$ when $B_G=\infty$.
For the fixed-data first-order algorithm with $W_n=I_{\dg}$ and constant batch sizes, this is precisely $\bar\Sigma_n$ in \eqref{def:bar-Sigma-n}.
}

Finally, let $\Sigma_\oo$ denote the asymptotic variance as both $n \to \infty$ and $t \to \infty$:
\begin{align*}
\Sigma_\oo=\lim_{t\to\infty}\Sigma_t
=\operatornamewithlimits{plim}_{n\to\infty}\Sigma_n
&=\frac1{B_g}G_\oo'W\Omega_\oo WG_\oo\\
&\quad+\frac1{B_gB_G}\mathbb E\big[
 (G(z_i,\theta_\oo)-G_\oo)'W\Omega_\oo W
 (G(z_i,\theta_\oo)-G_\oo)\big].
\end{align*}

{
Let $\mathcal F_{n,0}^\star$ contain $z_{1:n}$, $\theta_0^\star$, $W_n$, $\P_n$, and all preliminary draws used to construct them. Conditional on $\mathcal F_{n,0}^\star$, subsequent draws are independent and uniform on $z_{1:n}$. Define
\[
\mathcal F_{n,t}^\star
=\sigma\big(\mathcal F_{n,0}^\star,(\tilde z_i)_{1\le i\le B_t}\big),\qquad t\ge1.
\]
Then $\theta_{t-1}^\star$ is $\mathcal F_{n,t-1}^\star$-measurable and
$\mathbb E_n^\star[\xi_t(\theta_{t-1}^\star)\mid\mathcal F_{n,t-1}^\star]=0$.
For refinement, initial information includes the entire first-order run and the batches used for $\Phi_n$ and $W_{\mathrm{MB}}$; only the subsequent refinement batches enter $B_t$.
}

We collectively state the necessary assumptions in the following.

\begin{assumption}
\label{asm:appx:regularity}
The following holds.
\begin{enumerate}[({A\ref{asm:appx:regularity}}.1)]

\item
\label{asm:appx:initial}
For every $\eta>0$, there exist a nonrandom $M_\eta<\infty$, independent of $n$ and $\Ns$, and $\mathcal F_{n,0}^\star$-measurable events $A_{n,\eta}$ with $\liminf\mathbb P(A_{n,\eta})\ge1-\eta$, the limit being taken along the asymptotic sequence under consideration, such that on $A_{n,\eta}$
\begin{align*}
\|\theta_0^{\star}\| \le M_\eta, \quad \bar{Q}_{n,W}(\theta_0^{\star}) \le M_\eta \gs[0].
\end{align*}
When these bounds hold surely for some nonrandom $M$, one may take $A_{n,\eta}$ to be the whole space and $M_\eta=M$.

\item
\label{asm:appx:conditioning-and-weighting}
$\P_n \ppto \P$ and $W_n \ppto W$, where $\P>0$ and $W>0$ are non-random matrices;

\item
\label{asm:appx:batch-size}
As $t \to \infty$, $B_{G,t} \to B_G$ where either $B_G \in \mathbb{N}$ or $B_G = \infty;$

\item
\label{asm:appx:sample-regularity}

The following holds for some constants $c > 0$, $\delta > 0$, $M< \infty$, and $p \ge 1:$

% Assumption~\ref{asm:regularity} holds with \ref{asm:item:local-convex-Lojasiewicz}–\ref{asm:item:stability} replaced by the following conditions for some constants $c>0$, $\delta > 0$, $M < \infty$, and $p \ge 1$:
\begin{enumerate}[({A\ref{asm:appx:regularity}.4}.1)]

\item
\label{asm:appx:item:iid}
$(z_i)_{i=1}^n \subseteq \mathcal{Z}$ are i.i.d. draws from the population distribution $\mathbb{P};$

\item
\label{asm:appx:item:interior}
{
the domain and smoothness requirements of Assumption~\ref{assm:diff} hold, and $\theta_\oo\in\operatorname{int}(\Theta)$ is the unique solution of $g(\theta)=0$;
}

\item
\label{asm:appx:item:pd}
$G_\oo' G_\oo \ge c I_{\dtheta}$ and $\Omega_\oo \ge c I_{\dg};$

\item
\label{asm:appx:item:integrability}
It is assumed that $\mathbb{E}[\|g(z_i, \theta_\oo)\|^{2p}] < \infty$.
Moreover, there exists a measurable function $H : \mathcal{Z} \to \mathbb{R}$ such that $\|G(z,\theta)\|\le H(z)$ for all $z \in \mathcal{Z}$ and $\theta \in \Theta$, and $\mathbb{E}[H(z_i)^{2p}] < \infty;$

\item
\label{asm:appx:item:Lipschitz-G}
$\|G(z, \theta)-G(z, \tilde{\theta})\| \le L(z) \|\theta-\tilde{\theta}\|$ holds for any $z \in \mathcal{Z}$ and for all $\theta$ and $\tilde{\theta}$ in $\Theta$, where $L(\cdot)$ satisfies $\mathbb{E}[L(z_i)^{2p}] < \infty;$

\item
\label{asm:appx:item:local-convex-Lojasiewicz}

$\frac{\partial^2 \bar{Q}_{n,W}(\theta)}{\partial \theta \partial \theta'}  \ge c I_{\dtheta}$ for all $\|\theta - \hat{\theta}_{n,W}\|\le \delta$ \wpa;

\item
\label{asm:appx:item:away-from-zero}

$\bar{Q}_{n,W}(\theta) \ge c$ and $\|\frac{\partial \bar{Q}_{n,W}(\theta)}{\partial \theta}\|^2 \ge c$ for all $\|\theta - \hat{\theta}_{n,W} \| \ge \delta$ \wpa;

\item
\label{asm:appx:item:stability}

$\frac{\partial^2 \bar{Q}_{n,W}(\theta)}{\partial \theta \partial \theta'}  \le M I_{\dtheta}$ and $\frac{1}{n} \sum_{i=1}^n \|g(z_i, \theta)\|^{2p} \le M (\bar{Q}_{n,W}(\theta)^p + 1)$ for all $\theta \in \Theta$ \wpa.

\end{enumerate}

%\item
%$\theta_\oo \in \Theta$ is a unique solution to $G(\theta)' W g(\theta) = 0$.
%
%identification (unique stationary point).
%
%Rank condition on G and Omega.

\end{enumerate}
\end{assumption}

\subsection{Key Lemmas}\label{appx:key-lemmas}
We begin with the standard large-sample properties of the full-sample GMM estimator.
\begin{lemma}
\label{lem:consistency-and-normality}
Under Assumption~\ref{asm:appx:regularity}, $\hat{\theta}_{n,W} \pto \theta_\oo$
and, $$\sqrt{n}(\hat{\theta}_{n,W} - \theta_\oo) \dto \mathcal{N}(0,(G_\oo' W G_\oo)^{-1} G_\oo' W \Omega_\oo W G_\oo (G_\oo' W G_\oo)^{-1}).$$
Moreover, with $A_W=G_\oo'WG_\oo$,
\begin{align*}
\sqrt n(\hat\theta_{n,W}-\theta_\oo)&=-A_W^{-1}G_\oo'W\sqrt n\,\bar g_n(\theta_\oo)+o_p(1),\\
\sqrt n\,\bar g_n(\hat\theta_{n,W})&=(I_{\dg}-G_\oo A_W^{-1}G_\oo'W)\sqrt n\,\bar g_n(\theta_\oo)+o_p(1).
\end{align*}
\end{lemma}
%%%%%%%%%%%%%%%%%%%%

Next, we collect useful asymptotic bounds for the random quantities in our analysis, implied by Assumption~\ref{asm:appx:regularity}.

\begin{lemma}
\label{lem:sample-regularity}
Let Assumption~\ref{asm:appx:regularity} hold.
Then, the following holds \wpa:
\begin{align}
\label{eq:eq1}\underbar c I_{\dtheta} &\le \P_n  \le C I_{\dtheta}, \tag{\thesection.1}\\
\label{eq:eq2}\underbar c I_{\dg} &\le W_n  \le C I_{\dg}, \tag{\thesection.2}\\
\label{eq:eq3}\underbar c I_{\dtheta} &\le \bar{G}_n' W_n \bar{G}_n \le C I_{\dtheta},\tag{\thesection.3}\\
\label{eq:eq4}\underbar c I_{\dtheta} &\le \bar{\H}_n \le C I_{\dtheta},\tag{\thesection.4}\\
\label{eq:eq5}\|\bar{G}_n(\theta)\| & \le C \ \text{for all }  \theta \in \Theta,\tag{\thesection.5}\\
\label{eq:eq6}\|\bar{G}_n(\theta) - \bar{G}_n(\tilde{\theta})\| &\le C \|\theta - \tilde{\theta}\| \ \ \text{for all } \theta, \tilde{\theta} \in \Theta,\tag{\thesection.6}\\
\label{eq:eq7}\underbar c ( \|\theta - \hat{\theta}_{n,W}\|^2 \wedge \delta^2 ) &\le \bar{Q}_{n,W}(\theta)  \ \ \text{for all }  \theta \in \Theta,\tag{\thesection.7}\\
 \label{eq:eq8}\underbar c (\bar{Q}_{n,W}(\theta) \wedge 1) &\le \left\| \frac{\partial \bar{Q}_{n,W}(\theta)}{\partial \theta} \right\|^2 \ \ \text{for all }  \theta \in \Theta,\tag{\thesection.8}\\
\label{eq:eq9}\mathbb{E}_n^{\star}[\|\xi_t(\theta) - \xi_t(\hat{\theta}_{n,W})\|^2] &\le C \|\theta - \hat{\theta}_{n,W}\|^2 \ \ \text{for all } \theta \in \Theta \ \text{and}\ t \ge 1,\tag{\thesection.9}\\
\label{eq:eq10}\mathbb{E}_n^{\star}[\|\xi_t(\theta)\|^{2p}] & \le C (\bar{Q}_{n,W}(\theta)^p + 1) \ \ \text{for all } \theta\in \Theta \ \text{and}\ t \ge 1.\tag{\thesection.10}
\end{align}
\end{lemma}

{
For each $n$, let $E_n$ be the intersection of the events on which \eqref{eq:eq1}--\eqref{eq:eq10} and the assumed sample curvature and separation bounds hold, together with $\hat\theta_{n,W}\in\operatorname{int}(\Theta)$ and

\[
\bar g_n(\hat\theta_{n,W})'W_n\bar g_n(\hat\theta_{n,W})\le1,
\qquad \|\bar g_n(\theta_\oo)\|\le1,\qquad \|\hat\theta_{n,W}\|\le C,
\]
for a fixed $C>\|\theta_\oo\|+1$. These additional bounds and the interiority condition hold with probability approaching one by Lemma~\ref{lem:consistency-and-normality} and the moment law of large numbers.
We also include in $E_n$ the sample moment bounds $n^{-1}\sum_{i=1}^nH(z_i)^{2p}\le C$ and $n^{-1}\sum_{i=1}^nL(z_i)^{2p}\le C$, together with the moment-growth bound in Condition~\ref{asm:appx:item:stability}; these hold with probability approaching one by the law of large numbers and Assumption~\ref{asm:appx:regularity}.
}
Then, Assumption~\ref{asm:appx:regularity} and Lemma~\ref{lem:sample-regularity} ensure that $\mathbb{P}(E_n) \to 1$.

\emph{Localization of Condition~\ref{asm:appx:initial}.} In Lemmas~\ref{lem:stopping-time-and-tail}--\ref{lem:CLT-extension} and their proofs, $\eta>0$ is fixed, ``on the event $E_n$'' means on $E_n\cap A_{n,\eta}$ with $A_{n,\eta}$ from Condition~\ref{asm:appx:initial}, and absolute constants may depend on $\eta$. Every use of Condition~\ref{asm:appx:initial} in these proofs is pathwise given $\mathcal F_{n,0}^\star$. Because $\limsup\mathbb P((E_n\cap A_{n,\eta})^c)\le\eta$, the probability limits asserted in Lemmas~\ref{lem:control-approx-error}--\ref{lem:CLT-extension} and in the theorems follow by letting $\eta\downarrow0$ after the limits in $n$, $\Ns$, and $\N$ are taken.

{
The event $E_n$ is $\mathcal F_{n,0}^\star$-measurable. With random preliminary weights or preconditioners it need not be measurable with respect to the data alone.
}

Relative to the natural filtration $(\mathcal{F}_{n,t}^{\star})_{t \ge 0}$, we define a stopping time
\begin{equation*}
\label{eq:stopping-time}
\stt_{\loclevel}^{\star} = \inf\left\{ t \ge 0 : \bar{Q}_{n,W}(\theta_t^{\star}) \ge \loclevel \right\} \quad \text{for}\ \loclevel > 0,
\end{equation*}
where we set $\inf \varnothing \equiv \infty$.
We also define a tail-measurable time
\begin{equation}
\label{eq:tail-time}
\tmt_{\varepsilon}^{\star} = \inf \left\{ s \ge 0 : \sup_{t \ge s} \bar{Q}_{n,W}(\theta_t^{\star}) < \varepsilon \right\}\quad \text{for}\ \varepsilon > 0.
\end{equation}

The next lemma concerns the concentration and tightness of these random times conditional on the event $E_n$.

\begin{lemma}
\label{lem:stopping-time-and-tail}
Let Assumption~\ref{asm:appx:regularity} hold.
Then the following hold.
\begin{enumerate}

\item[(i)]
Let $n \in \mathbb{N}$ be fixed.
As $\N \to \infty$, $\|\theta_{\N}^{\star} - \hat{\theta}_{n,W}\| \psto 0$ on the event $E_n$.

\item [(ii)]
For any $\varepsilon > 0$, there exists an absolute constant $\loclevel < \infty$, independent of $n$ and $\Ns$, such that
\begin{equation*}
\sup_{n \in \mathbb{N}}\sup_{\Ns\ge0}\left\{\mathbb{P}_n^{\star}\left( \stt_{\loclevel}^{\star}  < \infty \right)\mathbbm{1}_{E_n}\right\}< \varepsilon.
\end{equation*}

\item [(iii)]
For any $\varepsilon > 0$, there exists an absolute constant $\tilde{T} \in \mathbb{N}$, independent of $n$ and $\Ns$, such that
\begin{equation*}
\sup_{n \in \mathbb{N}}\sup_{\Ns\ge0}\left\{\mathbb{P}_n^{\star}\left(  \tmt_{\varepsilon}^{\star} > \tilde{T} \right)\mathbbm{1}_{E_n}\right\} < \varepsilon.
\end{equation*}

\end{enumerate}
\end{lemma}

\phantom{Explain the implications of this lemma.}

%%%%%%%%%%%%%%%%%%%%
The next lemma establishes an explicit convergence rate of the Lyapunov function evaluated at stochastic iterates.
\begin{lemma}
\label{lem:sgd-conv-rate}
Let Assumption~\ref{asm:appx:regularity} hold.
For any $\loclevel > 0$, there exists an absolute constant $C_{\loclevel} > 0$ such that, on the event $E_n$,
\begin{equation*}
    \mathbb{E}_n^{\star}[\bar{Q}_{n,W}(\theta_t^{\star}) \I{\stt_{\loclevel}^{\star} \ge t}]  \le C_\loclevel \gs
\end{equation*}
for all $n \in \mathbb{N}$, $\Ns \ge 0$, and $t \ge 1$.
\end{lemma}

%%%%%%%%%%%%%%%%%%%%

To approximate the distribution of the stochastic approximation estimator, we introduce a first-order difference sequence $(\tto{t})_{t \ge 0}$, defined as follows:
\begin{equation}
\label{eq:appx:approximating-sequence}
\tto{t} =
\tto{t-1} -  \gs \P_n \bar{\H}_n (\tto{t-1} - \hat{\theta}_{n,W}) - \gs \P_n \xi_{t}(\theta_{t-1}^{\star}), \quad  \forall t \ge 1,
\end{equation}
with the same initial value $\tto{0} = \theta_0^{\star}$ as that of $(\theta_t^{\star})_{t \ge 0}$.

The following lemma shows that the partial sum trajectory $(\sum_{s=1}^t \theta_s^{\star})_{1\le t\le \N}$ of the stochastic iterates $(\theta_t^{\star})_{t \ge 1}$ can be uniformly approximated by the partial sum trajectory of $(\tto{t})_{t \ge 1}$.
%%%%%%%%%%%%%%%%%%%%

\begin{lemma}
\label{lem:control-approx-error}
Let Assumption~\ref{asm:appx:regularity} hold.
As $n \to \infty$, $\N \to \infty$, and $\N^{1-a}/n \to 0$, it holds,
for every $\varepsilon > 0$,
\begin{equation*}
  \sup_{\Ns \ge 0}\mathbb{P}\left[  \mathbb{P}_n^{\star} \left( \sup_{1\le t\le \N} \left\| \sum_{s=1}^t (\theta_s^{\star} - \tto{s})\right\| > \sqrt{\N} \varepsilon\right) \ge \varepsilon \right] \to 0.
\end{equation*}
\end{lemma}

%%%%%%%%%%%%%%%%%%%%

The next two lemmas establish that, under suitable moment conditions and growth conditions on the pair $(\Ns, \N)$ relative to each other, the CLT  and FCLT apply to the approximating sequence, thereby enabling distributional approximation of the stochastic approximation estimator.

%%%%%%%%%%%%%%%%%%%%
\begin{lemma}
\label{lem:FCLT-extension}
Let Assumption~\ref{asm:appx:regularity} hold for $p > (1-a)^{-1}$.
Then, as $n \to \infty$, $\N \to \infty$, $\Ns/\N = O(1)$, it holds
\begin{equation*}
 \frac{1}{\sqrt{\N}}  \sum_{t=1}^{\lfloor \N r \rfloor} (\tto{t} - \hat{\theta}_{n,W}) \wto \left( (G_\oo' W G_\oo)^{-1} \Sigma_\oo (G_\oo' W G_\oo)^{-1}\right)^{1/2}W(r).
\end{equation*}
Moreover, the following linear representation holds:
\begin{equation*}
\sup_{0\le r\le1}\left\|\frac{1}{\sqrt{\N}}\sum_{t=1}^{\lfloor \N r \rfloor}(\tto{t}-\hat\theta_{n,W})+\bar{\H}_n^{-1}\frac{1}{\sqrt{\N}}\sum_{s=1}^{\lfloor \N r \rfloor}\xi_s(\theta_{s-1}^\star)\right\|\pto0,
\end{equation*}
and the convergence is $\mathcal F_{n,0}^\star$-stable: for every $\mathcal F_{n,0}^\star$-measurable random vector $X_n$ with $X_n\dto X$,
\begin{equation*}
\left(X_n,\ \frac{1}{\sqrt{\N}}\sum_{s=1}^{\lfloor \N\cdot\rfloor}\xi_s(\theta_{s-1}^\star)\right)\wto\left(X,\ \Sigma_\oo^{1/2}W(\cdot)\right),
\end{equation*}
where the standard $\dtheta$-dimensional Wiener process $W(\cdot)$ is independent of $X$.
\end{lemma}
% {\color{red} The condition $\Ns^{a}/\N = o(1)$ has changed.

% Condition for Theorem 4 : $(T-N)^{1-a}/n \to 0$, $N/(T-N) = O(1)$, $N^a / n = O(1)$}

%%%%%%%%%%%%%%%%%%%%

\begin{lemma}
\label{lem:CLT-extension}
Let Assumption~\ref{asm:appx:regularity} hold for $p > 1$.
Then, as $n \to \infty$, $\N \to \infty$, $\Ns / \N = O(1)$, it holds
\begin{equation*}
\sqrt{\N}( \tto[\bar]{\N}   - \hat{\theta}_{n,W}  )\dto \mathcal{N}\left( 0, (G_\oo' W G_\oo)^{-1} \Sigma_\oo (G_\oo' W G_\oo)^{-1} \right),
\end{equation*}
where $\tto[\bar]{\N}=\N^{-1}\sum_{t=1}^{\N}\tto{t}$. Moreover,
\begin{equation*}
\sqrt{\N}( \tto[\bar]{\N}   - \hat{\theta}_{n,W}  )=-\bar{\H}_n^{-1}\frac{1}{\sqrt{\N}}\sum_{s=1}^{\N}\xi_s(\theta_{s-1}^\star)+o_p(1),
\end{equation*}
and the convergence is $\mathcal F_{n,0}^\star$-stable: for every $\mathcal F_{n,0}^\star$-measurable random vector $X_n$ with $X_n\dto X$, the pair $(X_n,\sqrt{\N}( \tto[\bar]{\N}   - \hat{\theta}_{n,W}))$ converges in distribution to $(X,\zeta)$, where $\zeta\sim\mathcal{N}( 0, (G_\oo' W G_\oo)^{-1} \Sigma_\oo (G_\oo' W G_\oo)^{-1})$ is independent of $X$.
\end{lemma}

%%%%%%%%%%%%%%%%%%%%

\paragraph{Lemmas for the Refinement Stage.}
For the second-order refinement in Section~\ref{sec:extensions}, we next verify Conditions~\ref{asm:appx:conditioning-and-weighting} and~\ref{asm:appx:initial}. As described in Section~\ref{appx:proof-setup}, the refinement takes the form \eqref{eq:appx:MB-GD-updating-rule} with $W_n=W_{\mathrm{MB}}${} (either~\eqref{def:Phi:W} or~\eqref{def:centered W}), $\P_n=(\Phi_n'W_{\mathrm{MB}}\Phi_n)^\dagger$, $\Ns=N$, and initial value $\theta_0^\star=\theta_N^*${} (or $\bar\theta_N^*$; see Lemma~\ref{lem:refinement-initial}). Thus $\gamma_0^\star=\gamma_0N^{-a}$, and $\hat\theta_{n,W}$ and $\bar Q_{n,W}$ are those of Section~\ref{sec:extensions}. Let $E_n^{(2)}$ denote the event $E_n$ constructed for this choice of $W_n$ and $\P_n$.

\begin{lemma}
\label{lem:refinement-weight}
Under the assumptions of Theorem~\ref{thm:asymp:normal:second:order}(i), as $n$, $N$, and $M_{\mathrm{MB}}\to\infty$,
\begin{equation*}
\Phi_n\pto G_\oo,\qquad W_{\mathrm{MB}}\pto\Omega_\oo^{-1},\qquad (\Phi_n'W_{\mathrm{MB}}\Phi_n)^\dagger\pto(G_\oo'\Omega_\oo^{-1}G_\oo)^{-1}.
\end{equation*}
Hence Condition~\ref{asm:appx:conditioning-and-weighting} holds for the refinement stage with $W=\Omega_\oo^{-1}$ and $\P=(G_\oo'\Omega_\oo^{-1}G_\oo)^{-1}$. Moreover, the conclusions of Lemmas~\ref{lem:consistency-and-normality} and~\ref{lem:sample-regularity} hold for the refinement stage, and $\mathbb P(E_n^{(2)})\to1$. These conclusions hold for $W_{\mathrm{MB}}$ in~\eqref{def:Phi:W} and for $W_{\mathrm{MB}}^c$ in~\eqref{def:centered W}, also when $\bar g_n(\bar\theta_N^*)$ in~\eqref{def:centered W} is replaced by an approximation $\hat g$ with $\hat g-\bar g_n(\bar\theta_N^*)\pto0$.
\end{lemma}

\begin{lemma}
\label{lem:refinement-initial}
Under the assumptions of Theorem~\ref{thm:asymp:normal:second:order}(i), Condition~\ref{asm:appx:initial} holds for the refinement stage: for every $\eta>0$, there exist a nonrandom $M_\eta<\infty$, independent of $n$ and $N$, and $\mathcal F_{n,0}^\star$-measurable events $A_{n,\eta}$ with $\liminf\mathbb P(A_{n,\eta})\ge1-\eta$ as $n$, $N$, and $M_{\mathrm{MB}}\to\infty$, such that on $A_{n,\eta}$
\begin{equation*}
\|\theta_N^*\|\le M_\eta,\qquad \bar Q_{n,W}(\theta_N^*)\le M_\eta\gamma_0N^{-a}.
\end{equation*}
The same holds with $\bar\theta_N^*$ in place of $\theta_N^*$, that is, when the refinement is initialized at $\bar\theta_N^*$.
\end{lemma}

%%%%%%%%%%%%%%%%%%%%

\paragraph{Other Technical Lemmas.}
\begin{lemma}
\label{lem:Robbins-Siegmund}
Suppose that $X_n, A_n, C_n$, and $D_n$ are finite, non-negative random variables, adapted to the filtration $\{ \mathcal{F}_n \}_{n =0}^\infty$, which satisfy
\begin{equation*}
\mathbb E [X_{n+1}|\mathcal{F}_n] \le (1 + A_n) X_n + C_n - D_n.
\end{equation*}
Then, on the event $\{\sum_{n=1}^\infty A_n < \infty,\ \sum_{n=1}^\infty C_n < \infty\}$, we have
\begin{equation*}
\sum_{n=1}^\infty D_n <\infty \quad  \text{ and } \quad X_n \to X
\end{equation*}
almost surely for some random variable $X$.
\end{lemma}

\begin{lemma}
\label{lem:recursive-bound}
Let $(\gamma_t)_{t\ge1}$ be a positive nonincreasing sequence; for example, $\gamma_t=\gamma_0t^{-a}$ or $\gamma_t=\gs$ with any offset $\Ns\ge0$.
Assume that $(v_n)_{n \ge 0}$ is a {nonnegative} sequence that satisfies for all $n \ge n_0$ ($n_0 \ge 0$) and for a given $\mu \in (0, 1/\gamma_{n_0+1})$:
\begin{equation*}
    v_{n+1} \le (1-\mu \gamma_{n+1}) v_n + C \gamma_{n+1},
\end{equation*}
then for all $n \ge n_0$,
\begin{equation*}
    v_n \le \max\{C,1\} (v_{n_0} + \mu^{-1}).
\end{equation*}

\end{lemma}
%%%%%%%%%%%%%%%%%%%%

For a $d$-dimensional matrix $A$, denote its eigenvalues by $\lambda_i(A)$, $i=1,\ldots,d$, counted with multiplicity.
We write $\mu_{\max}(A)$ and $\mu_{\min}(A)$ for $\max_{1\le i\le d} \re(\lambda_i(A))$ and $\min_{1\le i\le d} \re(\lambda_i(A))$, respectively.

\begin{lemma}
\label{lem:control-alpha-w}
Let $A$ be a $d$-dimensional matrix such that $\mu_{\min}(A) > 0$.
Then, there exists a constant $M_1(A,a,\gamma_0) <
 \infty$ that depend continuously on $A$, $a$, and $\gamma_0$, but not on $\Ns$, such that
\begin{equation*}
\|\alpha_{s}^{t}(A)\| \le M_1(A,a,\gamma_0)
\end{equation*}
for all $0 \le s \le t$ and $t \ge 1$,
where $\alpha_{s}^{t}(A)$ is defined as\footnote{We follow the convention that $\prod_{k=s+1}^s \cdot$ is an identity matrix.}
\begin{equation*}
\alpha_{s}^{t}(A) = \gs[s] \sum_{i=s}^t \prod_{k=s+1}^i (I_d - \gs[k] A).
\end{equation*}
Moreover, there exists a constant $M_2(A,a,\gamma_0) < \infty$ that depends continuously on $A$, $a$, and $\gamma_0$, but not on $\Ns$, such that, for all $t \ge 1$ and $\Ns \ge 0$,
\begin{equation*}
{\sum_{s=1}^t \|\alpha_{s}^{t}(A) - A^{-1}\| \le M_2(A,a,\gamma_0) (t+\Ns)^a.}
\end{equation*}
\end{lemma}

The next lemma is a martingale central limit theorem whose limit is independent of the initial information. It is applied to the stochastic approximation noise in Lemmas~\ref{lem:FCLT-extension} and~\ref{lem:CLT-extension}, and to the moment-batch and cluster noise in the proofs of Theorems~\ref{thm:Plug-in-and-online-J-tests} and~\ref{thm:clpop:joint}.

\begin{lemma}
\label{lem:stable-mclt}
For each $n$, let $(\mathcal G_{n,t})_{t\ge0}$ be a filtration, and let $(e_{n,s})_{s\ge1}$ be $\mathbb R^k$-valued random vectors such that $e_{n,s}$ is $\mathcal G_{n,s}$-measurable, $\mathbb E[\|e_{n,s}\|^2\mid\mathcal G_{n,s-1}]<\infty$, and $\mathbb E[e_{n,s}\mid\mathcal G_{n,s-1}]=0$. Put $S_n(r)=\N^{-1/2}\sum_{s=1}^{\lfloor\N r\rfloor}e_{n,s}$ for $r\in[0,1]$. Suppose that, as $n\to\infty$ and $\N\to\infty$,
\begin{enumerate}[(i)]
\item $\N^{-1}\sum_{s=1}^{\N}\mathbb E[\|e_{n,s}\|^2\mathbbm1\{\|e_{n,s}\|>\sqrt\N\,\delta\}\mid\mathcal G_{n,s-1}]\pto0$ for every $\delta>0$;
\item $\N^{-1}\sum_{s=1}^{\lfloor\N r\rfloor}\mathbb E[e_{n,s}e_{n,s}'\mid\mathcal G_{n,s-1}]\pto rV$ for every $r\in[0,1]$, where $V$ is a nonrandom positive semidefinite matrix.
\end{enumerate}
Then, for every $\mathcal G_{n,0}$-measurable random vector $X_n$ with $X_n\dto X$,
\begin{equation*}
\big(X_n,S_n(\cdot)\big)\wto\big(X,V^{1/2}W(\cdot)\big),
\end{equation*}
where the standard $k$-dimensional Wiener process $W(\cdot)$ is independent of $X$. If~(ii) holds only for $r=1$, then $(X_n,S_n(1))\dto(X,V^{1/2}Z)$, where $Z\sim\mathcal N(0,I_k)$ is independent of $X$.
\end{lemma}

\paragraph{Lemmas for the Fixed-Sample Case.}

\begin{lemma}
\label{lem:control-approx-error-fixed-n}
Let $z_{1:n}$ and $n \in \mathbb{N}$ be fixed.
Let $\varepsilon>0$ be arbitrary.
Under the assumptions of Theorem~\ref{thm:data:fixed}, as $N \to \infty$, it holds
\begin{equation*}
   \mathbb{P}_n^* \left( \sup_{1\le t\le N} \left\| \sum_{s=1}^t (\theta_s^* - \tto{s})\right\| > \sqrt{N} \varepsilon\right) \to 0.
\end{equation*}
\end{lemma}

\begin{lemma}
\label{lem:FCLT-fixed-n}
Let $z_{1:n}$ and $n \in \mathbb{N}$ be fixed.
Under the assumptions of Theorem~\ref{thm:data:fixed}, as $N \to \infty$,
\begin{equation*}
     \frac{1}{\sqrt{N}}  \sum_{t=1}^{\lfloor N r \rfloor} (\tto{t} - \hat{\theta}_n)\wto \left( \bar{\H}_n^{-1} \bar\Sigma_n \bar{\H}_n^{-1}\right)^{1/2}W(r).
\end{equation*}
\end{lemma}
The proofs of Lemmas~\ref{lem:control-approx-error-fixed-n} and~\ref{lem:FCLT-fixed-n} use two facts established in the first part of the proof of Theorem~\ref{thm:data:fixed}, which does not rely on these lemmas: $\nabla\bar Q_n(\hat\theta_n)=0$, and $\theta_t^*\to\hat\theta_n$ $\mathbb P_n^*$-almost surely.

\paragraph{Cluster Sampling.}
{
The following lemma verifies the stochastic-error conditions used to transfer the fixed-data results in Appendix~\ref{sec:setting:fixed} to the cluster sampling scheme in Appendix~\ref{sec:cluster}. Its proof is given in Appendix~\ref{appx:proofs-lems}.
}
For all $\theta \in \Theta$, write
\begin{equation*}
\bar{\Omega}^{\mathrm{cl}}_K(\theta) := \frac{1}{K} \sum_{k=1}^K \hat{g}_k(\theta) \hat{g}_k(\theta)',
\end{equation*}
so that $\bar{\Omega}^{\mathrm{cl}}_K = \bar{\Omega}^{\mathrm{cl}}_K(\hat{\theta}_n)$.

\begin{lemma}[Verification of Assumption~\ref{assm:regularity:data:fixed:2} under cluster sampling]
\label{lem:cluster:verify}
Let Assumptions \ref{assm:uniform} $($with clusters as sampling units$)$, \ref{assm:lr}, \ref{assm:diff}, \ref{assm:regularity:data:fixed:1}, and \ref{assm:cluster:reg} hold.
% [G7 patch] Removed the repeated convexity hypothesis and the unnecessary unit residual bound: convexity is part of Assumption~\ref{assm:diff}, and Condition~\ref{assm:cluster:item:stability} gives a finite residual bound by Jensen's inequality. The proof does not require that bound to be one (see steps (c)--(d) of the proof of Lemma~\ref{lem:cluster:verify}).
Then the stochastic errors $\xi_t(\theta)$ in \eqref{eq:cluster:xi} satisfy Assumption~\ref{assm:regularity:data:fixed:2}$:$ there exist finite constants $C$ and $M'$ such that, for all $\theta \in \Theta$ and $t \ge 1$,
\begin{align}
\mathbb{E}_n^*[\|\xi_t(\theta) - \xi_t(\hat{\theta}_n)\|^2] &\le C \|\theta - \hat{\theta}_n\|^2, \label{eq:cluster:eq9}\\
\mathbb{E}_n^*[\|\xi_t(\theta)\|^{2p}] &\le M' (\bar{Q}_n(\theta)^p + 1). \label{eq:cluster:eq10}
\end{align}
Moreover, the conditional variance evaluated along the iterate path satisfies
\begin{equation}\label{eq:cluster:var-limit}
\mathbb{E}_n^*[\xi_t(\theta_{t-1}^*)\xi_t(\theta_{t-1}^*)' \mid \mathcal{F}_{n,t-1}^*] \to \bar{\Sigma}^{\mathrm{cl}}_K \quad \text{as } \theta_{t-1}^* \to \hat{\theta}_n,
\end{equation}
{with $\bar\Sigma_K^{\mathrm{cl}}$ defined by the exact expression \eqref{eq:cluster:Sigma}.}
\end{lemma}

\paragraph{Debiased $J$-Test.}
{
The following lemma establishes the general testing result used by Theorem~\ref{thm:cluster:debiased-J}. It uses its own sampling and accuracy conditions, without requiring an efficient preliminary estimate or a computational FCLT.
Write $m=\dg$ and $r=m-\dtheta>0$. For an estimate $\tilde\theta$, define $\hat g=\bar g_n(\tilde\theta)$ and $\hat D=\bar G_n(\tilde\theta)$. Let $\hat V$ be a symmetric estimate of the covariance of the sample moment, and choose an $r\times m$ matrix $\hat H$ of full row rank satisfying $\hat H\hat D=0$. Set
\begin{equation}\label{eq:general:debiased-J}
\hat h=\hat H\hat g,\qquad \hat V_h=\hat H\hat V\hat H',\qquad
J_D(\tilde\theta)=\hat h'\hat V_h^{-1}\hat h.
\end{equation}
For a fixed partition $\hat D=(\hat D_1',\hat D_2')'$ with a nonsingular $d\times d$ block $\hat D_1$, one may take $\hat H=[-\hat D_2\hat D_1^{-1}\ \ I_r]$. The statistic is invariant to nonsingular changes of residual coordinates.

\begin{lemma}[Debiased $J$-Test at an Approximate Estimate]\label{lem:debiased-J:general}
Let $m,d$ be fixed and $K\to\infty$. Under the null, suppose an interior parameter $\theta_\oo$ satisfies the maintained moment restrictions and:
\begin{enumerate}[(i)]
\item $\sqrt K\,\bar g_n(\theta_\oo)\dto\mathcal N(0,\Omega)$ for a finite positive-semidefinite matrix $\Omega$.
\item On a fixed convex neighborhood of $\theta_\oo$, $\bar G_n(\theta)$ is Lipschitz with constant $O_p(1)$, and $\bar G_n(\theta_\oo)\pto D_0$ of full column rank.
\item $\tilde\theta-\theta_\oo=o_p(K^{-1/4})$, with probability taken over both the data and the algorithm.
\item $K\hat V\pto\Omega$, including evaluation at the random estimate $\tilde\theta$.
\item $\hat H\hat D=0$, $\hat H\pto H_0$ of full row rank $r$, and $H_0\Omega H_0'$ is positive definite.
\end{enumerate}
Then $J_D(\tilde\theta)\dto\chi^2_r$.
\end{lemma}

The residual form only requires $\hat V_h$ to be nonsingular, allowing a singular full covariance when the residual restrictions remain nondegenerate. When $\hat V>0$, it equals
\[
\hat g'\big[\hat V^{-1}-\hat V^{-1}\hat D
 (\hat D'\hat V^{-1}\hat D)^{-1}\hat D'\hat V^{-1}\big]\hat g.
\]
Indeed, with $B=\hat V^{-1/2}\hat D$, the rows of $\hat H\hat V^{1/2}$ span the orthogonal complement of the columns of $B$. Their orthogonal projector is $I-B(B'B)^{-1}B'$, yielding the identity.

More generally, if a scalar rate $a_K$ satisfies $a_K\bar g_n(\theta_\oo)\dto\mathcal N(0,\Omega)$ and $a_K^2\hat V\pto\Omega$, the same proof requires $a_K\|\tilde\theta-\theta_\oo\|^2=o_p(1)$. If the full-sample Jacobian is replaced by $\tilde D$ and residual coordinates annihilate $\tilde D$, the additional first-order term is negligible under
\[
\sqrt K\,\|\tilde D-\bar G_n(\tilde\theta)\|
             \|\tilde\theta-\theta_\oo\|=o_p(1),
\]
together with the corresponding convergence and rank conditions on the residual coordinates. A full-sample derivative avoids this approximation. The proof of the lemma is in Appendix~\ref{appx:proofs-lems}.
}

\subsection{Proofs of Corollaries~\ref{cor:inference:random} and~\ref{cor:plugin:eff}}\label{appx:proofs-cors}

\subsubsection{Proof of Corollary~\ref{cor:inference:random}}
Write $H=G_\oo'G_\oo$, $\Lambda=H^{-1}\Sigma_\oo H^{-1}$, $V_s=H^{-1}G_\oo'\Omega_\oo G_\oo H^{-1}$, $\tau_n=(n^{-1}+(NB_g)^{-1})^{-1}$, and $\lambda_n=\tau_n/n\in(0,1)$, so that $\tau_n/N=(1-\lambda_n)B_g$. The matrices $R\Lambda R'$ and $RV_sR'$ are positive definite, and $B_g\Lambda-V_s=H^{-1}(B_g\Sigma_\oo-G_\oo'\Omega_\oo G_\oo)H^{-1}\succeq0$.

Fix a subsequence along which $\lambda_n\to\lambda\in[0,1]$, and let $S_N(r)=N^{-1/2}\sum_{t=1}^{\lfloor Nr\rfloor}(\theta_t^*-\hat\theta_n)$. Under $H_0$,
\[
\sqrt{\tau_n}\,(R\bar\theta_N^*-c)=\sqrt{\lambda_n}\,R\sqrt n(\hat\theta_n-\theta_\oo)+\sqrt{(1-\lambda_n)B_g}\,RS_N(1),
\]
and, as in the proof of Corollary~\ref{cor:cluster:RS} in Section~\ref{appx:proofs-cluster} below,
\[
V_N(R)=\int_0^1R\{S_N(r)-rS_N(1)\}\{S_N(r)-rS_N(1)\}'R'\,dr+o_p(1).
\]
By Theorem~\ref{thm:FCLT} and the continuous mapping theorem, $T_{n,N}\dto Y'\mathcal J_\ell^{-1}Y$, where $\mathcal J_\ell=\int_0^1\bar W_\ell(r)\bar W_\ell(r)'dr$ and
\[
Y=\sqrt\lambda\,(B_gR\Lambda R')^{-1/2}(RV_sR')^{1/2}Z_\ell+\sqrt{1-\lambda}\,W_\ell(1).
\]
Here $Z_\ell$ is independent of $W_\ell$, and $W_\ell(1)$ is independent of the bridge $\bar W_\ell$. Hence $Y$ is independent of $\mathcal J_\ell$, and
\[
\operatorname{Var}(Y)=\lambda(B_gR\Lambda R')^{-1/2}RV_sR'(B_gR\Lambda R')^{-1/2}+(1-\lambda)I_\ell\preceq I_\ell,
\]
with $\operatorname{Var}(Y)$ positive definite. Let $Z\sim\mathcal N(0,I_\ell)$ be independent of $\mathcal J_\ell$. Conditional on $\mathcal J_\ell$, $Y'\mathcal J_\ell^{-1}Y$ and $Z'\mathcal J_\ell^{-1}Z$ are distributed as weighted sums of independent $\chi^2_1$ variables, with weights equal to the eigenvalues of $\mathcal J_\ell^{-1/2}\operatorname{Var}(Y)\mathcal J_\ell^{-1/2}$ and of $\mathcal J_\ell^{-1}$, respectively. Since $\mathcal J_\ell^{-1/2}\operatorname{Var}(Y)\mathcal J_\ell^{-1/2}\preceq\mathcal J_\ell^{-1}$, the ordered eigenvalues of the former do not exceed those of the latter. Coupling the two weighted sums through common $\chi^2_1$ variables, as in the proof of Corollary~\ref{cor:clpop:RS} below, shows that $Y'\mathcal J_\ell^{-1}Y$ is stochastically dominated by $Z'\mathcal J_\ell^{-1}Z$, whose $(1-\alpha)$ quantile is $\mathrm{cv}_{\ell,1-\alpha}$ by definition. Because $Y'\mathcal J_\ell^{-1}Y$ has a continuous distribution, $\lim\mathbb P(T_{n,N}>\mathrm{cv}_{\ell,1-\alpha})\le\alpha$ along the subsequence. Since every subsequence has a further subsequence along which $\lambda_n$ converges, the stated limsup bound follows.

Since the argument uses $N\asymp n$ only through subsequential limits $\lambda\in[0,1]$, the bound holds for any $n,N\to\infty$ satisfying the conditions of Theorem~\ref{thm:FCLT}, with equality when $\lambda=0$.

\subsubsection{Proof of Corollary~\ref{cor:plugin:eff}}
Put $A=G_\oo'\Omega_\oo^{-1}G_\oo$, $\tau=(n^{-1}+((T-N)B_g)^{-1})^{-1}$, and $\lambda_n=\tau/n\in(0,1)$, so that $\tau/(T-N)=(1-\lambda_n)B_g$. With $B_G=\infty$, Theorem~\ref{thm:asymp:normal:second:order}(i) gives the joint convergence of $\sqrt n(\hat\theta_{n,W}-\theta_\oo)$ and $\sqrt{T-N}(\bar\theta_T^*-\hat\theta_{n,W})$ to independent normal vectors with covariance matrices $A^{-1}$ and $B_g^{-1}A^{-1}$. Fix a subsequence along which $\lambda_n\to\lambda\in[0,1]$. Under $H_0$,
\begin{align*}
\sqrt\tau\,(R\bar\theta_T^*-c)&=\sqrt{\lambda_n}\,R\sqrt n(\hat\theta_{n,W}-\theta_\oo)+\sqrt{(1-\lambda_n)B_g}\,R\sqrt{T-N}(\bar\theta_T^*-\hat\theta_{n,W})\\
&\dto\mathcal N(0,RA^{-1}R'),
\end{align*}
because $\lambda RA^{-1}R'+(1-\lambda)B_g\cdot B_g^{-1}RA^{-1}R'=RA^{-1}R'$. The statistic in \eqref{def:Wald:eff:in-feasible} equals $\tau(R\bar\theta_T^*-c)'(RA^{-1}R')^{-1}(R\bar\theta_T^*-c)$, so it converges in distribution to $\chi^2_\ell$ along the subsequence. Since the limit does not depend on $\lambda$ and every subsequence has a further subsequence along which $\lambda_n$ converges, the convergence holds along the full sequence. For the feasible statistic \eqref{def:Wald:eff:feasible}, $\bar\theta_T^*\pto\theta_\oo$ and the local uniform laws of large numbers used in the proof of Lemma~\ref{lem:refinement-weight} give $\Phi_n(\bar\theta_T^*)\pto G_\oo$ and $W_n(\bar\theta_T^*)\pto\Omega_\oo^{-1}$, so the same limit follows by Slutsky's theorem.

\subsection{Proofs of Results for Fixed Data}\label{appx:proofs-fixed}

\subsubsection{Proof of Theorem~\ref{thm:data:fixed}}
We use $\hat\theta_n$ and $\bar Q_n$ for the identity-weight versions of $\hat\theta_{n,W}$ and $\bar Q_{n,W}$.

We first establish $\nabla\bar Q_n(\hat\theta_n)=0$ under the full assumptions of the theorem. Write $Q_t=\bar Q_n(\theta_t^*)$. Assumption~\ref{assm:diff} keeps the iterates in the convex domain. The upper Hessian bound, conditional unbiasedness, and the noise bound $\mathbb E_n^*[\|\xi_{t+1}(\theta_t^*)\|^2\mid\mathcal F_{n,t}^*]\le C(Q_t+1)$ give, by Taylor expansion, for all sufficiently large $t$,
\[
\mathbb E_n^*[Q_{t+1}\mid\mathcal F_{n,t}^*]
\le (1+C\gamma_{t+1}^2)Q_t
-\frac{\gamma_{t+1}}2\|\nabla\bar Q_n(\theta_t^*)\|^2
+C\gamma_{t+1}^2.
\]
Lemma~\ref{lem:Robbins-Siegmund} and $\sum_t\gamma_t^2<\infty$ imply
\[
\sum_{t\ge0}\gamma_{t+1}\|\nabla\bar Q_n(\theta_t^*)\|^2<\infty
\quad\mathbb P_n^*\text{-almost surely}.
\]
Since $\sum_t\gamma_t=\infty$, a subsequence has gradient norms tending to zero. For $d=\theta-\hat\theta_n$ with $\|d\|\le\delta$, constrained optimality and local strong convexity give $d'\nabla\bar Q_n(\theta)\ge c\|d\|^2$. Outside this ball, the gradient separation condition bounds the squared norm below by $c$. The subsequence therefore approaches $\hat\theta_n$, and continuity yields $\nabla\bar Q_n(\hat\theta_n)=0$.

We now establish consistency.
{Under Assumptions~\ref{assm:regularity:data:fixed:1} and~\ref{assm:regularity:data:fixed:2}{} and the stationarity just established, the bounds \eqref{eq:eq4}--\eqref{eq:eq10} hold with finite constants depending on the fixed sample. The drift bound uses the finite residual $\|\bar g_n(\hat\theta_n)\|$, without requiring it to be at most one. Condition~\ref{asm:appx:initial} holds with $A_{n,\eta}$ equal to the whole space, because $\|\theta_0^*\|\le M_0$ and Condition~\ref{assm:fixed:item:G-bounded} bound $\bar Q_n(\theta_0^*)$ by a constant depending on the fixed sample, while $\gamma_0^\star=\gamma_0$ for $\Ns=0$. Thus the proofs of Lemmas~\ref{lem:stopping-time-and-tail} and~\ref{lem:sgd-conv-rate} apply with $W_n=\P_n=I$ and $\Ns=0$, without invoking the random-sampling event $E_n$.}
The assertion that 
\begin{equation*}
\mathbb{P}_n^*\left( \theta_N^{*} \to \hat{\theta}_{n}\ \ \text{as}\ \ N \to \infty\right)=1
\end{equation*}
then follows from {the proof of} Lemma~\ref{lem:stopping-time-and-tail}(i).
As a result, the averaged estimator $\bar \theta_N^{*} = \frac{1}{N} \sum_{t=1}^N \theta_t^{*}$ also converges to the same limit $\mathbb{P}_n^{*}$-almost surely.
 We now establish the FCLT.
We write
\begin{equation*}
    \frac{1}{\sqrt{N}}\sum_{t=1}^{\lfloor N r \rfloor} (\theta_t^{*}-\hat\theta_n)= \frac{1}{\sqrt{N}}\sum_{t=1}^{\lfloor N r \rfloor} (\theta_t^{*}-\tto{t}) + \frac{1}{\sqrt{N}}\sum_{t=1}^{\lfloor N r \rfloor} (\tto{t}-\hat\theta_n).
\end{equation*}
For the first term, Lemma~\ref{lem:control-approx-error-fixed-n} implies that
$$
\sup_{r \in [0,1]} \left\|\frac{1}{\sqrt{N}}\sum_{t=1}^{\lfloor N r \rfloor} (\theta_t^{*}-\tto{t})\right\| = \frac{1}{\sqrt{N}}\sup_{1\le t\le N} \left\|\sum_{s=1}^{t} (\theta_s^{*}-\tto{s})\right\| \overset {\mathbb{P}^{*}} \to 0.
$$
For the second term, Lemma~\ref{lem:FCLT-fixed-n} establishes the FCLT.
Putting these together completes the proof.
%we have, as $N \to \infty$,
%$$ \frac{1}{\sqrt{N}}\sum_{t=1}^{\lfloor N r \rfloor} (\theta_t^{*}-\hat\theta_n)\wto \left( \bar{\H}_n^{-1} \bar \Sigma_n \bar{\H}_n^{-1}\right)^{1/2}W(r). $$

\subsubsection{Proof of Corollary~\ref{cor:Wald:fixed}}
Since $NB_g(B_gV_N(R))^{-1}=NV_N(R)^{-1}$, the statistic equals
\[
N(R\bar\theta_N^*-c)'V_N(R)^{-1}(R\bar\theta_N^*-c).
\]
The proof of Corollary~\ref{cor:cluster:RS} in Section~\ref{appx:proofs-cluster} then applies verbatim, with $\Lambda_n=\bar{\H}_n^{-1}\bar\Sigma_n\bar{\H}_n^{-1}$ and with Theorem~\ref{thm:data:fixed} in place of Theorem~\ref{thm:cluster:FCLT}.

%%%%%%%%%%%%%%%%%%%%%%%%%%%%%%%%%%%%%%%%%%%%%%%%%%%%%%%%%%%%%%%%%%%%%%%%%%
\subsection{Proofs of Results for Clustered Data}\label{appx:proofs-cluster}

\subsubsection{Proof of Theorem~\ref{thm:cluster:FCLT}}
By Lemma~\ref{lem:cluster:verify}, the cluster-resampled stochastic errors \eqref{eq:cluster:xi} satisfy Assumption~\ref{assm:regularity:data:fixed:2}, with steady-state conditional variance $\bar{\Sigma}^{\mathrm{cl}}_K$ by \eqref{eq:cluster:var-limit}.
The conclusion then follows from Theorem~\ref{thm:data:fixed} applied to the pseudo-sample $\{(\hat{g}_k, \hat{G}_k)\}_{k=1}^K$, with $\bar{\Sigma}_n$ replaced by $\bar{\Sigma}^{\mathrm{cl}}_K$.

\subsubsection{Proof of Corollary~\ref{cor:cluster:RS}}
{
Let $\Lambda_n=\bar{\H}_n^{-1}\bar\Sigma_K^{\mathrm{cl}}\bar{\H}_n^{-1}$ and
$S_N(r)=N^{-1/2}\sum_{t=1}^{\lfloor Nr\rfloor}(\theta_t^*-\hat\theta_n)$.

Theorem~\ref{thm:cluster:FCLT} implies $RS_N\rightsquigarrow^*Y$, where
$Y(r)=(R\Lambda_nR')^{1/2}W_\ell(r)$ in distribution. Under the null,
$\sqrt N(R\bar\theta_N^*-c)=RS_N(1)$, and
\[
V_N(R)=\int_0^1 R\{S_N(r)-rS_N(1)\}
                  \{S_N(r)-rS_N(1)\}'R'\,dr+o_{\mathbb P_n^*}(1).
\]
The continuous mapping theorem gives the limit
\[
Y(1)'\left[\int_0^1\{Y(r)-rY(1)\}\{Y(r)-rY(1)\}'\,dr\right]^{-1}Y(1).
\]
Since $R\Lambda_nR'$ is positive definite, its square root cancels. The Brownian-bridge integral is almost surely nonsingular, giving the stated limit.
}

\subsubsection{Proof of Theorem~\ref{thm:clpop:joint}}
{
We give the transfer argument explicitly, using the conditional bounds in the supporting lemmas of this appendix.

\emph{Bounds of Lemma~\ref{lem:sample-regularity}.} Fix $\eta>0$, and work on the events of Condition~\ref{assm:clpop:item:regularity}, intersected with events of probability approaching one on which the eigenvalues of $W_K$ and $P_K$ lie in $[\underbar c,C]$; such events exist by Condition~\ref{assm:clpop:item:weights}. On these events, the bounds \eqref{eq:eq1}--\eqref{eq:eq10} hold for the weighted cluster update, with cluster averages in place of observation averages, $Q_K$ in place of $\bar Q_{n,W}$, and noise $\widetilde D_t(\theta)'W_K\widetilde u_t(\theta)-\bar G_n(\theta)'W_K\bar g_n(\theta)$. Indeed, \eqref{eq:eq1}--\eqref{eq:eq2} hold by construction; \eqref{eq:eq5}--\eqref{eq:eq6} follow from Conditions~\ref{assm:cluster:item:envelope}--\ref{assm:cluster:item:Lipschitz} by Jensen's inequality; \eqref{eq:eq4} and \eqref{eq:eq7} follow from parts~(b) and~(c) of Condition~\ref{assm:clpop:item:regularity}, the latter through the variational inequality for the minimizer $\hat\theta_{n,W}$ over the convex set $\Theta$; and \eqref{eq:eq9}--\eqref{eq:eq10} follow from steps~(c) and~(d) of the proof of Lemma~\ref{lem:cluster:verify}, with constants multiplied by powers of $\|W_K\|$. The bounds \eqref{eq:eq3} and \eqref{eq:eq8} hold with probability approaching one, as in the proof of Lemma~\ref{lem:sample-regularity}, because $\hat\theta_{n,W}$ is consistent, and hence interior, by the sampling expansion below, and $\bar G_n(\hat\theta_{n,W})\pto D_0$ by Condition~\ref{assm:clpop:item:sampling} and \eqref{eq:eq6}. The constants depend on $\eta$ but not on $K$; as in Section~\ref{appx:key-lemmas}, the probability limits below are first obtained on these events and then follow by letting $\eta\downarrow0$.

\emph{Sampling expansion.} Since $Q_K(\theta_\oo)\le \tfrac12\bar g_n(\theta_\oo)'W_K\bar g_n(\theta_\oo)=O_p(K^{-1})$, bound \eqref{eq:eq7} gives $\hat\theta_{n,W}-\theta_\oo=O_p(K^{-1/2})$. Expanding the interior first-order condition and using the Jacobian bounds yields
\[
\sqrt K(\hat\theta_{n,W}-\theta_\oo)
=-J_W^{-1}D_0'W\sqrt K\bar g_n(\theta_\oo)+o_p(1).
\]
Consistency of $W_K$ suffices here, even if it depends on the preliminary algorithm. Also $q_K=\bar g_n(\hat\theta_{n,W})=O_p(K^{-1/2})$.

\emph{Conditional covariance.} Set $\widehat D=\bar G_n(\hat\theta_{n,W})$, $\Delta_k=D_k(\hat\theta_{n,W})-\widehat D$, and
$U_K=K^{-1}\sum_k u_k(\hat\theta_{n,W})u_k(\hat\theta_{n,W})'-q_Kq_K'$. Independence of the two batches gives the exact gradient-noise covariance
\begin{align*}
\Sigma_{K,t}
&=\frac1{B_g}\widehat D'W_KU_KW_K\widehat D\\
&\quad+\frac1{Kb_t}\sum_k\Delta_k'W_K
 \left(\frac{U_K}{B_g}+q_Kq_K'\right)W_K\Delta_k.
\end{align*}
The envelope and Lipschitz bounds allow evaluation at $\hat\theta_{n,W}$ instead of $\theta_\oo$. Consequently $U_K\pto\Omega$ and, uniformly in $t$,
\[
\Sigma_{K,t}
=\frac1{B_g}\left(D_0'W\Omega WD_0+\frac{\Xi_W}{b_t}\right)+o_p(1).
\]
Since $q_K=O_p(K^{-1/2})$, the terms involving $q_K$ vanish in this covariance limit. Condition~\ref{assm:clpop:item:batches} then gives $L^{-1}\sum_{t\le Lr}\Sigma_{K,t}\pto r\Sigma_W$ uniformly in $r$.

\emph{Path approximation.} Conditional on $\mathcal F_{K,0}$, the proofs of Lemmas~\ref{lem:stopping-time-and-tail} and \ref{lem:sgd-conv-rate} use the bounds \eqref{eq:eq1}--\eqref{eq:eq10} established above, together with the stated initialization, learning-rate, and conditional-noise conditions, rather than independence of the original observations. With $\stt_\loclevel$ denoting the first exit of $Q_K(\vartheta_t)$ above $\loclevel$, they give the same localized rate $\mathbb E[Q_K(\vartheta_t)\mathbf1\{\stt_\loclevel\ge t\}\mid\mathcal F_{K,0}]\le C_\loclevel(s+t+1)^{-a}$. Linearize the drift at $\hat\theta_{n,W}$ using its actual Hessian $H_K\pto J_W$. The proof of Lemma~\ref{lem:control-approx-error} bounds the normalized cumulative nonlinear remainder, after localization, by a quantity of order
\[
L^{1/2-a}+\|\bar g_n(\theta_\oo)\|L^{(1-a)/2}
=O_p\left(L^{1/2-a}+\sqrt{L^{1-a}/K}\right)=o_p(1).
\]
The initial-value contribution has order $O_p((s+1)^{a/2}/\sqrt L)=o_p(1)$. The matrix-product remainder has a localized $2p$th-moment bound of order $L^{1-(1-a)p}=o(1)$, as in Lemma~\ref{lem:FCLT-extension}. These arguments yield, uniformly in $r$,
\begin{equation}\label{eq:clpop:martingale-representation}
\frac1{\sqrt L}\sum_{t=1}^{\lfloor Lr\rfloor}(\vartheta_t-\hat\theta_{n,W})
=-H_K^{-1}\frac1{\sqrt L}\sum_{t=1}^{\lfloor Lr\rfloor}e_t+o_p(1),
\end{equation}
where
\[
e_t=\widetilde D_t(\vartheta_{t-1})'W_K\widetilde u_t(\vartheta_{t-1})
-\bar G_n(\vartheta_{t-1})'W_K\bar g_n(\vartheta_{t-1}).
\]
The cancellation $(P_KH_K)^{-1}P_K=H_K^{-1}$ explains the absence of $P$.

\emph{Joint convergence.} The $2p$th-moment bound gives the localized Lindeberg bound $O(L^{1-p})$. Mean-square continuity of the noise and the localized iterate rate permit replacing the predictable covariance at $\vartheta_{t-1}$ by $\Sigma_{K,t}$ in its time average. Hence the predictable bracket tends uniformly to $r\Sigma_W$. After the localization is removed by making its exceptional probability arbitrarily small, these are the conditions of Lemma~\ref{lem:stable-mclt}, with $K$ and $L$ in place of $n$ and $\N$ and with $V=\Sigma_W$. That lemma and representation \eqref{eq:clpop:martingale-representation} give the computational limit jointly with the sampling expansion, which is measurable with respect to $\mathcal F_{K,0}$, with an independent Brownian limit. This proves the joint result.
}

\subsubsection{Proof of Corollary~\ref{cor:clpop:RS}}
{
Put $A_R=RA_WR'$ and $C_R=RC_WR'$, both positive definite. Along any subsequence for which $\lambda_{K,L}=\tau_{K,L}/K\to\lambda\in[0,1]$, the scaled numerator is asymptotically normal with covariance
\[
M_R=\lambda A_R+(1-\lambda)B_gC_R\preceq B_gC_R.
\]
Theorem~\ref{thm:clpop:joint} gives $V_L(R)\dto C_R^{1/2}\mathcal J_\ell C_R^{1/2}$, where $\mathcal J_\ell=\int_0^1\overline{\mathbb B}_\ell(r)\overline{\mathbb B}_\ell(r)'dr$. The sampling normal and the computational Brownian endpoint are independent of this bridge. After normalization the statistic therefore has limit $Y'\mathcal J_\ell^{-1}Y$, with $Y$ independent of $\mathcal J_\ell$ and $\operatorname{Var}(Y)\preceq I_\ell$. Conditional on $\mathcal J_\ell$, the eigenvalues of $\mathcal J_\ell^{-1/2}\operatorname{Var}(Y)\mathcal J_\ell^{-1/2}$ do not exceed those of $\mathcal J_\ell^{-1}$. Coupling their weighted sums of independent $\chi^2_1$ variables proves stochastic dominance by $\mathcal T_\ell$. Equality follows when $M_R=B_gC_R$. Every subsequence has a further subsequence with a limit for $\lambda_{K,L}$, which proves the stated limsup without requiring a fixed ratio $L/K$.
}

\subsubsection{Proof of Corollary~\ref{cor:clpop:plugin}}
{
Put $A_R=RA_WR'$ and $C_R=RC_WR'$, both positive definite. Along any subsequence for which $\lambda_{K,L}=\tau_{K,L}/K\to\lambda\in[0,1]$, Theorem~\ref{thm:clpop:joint} gives
\[
\sqrt{\tau_{K,L}}(R\bar\vartheta_L-c)\dto\mathcal N(0,M_R),
\qquad M_R=\lambda A_R+(1-\lambda)B_gC_R>0.
\]
Since $\tau_{K,L}/L=(1-\lambda_{K,L})B_g$, consistency of the plug-in estimates gives
\[
\tau_{K,L}R\left(\frac{\widehat A_W}{K}+\frac{\widehat C_W}{L}\right)R'\pto M_R.
\]
Slutsky's theorem yields the stated $\chi^2_\ell$ limit. Every subsequence has a further subsequence with a limit for $\lambda_{K,L}$, so the conclusion does not require a fixed ratio $L/K$.
}

\subsubsection{Proof of Theorem~\ref{thm:cluster:debiased-J}}
{
Theorem~\ref{thm:clpop:joint} and $K/(LB_g)\to\tau<\infty$, with fixed $B_g$, imply
\[
\bar\vartheta_L-\theta_\oo=O_p(K^{-1/2}+L^{-1/2})
=O_p(K^{-1/2})=o_p(K^{-1/4}).
\]
The local Jacobian bounds in Assumption~\ref{assm:clpop} imply the Lipschitz condition in Lemma~\ref{lem:debiased-J:general} and $\bar\Phi\pto D_0$. Select a fixed nonsingular $\dtheta\times\dtheta$ row block of $D_0$ to construct $\hat H$ as in that lemma, with $\hat H\bar\Phi=0$ and $\hat H\pto H_0$ of full row rank $\dg-\dtheta$. Because $\Omega>0$, $H_0\Omega H_0'>0$. Set $\hat V=\hat\Omega_K(\bar\vartheta_L)/K$. The sampling CLT and the assumed covariance consistency verify the remaining conditions of the lemma, so its residual statistic has limit $\chi^2_{\dg-\dtheta}$.

The projection identity following the lemma shows that this residual statistic equals \eqref{eq:cluster:debiased-J}, since $\hat V^{-1}=K\bar W_K$. Thus it gives the asserted limit. In particular, the projected first-order estimation error vanishes without changing the statistic's normalization to account for computation.
}

\section{Proofs of Lemmas}\label{appx:proofs-lems}
%%%%%%%%%%%%%%%%%%%%

\subsection{Proof of Lemma~\ref{lem:consistency-and-normality}}
{
Write $\hat\theta=\hat\theta_{n,W}$, $q_0=\bar g_n(\theta_\oo)$, and $A=A_W=G_\oo'WG_\oo$.
Since $W_n\pto W>0$ and $\sqrt n q_0=O_p(1)$,
\[
0\le\bar Q_{n,W}(\theta_\oo)
\le\tfrac12 q_0'W_nq_0=O_p(n^{-1}).
\]
The assumed criterion separation outside the $\delta$-ball around $\hat\theta$ implies $\|\hat\theta-\theta_\oo\|<\delta$ with probability approaching one.
Since $\Theta$ is convex, constrained optimality gives $\nabla\bar Q_{n,W}(\hat\theta)'(\theta_\oo-\hat\theta)\ge0$. Taylor expansion and local strong convexity therefore imply, on that event,
\[
\bar Q_{n,W}(\theta_\oo)
\ge\nabla\bar Q_{n,W}(\hat\theta)'(\theta_\oo-\hat\theta)
+\frac c2\|\hat\theta-\theta_\oo\|^2
\ge\frac c2\|\hat\theta-\theta_\oo\|^2.
\]

Thus $\hat\theta-\theta_\oo=O_p(n^{-1/2})$.
Because $\theta_\oo\in\operatorname{int}(\Theta)$, this rate implies $\hat\theta\in\operatorname{int}(\Theta)$ with probability approaching one, and hence $\nabla\bar Q_{n,W}(\hat\theta)=0$ on that event.
The envelope and Lipschitz assumptions imply local uniform convergence of $\bar G_n$ to $G$, so
\[
D_n:=\int_0^1\bar G_n(\theta_\oo+u(\hat\theta-\theta_\oo))\,du\pto G_\oo,
\qquad \bar G_n(\hat\theta)\pto G_\oo.
\]
The exact moment expansion and first-order condition yield
\[
0=\bar G_n(\hat\theta)'W_n\{q_0+D_n(\hat\theta-\theta_\oo)\}.
\]
As $\bar G_n(\hat\theta)'W_nD_n\pto A>0$, this gives
\begin{equation}\label{eq:GMM:random-weight-expansion}
\sqrt n(\hat\theta_{n,W}-\theta_\oo)
=-A^{-1}G_\oo'W\sqrt n\bar g_n(\theta_\oo)+o_p(1).
\end{equation}
Substituting \eqref{eq:GMM:random-weight-expansion} into the exact moment expansion $\sqrt n\,\bar g_n(\hat\theta)=\sqrt n\,q_0+D_n\sqrt n(\hat\theta-\theta_\oo)$ and using $D_n\pto G_\oo$ gives the second representation.
The i.i.d. moment CLT proves the asserted covariance. Only $W_n\pto W$ is required here: preliminary algorithmic randomness enters the remainder, while the leading term is measurable from the sample. The result is convergence in probability and asymptotic normality; it does not assert almost-sure convergence for arbitrary random $W_n$.
}

\subsection{Proof of Lemma~\ref{lem:sample-regularity}}

\noindent \textit{Proof of \eqref{eq:eq1}:}
This is an immediate consequence of $\P_n \ppto \P > 0$ in {Condition}~\ref{asm:appx:conditioning-and-weighting}.
\medskip

\noindent \textit{Proof of \eqref{eq:eq2}:}
This is an immediate consequence of $W_n \ppto W > 0$ in {Condition}~\ref{asm:appx:conditioning-and-weighting}.
\medskip

\noindent \textit{Proof of \eqref{eq:eq3}:}
{
By Lemma~\ref{lem:consistency-and-normality}, $\hat\theta_{n,W}\pto\theta_\oo$. A local uniform law of large numbers and continuity of $G$ give $\bar G_n(\hat\theta_{n,W})\pto G_\oo$. Together with $W_n\pto W>0$, this implies $\bar G_n'W_n\bar G_n\pto G_\oo'WG_\oo>0$, establishing the eigenvalue bounds.
}
\medskip

\noindent \textit{Proof of \eqref{eq:eq4}:}
It suffices to establish that $\bar{\H}_n \ppto G_\oo' W G_\oo$.
In light of the proof of \eqref{eq:eq3}, this should follow from
\begin{equation*}
{\bar{\H}_n -\bar{G}_n'W_n \bar{G}_n = \sum_{j = 1}^{\dg} [W_n\bar g_n(\hat{\theta}_{n,W})]_j \frac{\partial^2 \bar{g}_{nj}(\hat{\theta}_{n,W})}{\partial \theta \partial \theta'} \ppto 0.}
\end{equation*}
It is straightforward to verify this since $\|\frac{\partial^2 \bar{g}_{nj}(\hat{\theta}_{n,W})}{\partial \theta \partial \theta'}\| \le C$ for $j = 1, \ldots, \dg$, by \eqref{eq:eq6}, $W_n \le C$ by \eqref{eq:eq2}, and
\begin{equation*}
\|\bar{g}_{nj}(\hat{\theta}_{n,W})\|^2 \le \underbar{c}^{-1} \bar{g}_n(\hat{\theta}_{n,W})' W_n \bar{g}_n(\hat{\theta}_{n,W}) \le \underbar{c}^{-1} \bar{g}_n(\theta_\oo)' W_n \bar{g}_n(\theta_\oo) \le \frac{C}{\underbar{c}} \|\bar{g}_n(\theta_\oo)\|^2 \ppto 0,
\end{equation*}
for each $j=1,\ldots, \dg$, where we use \eqref{eq:eq2}.

\medskip

\noindent \textit{Proof of \eqref{eq:eq5}:}
By assumption~\ref{asm:appx:item:integrability} and the SLLN (strong law of large numbers),
\begin{equation*}
\limsup_{n \to \infty} \sup_{\theta \in \Theta} \|\bar{G}_n(\theta)\| \le \lim_{n \to \infty}\frac{1}{n} \sum_{i=1}^n H(z_i) = \mathbb{E}[H(z_i)] < \infty\quad \text{a.s.}
\end{equation*}
% \medskip

\noindent \textit{Proof of \eqref{eq:eq6}:}
This follows from the fact that
\begin{equation*}
    \sup_{\theta \ne \tilde{\theta}}\frac{\|\bar{G}_n(\theta)-\bar{G}_n(\tilde{\theta})\|}{\|\theta - \tilde{\theta}\|}   \le \frac{1}{n} \sum_{i=1}^n L(z_i) \asto \mathbb{E}[L(z_i)] < \infty
\end{equation*}
where we use the SLLN.
This also implies
\begin{equation*}
\left\|\frac{\partial^2 \bar{g}_{nj}(\hat{\theta}_{n,W})}{\partial \theta \partial \theta'}\right\| \le C\quad \text{w.p.a.1 \ for }j=1,\ldots, \dg.
\end{equation*}

\noindent \textit{Proof of \eqref{eq:eq7}:}
We begin by writing
\begin{equation*}
    \bar{Q}_{n,W}(\theta) = \frac{1}{2}\left( \bar{g}_n(\theta)' W_n \bar{g}_n(\theta) - \bar{g}_n(\hat{\theta}_{n,W})' W_n \bar{g}_n(\hat{\theta}_{n,W})\right).
\end{equation*}
We divide into two cases.
On the event $\{\hat\theta_{n,W}\in\operatorname{int}(\Theta)\}$, whose probability tends to one by Lemma~\ref{lem:consistency-and-normality}, we have $\nabla\bar Q_{n,W}(\hat\theta_{n,W})=0$; this is used in the first case.
If $\theta$ satisfies $\|\theta - \hat{\theta}_{n,W}\| \le \delta$, then by Taylor's theorem, there exists $s \in (0,1)$ such that, for $\theta_s = s\theta + (1-s)\hat{\theta}_{n,W}$,
\begin{align}
\label{eq:second-order-expansion-Lyapunov}
    \bar{Q}_{n,W}(\theta) = \frac{1}{2}(\theta - \hat{\theta}_{n,W})' \frac{\partial^2 \bar{Q}_{n,W}(\theta_s)}{\partial \theta \partial \theta'} (\theta - \hat{\theta}_{n,W}).
\end{align}
By \ref{asm:appx:sample-regularity}, it follows
\begin{equation*}
       \bar{Q}_{n,W}(\theta) \ge c/2 \|\theta - \hat{\theta}_{n,W}\|^2 \ge c/2 \min\{\|\theta - \hat{\theta}_{n,W}\|^2, \delta^2\}
\end{equation*}
\wpa.

Otherwise, if $\|\theta - \hat{\theta}_{n,W}\| \ge \delta$, we have $\bar{Q}_{n,W}(\theta) \ge c$, thereby implying that
\begin{equation*}
\bar{Q}_{n,W}(\theta) \ge c \ge \underbar{c} \min\{\|\theta - \hat{\theta}_{n,W}\|^2, \delta^2\}
\end{equation*}
\wpa, where $\underbar{c} = \min\{c/2, c/\delta^2\}$.
This completes the proof of \eqref{eq:eq7}.
\medskip

\noindent \textit{Proof of \eqref{eq:eq8}:}
{
Let $d=\theta-\hat\theta_{n,W}$. On the event where the assumed curvature bounds hold and $\hat\theta_{n,W}\in\operatorname{int}(\Theta)$, so that $\nabla\bar Q_{n,W}(\hat\theta_{n,W})=0$, for $\|d\|\le\delta$ the whole segment $\hat\theta_{n,W}+ud$, $0\le u\le1$, is in the local strong-convexity region. Hence
\[
d'\nabla\bar Q_{n,W}(\theta)
=\int_0^1 d'\nabla^2\bar Q_{n,W}(\hat\theta_{n,W}+ud)d\,du
\ge c\|d\|^2.
\]
Cauchy--Schwarz and the upper Hessian bound give
\[
\|\nabla\bar Q_{n,W}(\theta)\|^2\ge c^2\|d\|^2,
\qquad \bar Q_{n,W}(\theta)\le\tfrac M2\|d\|^2.
\]
Thus the squared gradient is at least $(2c^2/M)\bar Q_{n,W}(\theta)$ throughout the entire $\delta$-ball. Outside this ball, {Condition}~\ref{asm:appx:item:away-from-zero} bounds the squared gradient below by $c$. Taking $\underbar c=\min\{2c^2/M,c\}$ proves \eqref{eq:eq8} without an uncovered intermediate region.
}
\medskip

\noindent \textit{Proof of \eqref{eq:eq9}:}
{
Write $\hat\theta=\hat\theta_{n,W}$ and
\begin{align*}
X_t(\theta)&=\tilde G_t(\theta)'W_n\tilde g_t(\theta),\\
X_t(\theta)-X_t(\hat\theta)
&=\tilde G_t(\theta)'W_n\{\tilde g_t(\theta)-\tilde g_t(\hat\theta)\}
 +(\tilde G_t(\theta)-\tilde G_t(\hat\theta))'W_n\tilde g_t(\hat\theta).
\end{align*}
The centered increment is exactly
\[
\xi_t(\theta)-\xi_t(\hat\theta)
=X_t(\theta)-X_t(\hat\theta)
 -\mathbb E_n^\star[X_t(\theta)-X_t(\hat\theta)].
\]
Consequently its conditional second moment is no greater than that of $X_t(\theta)-X_t(\hat\theta)$.
Let $d=\theta-\hat\theta$. The derivative envelope for the moment batch and the Lipschitz envelope for the Jacobian batch imply
\begin{align*}
\|\tilde g_t(\theta)-\tilde g_t(\hat\theta)\|
&\le\left\{\frac1{B_g}\sum_{j=1}^{B_g}
 H(\tilde z_{j+B_{t-1}+B_{G,t}})\right\}\|d\|,\\
\|\tilde G_t(\theta)-\tilde G_t(\hat\theta)\|
&\le\left\{\frac1{B_{G,t}}\sum_{j=1}^{B_{G,t}}
 L(\tilde z_{j+B_{t-1}})\right\}\|d\|.
\end{align*}
These inequalities follow by integration along the segment in $\Theta$. Jensen's inequality and the empirical second moments of $H$ and $L$ bound the squared batch envelopes uniformly in $t$ with probability approaching one. Also,
\[
\sup_{\theta\in\Theta}\mathbb E_n^\star\|\tilde G_t(\theta)\|^2\le C,
\qquad
\mathbb E_n^\star\|\tilde g_t(\hat\theta)\|^2
\le\left\{\frac1n\sum_i\|g(z_i,\hat\theta)\|^{2p}\right\}^{1/p}\le C,
\]
using $\bar Q_{n,W}(\hat\theta)=0$. Independence of the two batches conditional on $\mathcal F_{n,0}^\star$, together with $\|W_n\|\le C$, therefore proves \eqref{eq:eq9}.
}
\medskip

\noindent \textit{Proof of \eqref{eq:eq10}:}
{
Conditional centering and Jensen's inequality give
\begin{align*}
\mathbb E_n^\star\|\xi_t(\theta)\|^{2p}
&\le 2^{2p}\mathbb E_n^\star\|X_t(\theta)\|^{2p}\\
&\le C\left\{\frac1n\sum_iH(z_i)^{2p}\right\}
       \left\{\frac1n\sum_i\|g(z_i,\theta)\|^{2p}\right\}\\
&\le C\{\bar Q_{n,W}(\theta)^p+1\}
\end{align*}
with probability approaching one, uniformly in $\theta$ and $t$. The product bound uses independent moment and Jacobian batches and bounded $W_n$. The additive constant accounts for the possibly nonzero sample moment at the optimizer; no bound of $\|\bar g_n(\theta)\|^{2p}$ by $C\bar Q_{n,W}(\theta)^p$ is needed.
}

\subsection{Proof of Lemma~\ref{lem:stopping-time-and-tail}}
\noindent \textit{Part (i):}
Let us write 
\begin{equation*}
Q_t^{\star} = \bar{Q}_{n,W}(\theta_t^{\star}).
\end{equation*}
Conditional on the event $E_n$, by the second-order Taylor expansion, we have
\begin{align*}
Q_{t+1}^{\star} - Q_t^{\star} &=  \nabla \bar{Q}_{n,W}(\theta_t^{\star})^{\top}(\theta_{t+1}^{\star}-\theta_{t}^{\star}) + \frac{1}{2}(\theta_{t+1}^{\star}-\theta_{t}^{\star})^{\top}\frac{\partial^2 \bar{Q}_{n,W}(\tilde{\theta}_t^{\star})}{\partial \theta\partial \theta'}  (\theta_{t+1}^{\star}-\theta_{t}^{\star}),
\end{align*}
where $\tilde{\theta}_t^{\star}$ lies on the line segment $[\theta_t^{\star}, \theta_{t+1}^{\star}]$.
Using the fact that the Hessian of $\bar{Q}_{n, W}$ is uniformly bounded above by $M I_{\dtheta}$ and
\begin{equation*}
\theta_{t+1}^{\star}-\theta_{t}^{\star} = - \gs[t+1] \P_n \nabla \bar{Q}_{n,W}(\theta_t^{\star}) - \gs[t+1] \P_n \xi_{t+1}(\theta_{t}^{\star}),
\end{equation*}
we find that
\begin{align*}
Q_{t+1}^{\star} - Q_t^{\star} & \le - \underbar{c} \gs[t+1] \| \nabla \bar{Q}_{n,W}(\theta_t^{\star})\|^2 - \gs[t+1] \nabla \bar{Q}_{n,W}(\theta_t^{\star})^{\top}\P_n \xi_{t+1}(\theta_{t}^{\star}) \\
& \quad + C \gs[t+1]^2 (\|\nabla \bar{Q}_{n,W}(\theta_t^{\star})\|^2 + \|\xi_{t+1}(\theta_{t}^{\star})\|^2).
\end{align*}
where we use \eqref{eq:eq1}.
{
By Lemma~\ref{lem:sample-regularity}, on $E_n$,
$\|\nabla\bar Q_{n,W}(\theta_t^\star)\|^2\ge\underbar c(Q_t^\star\wedge1)$ and
\[
\mathbb E_n^\star\big[
\|\nabla\bar Q_{n,W}(\theta_t^\star)\|^2+
\|\xi_{t+1}(\theta_t^\star)\|^2
\mid\mathcal F_{n,t}^\star\big]\le C(Q_t^\star+1).
\]
Here the conditional noise bound follows from \eqref{eq:eq10} and Jensen's inequality. The gradient bound follows from bounded $\bar G_n$ and $W_n$ and the identity
$\bar g_n(\theta)'W_n\bar g_n(\theta)=2\bar Q_{n,W}(\theta)+\bar g_n(\hat\theta_{n,W})'W_n\bar g_n(\hat\theta_{n,W})$, with the last term bounded by the definition of $E_n$.
}
This leads to the basic inequality
\begin{equation}
\label{eq:recursion-Lyapunov}
\mathbb{E}_n^{\star}[Q_{t+1}^{\star} | \mathcal{F}^{\star}_{n,t}] \le Q_t^{\star} - \underbar{c} \gs[t+1] (Q_t^{\star}\wedge 1) + C \gs[t+1]^2(Q_t^{\star} + 1), \quad t \ge 0,
\end{equation}
which holds for all $n \in \mathbb{N}$ and $\Ns\ge0$ on the event $E_n$.

Let us define
\begin{equation*}
{\pi_t = \prod_{s=1}^t (1 + C \gs[s]^2),\quad t \ge 0,}
\end{equation*} 
and let $\pi_\infty = \lim_{t \to \infty} \pi_t < \infty$ denote its limit, which is uniformly bounded in $\Ns$ since $\sum_{s\ge 1}\gs[s]^2 \le \gamma_0^2 \sum_{s \ge 1} s^{-2a}$ is bounded by an absolute constant.
Let $\tilde{Q}_t^{\star} := Q_{t}^{\star}/\pi_t$.
Dividing each side of \eqref{eq:recursion-Lyapunov} by $\pi_{t+1}$, we find that
\begin{equation}
\label{eq:Vt-recursive-inequality}
\mathbb{E}_n^{\star}[\tilde{Q}_{t+1}^{\star} | \mathcal{F}^{\star}_{n,t}] \le \tilde{Q}_t^{\star} - \underbar{c} \gs[t+1] (\tilde{Q}_t^{\star} \wedge 1) + C\gs[t+1]^2 , \quad t \ge 0.
\end{equation}
(Recall that $\underbar{c}>0$ and $C>0$ can be different in each equation.)
By Lemma~\ref{lem:Robbins-Siegmund}, \eqref{eq:Vt-recursive-inequality} implies that 
\begin{equation*}
\mathbb{P}_n^{\star}\left(  \lim_{t\to\infty}\tilde{Q}_t^{\star} \text{ exists},\quad \sum_{t=1}^\infty \gs[t+1] (\tilde{Q}_t^{\star} \wedge 1 ) < \infty\right)  = 1.
\end{equation*}
These together imply that $\lim_{t\to\infty}\tilde{Q}_t^{\star} = 0$, yielding $\theta_{\N}^{\star} - \hat{\theta}_{n,W} \to 0$ $\mathbb{P}_n^{\star}$-a.s. as $\N \to\infty$ conditional on the event $E_n$ by \eqref{eq:eq7}.
\medskip

\noindent \textit{Part (ii):}
Define $W_t^{\star}$ by
\begin{equation*}
    W_t^{\star} = \tilde{Q}_t^{\star} + \underbar{c} \sum_{s=0}^{t-1} \gs[s+1] (\tilde{Q}_s^{\star} \wedge 1) - C \sum_{s=0}^{t-1} \gs[s+1]^2, \quad t \ge 0,
\end{equation*}
where $\underbar{c}>0$ and $C>0$ are the same as in \eqref{eq:Vt-recursive-inequality}.
Note that
\begin{equation*}
-C     \sum_{s=0}^{t-1} \gs[s+1]^2 \le W_t^{\star} \le \tilde{Q}_t^{\star} + \underbar{c} \sum_{s=0}^{t-1} \gs[s+1],
\end{equation*}
Taking expectations in \eqref{eq:Vt-recursive-inequality} and iterating gives $\mathbb{E}_n^{\star}[\tilde Q_t^{\star}]\le\tilde Q_0^{\star}+C\sum_{s\ge1}(\gs[s])^2<\infty$ by Condition~\ref{asm:appx:initial}. Hence, for all $t\ge 0$, $\mathbb{E}_n^{\star}[|W_t^{\star}|] < \infty$ on the event $E_n$.
By \eqref{eq:Vt-recursive-inequality}, it follows that $\mathbb{E}_n^{\star}[W_{t+1}^{\star} |\mathcal{F}_{n,t}^{\star}] \le W_t^{\star}$ on the event $E_n$, that is, $(W_t^{\star})_{t \ge 0}$ forms a supermartingale.

Conditional on the event $E_n$, consider a submartingale $(X_t^{\star}, \mathcal{F}_{n,t}^{\star}, \mathbb{P}_n^{\star})_{t \ge 0}$ defined by 
\begin{equation*}
X_t^{\star} := -W_{t}^{\star} + W_{0}^{\star},\quad t \ge 0.
\end{equation*} 
By the submartingale maximal inequality (see \citet[{Theorem~9.4.1}]{chungCourseProbabilityTheory2000}), we have, for all $n \in \mathbb{N}$, $\Ns \ge 0$, and $\N \in \mathbb{N}$,
\begin{equation*}
\mathbb{P}_n^{\star}\left( \min_{0 \le t \le \N} X_t^{\star} \le - \lambda \right) \le \lambda^{-1} \mathbb{E}_n^{\star}[(X_{\N}^{\star} \vee 0)], \quad \forall \lambda > 0
\end{equation*}
on the event $E_n$.
Note that
\begin{equation*}
X_{\N}^{\star} \le \tilde{Q}_0^{\star} + C \sum_{s=0}^{\infty} \gs[s+1]^2 \le C_1
\end{equation*}
by {Condition}~\ref{asm:appx:initial} with $C_1 > 0$ being an absolute constant.
Hence, for all $n \in \mathbb{N}$, $\Ns \ge 0$, and $\N \in \mathbb{N}$, it holds $\mathbb{E}_n^{\star}[(X_{\N}^{\star} \vee 0)] \le C_1$.
This allows us to choose an absolute constant {$\loclevel > C_1+C_1/\varepsilon$} such that
\begin{equation*}
\mathbb{P}_n^{\star}\left( \min_{0 \le t \le \N} X_t^{\star} \le - (\loclevel - C_1) \right) \mathbbm{1}_{E_n} < \varepsilon
\end{equation*}
for all $n \in \mathbb{N}$, $\Ns \ge 0$, and $\N \in \mathbb{ N} $.
We then observe that
\begin{equation*}
\max_{0 \le t \le \N} (-X_t^{\star}) \ge \max_{0 \le t \le \N} \tilde{Q}_t^{\star} - C \sum_{s =1}^\infty \gs[s]^2 - \tilde{Q}_0^{\star} \ge \max_{0\le t\le \N} \tilde{Q}_t^{\star} - C_1.
\end{equation*}
{
On $E_n$, this implies
\begin{equation*}
\mathbb{P}_n^{\star}\left( \max_{0 \le t \le \N} \tilde{Q}_t^{\star} \ge \loclevel \right) \le \mathbb{P}_n^{\star} \left(\max_{0 \le t \le \N} (-X_t^{\star}) \ge \loclevel - C_1 \right)\le \frac{C_1}{\loclevel-C_1}< \varepsilon.
\end{equation*}
Since $1\le\pi_t\le C_\pi:=\exp\{C\gamma_0^2\sum_{s=1}^\infty s^{-2a}\}$ uniformly in $t$ and $\Ns$, we have
\begin{equation*}
\mathbb{P}_n^{\star}(\stt_{C_\pi\loclevel}^{\star} \le \N)
\le \mathbb{P}_n^{\star}\left(\max_{0\le t\le\N}\tilde Q_t^{\star}\ge\loclevel\right)
\le \frac{C_1}{\loclevel-C_1}<\varepsilon.
\end{equation*}
Both $C_\pi$ and $\loclevel$ are independent of $n$, $\N$, and $\Ns$. Letting $\N\to\infty$ and relabeling $C_\pi\loclevel$ as $\loclevel$, we obtain, for all $n\in\mathbb N$ and $\Ns\ge0$,
\begin{equation*}
\mathbb{P}_n^{\star}(\stt_{\loclevel}^{\star} < \infty) \mathbbm{1}_{E_n} < \varepsilon.
\end{equation*}
}
This verifies part (ii).
\medskip

\noindent \textit{Part (iii):}
Choose positive integers $\tilde{T} < T$ and consider a submartingale $(W_{\tilde{T}}^{\star}-W_t^{\star}, \mathcal{F}_{n,t}^{\star}, \mathbb{P}_n^{\star})_{t = \tilde{T}}^{T}$.
By part (ii), we can find $\loclevel$ such that, for all $n \in \mathbb{N}$ and $\Ns \ge 0$, $\mathbb{P}_n^{\star}(\stt_{\loclevel}^{\star} < \infty) \mathbbm{1}_{E_n} < \varepsilon$.
By the submartingale maximal inequality (\citet*{chungCourseProbabilityTheory2000}), we have, for all $\varepsilon>0$,
\begin{equation*}
    \mathbb{P}_n^{\star}\left( \left. \min_{\tilde{T}\le t\le T} [W_{\tilde{T}}^{\star} - W_t^{\star}] < - \varepsilon \right| \mathcal{F}_{n, \tilde{T}}^{\star}\right) \mathbbm{1}_{ \mathcal{E}}\le \varepsilon^{-1} \mathbb{E}_n^{\star}[((W_{\tilde{T}}^{\star} - W_{T}^{\star}) \vee 0)|\mathcal{F}_{n, \tilde{T}}^{\star}] \mathbbm{1}_{ \mathcal{E}},
\end{equation*}
where the event $\mathcal{E} := \{\stt_{\loclevel}^{\star} \ge \tilde{T}+1\}$ is $\mathcal{F}_{n, \tilde{T}}^{\star}$-measurable since $\stt_{\loclevel}^{\star}$ is a stopping time with respect to $(\mathcal{F}_{n,t}^{\star})_{t\ge 0}$.
As $T$ tends to infinity, we have, on the event $E_n$,
\begin{align*}
   \max\{W_{\tilde{T}}^{\star} - W_{T}^{\star},0\} & \le 
   \tilde{Q}_{\tilde{T}}^{\star} + C \sum_{s=\tilde{T}}^{T-1} \gs[s+1]^2 \to \tilde{Q}_{\tilde{T}}^{\star} + C \sum_{s = \tilde{T}}^{\infty} \gs[s+1]^2 =: U_{0}
\end{align*}
By the monotone convergence theorem, we have, as $T\to\infty$,
\begin{equation*}
    \mathbb{P}_n^{\star}\left( \left. \inf_{\tilde{T}\le t < \infty} [W_{\tilde{T}}^{\star} - W_t^{\star}] <- \varepsilon \right|\mathcal{F}_{n, \tilde{T}}^{\star}\right) \mathbbm{1}_{ \mathcal{E}}\le \varepsilon^{-1} \mathbb{E}_n^{\star}[U_{0} | \mathcal{F}_{n, \tilde{T}}^{\star}]\mathbbm{1}_{ \mathcal{E}}.
\end{equation*}
By Lemma~\ref{lem:sgd-conv-rate}, which does not rely on part (iii), for all $n \in \mathbb{N}$ and $\Ns \ge 0$, it holds
\begin{equation*}
\mathbb{E}_n^{\star}\left [ U_{0}\I{\stt_{\loclevel}^{\star} \ge \tilde{T}+1} \right ] \le \mathbb{E}_n^{\star}\left[U_{0}\I{\stt_{\loclevel}^{\star} \ge \tilde{T}}\right] \le C_\loclevel \gs[\tilde{T}] + C \sum_{s \ge \tilde{T}} \gs[s+1]^2.
\end{equation*}
As a result, we can find $\tilde{T} < \infty$ independent of $n\in \mathbb{N}$ and $\Ns \ge 0$ such that for all $n \in \mathbb{N}$ and $\Ns \ge 0$, 
\begin{equation*}
\mathbb{E}_n^{\star}\left[  U_{0}\I{\stt_{\loclevel}^{\star} \ge \tilde{T}+1} \right]  \le C_\loclevel \gs[\tilde{T}] + C \sum_{s \ge \tilde{T}} \gs[s+1]^2 < (\varepsilon^2 \wedge \varepsilon)
\end{equation*} 
on the event $E_n$.
This implies that
\begin{equation*}
    \mathbb{P}_n^{\star}\left( \inf_{\tilde{T}\le t < \infty} [W_{\tilde{T}}^{\star} - W_t^{\star}] <- \varepsilon , \stt_{\loclevel}^{\star} \ge \tilde{T}+1\right) < \varepsilon,
\end{equation*}
and hence, for all $n \in \mathbb{N}$ and $\Ns \ge 0$,
\begin{equation*}
\mathbb{P}_n^{\star}\left( \inf_{\tilde{T}\le t < \infty} [W_{\tilde{T}}^{\star} - W_t^{\star}] <- \varepsilon \right) < \varepsilon + \mathbb{P}_n^{\star}(\stt_{\loclevel}^{\star} < \infty) < 2\varepsilon
\end{equation*}
on the event $E_n$.
Now, $\inf_{\tilde{T} \le t < \infty} [W_{\tilde{T}}^{\star} - W_t^{\star}] \ge - \varepsilon$ implies that for all $t \ge \tilde{T}$,
\begin{align*}
\tilde{Q}_t^{\star} &\le \tilde{Q}_{\tilde{T}}^{\star} - \underbar{c} \sum_{s=\tilde{T}}^{t-1} \gs[s+1] (\tilde{Q}_s^{\star} \wedge 1) + C \sum_{s=\tilde{T}}^{t-1} \gs[s+1]^2 + \varepsilon \le U_{0} + \varepsilon.
\end{align*}
Define an event $\mathcal{U} := \{ U_{0} \le \varepsilon \}$.
Since by Markov's inequality,
$$
\mathbb{P}_n^{\star}(\mathcal{U}^c) \le \varepsilon^{-1} \mathbb{E}_n^{\star}\left[U_0 \I{\stt_{\loclevel}^{\star} \ge \tilde{T}+1}\right] + {\mathbb{P}_n^{\star}}(\stt_{\loclevel}^{\star} < \infty) < 2 \varepsilon,
$$
we have
\begin{align*}
   \mathbb{P}_n^{\star}\left( \inf_{\tilde{T}\le t < \infty} [W_{\tilde{T}}^{\star} - W_t^{\star}] <- \varepsilon\right) &\ge \mathbb{P}_n^{\star}     \left( \left\{ \inf_{\tilde{T}\le t < \infty} [W_{\tilde{T}}^{\star} - W_t^{\star}] <- \varepsilon \right\} \cap \mathcal{U}\right) \\
   &\ge \mathbb{P}_n^{\star}\left( \left\{ \sup_{\tilde{T} \le t < \infty}\tilde{Q}_t^{\star} \le 2\varepsilon\right\}^c \cap \mathcal{U}\right) \\
   &\ge\mathbb{P}_n^{\star}\left( \sup_{\tilde{T} \le t < \infty}\tilde{Q}_t^{\star} > 2\varepsilon \right) - 2\varepsilon.
\end{align*}
This shows that, for all $n \in \mathbb{N}$ and $\Ns \ge 0$,
\begin{equation*}
        \mathbb{P}_n^{\star}\left( \sup_{\tilde{T} \le t < \infty}\tilde{Q}_t^{\star} > 2\varepsilon \right) < 4\varepsilon
\end{equation*}
on the event $E_n$.
Finally, $\{\tmt_\varepsilon^\star>\tilde T\}=\{\sup_{t\ge\tilde T}Q_t^\star\ge\varepsilon\}$, and $Q_t^{\star} \le \pi_\infty \tilde{Q}_t^{\star} \le C_\pi \tilde{Q}_t^{\star}$ with $C_\pi=\exp\left( C\gamma_0^2 \zeta(2a)\right)$, where $\zeta(2a) = \sum_{s \ge 1} s^{-2a}$ denotes the Riemann zeta function. Running the preceding argument with $\varepsilon$ replaced by $\varepsilon'=\min\{\varepsilon/4,\varepsilon/(3C_\pi)\}$ changes $\loclevel$ and $\tilde T$ but not their independence of $n$ and $\Ns$, and gives, on the event $E_n$,
\[
\mathbb P_n^\star(\tmt_\varepsilon^\star>\tilde T)\le\mathbb P_n^\star\Big(\sup_{t\ge\tilde T}\tilde Q_t^\star>2\varepsilon'\Big)<4\varepsilon'\le\varepsilon,
\]
because $Q_t^\star\ge\varepsilon$ implies $\tilde Q_t^\star\ge\varepsilon/C_\pi\ge3\varepsilon'$. This verifies part (iii).

%%%%%%%%%%%%%%%%%%%%
\subsection{Proof of Lemma~\ref{lem:sgd-conv-rate}}

Define an auxiliary stopping time
\begin{equation}
\label{eq:auxiliary-stopping-time}
\tilde{\stt}_{\loclevel}^{\star} := \inf \{t \ge 0: \tilde{Q}_t^{\star} \ge \loclevel\}
\end{equation}
for $\loclevel > 0$.
Since $\tilde{Q}_t^{\star} \le Q_t^{\star}$, we observe that
\begin{equation*}
\stt_{\loclevel}^{\star} \le \tilde{\stt}_{\pi_\infty \loclevel}^{\star}, \quad \forall \loclevel > 0.
\end{equation*}
Let $\bar{x}$ be a positive number.
Since $(x\wedge 1) \ge (1 \wedge \bar{x}^{-1}) x$ for all $x \in [0, \bar{x}]$, we find that 
\begin{equation*}
(\tilde{Q}_t^{\star} \wedge 1) \ge (\loclevel^{-1} \wedge 1) \tilde{Q}_t^{\star} \ge \underbar{c}_{\loclevel} \tilde{Q}_t^{\star}
\end{equation*}
conditional on $\mathcal{E}_{t+1} : = \{\tilde{\stt}_{\loclevel}^{\star} \ge t+1\}$, which is an $\mathcal{F}_{n,t}^{\star}$-measurable event.
Here, $\underbar{c}_{\loclevel} > 0$ denotes a generic absolute constant that may depend on $\loclevel$.
Taking the expectations of both sides of \eqref{eq:Vt-recursive-inequality} on $\mathcal{E}_{t+1}$, this implies
\begin{align*}
\mathbb{E}_n^{\star}\left[\tilde{Q}_{t+1}^{\star} \mathbbm{1}_{\mathcal{E}_{t+1}} \right] &\le \mathbb{E}_n^{\star}\left[\tilde{Q}_t^{\star}  \mathbbm{1}_{\mathcal{E}_{t+1}}\right](1 - \underbar{c}_{\loclevel} \gs[t+1]) + C \gs[t+1]^2\\
& \le \mathbb{E}_n^{\star}\left[\tilde{Q}_t^{\star}  \mathbbm{1}_{\mathcal{E}_{t}} \right](1 - \underbar{c}_{\loclevel} \gs[t+1]) + C \gs[t+1]^2
\end{align*}
for all $t \ge 0$ and some absolute constants $0 < \underbar{c}_{\loclevel} < 1/(2\gamma_0)$ and $C > 0$,
where we use the fact that $\mathcal{E}_{t+1} \subseteq \mathcal{E}_{t}$.
From this, we can also see that 
\begin{equation*}
\mathbb{E}_n^{\star}[\tilde{Q}_{t}^{\star} \mathbbm{1}_{\mathcal{E}_{t}} ]\le \mathbb{E}_n^{\star}[\tilde{Q}_{t-1}^{\star} \mathbbm{1}_{\mathcal{E}_{t}} ] + C \gs[t]^2 \le \loclevel + C\gamma_0^2
\end{equation*}
is uniformly bounded in $t \ge 1$.
Define a (random) sequence $v_t := \mathbb{E}_n^{\star}[\tilde{Q}_{t}^{\star} \mathbbm{1}_{\mathcal{E}_{t}}]/\gs[t]$, which must satisfy the following recursive bound:
\begin{align*}
v_{t+1} \le v_t (1- \underbar{c}_{\loclevel}\gs[t+1]) \frac{\gs[t]}{\gs[t+1]} + C \gs[t+1], \ \ \ t \ge 0.
\end{align*}
Using the fact that
\begin{equation*}
\gs[t]/\gs[t+1] \le (1 + 1/(t+\Ns))^a \le 1 + \frac{a}{t+\Ns}
\end{equation*}
we have
\begin{equation*}
 (1- \underbar{c}_{\loclevel}\gs[t+1]) \frac{\gs[t]}{\gs[t+1]} \le 1 - \frac{\underbar{c}_{\loclevel}\gs[t+1]}{2}
\end{equation*}
provided that $a/(t+\Ns) \le \underbar{c}_{\loclevel}\gs[t+1]/2$.
Since $\gs[t]/\gs[t+1]\le 2$, we deduce that $a/(t+\Ns) \le \underbar{c}_{\loclevel}\gs[t+1]/2$ is satisfied under
\begin{equation*}
\frac{a}{t+\Ns} \le \underbar{c}_{\loclevel} \gs[t]/4
\end{equation*}
i.e., 
\begin{equation*}
t \ge \mathcal{T}_{\loclevel}(\Ns) := \left( \left\lceil \left( \frac{4 a}{\underbar{c}_{\loclevel} \gamma_0}\right)^{1/(1-a)} \right\rceil -\Ns\right)\vee 0,
\end{equation*}
where we write $\mathcal{T} = \mathcal{T}_{\loclevel}(\Ns)$ for ease of notation.

We divide the analysis into two cases: (i) $\lceil ( \frac{4 a}{\underbar{c}_{\loclevel} \gamma_0})^{1/(1-a)} \rceil \ge \Ns + 1$, and (ii) $\lceil (\frac{4 a}{\underbar{c}_{\loclevel} \gamma_0})^{1/(1-a)} \rceil \le \Ns$.
In case (i), applying Lemma~\ref{lem:recursive-bound} to the sequence $(v_t)_{t \ge \mathcal{T}}$ satisfying
\[
v_{t+1} \le v_t \left(1 - \frac{\underbar{c}_{\loclevel}\,\gs[t+1]}{2}\right) + C\,\gs[t+1],
\]
we obtain
\[
v_t \le C \left(v_{\mathcal{T}} + \frac{2}{\underbar{c}_{\loclevel}}\right) \le C \left(\loclevel + C\gamma_0^2\right) \frac{\lceil ( \frac{4 a}{\underbar{c}_{\loclevel} \gamma_0})^{1/(1-a)} \rceil^a}{\gamma_0} + \frac{2 C}{\underbar{c}_{\loclevel}}
\]
for all $t \ge \mathcal{T}$ on the event $E_n$.
Note that the upper bound on the right-hand side does not depend on $(n, \Ns, \N)$ or the time index.
For the remaining time indices $t \le \mathcal{T}$, we simply observe that
\[
v_t = \frac{\mathbb{E}_n^{\star}\left[ \tilde{Q}_t^{\star} \mathbbm{1}_{\mathcal{E}_{t}}\right]}{\gs[t]} \le \frac{\loclevel + C\gamma_0^2}{\gs[\mathcal{T}]} \le \left( \loclevel + C\gamma_0^2 \right) \frac{\lceil ( \frac{4 a}{\underbar{c}_{\loclevel} \gamma_0})^{1/(1-a)} \rceil^a}{\gamma_0}
\]
for $t \ge 1$.
For $t = 0$, by assumption, $v_0 = \tilde{Q}_0^{\star} / \gs[0] = Q_0^{\star} / \gs[0] \le M$.
Hence, there exists an upper bound that does not depend on $(n,\Ns, \N)$ and applies to the entire sequence $(v_t)_{t\ge 1}$.
This completes the analysis of case (i).

In case (ii), we have $\mathcal{T} = 0$, so that
\[
v_{t+1} \le v_t \left(1 - \frac{\underbar{c}_{\loclevel}\,\gs[t+1]}{2}\right) + C\,\gs[t+1]
\]
holds for all $t\ge0$. Lemma~\ref{lem:recursive-bound} then implies
\[
v_t \le C\left( v_0 + \frac{2}{\underbar{c}_{\loclevel}}\right) \le C  \frac{\tilde{Q}_0^{\star}}{\gs[0]} + \frac{2C}{\underbar{c}_{\loclevel}}.
\]
By the assumption that $Q_0^{\star}\le M \gs[0]$ and the fact that $\tilde{Q}_0^{\star} = Q_0^{\star}$, we find that $v_t \le C M + \frac{2C}{\underbar{c}_{\loclevel}}$, providing an absolute upper bound for $v_t$ for all $t\ge 1$.
This completes the analysis of case (ii).

Putting these together, we have shown that
\begin{equation*}
\mathbb{E}_n^{\star}\left[ \tilde{Q}_t^{\star} \I{\tilde{\stt}_{\loclevel}^{\star} \ge t} \right]\le C_{\loclevel} \gs[t],\quad t \ge 0,
\end{equation*}
where $C_\loclevel$ is an absolute constant that depends on $\loclevel$.
{
To transfer the bound back to $Q_t^\star$, put
$C_\pi=\exp\{C\gamma_0^2\zeta(2a)\}$. Since $1\le\pi_t\le C_\pi$,
\[
\tilde Q_t^\star\le Q_t^\star\le C_\pi\tilde Q_t^\star,
\qquad
\{\stt_\loclevel^\star\ge t\}\subseteq\{\tilde\stt_\loclevel^\star\ge t\}.
\]
Therefore
\[
\mathbb E_n^\star[Q_t^\star\mathbbm1\{\stt_\loclevel^\star\ge t\}]
\le C_\pi\mathbb E_n^\star[\tilde Q_t^\star\mathbbm1\{\tilde\stt_\loclevel^\star\ge t\}]
\le C_\pi C_\loclevel\gs[t].
\]
Absorbing $C_\pi$ into the constant proves the asserted rate.
}

\subsection{Proof of Lemma~\ref{lem:control-approx-error}}
For convenience of future reference, we divide the proof into two parts.
\medskip

\noindent \textit{Part~(i):}
Let $d_t := \theta_t^{\star} - \tto{t}$ denote the approximation error.
Then, $d_t$ follows $d_t = d_{t-1} - \gs[t] \P_n \zeta_t$, where
\begin{equation*}
\zeta_t := \bar{G}_n(\theta_{t-1}^{\star})' W_n \bar{g}_n(\theta_{t-1}^{\star}) - \bar{\H}_n (\tto{t-1} - \hat{\theta}_{n,W}).
\end{equation*}
Let
\begin{equation*}
\kappa_t := \zeta_t - \bar{\H}_n d_{t-1} = \bar{G}_n(\theta_{t-1}^{\star})' W_n \bar{g}_n(\theta_{t-1}^{\star}) - \bar{\H}_n (\theta_{t-1}^{\star} - \hat{\theta}_{n,W}).
\end{equation*}
{
On $E_n$, let $d=\theta_{t-1}^\star-\hat\theta_{n,W}$ and $\theta(u)=\hat\theta_{n,W}+ud$. The integral Taylor formula gives
\[
\kappa_t=\int_0^1\{H_n(\theta(u))-H_n(\hat\theta_{n,W})\}d\,du,
\]
where the actual objective Hessian is
\[
H_n(\theta)=\bar G_n(\theta)'W_n\bar G_n(\theta)
 +\sum_{j=1}^{\dg}[W_n\bar g_n(\theta)]_j\nabla^2\bar g_{nj}(\theta).
\]
Put $q=\bar g_n(\hat\theta_{n,W})$ and $D=\bar G_n(\hat\theta_{n,W})$. For each point on the segment,
\begin{align*}
H_n(\theta(u))-H_n(\hat\theta_{n,W})
&=\bar G_n(\theta(u))'W_n\bar G_n(\theta(u))-D'W_nD\\
&\quad+\sum_j[W_n\{\bar g_n(\theta(u))-q\}]_j\nabla^2\bar g_{nj}(\theta(u))\\
&\quad+\sum_j[W_nq]_j\{\nabla^2\bar g_{nj}(\theta(u))-\nabla^2\bar g_{nj}(\hat\theta_{n,W})\}.
\end{align*}
The empirical Jacobian is bounded and Lipschitz, so its component Hessians are bounded. The first two terms have norm at most $C\|d\|$; bounding both Hessians in the last term gives $C\|q\|$. Therefore
\begin{align*}
\|\kappa_t\|&\le C\|d\|(\|d\|+\|q\|)\\
&\le C\{\|d\|^2+\|\bar g_n(\theta_\oo)\|\|d\|\},
\end{align*}
}
where we use $\|\bar{g}_n(\hat{\theta}_{n,W})\| \le C \|\bar{g}_n(\theta_\oo)\|$.
{
Let $\varepsilon_0=\underbar c\delta^2/2$, with $\underbar c$ from \eqref{eq:eq7}, and write $\tau^\star=\tmt_{\varepsilon_0}^\star$. If $s-1\ge\tau^\star$, then \eqref{eq:eq7} implies
$\|\theta_{s-1}^\star-\hat\theta_{n,W}\|^2\le Q_{s-1}^\star/\underbar c$.
Consequently,
\[
\|\kappa_s\|\le C\{Q_{s-1}^\star+
\|\bar g_n(\theta_\oo)\|\sqrt{Q_{s-1}^\star}\},\qquad s>\tau^\star.
\]

\noindent\textit{Part~(ii):}
The error recursion has $d_0=0$ and
$d_t=(I-\gs[t]\P_n\bar{\H}_n)d_{t-1}-\gs[t]\P_n\kappa_t$.
Thus
\[
\sum_{s=1}^t(\theta_s^\star-\tto{s})
=-\sum_{s=1}^t\alpha_s^t(\P_n\bar{\H}_n)\P_n\kappa_s.
\]
On $E_n$, the matrices $\P_n\bar{\H}_n$ lie in a compact subset of matrices with eigenvalues having positive real parts, since they are similar to $\P_n^{1/2}\bar{\H}_n\P_n^{1/2}$. Lemma~\ref{lem:control-alpha-w} therefore bounds all the coefficients in this display by an absolute constant.

Fix $\eta>0$. Lemma~\ref{lem:stopping-time-and-tail} supplies fixed $\loclevel,L<\infty$, independent of $n$ and $\Ns$, such that on $E_n$ the conditional probabilities of $\stt_\loclevel^\star<\infty$ and $\tau^\star>L$ are each at most $\eta$. On their complement, for $\N\ge L$,
\begin{align*}
\frac1{\sqrt\N}\sup_{t\le\N}\left\|\sum_{s=1}^t(\theta_s^\star-\tto{s})\right\|
&\le\frac C{\sqrt\N}\sum_{s=1}^L\|\kappa_s\|\\
&\quad+\frac C{\sqrt\N}\sum_{s=L+1}^{\N}
\{Q_{s-1}^\star+\|\bar g_n(\theta_\oo)\|\sqrt{Q_{s-1}^\star}\}.
\end{align*}
For each fixed $L$, the first sum has conditional expectation bounded by $C_L$ on $E_n$, uniformly in $n$ and $\Ns$. Indeed, the drift recursion bounds $\mathbb E_n^\star Q_s^\star$ at every fixed $s$, while \eqref{eq:eq10} bounds the second moments of the noise. The update recursion, bounded initial value, and bounded $\hat\theta_{n,W}$ then bound $\mathbb E_n^\star\|\theta_s^\star-\hat\theta_{n,W}\|^2$, and the bound on $\kappa_s$ in Part~(i) applies. Hence the first term vanishes by Markov's inequality.

For the second term, insert $\mathbbm1\{\stt_\loclevel^\star\ge s-1\}$ and apply Lemma~\ref{lem:sgd-conv-rate} and Cauchy--Schwarz. Its conditional expectation is at most
\begin{align*}
\frac{C_\loclevel}{\sqrt\N}\sum_{s=L+1}^{\N}
\{\gs[s-1]+\|\bar g_n(\theta_\oo)\|\sqrt{\gs[s-1]}\}
&\le C_\loclevel\{\N^{1/2-a}
 +\|\bar g_n(\theta_\oo)\|\N^{(1-a)/2}\}\\
&=o_p(1),
\end{align*}
uniformly in $\Ns$, because $a>1/2$, $\sqrt n\bar g_n(\theta_\oo)=O_p(1)$, and $\N^{1-a}/n\to0$. A second application of Markov's inequality, followed by $\eta\downarrow0$ and $\mathbb P(E_n^c)\to0$, establishes the conditional approximation asserted in the lemma.
}

\subsection{Proof of Lemma~\ref{lem:FCLT-extension}}
We use a decomposition analogous to \citet{SGMM:2023}.
However, in the presence of $n$ and $\Ns$, the argument differs in that (i) $\theta_0^{\star}$ and $\gs[t]$ can depend on $\Ns$, (ii) $\alpha_{s}^t(\P_n \bar{\mathcal{H}}_n)$ and $w_{s}^t(\P_n \bar{\mathcal{H}}_n)$ vary with $n$,  (iii) the Gaussian approximation holds as $n$, $\N$, and $\Ns$ shift simultaneously.

We begin by writing $\tfrac{1}{\sqrt{\N}}\sum_{t=1}^{\lfloor \N r \rfloor} (\tto{t} - \hat{\theta}_{n,W}) = I_0(r) + I_1(r) + I_2(r)$. {
Put $m_r=\lfloor\N r\rfloor$ and $A_n=\P_n\bar{\H}_n$. Then
}
\begin{align*}
I_0(r) & = {\frac{\{\alpha_0^{m_r}(A_n)-\gs[0]I_{\dtheta}\}(\theta_0^\star-\hat\theta_{n,W})}{\sqrt\N\,\gs[0]}},\\
      I_1(r) & = -\bar{\H}_n^{-1}\frac{1}{\sqrt{\N}}\sum_{s=1}^{\lfloor \N r \rfloor} \xi_s(\theta_{s-1}^{\star}),\\
      I_2(r) & = \frac{1}{\sqrt{\N}}\sum_{s=1}^{\lfloor \N r \rfloor} w_s^{\lfloor \N r \rfloor}(\P_n \bar{\H}_n)  \P_n \xi_s(\theta_{s-1}^{\star}).
\end{align*}
{
Here $\alpha_s^t$ is defined in Lemma~\ref{lem:control-alpha-w}, and
$w_s^t(A_n)=A_n^{-1}-\alpha_s^t(A_n)$.
As shown in the preceding proof, on $E_n$ the matrices $A_n$ lie in a compact subset of matrices with eigenvalues having positive real parts. The same lemma therefore gives
\[
\sup_{0\le s\le t}\|\alpha_s^t(A_n)\|\le C,
\qquad\sum_{s=1}^t\|w_s^t(A_n)\|\le C(t+\Ns)^a.
\]
}

Since $\tfrac{1}{\sqrt \N \gs[0]} \le C \tfrac{(\Ns+1)^{a}}{\sqrt \N}$, {$\sup_{0\le r\le1}\|\alpha_0^{m_r}(A_n)-\gs[0]I_{\dtheta}\|\le C$}, and
\begin{align*}
(\|\theta_0^{\star} - \hat{\theta}_{n,W}\|\wedge 1)^2 \le C Q_0^{\star} \le C \gs[0] \le C (\Ns+1)^{-a},
\end{align*}
where $\|\theta_0^{\star}-\hat\theta_{n,W}\|\le C$ on $E_n$ by Condition~\ref{asm:appx:initial} and $\|\hat\theta_{n,W}\|\le C$, so that $\sup_{0\le r\le1}\|I_0(r)\|\le C\{(\Ns+1)^{a}/\N\}^{1/2}$ on $E_n$, we have, for every $\varepsilon > 0$,
\[
\mathbb{P}_n^{\star}\left( \sup_{0 \le r \le 1} \|I_0(r)\| \ge \varepsilon \right) \mathbbm 1_{E_n} \pto 0
\]
as $n, \N \to \infty$ and $\Ns/\N = O(1)$, which implies $\Ns^{a}/\N \to 0$ because $a<1$.

For $I_1(r)$, we verify that $I_1(r) \wto \left( (G_\oo'W G_\oo)^{-1} \Sigma_\oo (G_\oo' W G_\oo)^{-1}\right)^{1/2} W(r)$ at the end of this proof.

To address the last term $\sup_{0\le r \le 1}\|I_2(r)\|\ppto 0$, we first note that, for all $\varepsilon > 0$ and $\loclevel>0$,
\begin{align*}
& \p\left( \sup_{r \in [0,1]}\|I_2(r)\| > \varepsilon\right) \\
\le &\ \p\left( \sup_{r \in [0,1]}\|I_2(r)\| > \varepsilon, E_n\right) + \mathbb{P}(E_n^c) \\
\le &\ \p\left( \sup_{r \in [0,1]}\|I_2(r)\| > \varepsilon, E_n, \stt_{\loclevel}^{\star} = \infty\right) + \p(E_n, \stt_{\loclevel}^{\star} < \infty) + \mathbb{P}(E_n^c) \\
\le &\ \p\left( \sup_{r \in [0,1]}\|I_2(r)\| > \varepsilon, E_n, \stt_{\loclevel}^{\star} = \infty\right) + \mathbb{E}\left[ \mathbb{P}_n^{\star}(\stt_{\loclevel}^{\star} < \infty) \mathbbm{1}_{E_n}\right] + \mathbb{P}(E_n^c).
\end{align*}
By Lemma~\ref{lem:stopping-time-and-tail}, for every $\varepsilon > 0$, we can find $\loclevel > 0$ independent of $\Ns$ such that $\sup_{n \in \mathbb{N}}\mathbb{P}_n^{\star}(\stt_{\loclevel}^{\star} < \infty) \mathbbm{1}_{E_n} < \varepsilon$.
Moreover, $\mathbb{P}(E_n^c)\to 0$ as $n \to \infty$.
Thus, it suffices to show that, for each given $\loclevel>0$, $\sup_{r \in [0,1]}\|I_2(r)\| \ppto 0$ on the event $E_n \cap \{\stt_{\loclevel}^{\star} = \infty\}$.
Specifically, we establish that, for all $n \in \mathbb{N}$ and $\N \in \mathbb{N}$,
\begin{equation*}
   \mathbb{E}_n^{\star} \left[ \sum_{t=1}^{\N} \left\|\sum_{s=1}^{t} w_s^{t}(\P_n \bar{\H}_n) \P_n \xi_s(\theta_{s-1}^{\star})\I{\stt_{\loclevel}^{\star} \ge s}\right\|^{2p}  \right] \mathbbm{1}_{E_n} \le C_{\loclevel} (\N+\Ns)^{ap+1}
\end{equation*}
for an absolute constant $C_\loclevel>0$ that depends only on $\loclevel$.
This bound is sufficient, since conditional on the event $E_n \cap \{\stt_{\loclevel}^{\star} = \infty\}$,
\begin{align*}
\sup_{0\le r \le 1}\|I_2(r)\|^{2p} & \le \N^{-p}\sup_{1\le t \le \N} \left\|\sum_{s=1}^{t} w_s^{t}(\P_n \bar{\H}_n) \P_n \xi_s(\theta_{s-1}^{\star})\I{\stt_{\loclevel}^{\star} \ge s}\right\|^{2p} \mathbbm{1}_{E_n}\\
& \le \N^{-p} \sum_{t=1}^\N \left\|\sum_{s=1}^{t} w_s^{t}(\P_n \bar{\H}_n) \P_n \xi_s(\theta_{s-1}^{\star}) \I{\stt_{\loclevel}^{\star} \ge s} \right\|^{2p} \mathbbm{1}_{E_n},
\end{align*}
which must be on the order of $O_{\p}(\N^{1-(1-a)p}) = o_{\p}(1)$ under $\Ns/\N = O(1)$ and $p > (1-a)^{-1}$.

By applying Burkholder's inequality to the martingale difference partial sum, $\sum_{s=1}^{t} w_s^{t}(\P_n \bar{\H}_n) \P_n \xi_s(\theta_{s-1}^{\star})\I{\stt_{\loclevel}^{\star} \ge s}$, we observe that for all $t \in \mathbb{N}$,
\begin{align*}
& \mathbb{E}_n^{\star}\left\|\sum_{s=1}^{t} w_s^{t}(\P_n \bar{\H}_n) \P_n \xi_s(\theta_{s-1}^{\star}) \I{\stt_{\loclevel}^{\star} \ge s} \right\|^{2p} \\
\le &\ C_p \mathbb{E}_n^{\star} \left[ \left( \sum_{s=1}^t \|w_s^t(\P_n \bar{\H}_n)\P_n \|^2 \|\xi_s(\theta^{\star}_{s-1})\|^2 \I{\stt_{\loclevel}^{\star} \ge s}\right)^{p}\right].
\end{align*}
Here, $C_p$ denotes an absolute constant depending only on $p$.
Applying Hölder's inequality
$$
\sum_s |a_sb_s| \le \left( \sum_s |a_s|^\alpha\right)^{1/\alpha} \left( \sum_s |b_s|^\beta\right)^{1/\beta}
$$
with $(a_s,b_s) = (\|w_s^t(\P_n \bar{\H}_n)\P_n \|^{2-2/p}, \|w_s^t(\P_n \bar{\H}_n)\P_n \|^{2/p}\|\xi_s(\theta_{s-1}^{\star})\|^2\I{\stt_{\loclevel}^{\star} \ge s})$ and $(\alpha,\beta) = (p/(p-1), p)$, we get
\begin{align*}
& \sum_{s=1}^t \|w_s^t(\P_n \bar{\H}_n)\P_n \|^2 \|\xi_s(\theta^{\star}_{s-1})\|^2\I{\stt_{\loclevel}^{\star} \ge s} \\
\le &\ \left( \sum_{s=1}^t \|w_s^t(\P_n \bar{\H}_n)\P_n \|^2 \right)^{\frac{p-1}{p}} \left( \sum_{s=1}^t \|w_s^t(\P_n \bar{\H}_n) \P_n\|^2 \|\xi_s(\theta^{\star}_{s-1})\|^{2p}\I{\stt_{\loclevel}^{\star} \ge s}\right)^{\frac{1}{p}}.
\end{align*}
This implies that, for all $t \in \mathbb{N}$,
\begin{align*}
   & \mathbb{E}_n^{\star} \left[ \left( \sum_{s=1}^t \|w_s^t(\P_n \bar{\H}_n)\P_n \|^2 \|\xi_s(\theta^{\star}_{s-1})\|^2 \I{\stt_{\loclevel}^{\star} \ge s}\right)^{p}\right] \\
\le &\ \left( \sum_{s=1}^t \|w_s^t(\P_n \bar{\H}_n) \P_n\|^2\right) ^{p-1} \sum_{s=1}^t \|w_s^t(\P_n \bar{\H}_n) \P_n\|^2 \mathbb{E}_n^{\star}\left[ \|\xi_s(\theta^{\star}_{s-1})\|^{2p}\I{\stt_{\loclevel}^{\star} \ge s}\right]\\
\le &\ \left( \sum_{s=1}^t \|w_s^t(\P_n \bar{\H}_n) \P_n\|^2\right) ^{p-1} \\
&\quad\times \sum_{s=1}^t \|w_s^t(\P_n \bar{\H}_n) \P_n\|^2 \mathbb{E}_n^{\star}\left[ \mathbb{E}_n^{\star}[\|\xi_s(\theta^{\star}_{s-1})\|^{2p}|\mathcal{F}_{n,s-1}^{\star}]\I{\stt_{\loclevel}^{\star} \ge s}\right]\\
\le &\ C \left( \sum_{s=1}^t \|w_s^t(\P_n \bar{\H}_n) \P_n\|^2\right) ^{p-1} \sum_{s=1}^t \|w_s^t(\P_n \bar{\H}_n) \P_n\|^2 \mathbb{E}_n^{\star}\left[ (Q_{s-1}^{\star}{}^p + 1)\I{\stt_{\loclevel}^{\star} \ge s}\right]\\
\le &\ C\left( \sum_{s=1}^t \|w_s^t(\P_n \bar{\H}_n) \P_n\|^2\right) ^{p} (\loclevel^p + 1)
\end{align*}
on the event $E_n$,
where we use the fact that 
$$
\mathbb{E}_n^{\star}\left[ \|\xi_s(\theta^{\star}_{s-1})\|^{2p}|\mathcal{F}_{n,s-1}^{\star}\right] = \left.\mathbb{E}_n^{\star}[\|\xi_s(\theta)\|^{2p}]\right|_{\theta = \theta_{s-1}^{\star}} \le M (Q_{s-1}^{\star}{}^p + 1)
$$
on the event $E_n$, and $Q_{s-1}^{\star} \le \loclevel$ conditional on $\{\stt_{\loclevel}^{\star} \ge s\}$.
Uniformly in $s \in \mathbb{N}$, $t \ge s$, and $n \in \mathbb{N}$, $\|w_{s}^t(\P_n \bar{\H}_n)\P_n \| \le \|\alpha_s^t(\P_n \bar{\H}_n)\P_n \| + \|\bar{\H}_n^{-1}\| \le C$ on the event $E_n$, whence it follows, for all $t \ge 1$ and $n \in \mathbb{N}$,
$$
\sum_{s=1}^t \|w_s^t(\P_n \bar{\H}_n) \P_n\|^2\le C \sum_{s=1}^t \|w_s^t(\P_n \bar{\H}_n) \P_n\| \le C (t+\Ns)^a,
$$
{
Summing these bounds gives
\begin{align*}
&\mathbb E_n^\star\left[\sum_{t=1}^{\N}\left\|
 \sum_{s=1}^t w_s^t(A_n)\P_n\xi_s(\theta_{s-1}^\star)
 \mathbbm1\{\stt_\loclevel^\star\ge s\}\right\|^{2p}\right]\\
&\qquad\le C_\loclevel\sum_{t=1}^{\N}\left(\sum_{s=1}^t\|w_s^t(A_n)\P_n\|^2\right)^p\\
&\qquad\le C_\loclevel\sum_{t=1}^{\N}(t+\Ns)^{ap}
\le C_\loclevel(\N+\Ns)^{ap+1}.
\end{align*}
}
This concludes $\sup_{0\le r \le 1}\|I_2(r)\|\overset {\p} \to 0$, as discussed before.

{
Finally, put $e_s=\xi_s(\theta_{s-1}^\star)$. For the martingale FCLT it suffices to verify, as $n,\N\to\infty$,
\begin{align}
\label{eq:FCLT-first}
\frac1\N\sum_{s=1}^{\N}\mathbb E_n^\star
[\|e_s\|^2\mathbbm1\{\|e_s\|>\sqrt\N\,\delta\}\mid\mathcal F_{n,s-1}^\star]
&\pto0,\qquad \delta>0,\\
\label{eq:FCLT-second}
\frac1\N\sum_{s=1}^{\lfloor\N r\rfloor}\mathbb E_n^\star
[e_se_s'\mid\mathcal F_{n,s-1}^\star]
&\pto r\Sigma_\oo,\qquad r\in[0,1].
\end{align}
By Lemma~\ref{lem:stable-mclt}, applied with $\mathcal G_{n,t}=\mathcal F_{n,t}^\star$, for which $\mathbb E[\cdot\mid\mathcal F_{n,s-1}^\star]=\mathbb E_n^\star[\cdot\mid\mathcal F_{n,s-1}^\star]$, and with $V=\Sigma_\oo$, these conditions imply that $\N^{-1/2}\sum_{s=1}^{\lfloor\N\cdot\rfloor}e_s$ converges $\mathcal F_{n,0}^\star$-stably to $\Sigma_\oo^{1/2}W(\cdot)$. This covers the sample and all preliminary algorithmic draws, without requiring the initial sigma-algebras to be nested.

Together with $\bar{\H}_n\pto G_\oo'WG_\oo$, this gives the asserted limit for $I_1$.

\noindent\textit{Proof of \eqref{eq:FCLT-first}:}
For $p>1$, on $E_n$ the moment bound \eqref{eq:eq10} implies
\begin{align*}
\mathbb E_n^\star[\|e_s\|^2\mathbbm1\{\|e_s\|>\sqrt\N\,\delta\}
 \mid\mathcal F_{n,s-1}^\star]
&\le\N^{1-p}\delta^{2(1-p)}\mathbb E_n^\star[
 \|e_s\|^{2p}\mid\mathcal F_{n,s-1}^\star]\\
&\le C\N^{1-p}\delta^{2(1-p)}\{(Q_{s-1}^\star)^p+1\}.
\end{align*}
After inserting the predictable indicator $\mathbbm1\{\stt_\loclevel^\star\ge s\}$ in each summand, the average is bounded by
$C\N^{1-p}\delta^{2(1-p)}(\loclevel^p+1)\to0$ for each fixed $\loclevel$. The probability of ever hitting the level $\loclevel$ can be made arbitrarily small, uniformly in $n$ and $\Ns$, by Lemma~\ref{lem:stopping-time-and-tail}. This proves \eqref{eq:FCLT-first} after removing the localization.
}

\noindent \textit{Proof of \eqref{eq:FCLT-second}:}
Recall the definition of $\Sigma_{n,t}(\theta)$: $\Sigma_{n,t}(\theta) = \mathbb{E}_n^{\star}[\xi_t(\theta)\xi_t(\theta)']$, which depends on both $n \in \mathbb{N}$ and $t\ge 1$.
We have also defined the following matrices:
\begin{equation*}
\Sigma_{n,t} := \Sigma_{n,t}(\hat{\theta}_{n,W}),\quad \Sigma_t := \operatornamewithlimits{plim}_{n\to\infty} \Sigma_{n,t},\quad \Sigma_\oo := \operatornamewithlimits{lim}_{t\to\infty} \Sigma_t.
\end{equation*}
We begin by writing
\begin{align*}
\frac{1}{\N} \sum_{s=1}^{\lfloor \N r \rfloor} \mathbb{E}_n^{\star} \left[  \left. \xi_s(\theta^{\star}_{s-1}) \xi_s(\theta^{\star}_{s-1})'\right| \mathcal{F}^{\star}_{n,s-1}\right] 
&= \frac{1}{\N} \sum_{s=1}^{\lfloor \N r \rfloor} \Sigma_{n,s}(\theta_{s-1}^{\star})\\
&= \frac{1}{\N} \sum_{s=1}^{\lfloor \N r \rfloor} \Sigma_s + \frac{1}{\N} \sum_{s=1}^{\lfloor \N r \rfloor} [\Sigma_{n,s}-\Sigma_s]\\
&\quad +\frac{1}{\N} \sum_{s=1}^{\lfloor \N r \rfloor} [\Sigma_{n,s}(\theta_{s-1}^{\star})-\Sigma_{n,s}].
\end{align*}
We show that the latter two terms are negligible.
Observe that
\begin{align*}
\xi_t(\hat\theta_{n,W})
&=\tilde G_t(\hat\theta_{n,W})'W_n\tilde g_t(\hat\theta_{n,W})\\
&=\big(\tilde G_t(\hat\theta_{n,W})-\bar G_n\big)'W_n\tilde g_t(\hat\theta_{n,W})\\
&\quad+\bar G_n'W_n\big(\tilde g_t(\hat\theta_{n,W})-\bar g_n(\hat\theta_{n,W})\big).
\end{align*}
Since $\tilde{G}_t(\hat{\theta}_{n,W}) - \bar{G}_n$ is centered at zero and independent of $\tilde{g}_t(\hat{\theta}_{n,W})$ conditional on {$\mathcal F_{n,0}^{\star}$}, these two terms are uncorrelated to each other, yielding
{
\begin{align*}
\Sigma_{n,t}
&=\frac1{B_g}\bar G_n'W_n\tilde\Omega_nW_n\bar G_n\\
&\quad+\frac1{nB_{G,t}}\sum_{i=1}^n
 \Delta_i'W_n\left(\frac{\tilde\Omega_n}{B_g}+q_nq_n'\right)W_n\Delta_i,
\end{align*}
where $q_n=\bar g_n(\hat\theta_{n,W})$ and
$\Delta_i=G(z_i,\hat\theta_{n,W})-\bar G_n$.
The empirical moments converge to their population limits, $q_n=o_p(1)$,
and $B_{G,t}^{-1}\le1$ uniformly in $t$. Thus the exact expression gives
}
\begin{equation*}
\sup_{t \ge 1}\|\Sigma_{n,t} - \Sigma_t \| \pto 0\quad \text{as}\ \ \ n \to \infty.
\end{equation*}
In turn, this leads to {$\frac{1}{\N} \sum_{s=1}^{\lfloor \N r \rfloor} [\Sigma_{n,s} - \Sigma_s]  \pto 0$} as $n \to \infty$, addressing the second term.
To show that the last term converges in probability to zero, it suffices to show
\begin{equation*}
\frac{1}{\N} \sum_{s=1}^{\lfloor \N r \rfloor} [\Sigma_{n,s}(\theta_{s-1}^{\star})-\Sigma_{n,s}] \I{\stt_{\loclevel}^{\star} \ge s} \mathbbm{1}_{E_n} \ppto 0
\end{equation*}
for any given $\loclevel$.
{
Fresh independent draws allow us to apply \eqref{eq:eq9}--\eqref{eq:eq10} conditional on $\mathcal F_{n,s-1}^{\star}$ at the measurable iterate $\theta_{s-1}^{\star}$. Cauchy--Schwarz gives
\begin{align*}
&\|\Sigma_{n,s}(\theta_{s-1}^{\star})-\Sigma_{n,s}\|\\
&\quad\le
\Big(\mathbb E_n^{\star}[\|\xi_s(\theta_{s-1}^{\star})-\xi_s(\hat\theta_{n,W})\|^2
\mid\mathcal F_{n,s-1}^{\star}]\Big)^{1/2}\\
&\qquad\times
\Big(2\mathbb E_n^{\star}[\|\xi_s(\theta_{s-1}^{\star})\|^2+
\|\xi_s(\hat\theta_{n,W})\|^2\mid\mathcal F_{n,s-1}^{\star}]\Big)^{1/2}.
\end{align*}
On $E_n\cap\{\stt_{\loclevel}^{\star}\ge s\}$, the first factor is at most $C\|\theta_{s-1}^{\star}-\hat\theta_{n,W}\|$, and the second is bounded by a constant $C_{\loclevel}$, since $Q_{s-1}^{\star}<\loclevel$.
The same second-moment bounds also bound the covariance difference by $C_{\loclevel}$. Thus, uniformly in $n$, $s\ge1$, and $\Ns$,
\begin{align*}
&\|\Sigma_{n,s}(\theta_{s-1}^{\star})-\Sigma_{n,s}\|
\I{\stt_{\loclevel}^{\star}\ge s}\mathbbm1_{E_n}\\
&\quad\le C_{\loclevel}(\|\theta_{s-1}^{\star}-\hat\theta_{n,W}\|\wedge1)
\I{\stt_{\loclevel}^{\star}\ge s}\mathbbm1_{E_n}.
\end{align*}
Thus, as $n\to\infty$ and $\N \to\infty$,
\begin{align*}
   & \mathbb{E}\mathbb{E}_n^{\star} \frac{1}{\N} \sum_{s=1}^{\lfloor \N r \rfloor} \left\|\Sigma_{n,s}(\theta_{s-1}^{\star})-\Sigma_{n,s}\right\| \I{\stt_{\loclevel}^{\star} \ge s} \mathbbm{1}_{E_n} \\
   \le &\ \mathbb{E}\frac{C_{\loclevel}}{\N} \sum_{s=1}^{\lfloor \N r \rfloor} \mathbb{E}_n^{\star}[(\|\theta_{s-1}^{\star}-\hat{\theta}_{n,W}\|\wedge 1)\I{\stt_{\loclevel}^{\star} \ge s-1}] \mathbbm{1}_{E_n}\\
   \le &\ \mathbb{E}\frac{C_{\loclevel}}{\N} \sum_{s=1}^{\N} \mathbb{E}_n^{\star}[\sqrt{Q_{s-1}^{\star}} \I{\stt_{\loclevel}^{\star} \ge s-1}] \mathbbm{1}_{E_n}   \le \frac{1}{\N} \sum_{s=1}^{\N} \sqrt{C_{\loclevel}\gs[s-1]} \to 0
\end{align*}
by \eqref{eq:eq7} and Lemma~\ref{lem:sgd-conv-rate}, with $C_{\loclevel}$ allowed to change between inequalities. The convergence is uniform in $\Ns \ge 0$.
This verifies
\begin{equation*}
\frac{1}{\N} \sum_{s=1}^{\lfloor \N r \rfloor} [\Sigma_{n,s}(\theta_{s-1}^{\star})-\Sigma_{n,s}] \I{\stt_{\loclevel}^{\star} \ge s} \mathbbm{1}_{E_n} \ppto 0
\end{equation*}
uniformly in $\Ns$.
}
Now, \eqref{eq:FCLT-second} follows from the fact that $\frac{1}{\N} \sum_{s=1}^{\lfloor \N r \rfloor} \Sigma_s \to r \Sigma_\oo$
since the deterministic sequence $(\Sigma_t)_{t \ge 1}$ converges to $\Sigma_\oo$ as $t \to \infty$.
Finally, the decomposition $I_0+I_1+I_2$, with $\sup_r\|I_0(r)\|+\sup_r\|I_2(r)\|=o_p(1)$ and $I_1(r)=-\bar{\H}_n^{-1}\N^{-1/2}\sum_{s\le\lfloor\N r\rfloor}\xi_s(\theta_{s-1}^\star)$, is the asserted linear representation, and the stable convergence of the martingale part, obtained above from Lemma~\ref{lem:stable-mclt}, yields the asserted $\mathcal F_{n,0}^\star$-stable convergence.
%%%%%%%%%%%%%%%%%%%%
\subsection{Proof of Lemma~\ref{lem:CLT-extension}}
We note that Lemma~\ref{lem:CLT-extension} arises as a special case of the approximation in Lemma~\ref{lem:FCLT-extension} at $r = 1$.
Relative to the proof of Lemma~\ref{lem:FCLT-extension}, the relaxed assumption of $p > 1$ only changes the way we address $I_1(1)$ and $I_2(1)$.

For $I_1(1)$, we first observe that the Lindeberg condition \eqref{eq:FCLT-first} and the bracket condition \eqref{eq:FCLT-second} remain valid, as the arguments in the proof of Lemma~\ref{lem:FCLT-extension} continue to work under $p > 1$.
As a result, the CLT applies to $I_1(r)$ with $r = 1$, yielding
\begin{equation*}
    I_1(1) \dto \mathcal{N}(0,(G_\oo' W G_\oo)^{-1}\Sigma_\oo    (G_\oo' W G_\oo)^{-1})
\end{equation*}
as $n, \N \to \infty$.

{
For $I_2(1)$, conditional martingale orthogonality and \eqref{eq:eq10} give
\begin{align*}
&\mathbb E_n^\star\left\|\sum_{s=1}^{\N}
 w_s^{\N}(\P_n\bar{\H}_n)\P_n\xi_s(\theta_{s-1}^\star)
 \mathbbm1\{\stt_\loclevel^\star\ge s\}\right\|^2\\
&\quad\le C_\loclevel\sum_{s=1}^{\N}\|w_s^{\N}(\P_n\bar{\H}_n)\P_n\|^2
\le C_\loclevel(\N+\Ns)^a
\end{align*}
on $E_n$. The stopping indicators are predictable, and the coefficient matrices are $\mathcal F_{n,0}^\star$-measurable.
}

As the learning rate exponent $a < 1$, we have, for all $\loclevel>0$,
\begin{equation*}
\sup_{n \in \mathbb{N}}\mathbb{E}_n^{\star} [\|I_2(1)\|^2 \I{\stt_{\loclevel}^{\star} = \infty}] \mathbbm{1}_{E_n} \le C_\loclevel (\N+\Ns)^{a}/\N \to 0
\end{equation*}
as $\N \to \infty$ and $\Ns/\N = O(1)$.
This implies that $I_2(1) \ppto 0$.
Together with $I_0(1)=o_p(1)$, this gives the asserted linear representation at $r=1$, and Lemma~\ref{lem:stable-mclt} at $r=1$, whose conditions hold under $p>1$, gives the asserted stable convergence.

%%%%%%%%%%%%%%%%%%%%
\subsection{Proof of Lemma~\ref{lem:refinement-weight}}
We verify the consistency of $\Phi_n$ and $W_{\mathrm{MB}}$ in \eqref{def:Phi:W} as $n\to\infty$, $M_{\mathrm{MB}} \to \infty$, and $N \to \infty$. By Theorem~\ref{thm:consistency}, $\bar\theta_N^*\pto\theta_\oo$, and local uniform laws of large numbers, implied by the envelope and Lipschitz conditions in Assumption~\ref{asm:regularity}, give $\Phi_n\pto G_\oo$, $C_n(\bar\theta_N^*)\pto\Omega_\oo$, and $\bar g_n(\bar\theta_N^*)\pto0$, so the term $B_g\bar g_n(\bar\theta_N^*)\bar g_n(\bar\theta_N^*)'$ in the conditional mean displayed before~\eqref{def:centered W} vanishes. Because the batches defining $W_{\mathrm{MB}}$ are drawn after $\bar\theta_N^*$ is computed, a conditional weak law of large numbers, which holds under Condition~\ref{asm:item:stability} with $p>1$, shows that the matrix inside the generalized inverse in \eqref{def:Phi:W} converges in probability to $\Omega_\oo>0$; continuity of the Moore--Penrose inverse at nonsingular matrices then gives $W_{\mathrm{MB}}\pto\Omega_\oo^{-1}$. For $W_{\mathrm{MB}}^c$, the matrix inside the generalized inverse in~\eqref{def:centered W} has conditional mean exactly $C_n(\bar\theta_N^*)$, so the same argument gives $W_{\mathrm{MB}}^c\pto\Omega_\oo^{-1}$. Replacing $\bar g_n(\bar\theta_N^*)$ by $\hat g$ adds $B_g(\hat g-\bar g_n)(\hat g-\bar g_n)'-B_g(\tilde g^\circ-\bar g_n)(\hat g-\bar g_n)'-B_g(\hat g-\bar g_n)(\tilde g^\circ-\bar g_n)'$ to that matrix, where $\bar g_n=\bar g_n(\bar\theta_N^*)$ and $\tilde g^\circ=M_{\mathrm{MB}}^{-1}\sum_{\ell=1}^{M_{\mathrm{MB}}}\tilde g_\ell(\bar\theta_N^*)$; this is $o_p(1)$ because $B_g$ is fixed and $\hat g-\bar g_n=o_p(1)$.
By continuous mapping theorem, it is evident that $\P_n = (\Phi_n' W_{\mathrm{MB}} \Phi_n)^{\dagger} \ppto \P = (G_\oo'\Omega_\oo^{-1}G_\oo)^{-1}$.
The proofs of Lemmas~\ref{lem:consistency-and-normality} and~\ref{lem:sample-regularity}, and the argument following Lemma~\ref{lem:sample-regularity} that gives $\mathbb P(E_n)\to1$, do not use Condition~\ref{asm:appx:initial}. For the refinement stage, Condition~\ref{asm:appx:conditioning-and-weighting} has just been verified, and Conditions~\ref{asm:appx:batch-size}--\ref{asm:appx:sample-regularity} are imposed by the assumptions of Theorem~\ref{thm:asymp:normal:second:order}. Hence the conclusions of Lemmas~\ref{lem:consistency-and-normality} and~\ref{lem:sample-regularity} hold for the refinement stage, and $\mathbb P(E_n^{(2)})\to1$.

%%%%%%%%%%%%%%%%%%%%
\subsection{Proof of Lemma~\ref{lem:refinement-initial}}
In this proof, $E_n$, $\bar Q_n$, and $\stt^*_{\loclevel}$ refer to the first-order stage ($W_n=I_{\dg}$, $\P_n=I_{\dtheta}$, $\Ns=0$), for which Assumption~\ref{asm:appx:regularity} holds, as verified in the proof of Theorem~\ref{thm:consistency}, whereas $\hat\theta_{n,W}$ and $\bar Q_{n,W}$ refer to the refinement stage.
First of all, the initial value satisfies $\theta_N^* = O_{\mathrm P}(1)$ because $\theta_N^*$ is consistent for $\theta_\oo$ by Theorem~\ref{thm:consistency}.
Thus, we can find $M < \infty$ such that $\|\theta_N^*\|\le M$ with probability greater than $1-\varepsilon$ for any given $\varepsilon > 0$.
On the other hand, Lemma~\ref{lem:stopping-time-and-tail} implies that, for any given $\varepsilon>0$, there exists $\loclevel > 0$ such that
\begin{equation*}
\sup_{n \in \mathbb{N}}\mathbb{P}_n^*(\stt_{\loclevel}^* < \infty) \mathbbm{1}_{E_n} < \varepsilon.
\end{equation*}
Moreover, Lemma~\ref{lem:sgd-conv-rate} implies that, for such a $\loclevel>0$, there exists $C_\loclevel < \infty$ such that
\begin{equation*}
\mathbb{E}_n^*[\bar{Q}_{n}(\theta_N^*)\I{\stt_{\loclevel}^* \ge N}] \mathbbm{1}_{E_n} \le C_{\loclevel} \gamma_0 N^{-a}, \quad \forall n,N \in \mathbb{N}.
\end{equation*}
Let $\mathcal{E}_{N,n} := \{\stt_{\loclevel}^* \ge N\} \cap E_n$ so that
\begin{equation*}
\sup_{N \in \mathbb{N}} \p(\mathcal{E}_{N,n}^c) \le \mathbb{E}[\mathbb{P}_n^*(\stt_{\loclevel}^* < \infty) \mathbbm{1}_{E_n}] + \mathbb{P}(E_n^c) \le \varepsilon + \mathbb{P}(E_n^c), \quad \forall n \in \mathbb{N}.
\end{equation*}
We choose $n_0$ such that $\mathbb{P}(E_n^c) < \varepsilon$ for all $n \ge n_0$, and hence $\sup_{N \in \mathbb{N}} \p(\mathcal{E}_{N,n}^c) < 2 \varepsilon$ for all $n \ge n_0$.
Taking $\varepsilon<1/2$ and absorbing $\p(\mathcal E_{N,n})^{-1}\le(1-2\varepsilon)^{-1}$ into $C_{\loclevel}$, we have
\begin{equation*}
\mathbb{E}[\bar{Q}_{n}(\theta_{N}^*) | \mathcal{E}_{N,n}] = \mathbb{E}\left[ \mathbb{E}_n^*[\bar{Q}_{n}(\theta_N^*)\I{\stt_{\loclevel}^* \ge N}]\mathbbm{1}_{E_n} \right]/\p (\mathcal{E}_{N,n}) \le C_{\loclevel}\gamma_0 N^{-a}.
\end{equation*}
We can therefore find an absolute constant $C_1 = C_1(\loclevel, \varepsilon) < \infty$ such that
\begin{equation*}
\p\left( \bar{Q}_{n}(\theta_N^*) \le C_1 \gamma_0 N^{-a}, \mathcal{E}_{N,n}\right) \ge 1 - 3\varepsilon\quad \text{for all}\ n\ge n_0\text{ and } N \in \mathbb{N}.
\end{equation*}
To complete the proof, we now show that $\bar Q_{n,W}(\theta_N^*)\le C(\gamma_0N^{-a}+n^{-1})$ on an event of arbitrarily high probability.
As $N \to \infty$, on the event $\mathcal{E}_{N,n} \subseteq E_n$, we have
\begin{equation*}
\underbar{c} (\|\theta_N^* - \hat{\theta}_{n}\|^2 \wedge \delta^2) \le \bar{Q}_{n}(\theta_N^*) \le C_1 \gamma_0 N^{-a} < \underbar{c}\delta^2,
\end{equation*}
and hence $\|\theta_N^* - \hat{\theta}_{n}\|^2 \le (C_1/\underbar{c}) \gamma_0 N^{-a}$.
Since $\sqrt{n}\|\hat{\theta}_{n} - \hat{\theta}_{n,W}\|= O_{\mathbb{P}}(1)$ by Lemma~\ref{lem:consistency-and-normality} for the two stages (see Lemma~\ref{lem:refinement-weight}),
\begin{equation*}
\mathbb{P}\left( \|\hat{\theta}_{n} - \hat{\theta}_{n,W}\|^2 \le C_2/n\right) \ge 1-\varepsilon
\end{equation*}
for some absolute constant $C_2 = C_2(\varepsilon)$.
Since $\bar{Q}_{n,W}(\theta) \le C \|\theta - \hat{\theta}_{n,W}\|^2$ for all $\theta \in \Theta$ on $E_n^{(2)}$, we have
\begin{equation*}
\bar{Q}_{n,W}(\theta_N^*) \le C  \|\theta_N^* - \hat{\theta}_{n,W}\|^2 \le C (\|\theta_N^* - \hat{\theta}_{n}\|^2 + \|\hat{\theta}_{n} - \hat{\theta}_{n,W}\|^2) \le C(\gamma_0 N^{-a} + n^{-1})
\end{equation*}
on the event $\mathcal{E}_{N,n} \cap \{\bar Q_n(\theta_N^*)\le C_1\gamma_0N^{-a}\}\cap{} \{\|\hat{\theta}_{n} - \hat{\theta}_{n,W}\|^2 \le C_2/n\}{}\cap E_n^{(2)}$ that has probability at least $1-4\varepsilon{}-\mathbb P((E_n^{(2)})^c)$.
Under $N^a/n=O(1)$, this gives $\bar Q_{n,W}(\theta_N^*)\le C\gamma_0N^{-a}=C\gamma_0^\star$. Hence, for every $\varepsilon\in(0,1/2)$, the $\mathcal F_{n,0}^\star$-measurable event
\[
A_{n,5\varepsilon}:=\mathcal E_{N,n}\cap\{\bar Q_n(\theta_N^*)\le C_1\gamma_0N^{-a}\}\cap\{\|\hat\theta_n-\hat\theta_{n,W}\|^2\le C_2/n\}\cap\{\|\theta_N^*\|\le M\}\cap E_n^{(2)}
\]
satisfies $\liminf\mathbb P(A_{n,5\varepsilon})\ge1-5\varepsilon$ by Lemma~\ref{lem:refinement-weight}, and on it $\|\theta_N^*\|\le M$ and $\bar Q_{n,W}(\theta_N^*)\le C\gamma_0^\star$. This verifies Condition~\ref{asm:appx:initial} for the refinement stage with $\eta=5\varepsilon$ and $M_\eta=\max\{M,C\}$.

For the initial value $\bar\theta_N^*$, which also lies in $\Theta$ and is $\mathcal F_{n,0}^\star$-measurable, the quadratic bound on $E_n^{(2)}$ gives $\bar Q_{n,W}(\bar\theta_N^*)\le C(\|\bar\theta_N^*-\hat\theta_n\|^2+\|\hat\theta_n-\hat\theta_{n,W}\|^2)$. Here $\|\bar\theta_N^*-\hat\theta_n\|^2=O_p(N^{-1})$ by Theorem~\ref{thm:asymp:normal}, whose rate condition $N^{1-a}/n\to0$ is implied by $N^a/n=O(1)$, and $\|\hat\theta_n-\hat\theta_{n,W}\|^2=O_p(n^{-1})$. Since $N^{-1}\le N^{-a}$ and $n^{-1}=O(N^{-a})$, we obtain $\bar Q_{n,W}(\bar\theta_N^*)=O_p(N^{-a})$, while $\|\bar\theta_N^*\|=O_p(1)$ by Theorem~\ref{thm:consistency}. Hence, for every $\eta>0$, there are $M$ and $C$ such that $\{\|\bar\theta_N^*\|\le M\}\cap\{\bar Q_{n,W}(\bar\theta_N^*)\le C\gamma_0N^{-a}\}\cap E_n^{(2)}$ has probability at least $1-\eta$ for all large $n$ and $N$, which verifies Condition~\ref{asm:appx:initial} for this initial value.

%%%%%%%%%%%%%%%%%%%%

\subsection{Proof of Lemma~\ref{lem:Robbins-Siegmund}}
See Lemma~5.2.2 in \citet{benveniste2012adaptive}.
% \citet[p. 344]{benveniste2012adaptive}.
%%%%%%%%%%%%%%%%%%%%
\subsection{Proof of Lemma~\ref{lem:recursive-bound}}
{
Since $0<\mu\gamma_{n+1}<1$ for $n\ge n_0$, the recursion gives
\[
v_{n+1}\le(1-\mu\gamma_{n+1})v_n+\mu\gamma_{n+1}(C/\mu)
\le\max\{v_n,C/\mu\}.
\]
Induction yields $v_n\le\max\{v_{n_0},C/\mu\}\le\max\{C,1\}(v_{n_0}+\mu^{-1})$, where the last inequality uses $v_{n_0}\ge0$.
}

%%%%%%%%%%%%%%%%%%%%
\subsection{Proof of Lemma~\ref{lem:control-alpha-w}}
{
Let $V=V(A)>0$ solve $A'V+VA=I_d$, and put $L=\lambda_{\max}(V)$ and $\ell=\lambda_{\min}(V)$. In the norm $\|x\|_V^2=x'Vx$,
\[
\|(I_d-\gamma A)x\|_V^2
\le\left(1-\frac\gamma L+\frac{\|A\|^2L}{\ell}\gamma^2\right)\|x\|_V^2.
\]
Taking square roots, applying $1+u\le e^u$, and using equivalence of norms gives
\begin{align}\label{eq:each-term-bound}
\left\|\prod_{j=s+1}^i(I_d-\gs[j]A)\right\|
&\le\sqrt{\frac L\ell}\exp\left\{-\frac12\sum_{j=s+1}^i
 \left(\frac{\gs[j]}L-\frac{\|A\|^2L}{\ell}\gs[j]^2\right)\right\}\nonumber\\
&\le\kappa\exp\left\{-\lambda\sum_{j=s+1}^i\gs[j]\right\},
\end{align}
where
\[
\lambda=\frac1{2L},\qquad
\kappa=\sqrt{\frac L\ell}\exp\left\{\frac{\|A\|^2L}{2\ell}\gamma_0^2\zeta(2a)\right\},
\qquad \zeta(2a)=\sum_{j=1}^\infty j^{-2a}.
\]
The matrices $V(A)$ and these constants depend continuously on $A$ (and on $a,\gamma_0$ where applicable).

To bound $\alpha_s^t$, put $b=s+\Ns+1\ge1$. Monotonicity gives
$\sum_{j=s+1}^i\gs[j]\ge\gamma_0\int_{s+1}^{i+1}(u+\Ns)^{-a}\,du$.
Comparing the resulting decreasing sum with its integral and using
$z=(x+\Ns+1)^{1-a}-b^{1-a}$ yields
\begin{align*}
\|\alpha_s^t\|
&\le\kappa\gs[s]\left[1+\frac1{1-a}\int_0^\infty
 e^{-\lambda\gamma_0z/(1-a)}(z+b^{1-a})^{a/(1-a)}\,dz\right]\\
&\le\kappa\gamma_0\left[1+\frac{2^a}{1-a}\int_0^\infty
 e^{-\lambda\gamma_0z/(1-a)}(1+z)^{a/(1-a)}\,dz\right]=:M_1.
\end{align*}
Here $\gs[s]b^a\le2^a\gamma_0$, including $s=\Ns=0$ under the stated convention for $\gs[0]$. The integral is finite and depends continuously on its positive arguments.

For the second assertion, write $w_s^t=\alpha_s^t-A^{-1}$ and
\[
\beta_s^t=\sum_{i=s}^t(\gs[s]-\gs[i+1])
                    \prod_{k=s+1}^i(I_d-\gs[k]A).
\]
Telescoping the matrix products gives
$\beta_s^t=w_s^t+A^{-1}\prod_{k=s+1}^{t+1}(I_d-\gs[k]A)$.
For $0\le x\le y\le\gamma_0$,
$y-x\le e^{\lambda\gamma_0}(e^{-\lambda x}-e^{-\lambda y})/\lambda$.
Consequently, summing the scalar exponential differences over $1\le s\le i\le t$ gives
\begin{align*}
\sum_{s=1}^t\|\beta_s^t\|
&\le\frac{\kappa e^{\lambda\gamma_0}}\lambda
 \sum_{s=1}^t\left[
 e^{-\lambda\sum_{k=s+1}^{t+1}\gs[k]}
 -e^{-\lambda\sum_{k=1}^{s}\gs[k]}\right]\\
&\le\frac{\kappa e^{\lambda\gamma_0}}\lambda R_t,
\qquad R_t:=\sum_{s=1}^t e^{-\lambda\sum_{k=s+1}^{t+1}\gs[k]}.
\end{align*}
Since every summand in the inner sum is at least $\gs[t+1]$,
\[
R_t\le\sum_{j=1}^t e^{-\lambda j\gs[t+1]}
\le\frac{e^{\lambda\gamma_0}}{\lambda\gs[t+1]}
\le\frac{2^ae^{\lambda\gamma_0}}{\lambda\gamma_0}(t+\Ns)^a.
\]
It follows that
\[
\sum_{s=1}^t\|w_s^t\|
\le\kappa\left(\frac{e^{\lambda\gamma_0}}\lambda+\|A^{-1}\|\right)
\frac{2^ae^{\lambda\gamma_0}}{\lambda\gamma_0}(t+\Ns)^a
=:M_2(t+\Ns)^a.
\]
Both constants have the continuity and uniformity properties asserted in the lemma.
}

%%%%%%%%%%%%%%%%%%%%
\subsection{Proof of Lemma~\ref{lem:stable-mclt}}
Fix $0\le r_1<\cdots<r_m\le1$, let $S_n=(S_n(r_1)',\ldots,S_n(r_m)')'$, and let $C$ be the matrix with blocks $C_{ij}=\min(r_i,r_j)V$. For a fixed vector $v$, put $\Delta_n=\mathbb E[e^{\mathrm iv'S_n}\mid\mathcal G_{n,0}]-e^{-v'Cv/2}$. We first show that $\mathbb E|\Delta_n|\to0$.

Let $Y_n$ be $\mathcal G_{n,0}$-measurable with $1/3\le Y_n\le3$ and $\mathbb E[Y_n]=1$, and let $\mathbb Q_n$ be the probability measure with $d\mathbb Q_n/d\mathbb P=Y_n$. Because $Y_n$ is $\mathcal G_{n,0}$-measurable, conditional expectations given $\mathcal G_{n,s-1}$, $s\ge1$, are the same under $\mathbb Q_n$ and $\mathbb P$. Hence $(e_{n,s})$ is a martingale-difference array under $\mathbb Q_n$, and the random variables in conditions~(i) and~(ii) are unchanged; because $\mathbb Q_n\le3\mathbb P$, they converge in $\mathbb Q_n$-probability to the same limits. Write $v'S_n=\N^{-1/2}\sum_s c_{n,s}'e_{n,s}$, where $c_{n,s}=\sum_{j:\,s\le\lfloor\N r_j\rfloor}v_j$ is nonrandom and bounded. By~(i) and~(ii), this scalar array satisfies the conditional Lindeberg condition, and its predictable variance converges in $\mathbb Q_n$-probability to $v'Cv$. The martingale central limit theorem, which requires no nesting of the filtrations because the limiting variance is nonrandom, therefore gives $\mathbb E_{\mathbb Q_n}e^{\mathrm iv'S_n}\to e^{-v'Cv/2}$. Since $\mathbb E_{\mathbb Q_n}e^{\mathrm iv'S_n}=\mathbb E[Y_n\Delta_n]+e^{-v'Cv/2}$, this means $\mathbb E[Y_n\Delta_n]\to0$.

Take $Y_n=1$, and then $Y_n=\{2+\operatorname{sgn}(\operatorname{Re}\Delta_n)\}/\mathbb E[2+\operatorname{sgn}(\operatorname{Re}\Delta_n)]$, whose normalizing constant lies in $[1,3]$. This gives $\mathbb E[\operatorname{Re}\Delta_n]\to0$ and $2\mathbb E[\operatorname{Re}\Delta_n]+\mathbb E|\operatorname{Re}\Delta_n|\to0$, so $\mathbb E|\operatorname{Re}\Delta_n|\to0$. The imaginary part is handled in the same way, and hence $\mathbb E|\Delta_n|\to0$.

Now let $X_n$ be $\mathcal G_{n,0}$-measurable with $X_n\dto X$. For fixed vectors $u$ and $v$,
\[
\mathbb E\,e^{\mathrm i(u'X_n+v'S_n)}
=\mathbb E\big[e^{\mathrm iu'X_n}\big]e^{-v'Cv/2}+\mathbb E\big[e^{\mathrm iu'X_n}\Delta_n\big]
\to\mathbb E\big[e^{\mathrm iu'X}\big]e^{-v'Cv/2}.
\]
Hence, with $W$ independent of $X$, the finite-dimensional distributions of $(X_n,S_n(\cdot))$ converge to those of $(X,V^{1/2}W(\cdot))$. Conditions~(i) and~(ii) also give the martingale functional central limit theorem for $S_n(\cdot)$, so $S_n(\cdot)$ is asymptotically tight. Together with the tightness of $(X_n)$, this proves the asserted weak convergence. If~(ii) holds only for $r=1$, the same argument with $m=1$ and $r_1=1$ gives the final claim. The argument allows initial sigma-algebras $\mathcal G_{n,0}$ that are not nested in $n$, such as those generated by the sample and all preliminary algorithmic draws.

\subsection{Proof of Lemma~\ref{lem:control-approx-error-fixed-n}}
We follow the same line of arguments as in the proof of Lemma~\ref{lem:control-approx-error} with the following modifications.
Let $n \in \mathbb{N}$ be given and fixed and let $\Ns = 0$.
In this proof, we use the symbol $*$ rather than $\star$, and write $\gamma_t = \gamma_0 t^{-a}$ for the learning rate.
The conditioning and weighting matrices are set to identity matrices and we write $\hat{\theta}_{n}$ for the target estimator.

{The fixed-sample bounds used in the proof of Theorem~\ref{thm:data:fixed} apply here as well.}
Additionally, we assume that the Hessian $\theta \mapsto \frac{\partial^2 \bar{g}_{nj}}{\partial \theta \partial \theta'}(\theta)$ is Lipschitz continuous on the $\delta$-neighborhood of $\hat{\theta}_{n}$ for each $j=1,\ldots, \dg$ in {Condition}~\ref{assm:fixed:item:g-Hessian-Lip}.
Let $(\bar{L}_j)_{j=1}^{\dg}$ denote the Lipschitz constants for these Hessians.
Let $\varepsilon_0=\underbar c\delta^2/2$, with $\underbar c$ from~\eqref{eq:eq7}, and let $\tmt^*=\tmt_{\varepsilon_0}^*$ be the tail time in~\eqref{eq:tail-time} specialized to the first-order algorithm. The fixed-sample version of Lemma~\ref{lem:stopping-time-and-tail} gives $\tmt^*<\infty$ almost surely. For $t>\tmt^*$, \eqref{eq:eq7} places $\theta_{t-1}^*$ within the $\delta$-neighborhood, where the Taylor expansion in the proof of Lemma~\ref{lem:control-approx-error} yields
\begin{equation*}
\|\kappa_t\| \le C \left(  \|\theta_{t-1}^{*} - \hat{\theta}_{n}\|^2 + \|\bar{g}_n(\hat{\theta}_{n})\| \sqrt{ \sum_{j=1}^{\dg} \bar{L}_j^2 } \|\theta_{t-1}^{*} - \hat{\theta}_{n}\|^2 \right) \le C \|\theta_{t-1}^{*} - \hat{\theta}_{n}\|^2
\end{equation*}
under fixed $n$ with $\|\bar{g}_n(\hat\theta_n)\|$ fixed.
On $\{\tmt^{*} < t\}$, this implies $\|\kappa_t\| \le C Q_{t-1}^{*}$.

We write, for $1\le t \le N$,
\begin{equation*}
\sum_{s=1}^t d_s = - \sum_{s=1}^t \alpha_{s}^t \kappa_s    
\end{equation*}
{
where $\alpha_s^t := \gamma_s \sum_{i=s}^t \prod_{k = s+1}^i (I_\dtheta - \gamma_k \bar{\H}_n)$ is uniformly bounded in $s \ge 1$ and $t \ge s$.
}
This yields
\begin{equation*}
\sup_{1\le t \le N} \left\|\sum_{s=1}^t (\theta_s^* - \tto{s}) \right\| \le C \sum_{t=1}^N \|\kappa_t\| \le C \sum_{t =1}^{\tmt^{*}} \|\kappa_t\| + C \sum_{t = \tmt^{*}+1}^N Q_{t-1}^{*}.
\end{equation*}

Following the argument in the proof of Lemma~\ref{lem:control-approx-error}, we can show that, for every $\varepsilon > 0$,
\begin{equation*}
\mathbb{P}_n^{*} \left( \frac{1}{\sqrt{N}}\sum_{t =1}^{\tmt^{*}} \|\kappa_t\| > \varepsilon \right) \to 0 \quad \text{as}\ \ \ N \to \infty,
\end{equation*} 
which addresses the first term.
For the second term, we have 
\begin{align*}
\mathbb{P}_n^{*} \left( \frac{1}{\sqrt{N}}\sum_{t = \tmt^{*}+1}^N Q_{t-1}^{*} > \varepsilon\right) &\le \mathbb{P}_n^{*} \left( \frac{1}{\sqrt{N}}\sum_{t = 1}^N Q_{t-1}^{*} > \varepsilon\right)\\
&\le \frac{C_{\loclevel}}{ \varepsilon \sqrt{N}} \sum_{t=1}^N \gamma_{t-1} + \mathbb{P}_n^{*}(\stt_{\loclevel}^{*} < \infty).
\end{align*}
For arbitrary $\eta > 0$, we choose $\loclevel$ to satisfy $\mathbb{P}_n^{*}(\stt_{\loclevel}^{*} < \infty) < \eta$.
Then, as $N \to \infty$, 
\begin{equation*}
\limsup_{N\to\infty}\mathbb{P}_n^{*} \left( \frac{1}{\sqrt{N}}\sum_{t = \tmt^{*}+1}^N Q_{t-1}^{*} > \varepsilon\right) \le \eta ,\quad \forall \eta > 0.
\end{equation*}
Putting these together, we conclude that 
\begin{equation*}
\lim_{N \to \infty} \mathbb{P}_n^{*}\left( \frac{1}{\sqrt{N}} \sup_{1\le t \le N} \left\|\sum_{s=1}^t (\theta_s^* - \tto{s}) \right\| > \varepsilon \right) = 0.
\end{equation*} 
%%%%%%%%%%%%%%%%%%%%
\subsection{Proof of Lemma~\ref{lem:FCLT-fixed-n}}
We adapt the proof of Lemma~\ref{lem:FCLT-extension} to the case with fixed data.
Consider the decomposition {$\tfrac{1}{\sqrt{N}}\sum_{t=1}^{\lfloor N r \rfloor} (\tto{t} - \hat{\theta}_{n}) = I_0(r) + I_1(r) + I_2(r)$}, where
\begin{align*}
      I_0(r) &= {\frac{\{\alpha_0^{\lfloor Nr\rfloor}-\gamma_0 I_{\dtheta}\}(\theta_0^* -\hat\theta_n)}{\sqrt N\,\gamma_0}},\\
      I_1(r) &= -\bar{\H}_n^{-1}\frac{1}{\sqrt{N}}\sum_{s=1}^{\lfloor N r \rfloor}  \xi_s(\theta_{s-1}^{*}),\\
      I_2(r) &= \frac{1}{\sqrt{N}}\sum_{s=1}^{\lfloor N r \rfloor} w_s^{\lfloor N r \rfloor} \xi_s(\theta_{s-1}^{*}),
\end{align*}
where $\alpha_s^t = \gamma_s \sum_{i=s}^t \prod_{k=s+1}^i (I_\dtheta - \gamma_k \bar{\H}_n)$ and $w_s^t = \bar{\H}_n^{-1} - \alpha_s^t$ for $1 \le s \le t$.
Under Assumption~\ref{assm:regularity:data:fixed:1}, we have $\sup_{s\le t} \|\alpha_s^t\| \le C$ and $\sum_{s=1}^t \|w_s^t\| \le Ct^a$ for all $t \ge 1$ as in the proof of Lemma~\ref{lem:FCLT-extension}.

{By the same fixed-sample bounds,} $\sup_{r\in[0,1]} \|I_0(r)\| \psto 0$ and $\sup_{r\in[0,1]} \|I_2(r)\| \psto 0$ continue to hold.

To establish the FCLT for $I_1(r)$, we need to verify the following conditions:
as $N \to \infty$,
\begin{align}
\label{eq:FCLT-first-fixed-n}
   \frac{1}{N} \sum_{s=1}^{N} \mathbb{E}_n^{*} \left[  \left. \|\xi_s(\theta^{*}_{s-1})\|^2 \I{\|\xi_s(\theta^{*}_{s-1})\|^2 > N \delta} \right| \mathcal{F}^{*}_{n,s-1}\right] & \psto 0\ \ \text{for any} \  \delta > 0,\\
\label{eq:FCLT-second-fixed-n}   \frac{1}{N} \sum_{s=1}^{\lfloor N r \rfloor} \mathbb{E}_n^{*} \left[  \left. \xi_s(\theta^{*}_{s-1}) \xi_s(\theta^{*}_{s-1})'\right| \mathcal{F}^{*}_{n,s-1}\right] & \psto r \bar\Sigma_n  \ \ \text{for any}\   r \in [0,1],
\end{align}
{
Here $\bar\Sigma_n$ is the exact covariance in \eqref{def:bar-Sigma-n}.
The $2p$th-moment bound and localization give \eqref{eq:FCLT-first-fixed-n}.
Fresh draws, {Condition}~\ref{assm:fixed:item:5}, Cauchy--Schwarz, and local moment bounds make the conditional covariance function continuous at $\hat\theta_n$.
Almost-sure iterate convergence and Ces\`aro averaging then give \eqref{eq:FCLT-second-fixed-n}.
Together with the negligible $I_0$ and $I_2$ terms, these conditions prove
\[
\left\{\frac1{\sqrt N}\sum_{t=1}^{\lfloor Nr\rfloor}(\tto{t}-\hat\theta_n)\right\}_{r\in[0,1]}
\wto(\bar{\H}_n^{-1}\bar\Sigma_n\bar{\H}_n^{-1})^{1/2}W(r).
\]
}

%%%%%%%%%%%%%%%%%%%%
\subsection{Proof of Lemma~\ref{lem:cluster:verify}}
Throughout, $C$ denotes a finite constant that may change from line to line and depends only on the constants $\bar{C}$, $M$, and $p$ in the maintained assumptions.
For $p > (1-a)^{-1} > 2$, the power-mean inequality and \ref{assm:cluster:item:stability} give, since $\bar{Q}_n(\hat{\theta}_n) = 0$,
\begin{equation}\label{eq:cluster:g-second-moment}
\frac{1}{K} \sum_{k=1}^K \|\hat{g}_k(\hat{\theta}_n)\|^2 \le \left( \frac{1}{K} \sum_{k=1}^K \|\hat{g}_k(\hat{\theta}_n)\|^{2p} \right)^{1/p} \le M^{1/p},
\end{equation}
and, similarly, $K^{-1}\sum_{k} \hat{H}_k^2 \le M^{1/p}$ by \ref{assm:cluster:item:envelope}.

\medskip
\noindent \textit{Step (a): unbiasedness and the martingale difference property.}
Let $k_1^*, \ldots, k_{B_G}^*$ and $h_1^*, \ldots, h_{B_g}^*$ denote the cluster indices in the Jacobian batch and the moment batch of iteration $t$, respectively; all are i.i.d.\ uniform on $\{1,\ldots,K\}$, the two batches are mutually independent, and both are independent of $\mathcal{F}_{n,t-1}^*$.
Then
\begin{equation*}
\mathbb{E}_n^*[\tilde{G}_t(\theta)' \tilde{g}_t(\theta) \mid \mathcal{F}_{n,t-1}^*]
= \mathbb{E}_n^*[\tilde{G}_t(\theta)]' \, \mathbb{E}_n^*[\tilde{g}_t(\theta)]
= \bar{G}_n(\theta)' \bar{g}_n(\theta),
\end{equation*}
by \eqref{eq:cluster:identity}, so $\mathbb{E}_n^*[\xi_t(\theta) \mid \mathcal{F}_{n,t-1}^*] = 0$ for every fixed $\theta$, and, by the same computation applied conditionally, $\mathbb{E}_n^*[\xi_t(\theta_{t-1}^*) \mid \mathcal{F}_{n,t-1}^*] = 0$ since $\theta_{t-1}^*$ is $\mathcal{F}_{n,t-1}^*$-measurable.

\medskip
{
\noindent \textit{Step (b): exact conditional variance.}
For fixed $\theta$, abbreviate $q=\bar g_n(\theta)$, $D=\bar G_n(\theta)$,
$C=\bar\Omega_K^{\mathrm{cl}}(\theta)-qq'$, and
$\Delta_k(\theta)=\hat G_k(\theta)-D$; here $C$ denotes a matrix.
Then
\begin{equation}\label{eq:cluster:xi-decomp}
\xi_t(\theta)=D'(\tilde g_t(\theta)-q)
 +(\tilde G_t(\theta)-D)'\tilde g_t(\theta).
\end{equation}
The terms are uncorrelated: conditional on the moment batch, the centered Jacobian batch has mean zero. Uniform independent draws give
$\operatorname{Var}_n^*(\tilde g_t)=C/B_g$ and
$\mathbb E_n^*[\tilde g_t\tilde g_t']=C/B_g+qq'$. Hence
\begin{align}\label{eq:cluster:exact-variance}
\mathbb E_n^*[\xi_t(\theta)\xi_t(\theta)']
=\frac{1}{B_g}D'CD
 +\frac{1}{KB_G}\sum_{k=1}^K
 \Delta_k(\theta)'\left(\frac C{B_g}+qq'\right)\Delta_k(\theta).
\end{align}
At $\theta=\hat\theta_n$ this is exactly \eqref{eq:cluster:Sigma}.
For continuity, Cauchy--Schwarz gives
\begin{align*}
&\big\|\mathbb E_n^*[\xi_t(\theta)\xi_t(\theta)']
 -\mathbb E_n^*[\xi_t(\hat\theta_n)\xi_t(\hat\theta_n)']\big\|\\
&\quad\le
 \big(\mathbb E_n^*\|\xi_t(\theta)-\xi_t(\hat\theta_n)\|^2\big)^{1/2}
 \big(2\mathbb E_n^*\|\xi_t(\theta)\|^2
       +2\mathbb E_n^*\|\xi_t(\hat\theta_n)\|^2\big)^{1/2}.
\end{align*}
The first factor tends to zero by \eqref{eq:cluster:eq9}, verified in step (c), and the second is locally bounded by \eqref{eq:cluster:eq10}, verified in step (d). Fresh independent draws allow the same calculation conditional on $\mathcal F_{n,t-1}^*$ at its measurable iterate $\theta_{t-1}^*$, proving \eqref{eq:cluster:var-limit}.
}

\medskip
\noindent \textit{Step (c): proof of \eqref{eq:cluster:eq9} $($mean-square continuity$)$.}
Following the proof of \eqref{eq:eq9}, observe that
\begin{align*}
\xi_t(\theta)-\xi_t(\hat\theta_n)
&=\tilde G_t(\theta)'\big(\tilde g_t(\theta)-\tilde g_t(\hat\theta_n)\big)\\
&\quad+\big(\tilde G_t(\theta)-\tilde G_t(\hat\theta_n)\big)'\tilde g_t(\hat\theta_n)\\
&\quad-\big(\bar G_n(\theta)'\bar g_n(\theta)-\bar G_n'\bar g_n(\hat\theta_n)\big),
\end{align*}
and the last (nonrandom) term is the $\mathbb{E}_n^*$-mean of the sum of the first two, so it suffices to bound the second moments of the first two terms.
By the independence of the two cluster batches and the Cauchy--Schwarz inequality,
\begin{align*}
\mathbb{E}_n^*[\|\xi_t(\theta) - \xi_t(\hat{\theta}_n)\|^2]
\le C \Big( & \mathbb{E}_n^*[\|\tilde{G}_t(\theta)\|^2] \, \mathbb{E}_n^*[\|\tilde{g}_t(\theta) - \tilde{g}_t(\hat{\theta}_n)\|^2] \\
& + \mathbb{E}_n^*[\|\tilde{G}_t(\theta) - \tilde{G}_t(\hat{\theta}_n)\|^2] \, \mathbb{E}_n^*[\|\tilde{g}_t(\hat{\theta}_n)\|^2] \Big).
\end{align*}
We bound the four factors in turn.
First, by convexity of $\|\cdot\|^2$ and the envelope condition \ref{assm:cluster:item:envelope},
\begin{equation*}
\sup_{\theta \in \Theta} \mathbb{E}_n^*[\|\tilde{G}_t(\theta)\|^2]
\le \sup_{\theta \in \Theta} \frac{1}{K} \sum_{k=1}^K \|\hat{G}_k(\theta)\|^2 \le \frac{1}{K} \sum_{k=1}^K \hat{H}_k^2 \le M^{1/p}.
\end{equation*}
Second, by \eqref{eq:cluster:g-second-moment},
\begin{equation*}
\mathbb{E}_n^*[\|\tilde{g}_t(\hat{\theta}_n)\|^2]
\le \frac{1}{K} \sum_{k=1}^K \|\hat{g}_k(\hat{\theta}_n)\|^2 \le M^{1/p}.
\end{equation*}
Third, by the Lipschitz condition \ref{assm:cluster:item:Lipschitz},
$\|\tilde{G}_t(\theta) - \tilde{G}_t(\hat{\theta}_n)\| \le \big( B_G^{-1} \sum_{b=1}^{B_G} L_{k_b^*} \big) \|\theta - \hat{\theta}_n\|$, so that, by convexity of $x \mapsto x^2$ and $K^{-1}\sum_k L_k^2 \le M$,
\begin{equation*}
\mathbb{E}_n^*[\|\tilde{G}_t(\theta) - \tilde{G}_t(\hat{\theta}_n)\|^2]
\le \mathbb{E}_n^*\Big[ \frac{1}{B_G} \sum_{b=1}^{B_G} L_{k_b^*}^2 \Big] \|\theta - \hat{\theta}_n\|^2
= \frac{1}{K} \sum_{k=1}^K L_k^2 \, \|\theta - \hat{\theta}_n\|^2
\le M \|\theta - \hat{\theta}_n\|^2.
\end{equation*}
Fourth, since $g(z,\cdot)$ is continuously differentiable by Assumption~\ref{assm:diff}, $\partial \hat{g}_k(\theta) / \partial \theta' = \hat{G}_k(\theta)$, and, for $\theta_s := \hat{\theta}_n + s(\theta - \hat{\theta}_n)$, $s \in [0,1]$ (contained in $\Theta$ by convexity),
\begin{equation*}
\|\hat{g}_k(\theta) - \hat{g}_k(\hat{\theta}_n)\|
= \Big\| \int_0^1 \hat{G}_k(\theta_s)\, ds \cdot (\theta - \hat{\theta}_n) \Big\|
\le \hat{H}_k \|\theta - \hat{\theta}_n\|
\end{equation*}
by the envelope condition \ref{assm:cluster:item:envelope}. Averaging over the moment batch and using convexity of $\|\cdot\|^2$,
\begin{equation*}
\mathbb{E}_n^*[\|\tilde{g}_t(\theta) - \tilde{g}_t(\hat{\theta}_n)\|^2]
\le \frac{1}{K}\sum_{k=1}^K \hat{H}_k^2 \, \|\theta - \hat{\theta}_n\|^2
\le M^{1/p} \|\theta - \hat{\theta}_n\|^2 .
\end{equation*}
Combining the four bounds yields \eqref{eq:cluster:eq9} with a constant $C$ that does not depend on $t$, $\theta$, or the batch sizes.

\medskip
\noindent \textit{Step (d): proof of \eqref{eq:cluster:eq10} $($the $2p$th-moment bound$)$.}
Following the proof of \eqref{eq:eq10}, there exists a constant $C_p$ such that
\begin{equation*}
\mathbb{E}_n^*[\|\xi_t(\theta)\|^{2p}]
\le C_p \Big( \mathbb{E}_n^*[\|\tilde{G}_t(\theta)\|^{2p}] \, \mathbb{E}_n^*[\|\tilde{g}_t(\theta)\|^{2p}]
+ \|\bar{G}_n(\theta)\|^{2p} \|\bar{g}_n(\theta)\|^{2p} \Big),
\end{equation*}
where the factorization of the first term uses the independence of the two batches.
By convexity of $\|\cdot\|^{2p}$, \ref{assm:cluster:item:stability}, and \ref{assm:cluster:item:envelope}, for all $\theta \in \Theta$ and $t \ge 1$,
\begin{align*}
\mathbb E_n^*[\|\tilde g_t(\theta)\|^{2p}]
&\le\frac1K\sum_{k=1}^K\|\hat g_k(\theta)\|^{2p}
 \le M\big(\bar Q_n(\theta)^p+1\big),\\
\mathbb E_n^*[\|\tilde G_t(\theta)\|^{2p}]
&\le\frac1K\sum_{k=1}^K\hat H_k^{2p}\le M.
\end{align*}
Finally, $\sup_{\theta \in \Theta} \|\bar{G}_n(\theta)\| \le M$ by \ref{assm:fixed:item:G-bounded}, and, by \eqref{eq:cluster:identity}, Jensen's inequality, and \ref{assm:cluster:item:stability},
\begin{equation*}
\|\bar{g}_n(\theta)\|^{2p} = \Big\| \frac{1}{K}\sum_{k=1}^K\hat g_k(\theta)\Big\|^{2p} \le \frac{1}{K}\sum_{k=1}^K\|\hat g_k(\theta)\|^{2p}\le M \big( \bar{Q}_n(\theta)^p + 1 \big).
\end{equation*}
Combining the displays gives \eqref{eq:cluster:eq10} with a constant $M'$ independent of $t$ and $\theta$, completing the proof.

\subsection{Proof of Lemma~\ref{lem:debiased-J:general}}
{
Set $\delta=\tilde\theta-\theta_\oo$. A backward Taylor expansion using $\hat D=\bar G_n(\tilde\theta)$ gives, with probability tending to one,
\[
\hat g=\bar g_n(\theta_\oo)+\hat D\delta+R_K,
\qquad \|R_K\|\le L_K\|\delta\|^2/2,\quad L_K=O_p(1).
\]
Since $\hat H\hat D=0$, $\hat H=O_p(1)$, and
$\sqrt K\|\delta\|^2=o_p(1)$,
\[
\sqrt K\,\hat h
=\hat H\sqrt K\,\bar g_n(\theta_\oo)+o_p(1)
\dto\mathcal N(0,H_0\Omega H_0').
\]
Also $K\hat V_h=\hat H(K\hat V)\hat H'\pto H_0\Omega H_0'$. Studentization proves the claim. Slutsky's theorem applies even when $\tilde\theta$, $\hat H$, and the sample moments use the same data; independence of the preliminary estimate and the moments is unnecessary.
}

%\clearpage
\section{EASI Calibration and Monte Carlo Details}\label{app:easi:details}

This appendix provides the calibration, simulation construction, and implementation details for Section~\ref{sec:MC}, together with the complete versions of its shortened tables. The common learning-rate calibration rule is given in Appendix~\ref{subsec:Selection_gamma0}.

\subsection{Calibration Sample and Empirical Estimates}\label{app:easi:calibration}

The empirical analysis of \citet{EASI} examines annual expenditures across $J = 9$ categories:
food at home, food out, rent, household operation, household furnishing, clothing, transportation operation, recreation, and personal care.
The dataset incorporates price information spanning 12 years across 4 Canadian regions, resulting in 48 distinct price vectors.
The model includes $L = 5$ demographic characteristics: age minus 40, binary indicators for male, non-ownership of a car, and receipt of social assistance, along with a linear time trend (calendar year minus 1986).
The dataset comprises $n=4,847$ observations of single-person households renting their homes, aged 25 to 64, with positive expenditures on rent, recreation, and transportation. For further details on the estimation sample, refer to \citet{EASI}, Section II.B and Table 1.

As is standard practice, \citet{EASI} excluded the last demand equation (personal care) to satisfy the adding-up constraint for the budget shares and estimated eight share equations. Consequently, only $J-1 = 8$ log prices, expressed relative to the omitted category, need to be included. This results in a total of $[6 + 2L + (L+2)(J-1)](J-1)  = 576$ unknown parameters, which will be further reduced to 380 if Slutsky symmetry is imposed by assuming the symmetry of $\{ \mathbf{A}_l: l=0,\ldots,L \}$, and $\mathbf{B}$.
The instruments in~\eqref{GMM:inst:simple} follow \citet{Pendakur:2015}. The original instruments in \citet{EASI} require preliminary estimation steps. Without symmetry the specification used here is just-identified; our experiments impose symmetry and use the overidentified model.

As in \citet{EASI}, we evaluate the estimated budget shares at $\mathbf{p} = \bm{0}_J$, $\bm{z} = \bm{0}_L$, and $\bm{\varepsilon} = \bm{0}_J$,
where $\bm{0}_M $ denotes the zero vector of dimension $M$:
$$
x \mapsto \widehat{\mathbf{w}}(x) =\sum_{r=0}^5 \widehat{\mathbf{b}}_r x^r,
$$
where $\{\widehat{\mathbf{b}}_r\}_{r=0}^5$ are the GMM estimators of $\{\mathbf{b}_r\}_{r=0}^5$. As a function of $x$, it corresponds to the Engel curves for the reference individual facing the log prices of $\mathbf{p} = \bm{0}_J$. Here,
the reference individual is defined as a 40-year-old, car-owning female in 1986 who did not receive social assistance (i.e., $\bm{z} = \bm{0}_L$) with $\bm{\varepsilon} = \bm{0}_J$.

\begin{figure}[!htbp]
	\begin{center}
		\includegraphics[scale=0.50]{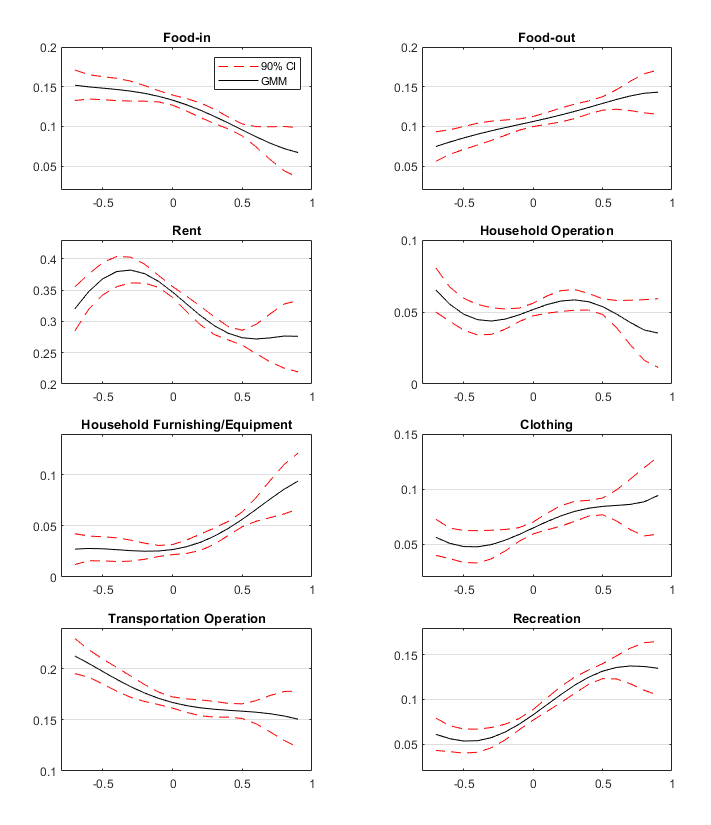}
	\end{center}
        \caption{Estimated Engel Curves}
	\label{figure1-GMM-original}
\end{figure}

Figure \ref{figure1-GMM-original} shows the estimated Engel curves along with 90\% pointwise confidence intervals.
The domain of the curves is restricted to [-0.7, 0.9], which is well within the support of $x$.
The estimated Engel curves presented in Figure \ref{figure1-GMM-original} differ slightly from those in \citet{EASI}, as the latter are based on nonlinear 3SLS and use a different set of instruments as described above. For instance, the slope of the household operation Engel curve becomes negative when the log nominal expenditure, $x$, is large, whereas their estimate remains nearly a positive linear segment in that range. We also find that the confidence intervals based on the GMM tend to be wider than those of the 3SLS particularly when $x$ is large.

For the reference individual at $x=0$, \citet[p. 855]{EASI} report a compensated own-price semi-elasticity of the rent budget share of 0.063. Using the instruments of \citet{Pendakur:2015}, our estimate is 0.053, with a 90\% confidence interval of $[0.010,0.096]$. This estimate supplies the rent-effect calibration used in Section~\ref{sec:MC}.

Obtaining these empirical results took 27 minutes on the laptop described in Appendix~\ref{app:easi:implementation}.

\subsection{Simulation Data Generation}\label{app:easi:dgp}

Because of the simultaneity of $\mathbf{w}$ and $y$, it is challenging to simulate $\mathbf{w}$ while allowing for non-zero higher-order polynomial terms of $y$. As the main goal of the experiment is to compare the standard GMM with our stochastic approximation approach in terms of inference precision and computational cost in a well-identified setting, we simplify the data generating process in the experiments as follows.

We set the true parameter values at the GMM estimates described in Appendix~\ref{app:easi:calibration}.
We obtain the sample variance of the regression residuals, denoted by $\widehat{\sigma}_{j}^2$, where $j=1,\ldots,J-1$.
We compute the sample average of observed budget shares in the original calibration sample, denoted by $\overline{\bm{w}}_{\textrm{sample}}$. We normalize the scale of the demographic characteristics, $\bm{z}_i$,  by dividing each component by its maximum absolute value.

To simulate observations, we draw $\left\{\left(x_i, \mathbf{p}_i, \bm{z}_i\right), i=1, \ldots, n\right\}$ with replacement from the original dataset, with $\bm{z}_i$ normalized as described above, where $n$ can be larger than the original sample size.
We then generate $y_i$ by replacing $\mathbf{w}$ in \eqref{y:eq} with $\overline{\bm{w}}_{\textrm{sample}}$:
\begin{align}\label{y:eq:sim}
y_i=\frac{x_i-\mathbf{p}_i^{\prime} \overline{\bm{w}}_{\textrm{sample}}+\frac{1}{2} \sum_{l=0}^L z_{l,i} \mathbf{p}_i^{\prime} \widehat{\mathbf{A}}_l \mathbf{p}_i}{1-\frac{1}{2} \mathbf{p}_i^{\prime} \widehat{\mathbf{B}} \mathbf{p}_i},
\end{align}
where $\widehat{\mathbf{A}}_l$ and $\widehat{\mathbf{B}}$ are the GMM estimates obtained in the preliminary step.
Finally, for each $i=1,\ldots,n$, we generate $\mathbf{w}_i$ by
\begin{align}\label{EASI:eq:sim}
\mathbf{w}_i=\sum_{r=0}^5 \widehat{\mathbf{b}}_r y_i^r+\widehat{\bf{C}} \bm{z}_i+\widehat{\bm{D}} \bm{z}_i y_i
+\sum_{l=0}^L z_{l,i} \widehat{\bm{A}}_l \mathbf{p}_i+\widehat{\bm{B}} \mathbf{p}_i y_i+\bm{\varepsilon}_i,
\end{align}
where, for $j=1,\ldots,J-1$, the $j$th component of $\bm{\varepsilon}_i$ is randomly drawn from $N\left(0, \hat{\sigma}_{j}^2\right)$
and these $J-1$ components are mutually independent.
The ninth share is recovered by adding-up.
By construction, simulated utility is independent of the generated preference shocks.
The simulation design does not guarantee that each element of $\mathbf{w}_i$ lies between 0 and 1; the moment conditions do not impose that restriction.

\subsection{Implementation and Computational Benchmarks}\label{app:easi:implementation}

Let $\theta$ denote the vectorization of model parameters $\{ \mathbf{b}_r: r=0,\ldots,5 \}$,
$\mathbf{C}$,
$\mathbf{D}$,
$\{ \mathbf{A}_l: l=0,\ldots,L \}$,
and
$\mathbf{B}$ under the Slutsky symmetry restrictions. For the experiments with $n\le100{,}000$, our procedure follows the steps outlined below. First, we apply the warm-start algorithm described in Algorithm \ref{alg:sequential-reshuffle} with $\theta_{0,\mathrm{ws}}^\ast=\bm{0}_{380}$. We set $\gamma_{0,\mathrm{ws}} = 0.1$ and $a=0.501$. Second, we use our stochastic approximation, based on Algorithm \ref{alg:mini-batch}, to obtain $\theta_N^\ast$ and $\bar{\theta}_N^\ast$. In both steps, we employ the weighting matrix,
\[
W_{2SLS} = I_{J-1} \otimes \left(\frac{1}{n} \sum_{i=1}^{n} \bm{q}_i \bm{q}_i^\prime \right)^{-1},
\]
assuming the independence of unobserved preference shocks across equations. The initial step size $\gamma_0$ is selected based on the rule-of-thumb method proposed in Appendix~\ref{subsec:Selection_gamma0}. We use $s_0=5$ and $8$ for $n=20,000$, which yield average values of $\gamma_0 = 0.082$ and $0.051$, respectively. For $n=50,000$ and $100,000$, we use $s_0=3$ and $8$, resulting in average values of $\gamma_0 = 0.136$ and $0.051$, respectively.
Finally, we apply the refinement step for efficient estimation, as proposed in Section \ref{sec:extensions}. We initialize the refinement stage at $\theta_N^\ast$ and use the mini-batch based centered optimal weight in (\ref{def:centered W}), with $M_{\mathrm{MB}}=50,000$, under the same independence restriction.

The batch sizes are $B_G=B_g=512$ in the warm-start and first-order stages. During refinement, $B_g=512$ and $B_{G,t}=512+\lfloor\log(t-N)\rfloor$ for $t>N$, with the local refinement index starting at one. Averaging and the RS accumulators restart at refinement, as described in Section~\ref{sec:extensions}.

For $n=20{,}000$, $50{,}000$, and $100{,}000$, the warm-start epoch counts are $E_{\mathrm{ws}}=25$, $4$, and $1$, respectively. These give approximately 37,000--38,000 updates, with count $\lfloor n/B_{\mathrm{ws}}\rfloor(\lfloor n/B_{\mathrm{ws}}\rfloor-1)E_{\mathrm{ws}}$. We use $(N,T)=(20{,}000,40{,}000)$ and $(20{,}000,60{,}000)$ at each of these sample sizes.

For the million-observation exercise, Table~\ref{tab:easi:million-full} reports each method's subsample sizes, tuning values, and iteration budgets. The refined-weight first-order method omits the curvature adjustment in~\eqref{def:SGD:eff}; the unrefined method keeps the initial 2SLS weight.

The simulations use 500 replications on the University of Connecticut HPC cluster. Its compute nodes include Intel Skylake and AMD EPYC 64-core and 128-core processors, with 34--126 cores and 187--503 GB of RAM. Times are measured separately using five replications on a laptop with an Apple M1 Pro chip (8-core CPU, 14-core GPU), 16 GB of RAM, and macOS Ventura 13.5. Both MATLAB SLIM and the Stata GMM benchmark use the same pregenerated datasets. The timing measurements for the Engel-curve and rent-effect exercises are reported in the tables. Engel-curve RIMSE integrates over $[-0.7,0.9]$ using a grid with spacing 0.1.

\subsection{Complete Estimation and Inference Results}\label{app:easi:results}

Table~\ref{tab:easi:million-full} gives the complete million-observation results and settings corresponding to Table~\ref{tab:sim4}. Table~\ref{tab:easi:rent-full} adds bias and standard deviation to the rent-effect results in Table~\ref{tab:sim2}, retaining every learning-rate and iteration configuration.

\begin{table}[!htbp]
\caption{Complete Simulation Results when $n=$ 1,000,000}
\begin{center}
\begin{tabular}{ccccccc}
\hline\hline
 & \multicolumn{2}{c}{First-order SLIM}&  \multicolumn{2}{c}{First-order SLIM}&  \multicolumn{2}{c}{Second-order SLIM} \\
 & \multicolumn{2}{c}{(2SLS Weight)}& \multicolumn{2}{c}{(Refined Weight)}&   \\
 & (1) & (2) & (3)  & (4) & (5) &(6) \\ \hline
\hline
 $N$ &  300,000  &  500,000  & 20,000  & 20,000 & 20,000 & 20,000 \\
 $T$ &           &           & 120,000 & 140,000 & 40,000 & 60,000 \\
\hline
$n_{\mathrm{ws}}$ & 150,000 & 150,000 & 100,000 & 100,000  & 100,000 & 100,000 \\
$n_{\mathrm{w}}$ & 150,000 & 150,000 & 100,000 & 100,000  & 100,000 & 100,000 \\
$(\gamma_{0,\mathrm{ws}},s_0)$ & (0.4, 1) & (0.4, 1) & (0.1, 3)  & (0.1, 3) & (0.1, 3)  & (0.1, 3)  \\
\hline
\multicolumn{7}{c}{Panel A. Engel Curves } \\
RIMSE & 0.020 & 0.020 & 0.030 & 0.030 & 0.006 & 0.005  \\
\hline
\multicolumn{7}{c}{Panel B. Price Effect on Rent Budget Share } \\
Bias &-0.001 & -0.001 & 0.001 & 0.001 & -0.001 & -0.001  \\
SD & 0.005 & 0.005 & 0.003 & 0.003 & 0.003 & 0.003  \\
RMSE & 0.005 & 0.005 & 0.003 & 0.003 & 0.003 & 0.003  \\
Coverage Probability & 0.866 & 0.924 & 0.974& 0.968& 0.944 & 0.968 \\
 CI Length &  0.024 & 0.035 & 0.030 & 0.031 &  0.018 & 0.019 \\
\hline
\multicolumn{7}{c}{Panel C. Computational Cost (Hours)} \\
 & 4.98 & 7.74  & 2.09 & 2.38 &  1.11 & 1.38 \\
\hline \hline
\multicolumn{7}{l}{$E_{\mathrm{ws}}=1$ all  methods. $n_{\mathrm{ws}}$ is the number of observations used in the warm-start} \\
\multicolumn{7}{l}{stage. $n_{\mathrm{w}}$ denotes the number of observations used for the initial weight, $\Phi_n$ and}\\ \multicolumn{7}{l}{$\bar{g}_n\bigl(\bar{\theta}_N^\ast\bigr)$ in the centered weight in the refinement stage. ``Coverage Probability" and}\\
\multicolumn{7}{l}{ ``CI Length" in Panel B are based on 95\% confidence.}
\end{tabular}
\end{center}
\label{tab:easi:million-full}
\end{table}

\begin{table}[!htbp]
\caption{Complete Simulation Results for the Price Effect on the Rent Budget Share}
\begin{center}
\setlength{\tabcolsep}{4.5pt}
\begin{tabular}{cccccccccc}
\hline\hline
  & Full-Sample &  \multicolumn{8}{c}{SLIM} \\
  & GMM &  \multicolumn{8}{c}{} \\
 & (1) & (2) & (3) & (4) & (5)  & (6)  & (7)  & (8)  & (9) \\
 \hline
 Sample Size   & &  \multicolumn{8}{c}{($N,T$)} \\
$(n)$ & & \multicolumn{4}{c}{(20,000, \ 40,000)} & \multicolumn{4}{c}{(20,000, \ 60,000)} \\
 \hline
 & & \multicolumn{8}{c}{($s_0,E_{\mathrm{ws}}$)} \\
 \ 20,000 &       & \multicolumn{2}{c}{$(5, 25)$} & \multicolumn{2}{c}{$(8, 25)$}& \multicolumn{2}{c}{$(5, 25)$}& \multicolumn{2}{c}{$(8, 25)$}\\
 \ 50,000 &       & \multicolumn{2}{c}{$(3,\ 4)$} & \multicolumn{2}{c}{$(8,\ 4)$} & \multicolumn{2}{c}{$(3,\ 4)$} & \multicolumn{2}{c}{$(8, \ 4)$}\\
100,000 &       &   \multicolumn{2}{c}{$(3,\ 1)$} & \multicolumn{2}{c}{$(8,\ 1)$} & \multicolumn{2}{c}{$(3, \ 1)$} & \multicolumn{2}{c}{$(8, \ 1)$}\\
\hline
\multicolumn{10}{c}{Panel A. Bias} \\
\ 20,000 & 0.001  &  \multicolumn{2}{c}{ 0.000} & \multicolumn{2}{c}{0.001 } & \multicolumn{2}{c}{0.000 } & \multicolumn{2}{c}{ 0.000} \\
\ 50,000 &  0.000 &  \multicolumn{2}{c}{ 0.000 } & \multicolumn{2}{c}{ 0.000} & \multicolumn{2}{c}{ 0.000}  & \multicolumn{2}{c}{ 0.000}  \\
100,000 &  NA   &  \multicolumn{2}{c}{0.001 } & \multicolumn{2}{c}{ 0.000} & \multicolumn{2}{c}{0.001}  & \multicolumn{2}{c}{ 0.000}  \\
 \hline
\multicolumn{10}{c}{Panel B. SD} \\
\ 20,000 & 0.019  &  \multicolumn{2}{c}{ 0.019 } & \multicolumn{2}{c}{ 0.018 } & \multicolumn{2}{c}{ 0.019 } & \multicolumn{2}{c}{ 0.019 }  \\
\ 50,000 &  0.013 &  \multicolumn{2}{c}{0.013 } & \multicolumn{2}{c}{0.012 } & \multicolumn{2}{c}{ 0.013}  & \multicolumn{2}{c}{0.012 }  \\
100,000 &  NA   &  \multicolumn{2}{c}{ 0.009 } & \multicolumn{2}{c}{ 0.009} & \multicolumn{2}{c}{0.009 }  & \multicolumn{2}{c}{0.009 }  \\
\hline
\multicolumn{10}{c}{Panel C. RMSE} \\
\ 20,000 & 0.019  &  \multicolumn{2}{c}{ 0.019 } & \multicolumn{2}{c}{ 0.018 } & \multicolumn{2}{c}{ 0.019 } & \multicolumn{2}{c}{ 0.019 }  \\
\ 50,000 &  0.013 &  \multicolumn{2}{c}{0.013 } & \multicolumn{2}{c}{0.012 } & \multicolumn{2}{c}{ 0.013}  & \multicolumn{2}{c}{0.013 }  \\
100,000 &  NA   &  \multicolumn{2}{c}{ 0.009 } & \multicolumn{2}{c}{ 0.009} & \multicolumn{2}{c}{0.009 }  & \multicolumn{2}{c}{0.009 }  \\
\hline
\multicolumn{10}{c}{Panel D. Coverage Probability (Nominal Level: 0.95)} \\
        &       & \underline{RS} &  \underline{PI} &  \underline{RS} &  \underline{PI} &  \underline{RS} &  \underline{PI} &  \underline{RS} &  \underline{PI} \\
\ 20,000  & 0.962 &  0.974   & 0.966    &   0.972  & 0.972  & 0.978 & 0.962 & 0.978 & 0.968 \\
\ 50,000  & 0.948 & 0.944 & 0.946 & 0.924 & 0.950 & 0.970 & 0.948  & 0.954 & 0.950 \\
100,000 & NA    & 0.954 & 0.952  & 0.940 & 0.960 &  0.960     &   0.954    &  0.954  & 0.956   \\
\hline
\multicolumn{10}{c}{Panel E. CI Length (Nominal Level: 0.95)} \\
        &       & \underline{RS} &  \underline{PI} &  \underline{RS} &  \underline{PI} &  \underline{RS} &  \underline{PI} &  \underline{RS} &  \underline{PI} \\
\ 20,000 & 0.077  & 0.202  & 0.077  & 0.237   & 0.077  &0.179 & 0.077 & 0.220 & 0.077\\
\ 50,000 & 0.049  & 0.079 & 0.049 & 0.085 & 0.049 & 0.077 & 0.049  & 0.085 & 0.049\\
100,000 & NA    & 0.054 & 0.035 & 0.057 & 0.035 & 0.056      &   0.035     & 0.059 & 0.035      \\
\hline
\multicolumn{10}{c}{Panel F. Computational Cost (Hours)} \\
        &       & \underline{RS} &  \underline{PI} &  \underline{RS} &  \underline{PI} &  \underline{RS} &  \underline{PI} &  \underline{RS} &  \underline{PI} \\
\ 20,000   & \ 1.93       &  1.05  &  1.04  &  1.03  &  1.05  &  1.33  &  1.29  &   1.36 &  1.30  \\
\ 50,000  & \ 7.84       &  1.06  &  1.06  &  1.07  & 1.06   &  1.34  &  1.36  & 1.34   &  1.34  \\
100,000    & 18.93$^\ast$ & 1.07   & 1.09   & 1.07   & 1.08   &  1.38  &  1.38  &   1.36 &  1.37   \\
\hline \hline
\multicolumn{10}{l}{$\ast$ This exceeds our computational budget.}
\end{tabular}
\end{center}
\label{tab:easi:rent-full}
\end{table}

%\clearpage

\section{Complete Coefficient Results for the LinkedIn Application}\label{app:linkedin:results}

Tables~\ref{tab:linkedin:full-levels} and~\ref{tab:linkedin:full-demographics} report the full model. The industry and occupation intercepts are relative to the stated reference industry and occupation, and demographic terms are industry interactions relative to the reference industry. %The farming, fishing, and forestry coefficient is retained with its uncertainty. No parameter is removed on the basis of its precision, and both inference methods are reported for every coefficient.

% Generated by Code/linkedin/build_paper_application_outputs.R; do not hand-edit numbers.
\begin{table}[!htbp]
\centering
\caption{Complete destination-choice estimates: destination propensity and intercepts}\label{tab:linkedin:full-levels}
\begingroup
\footnotesize
\setstretch{1}\setlength{\tabcolsep}{3pt}\renewcommand{\arraystretch}{1.18}
\begin{tabularx}{\textwidth}{@{}Xrrr@{}}\toprule
Coefficient & Estimate & Plug-in 95\% CI & RS 95\% CI \\
\midrule
Destination propensity & $-1.2957$ & $[-1.3001, -1.2912]$ & $[-1.3043, -1.2870]$ \\
\addlinespace\multicolumn{4}{@{}l}{Industry intercepts}\\
Natural resources, utilities, construction & $0.1377$ & $[0.1197, 0.1557]$ & $[0.0521, 0.2233]$ \\
Manufacturing & $0.0704$ & $[0.0640, 0.0768]$ & $[0.0315, 0.1093]$ \\
Trade and transportation & $-0.0263$ & $[-0.0333, -0.0193]$ & $[-0.0390, -0.0136]$ \\
Education, health, public services & $-0.2727$ & $[-0.2828, -0.2625]$ & $[-0.3458, -0.1996]$ \\
Consumer and other services & $-0.2032$ & $[-0.2127, -0.1936]$ & $[-0.2383, -0.1681]$ \\
\addlinespace\multicolumn{4}{@{}l}{Occupation intercepts}\\
Business and financial operations & $-0.0296$ & $[-0.0333, -0.0259]$ & $[-0.0580, -0.0012]$ \\
Computer and mathematical & $-0.0653$ & $[-0.0706, -0.0601]$ & $[-0.1041, -0.0266]$ \\
Architecture and engineering & $-0.0567$ & $[-0.0631, -0.0503]$ & $[-0.0770, -0.0364]$ \\
Life, physical, and social science & $-0.2307$ & $[-0.2387, -0.2227]$ & $[-0.3026, -0.1588]$ \\
Community and social service & $-0.1370$ & $[-0.1445, -0.1294]$ & $[-0.1912, -0.0827]$ \\
Legal & $-0.1165$ & $[-0.1233, -0.1096]$ & $[-0.1594, -0.0736]$ \\
Education, training, and library & $-0.1544$ & $[-0.1611, -0.1477]$ & $[-0.2068, -0.1021]$ \\
Arts, design, entertainment, sports, media & $-0.0088$ & $[-0.0112, -0.0065]$ & $[-0.0263, 0.0086]$ \\
Healthcare practitioners and technical & $-0.0929$ & $[-0.0998, -0.0859]$ & $[-0.1307, -0.0550]$ \\
Healthcare support & $-0.0787$ & $[-0.0851, -0.0722]$ & $[-0.1026, -0.0548]$ \\
Protective service & $-0.1424$ & $[-0.1523, -0.1324]$ & $[-0.2019, -0.0829]$ \\
Food preparation and serving & $-0.1066$ & $[-0.1132, -0.0999]$ & $[-0.1494, -0.0637]$ \\
Building and grounds maintenance & $-0.4021$ & $[-0.4731, -0.3310]$ & $[-0.6282, -0.1759]$ \\
Personal care and service & $-0.1548$ & $[-0.1627, -0.1468]$ & $[-0.2041, -0.1054]$ \\
Sales and related & $-0.0012$ & $[-0.0034, 0.0010]$ & $[-0.0150, 0.0126]$ \\
Office and administrative support & $-0.0217$ & $[-0.0241, -0.0192]$ & $[-0.0365, -0.0068]$ \\
Farming, fishing, and forestry & $-0.7969$ & $[-1.4009, -0.1930]$ & $[-1.4436, -0.1502]$ \\
Construction and extraction & $-0.1975$ & $[-0.2127, -0.1824]$ & $[-0.2684, -0.1267]$ \\
Installation, maintenance, and repair & $-0.1179$ & $[-0.1275, -0.1083]$ & $[-0.1553, -0.0805]$ \\
Production & $-0.1212$ & $[-0.1285, -0.1139]$ & $[-0.1505, -0.0920]$ \\
Transportation and material moving & $-0.0558$ & $[-0.0615, -0.0502]$ & $[-0.0691, -0.0426]$ \\
\bottomrule\end{tabularx}
\par\smallskip
\begin{minipage}{\textwidth}\footnotesize Notes: Reference industry: information, finance, and business; reference occupation: management. The destination measure is standardized. Estimates and both intervals use the same averaged second-order SLIM estimator. Plug-in intervals include sampling and computational uncertainty. RS intervals are pointwise.\end{minipage}
\endgroup
\end{table}
% Generated by Code/linkedin/build_paper_application_outputs.R; do not hand-edit numbers.
\begin{table}[!htbp]
\centering
\caption{Complete destination-choice estimates: demographic interactions}\label{tab:linkedin:full-demographics}
\begingroup
\footnotesize
\setstretch{1}\setlength{\tabcolsep}{3pt}\renewcommand{\arraystretch}{1.18}
\begin{tabularx}{\textwidth}{@{}Xrrr@{}}\toprule
Coefficient & Estimate & Plug-in 95\% CI & RS 95\% CI \\
\midrule
\addlinespace\multicolumn{4}{@{}l}{Natural resources, utilities, construction $\times$}\\
Female probability & $-0.4660$ & $[-0.5072, -0.4247]$ & $[-0.7496, -0.1824]$ \\
Asian/Pacific Islander probability & $-0.6211$ & $[-0.6959, -0.5463]$ & $[-0.9071, -0.3350]$ \\
Hispanic probability & $0.1028$ & $[0.0595, 0.1460]$ & $[-0.0149, 0.2204]$ \\
Black probability & $-0.1908$ & $[-0.2316, -0.1500]$ & $[-0.2914, -0.0901]$ \\
Other race/ethnicity probability & $0.1046$ & $[-0.0251, 0.2342]$ & $[-0.0568, 0.2659]$ \\
\addlinespace\multicolumn{4}{@{}l}{Manufacturing $\times$}\\
Female probability & $-0.1135$ & $[-0.1246, -0.1025]$ & $[-0.1731, -0.0539]$ \\
Asian/Pacific Islander probability & $-0.1107$ & $[-0.1332, -0.0881]$ & $[-0.1543, -0.0670]$ \\
Hispanic probability & $-0.0057$ & $[-0.0225, 0.0112]$ & $[-0.0333, 0.0220]$ \\
Black probability & $-0.2488$ & $[-0.2700, -0.2276]$ & $[-0.3634, -0.1342]$ \\
Other race/ethnicity probability & $-0.5148$ & $[-0.5914, -0.4383]$ & $[-0.7506, -0.2790]$ \\
\addlinespace\multicolumn{4}{@{}l}{Trade and transportation $\times$}\\
Female probability & $0.0281$ & $[0.0133, 0.0430]$ & $[0.0175, 0.0388]$ \\
Asian/Pacific Islander probability & $-0.2650$ & $[-0.2988, -0.2313]$ & $[-0.3925, -0.1376]$ \\
Hispanic probability & $0.0615$ & $[0.0363, 0.0866]$ & $[0.0180, 0.1049]$ \\
Black probability & $-0.0051$ & $[-0.0226, 0.0123]$ & $[-0.0458, 0.0355]$ \\
Other race/ethnicity probability & $-0.1952$ & $[-0.2686, -0.1219]$ & $[-0.3562, -0.0342]$ \\
\addlinespace\multicolumn{4}{@{}l}{Education, health, public services $\times$}\\
Female probability & $0.2959$ & $[0.2777, 0.3140]$ & $[0.1841, 0.4076]$ \\
Asian/Pacific Islander probability & $-0.2898$ & $[-0.3195, -0.2601]$ & $[-0.3901, -0.1895]$ \\
Hispanic probability & $0.1659$ & $[0.1421, 0.1898]$ & $[0.0788, 0.2530]$ \\
Black probability & $0.1434$ & $[0.1269, 0.1599]$ & $[0.0998, 0.1871]$ \\
Other race/ethnicity probability & $0.2067$ & $[0.1494, 0.2641]$ & $[0.1095, 0.3040]$ \\
\addlinespace\multicolumn{4}{@{}l}{Consumer and other services $\times$}\\
Female probability & $0.2393$ & $[0.2176, 0.2611]$ & $[0.1436, 0.3350]$ \\
Asian/Pacific Islander probability & $-0.4315$ & $[-0.4789, -0.3841]$ & $[-0.6480, -0.2150]$ \\
Hispanic probability & $0.1225$ & $[0.0959, 0.1490]$ & $[0.0587, 0.1862]$ \\
Black probability & $0.0848$ & $[0.0624, 0.1072]$ & $[0.0376, 0.1320]$ \\
Other race/ethnicity probability & $0.3199$ & $[0.2328, 0.4070]$ & $[0.2216, 0.4182]$ \\
\bottomrule\end{tabularx}
\par\smallskip
\begin{minipage}{\textwidth}\footnotesize Notes: Interactions are relative to information, finance, and business. Probabilities enter in 0--1 units; White is the omitted race/ethnicity category. Other combines Native American and multiple-race probabilities. Plug-in intervals include sampling and computational uncertainty; RS intervals are pointwise.\end{minipage}
\endgroup
\end{table}

%\clearpage

\section{Convergence with Multiple Stationary Points}\label{appx:multi-local}
%%%%%%%%%%%%%%%%%%%%%%%%%%%%%%%%%%%%%%%%%%%%%%%%%%%%%%%%%%%%%%%%%%%%%%%%%%

This appendix extends Theorem~\ref{thm:consistency} to allow the sample GMM criterion to possess multiple stationary points. Condition~\ref{asm:item:id} requires the gradient of the sample Lyapunov function to be bounded away from zero whenever $\|\theta - \hat{\theta}_n\| \ge \delta$, and Condition~\ref{asm:item:local-convex-Lojasiewicz} makes the criterion strongly convex within distance $\delta$ of $\hat{\theta}_n$. Together, they make $\hat{\theta}_n$ the only stationary point of the sample criterion. In overidentified nonlinear models, however, the GMM criterion may possess multiple local minima. The following assumption relaxes Conditions~\ref{asm:item:local-convex-Lojasiewicz} and \ref{asm:item:id} to accommodate this possibility.

Convergence of stochastic approximation to the set of stationary points of a nonconvex objective is classical \citep{bertsekas2000gradient,kushner2003stochastic}.

{\subsection{Convergence to a Stationary Point}}

\begin{assumption}[Multiple Stationary Points]\label{asm:multi-local}
	There exist an integer $m \ge 1$, constants $c > 0$ and $\delta > 0$, nonrandom points $\theta_\oo^{(1)}, \ldots, \theta_\oo^{(m)} \in \Theta$ with $\theta_\oo^{(1)} = \theta_\oo$, and $\sigma(z_{1:n})$-measurable random vectors $\hat{\theta}_n^{(1)}, \ldots, \hat{\theta}_n^{(m)}$ with $\hat{\theta}_n^{(1)} = \hat{\theta}_n$, such that $\min_{j \ne k} \|\theta_\oo^{(j)} - \theta_\oo^{(k)}\| \ge 6\delta$, $\{\theta : \|\theta - \theta_\oo^{(j)}\| \le 3\delta\} \subseteq \Theta$ for each $j$, and the following hold:
	\begin{enumerate}[({A\ref{asm:multi-local}}.1)]
		\item \label{asm:item:multi:consistency}
		$\max_{1 \le j \le m} \|\hat{\theta}_n^{(j)} - \theta_\oo^{(j)}\| = \oP(1)$ as $n \to \infty$.

		\item \label{asm:item:multi:critical}
		For each $j$, $\hat{\theta}_n^{(j)}$ is the unique solution of $\bar{G}_n(\theta)'\bar{g}_n(\theta) = 0$ in $\{\theta : \|\theta - \hat{\theta}_n^{(j)}\| \le \delta\}$ \wpa.

		\item \label{asm:item:multi:value-sep}
		For all $\|\theta - \hat{\theta}_n\| \ge \delta$, $\bar{Q}_n(\theta) \ge c$ \wpa.

		\item \label{asm:item:multi:grad-sep}
		For all $\theta \in \Theta$ with $\min_{1 \le j \le m}\|\theta - \hat{\theta}_n^{(j)}\| \ge \delta$, $\|\bar{G}_n(\theta)'\bar{g}_n(\theta)\|^2 \ge c$ \wpa.
	\end{enumerate}
\end{assumption}

Assumption~\ref{asm:multi-local} replaces the two-region structure of Conditions~\ref{asm:item:local-convex-Lojasiewicz}--\ref{asm:item:id} with an $(m+1)$-region structure. Condition~\ref{asm:item:multi:critical} requires each $\hat{\theta}_n^{(j)}$ to be an isolated stationary point of the sample Lyapunov function; no curvature condition is imposed, so the $\hat{\theta}_n^{(j)}$ may be local minima, saddle points, or local maxima. By Condition~\ref{asm:item:multi:consistency}, they consistently estimate distinct population stationary points, the first of which is the true parameter $\theta_\oo$. {Condition~\ref{asm:item:multi:value-sep} separates the global minimizer $\hat{\theta}_n$ in criterion value, as in Condition~\ref{asm:item:id}. {Consequently, $\bar{Q}_n(\hat{\theta}_n^{(j)}) \ge c$ \wpa\ for every $j \ne 1$.}} The substantive relaxation is Condition~\ref{asm:item:multi:grad-sep}: the gradient must be bounded away from zero only outside the $\delta$-neighborhoods of the $m$ stationary points rather than outside a single neighborhood of $\hat{\theta}_n$. {When $m = 1$, Assumption~\ref{asm:multi-local} implies Condition~\ref{asm:item:id} and requires $\hat{\theta}_n$ to be an isolated stationary point. This is weaker than imposing the local strong convexity in Condition~\ref{asm:item:local-convex-Lojasiewicz} together with Condition~\ref{asm:item:id}.}

\begin{theorem}[Convergence to a Stationary Point]
	\label{thm:consistency:local}
	Let Assumptions \ref{assm:uniform}, \ref{assm:lr}, and \ref{assm:diff} hold, and let Assumption \ref{asm:regularity} hold with Conditions~\ref{asm:item:local-convex-Lojasiewicz} and \ref{asm:item:id} replaced by Assumption~\ref{asm:multi-local}. Then there exists a sequence of events $E_n \in \mathcal{F}_n$ with $\mathbb{P}(E_n) \to 1$ such that, on $E_n$,
	\begin{equation*}
		\mathbb{P}_n^*\left( \lim_{N \to \infty} \theta_N^* = \lim_{N \to \infty} \bar{\theta}_N^* = \hat{\theta}_n^{(\mathcal{J})} \ \text{ for some random index } \mathcal{J} \in \{1, \ldots, m\} \right) = 1.
	\end{equation*}
	Consequently, for every $\varepsilon > 0$,
	\begin{equation*}
		\lim_{n \to \infty} \limsup_{N \to \infty}\, \mathbb{P}\left( \min_{1 \le j \le m} \|\theta_N^* - \theta_\oo^{(j)}\| > \varepsilon \right) = 0,
	\end{equation*}
	and the same holds with $\theta_N^*$ replaced by $\bar{\theta}_N^*$.
\end{theorem}

Theorem~\ref{thm:consistency:local} states that the stochastic path cannot oscillate between the neighborhoods of distinct stationary points indefinitely: the iterates settle near a single stationary point and converge to it. In the leading case in which every stationary point is a local minimizer, the theorem states that the algorithm converges to one of the local optimizers. The index $\mathcal{J}$ is random, determined by the initial value and the realized path of mini-batch draws, and the theorem places no restriction on which limit is selected. {The proof is given in Section~\ref{appx:proofs-multi-local}.}

\begin{remark}[Relation to Theorem~\ref{thm:consistency}]
	{Theorem~\ref{thm:consistency:local} establishes conditional convergence to a stationary point on $E_n$ for each $n$. It neither guarantees selection of the global minimizer nor provides a convergence bound uniform in $n$.} The $O_p(N^{-a/2})$ rate for the iterates requires additional local structure. If $\bar{Q}_n$ is strongly convex on a neighborhood of $\hat{\theta}_n^{(\mathcal{J})}$, as in Condition~\ref{asm:item:local-convex-Lojasiewicz}, then the recentred criterion $\bar Q_n-\bar Q_n(\hat{\theta}_n^{(\mathcal{J})})$ satisfies a local {\L}ojasiewicz-type inequality, and the arguments of Lemmas~\ref{lem:stopping-time-and-tail} and~\ref{lem:sgd-conv-rate} could be adapted using stopping times that localize the iterates to that neighborhood. Such a localized rate result, and a basin-conditional distribution theory including the analogue of Theorem~\ref{thm:asymp:normal} centered at $\hat{\theta}_n^{(\mathcal{J})}$, are left for future research.
\end{remark}

{
\subsection{Interpreting the Debiased $J$-Test at a Stationary Solution}
The value separation in Assumption~\ref{asm:multi-local} also permits a statistical diagnostic under correct specification. Here $m$ remains the number of stationary points and $r=\dg-\dtheta>0$. For numerical approximations $\tilde\theta_n^{(j)}$ to the points $\hat\theta_n^{(j)}$, define
\begin{align*}
\hat g_{n,j}&=\bar g_n(\tilde\theta_n^{(j)}),\qquad
\hat D_{n,j}=\bar G_n(\tilde\theta_n^{(j)}),\qquad
\hat H_{n,j}\hat D_{n,j}=0,\\
\hat h_{n,j}&=\hat H_{n,j}\hat g_{n,j},\qquad
\hat V_{h,n,j}=\hat H_{n,j}\hat V_{n,j}\hat H_{n,j}',\qquad
J_{D,n}^{(j)}=\hat h_{n,j}'\hat V_{h,n,j}^{-1}\hat h_{n,j}.
\end{align*}
The matrix $\hat V_{n,j}$ is in sample-mean covariance units, as in \eqref{eq:cluster:terminal-cov}; no outer $n$ multiplies this statistic. With positive-definite $\hat V_{n,j}$ it equals the projected-weight statistic in Section~\ref{subsec:J-test}, with test weight $(n\hat V_{n,j})^{-1}$.

\begin{proposition}[Screening separated stationary points]\label{prop:multi:J-screening}
Maintain Assumption~\ref{asm:multi-local} with fixed $m$ and $g(\theta_\oo)=0$. Suppose:
\begin{enumerate}[(i)]
\item On neighborhoods of all $\theta_\oo^{(j)}$, the sample moments and Jacobians converge uniformly in probability to continuous $g(\theta)$ and $G(\theta)$. Each $D_j=G(\theta_\oo^{(j)})$ has full column rank, and $\tilde\theta_n^{(j)}-\hat\theta_n^{(j)}=o_p(1)$.
\item Each $\hat H_{n,j}$ has $r$ rows, satisfies the displayed annihilation identity, and converges to a full-row-rank matrix $H_j$. Each symmetric $\hat V_{h,n,j}$ is positive definite with probability tending to one. For $j\ne1$, $\lambda_{\max}(\hat V_{h,n,j})\pto0$.
\item At $j=1$, the conditions of Lemma~\ref{lem:debiased-J:general} hold with each observation treated as a cluster, $K=n$, and the corresponding moments, Jacobian, and covariance estimates. In particular, $\tilde\theta_n^{(1)}-\theta_\oo=o_p(n^{-1/4})$.
\end{enumerate}
Then $J_{D,n}^{(1)}\dto\chi^2_r$, whereas $J_{D,n}^{(j)}\pto\infty$ for every $j\ne1$. A test with a fixed $\chi^2_r$ critical value therefore rejects all the spurious candidates simultaneously with probability tending to one.
\end{proposition}

The proof is given in Section~\ref{appx:proofs-multi-local}. The power argument only needs covariance shrinkage at a spurious point, not a computational CLT or a null covariance formula valid there. Writing $\mu_j=g(\theta_\oo^{(j)})$ and $D_j=G(\theta_\oo^{(j)})$, it also permits different optimization and testing weights: if population stationarity is $D_j'A\mu_j=0$ for a positive-definite optimization weight $A$, the contradiction becomes $\mu_j'A\mu_j=0$, provided the corresponding separation and convergence conditions hold. No efficiency of the optimization weight is required.

The proposition assumes the stated joint sample-size and iteration accuracy; the convergence theorem alone does not supply it or establish a null distribution conditional on a selected stationary point. Rejection can reflect either invalid moment restrictions or convergence to a spurious stationary point, and the test alone does not distinguish these explanations. Nonrejection does not establish global optimality. At the correct solution, the test still rejects with limiting probability $\alpha$ at level $\alpha$.
}

Proposition~\ref{prop:multi:J-screening} gives the interpretation at a stationary solution under its separate conditions. Its power argument extends to clusters when the corresponding stationary-point limits, separation, and shrinking residual covariance hold; the null limit at the correct solution follows from Lemma~\ref{lem:debiased-J:general} under its cluster sampling and accuracy conditions.

\begin{comment}
    \begin{remark}[Saddle Points]
	Because Assumption~\ref{asm:multi-local} imposes no curvature at the stationary points, the limit $\hat{\theta}_n^{(\mathcal{J})}$ may in principle be a saddle point or a local maximum. Establishing that the iterates avoid such unstable stationary points requires avoidance-of-traps arguments \citep{pemantle1990nonconvergence,brandiere1996traps}, which exploit excitation of the noise in the unstable directions. Such excitation is natural in our setting: $\bar{g}_n$ is bounded away from zero at every spurious stationary point by Condition~\ref{asm:item:multi:value-sep}, so the mini-batch noise $\xi_t$ does not degenerate there. A formal treatment, which would sharpen the conclusion of Theorem~\ref{thm:consistency:local} to convergence to a local minimizer, is left for future work.
\end{remark}

\end{comment}

\subsection{Proofs of Results for Multiple Stationary Points}\label{appx:proofs-multi-local}

\subsubsection{Proof of Theorem~\ref{thm:consistency:local}}
{
The proof reuses the supporting lemmas in Appendix~\ref{appx:lemmas-proofs}. It replaces the global {\L}ojasiewicz-type inequality \eqref{eq:eq8}, which fails in the presence of multiple stationary points, with a crossing argument: summability of the weighted gradient norms prevents infinitely many crossings between distinct stationary-point neighborhoods; applying the argument at arbitrarily small radii gives convergence.
}

\begin{proof}
	Throughout the proof, we consider the first-order algorithm with $\P_n = I_{\dtheta}$ and $W_n = I_{\dg}$, write $Q_t := \bar{Q}_n(\theta_t^*)$ and ${\nabla \bar{Q}_n(\theta) = \bar{G}_n(\theta)'\bar{g}_n(\theta)}$, and let $C < \infty$ denote an absolute constant that may change from line to line. The stochastic errors are $\xi_t(\theta) := \tilde{G}_t(\theta)'\tilde{g}_t(\theta) - \bar{G}_n(\theta)'\bar{g}_n(\theta)$, so that the update rule reads
	\begin{equation}
		\label{eq:update-local}
		\theta_t^* = \theta_{t-1}^* - \gamma_t \left( {\nabla \bar{Q}_n(\theta_{t-1}^*)} + \xi_t(\theta_{t-1}^*) \right), \quad t \ge 1,
	\end{equation}
	and $\mathbb{E}_n^*[\xi_t(\theta_{t-1}^*) \,|\, \mathcal{F}_{n,t-1}^*] = 0$.

	Define $E_n \in \mathcal{F}_n$ as the event on which the following statements hold: \eqref{eq:eq1}--\eqref{eq:eq6}, \eqref{eq:eq9}, and \eqref{eq:eq10} of Lemma~\ref{lem:sample-regularity}; $\|\bar{g}_n(\hat{\theta}_n)\| \le 1$; {the upper Hessian bound in Condition~\ref{asm:item:stability};} the statements of Conditions~\ref{asm:item:multi:critical}, \ref{asm:item:multi:value-sep}, and \ref{asm:item:multi:grad-sep}; and $\max_{1 \le j \le m}\|\hat{\theta}_n^{(j)} - \theta_\oo^{(j)}\| \le \delta/2$.
	We claim $\mathbb{P}(E_n) \to 1$.
	Conditions~\ref{asm:item:multi:critical}--\ref{asm:item:multi:grad-sep} {and the upper Hessian bound} hold \wpa\ by assumption, $\max_{j}\|\hat{\theta}_n^{(j)} - \theta_\oo^{(j)}\| = \oP(1)$ by Condition~\ref{asm:item:multi:consistency}, and $\|\bar{g}_n(\hat{\theta}_n)\|^2 \le \|\bar{g}_n(\theta_\oo)\|^2 \asto 0$ since $\hat{\theta}_n$ minimizes $\theta \mapsto \|\bar{g}_n(\theta)\|^2$.
	For the conclusions of Lemma~\ref{lem:sample-regularity}, {bounds \eqref{eq:eq7} and \eqref{eq:eq8} are not used below}: conclusions \eqref{eq:eq1}--\eqref{eq:eq2} are trivial here; \eqref{eq:eq5}, \eqref{eq:eq6}, \eqref{eq:eq9}, and \eqref{eq:eq10} rely only on Conditions~\ref{asm:item:iid}--\ref{asm:item:Lipschitz-G} and \ref{asm:item:stability}. {Bounds \eqref{eq:eq3}--\eqref{eq:eq4} follow from consistency of $\hat{\theta}_n$ under Condition~\ref{asm:item:multi:consistency}, the local uniform convergence of $\bar G_n$, the population rank condition, and the vanishing sample moments at $\hat{\theta}_n$ together with the bounded second derivatives of the sample moments.}
	Hence $\mathbb{P}(E_n) \to 1$.

	Three implications on $E_n$ are recorded for later use. First, since $\min_{j \ne k}\|\theta_\oo^{(j)} - \theta_\oo^{(k)}\| \ge 6\delta$ and $\max_j \|\hat{\theta}_n^{(j)} - \theta_\oo^{(j)}\| \le \delta/2$, we have $\min_{j \ne k}\|\hat{\theta}_n^{(j)} - \hat{\theta}_n^{(k)}\| \ge 5\delta$, so the balls $\mathcal{B}_j := \{\theta : \|\theta - \hat{\theta}_n^{(j)}\| \le 2\delta\}$, $j = 1, \ldots, m$, are disjoint and pairwise separated by at least $\delta$. Second, $\mathcal{B}_j \subseteq \Theta$ for each $j$ because $\{\theta : \|\theta - \theta_\oo^{(j)}\| \le 3\delta\} \subseteq \Theta$. Third, $\hat{\theta}_n^{(j)}$ is the unique stationary point of $\bar{Q}_n$ in $\mathcal{B}_j$: Condition~\ref{asm:item:multi:critical} excludes further stationary points within distance $\delta$ of $\hat{\theta}_n^{(j)}$, while any $\theta \in \mathcal{B}_j$ with $\|\theta - \hat{\theta}_n^{(j)}\| \ge \delta$ satisfies $\min_{k}\|\theta - \hat{\theta}_n^{(k)}\| \ge \delta$, because the centers $\hat{\theta}_n^{(k)}$ with $k \ne j$ are at distance at least $3\delta$ from $\theta$, so $\bar{G}_n(\theta)'\bar{g}_n(\theta) \ne 0$ by Condition~\ref{asm:item:multi:grad-sep}.
	In the remainder of the proof, we fix $n$, condition on the event $E_n$, and establish that
	\begin{equation}
		\label{eq:as-conv-local}
		\mathbb{P}_n^*\left( \lim_{N \to \infty} \theta_N^* = \hat{\theta}_n^{(\mathcal{J})} \ \text{ for some random } \mathcal{J} \in \{1, \ldots, m\}\right) = 1;
	\end{equation}
	all almost-sure statements below are under $\mathbb{P}_n^*$.

	We first derive a stochastic recursion for the Lyapunov value $Q_t$. By Assumption~\ref{assm:diff} and the second-order Taylor expansion, together with $\frac{\partial^2 \bar{Q}_n(\theta)}{\partial \theta \partial \theta'} \le M I_{\dtheta}$ from Condition~\ref{asm:item:stability} and the update rule \eqref{eq:update-local},
	\begin{align*}
		Q_t - Q_{t-1} & \le - \gamma_t \left( {\|\nabla \bar{Q}_n(\theta_{t-1}^*)\|^2} + \nabla \bar{Q}_n(\theta_{t-1}^*)' \xi_t(\theta_{t-1}^*) \right) \\
		& \quad + \frac{M}{2} \gamma_t^2 \left( {2} \|\nabla \bar{Q}_n(\theta_{t-1}^*)\|^2 + 2\|\xi_t(\theta_{t-1}^*)\|^2 \right).
	\end{align*}
	On $E_n$, $\|\nabla \bar{Q}_n(\theta)\|^2 \le {\|\bar{G}_n(\theta)\|^2} \|\bar{g}_n(\theta)\|^2 \le C (\bar{Q}_n(\theta) + 1)$ for all $\theta \in \Theta$, using \eqref{eq:eq5} and $\|\bar{g}_n(\theta)\|^2 = {2}\bar{Q}_n(\theta) + \|\bar{g}_n(\hat{\theta}_n)\|^2 \le {2}\bar{Q}_n(\theta) + 1$, and $\mathbb{E}_n^*[\|\xi_t(\theta)\|^2] \le \mathbb{E}_n^*[\|\xi_t(\theta)\|^{2p}]^{1/p} \le C(\bar{Q}_n(\theta) + 1)$ by \eqref{eq:eq10}.
	Taking conditional expectations therefore yields
	\begin{equation}
		\label{eq:recursion-Lyapunov-local}
		\mathbb{E}_n^*[Q_t \,|\, \mathcal{F}_{n,t-1}^*] \le Q_{t-1} - \frac{\gamma_t}{2} \|\nabla \bar{Q}_n(\theta_{t-1}^*)\|^2 + C \gamma_t^2 (Q_{t-1} + 1), \quad t \ge 1.
	\end{equation}
	Let $\pi_t := \prod_{s=1}^t (1 + C\gamma_s^2)$ with the constant $C$ from \eqref{eq:recursion-Lyapunov-local}, and note $\pi_\infty := \lim_t \pi_t < \infty$ since $\sum_s \gamma_s^2 < \infty$ under Assumption~\ref{assm:lr}. Setting $\tilde{Q}_t := Q_t / \pi_t$, dividing \eqref{eq:recursion-Lyapunov-local} by $\pi_t$ gives
	\begin{equation*}
		\mathbb{E}_n^*[\tilde{Q}_t \,|\, \mathcal{F}_{n,t-1}^*] \le \tilde{Q}_{t-1} - \frac{\gamma_t}{2\pi_\infty} \|\nabla \bar{Q}_n(\theta_{t-1}^*)\|^2 + C \gamma_t^2.
	\end{equation*}
	By Lemma~\ref{lem:Robbins-Siegmund}, $\mathbb{P}_n^*$-almost surely, $\lim_t \tilde{Q}_t$ exists and
	\begin{equation}
		\label{eq:grad-summability}
		\sum_{t=1}^\infty \gamma_t \|\nabla \bar{Q}_n(\theta_{t-1}^*)\|^2 < \infty.
	\end{equation}
	Since $\pi_t \to \pi_\infty \in (0, \infty)$, $\lim_t Q_t$ exists and $S := \sup_{t \ge 0} Q_t < \infty$, almost surely.

	Next, we show that the noise series converges and that the increments of the iterates vanish. Define the stopping time $\stt_{\loclevel}^* := \inf\{t \ge 0 : Q_t \ge \loclevel\}$ for $\loclevel > 0$, and the $\mathbb{R}^{\dtheta}$-valued martingale
	\begin{equation*}
		M_T^{(\loclevel)} := \sum_{t=1}^T \gamma_t \I{\stt_{\loclevel}^* \ge t}\, \xi_t(\theta_{t-1}^*), \quad T \ge 1,
	\end{equation*}
	which is indeed a martingale with respect to $(\mathcal{F}_{n,T}^*)_{T \ge 1}$ because $\{\stt_{\loclevel}^* \ge t\} \in \mathcal{F}_{n,t-1}^*$ and $\xi_t(\theta_{t-1}^*)$ is a martingale difference. On $\{\stt_{\loclevel}^* \ge t\}$, $Q_{t-1} < \loclevel$, so, by \eqref{eq:eq10}, the $t$th increment of $M_T^{(\loclevel)}$ has second moment at most $C(\loclevel+1)\gamma_t^2$. Since martingale increments are orthogonal,
	\begin{equation*}
		\sup_{T \ge 1} \mathbb{E}_n^*\big[\|M_T^{(\loclevel)}\|^2\big] \le C(\loclevel+1) \sum_{t=1}^\infty \gamma_t^2 < \infty.
	\end{equation*}
	By Doob's martingale convergence theorem, $M_T^{(\loclevel)}$ converges almost surely as $T \to \infty$, and on the event $\{\stt_{\loclevel}^* = \infty\}$ it coincides with $\sum_{t \le T} \gamma_t \xi_t(\theta_{t-1}^*)$. Since $\sup_t Q_t < \infty$ almost surely, $\mathbb{P}_n^*(\bigcup_{\loclevel \in \mathbb{N}} \{\stt_{\loclevel}^* = \infty\}) = 1$, whence
	\begin{equation}
		\label{eq:noise-series}
		\sum_{t=1}^\infty \gamma_t \xi_t(\theta_{t-1}^*) \ \text{ converges in } \mathbb{R}^{\dtheta} \ \text{ almost surely.}
	\end{equation}
	In particular, the tail sums $r_s := \sup_{u > s} \| \sum_{t=s+1}^u \gamma_t \xi_t(\theta_{t-1}^*) \|$ satisfy $r_s \to 0$ almost surely as $s \to \infty$. Moreover, $\sup_t \|{\nabla \bar{Q}_n(\theta_t^*)}\| \le C(S + 1)^{1/2} =: C_S < \infty$ almost surely, so by \eqref{eq:update-local},
	\begin{equation}
		\label{eq:vanishing-increments}
		\|\theta_t^* - \theta_{t-1}^*\| \le \gamma_t C_S + \gamma_t \|\xi_t(\theta_{t-1}^*)\| \to 0 \ \text{ almost surely,}
	\end{equation}
	where $\gamma_t \xi_t(\theta_{t-1}^*) \to 0$ almost surely because it is the increment of the convergent series \eqref{eq:noise-series}.

	We now show that the iterates eventually remain in a single ball $\mathcal{B}_j$. Let $D_t := \min_{1 \le j \le m} \|\theta_t^* - \hat{\theta}_n^{(j)}\|$, which is $1$-Lipschitz in $\theta_t^*$.
	First, $D_t < \delta$ infinitely often, almost surely. Indeed, on the event that $D_t \ge \delta$ for all $t \ge T_0$ for some $T_0 < \infty$, Condition~\ref{asm:item:multi:grad-sep} gives $\|\nabla \bar{Q}_n(\theta_t^*)\|^2 = {\|\bar{G}_n(\theta_t^*)'\bar{g}_n(\theta_t^*)\|^2 \ge c}$ for all $t \ge T_0$, so $\sum_t \gamma_{t+1} \|\nabla \bar{Q}_n(\theta_t^*)\|^2 \ge {c} \sum_{t > T_0} \gamma_t = \infty$ by Assumption~\ref{assm:lr}, contradicting \eqref{eq:grad-summability}; hence that event is $\mathbb{P}_n^*$-null.
	Second, almost surely, there are only finitely many disjoint intervals $[s_i, u_i]$, $i \ge 1$, with $D_{s_i} \le \delta$, $D_{u_i} \ge 2\delta$, and $D_t > \delta$ for $s_i < t < u_i$. To see this, suppose such intervals exist for all $i \ge 1$ on an event of positive probability; on that event $s_i \to \infty$. Since $D_t$ is $1$-Lipschitz, \eqref{eq:update-local} and the triangle inequality give
	\begin{equation*}
		\delta \le D_{u_i} - D_{s_i} \le \|\theta_{u_i}^* - \theta_{s_i}^*\| \le C_S \sum_{t = s_i + 1}^{u_i} \gamma_t + \Big\| \sum_{t = s_i + 1}^{u_i} \gamma_t \xi_t(\theta_{t-1}^*) \Big\| \le C_S \sum_{t = s_i + 1}^{u_i} \gamma_t + r_{s_i}.
	\end{equation*}
	For $i$ large enough that $r_{s_i} \le \delta/2$ and $\gamma_{s_i + 1} \le \delta / (4 C_S)$, this yields $\sum_{t = s_i + 2}^{u_i} \gamma_t \ge \delta/(4 C_S)$. Since $D_{t-1} > \delta$ for $t \in \{s_i + 2, \ldots, u_i\}$, Condition~\ref{asm:item:multi:grad-sep} implies
	\begin{equation*}
		\sum_{t = s_i + 2}^{u_i} \gamma_t \|\nabla \bar{Q}_n(\theta_{t-1}^*)\|^2 \ge %{c} \cdot \frac{\delta}{4 C_S}
		%=
		{\frac{c \delta}{4 C_S}}
	\end{equation*}
	for every such $i$, contradicting \eqref{eq:grad-summability}.
	Third, combining the two previous claims, almost surely there exists $T_1 < \infty$ such that $D_t < 2\delta$ for all $t \ge T_1$: if $D_t \ge 2\delta$ for arbitrarily large $t$ while $D_t < \delta$ infinitely often, we could extract infinitely many disjoint crossing intervals as above. For $t \ge T_1$, let $j_t$ denote the unique index with $\theta_t^* \in \mathcal{B}_{j_t}$, which is well defined because the balls $\mathcal{B}_j$ are disjoint. Since the balls are separated by at least $\delta$, a change of index requires $\|\theta_t^* - \theta_{t-1}^*\| \ge \delta$, which by \eqref{eq:vanishing-increments} occurs only finitely often. Hence, almost surely, there exist $T_2 \ge T_1$ and a random index $\mathcal{J} \in \{1, \ldots, m\}$ such that $\theta_t^* \in \mathcal{B}_{\mathcal{J}}$ for all $t \ge T_2$.

	It remains to establish convergence within the ball $\mathcal{B}_{\mathcal{J}}$. In what follows, we work on the almost-sure event on which \eqref{eq:grad-summability}--\eqref{eq:vanishing-increments} and the localization above hold, and let $\mathcal{A}$ denote the set of accumulation points of $(\theta_t^*)_{t \ge T_2}$, which is nonempty, compact, and contained in $\mathcal{B}_{\mathcal{J}}$ since the sequence is bounded and eventually confined to the closed ball $\mathcal{B}_{\mathcal{J}}$.
	We first show $\hat{\theta}_n^{(\mathcal{J})} \in \mathcal{A}$. Since $\sum_t \gamma_t \|\nabla \bar{Q}_n(\theta_{t-1}^*)\|^2 < \infty$ by \eqref{eq:grad-summability} while $\sum_t \gamma_t = \infty$ by Assumption~\ref{assm:lr}, we have $\liminf_{t \to \infty} \|\nabla \bar{Q}_n(\theta_t^*)\| = 0$. Extract a subsequence along which $\nabla \bar{Q}_n(\theta_t^*) \to 0$ and, by compactness, $\theta_t^* \to \bar{\theta} \in \mathcal{A}$ along a further subsequence. Since $\nabla \bar{Q}_n$ is continuous under Assumption~\ref{assm:diff}, $\nabla \bar{Q}_n(\bar{\theta}) = 0$, and $\bar{\theta} = \hat{\theta}_n^{(\mathcal{J})}$ because $\hat{\theta}_n^{(\mathcal{J})}$ is the unique stationary point of $\bar{Q}_n$ in $\mathcal{B}_{\mathcal{J}}$, as recorded at the beginning of the proof.
	Next, we show $\mathcal{A} = \{\hat{\theta}_n^{(\mathcal{J})}\}$. Suppose instead that some $\bar{\theta}' \in \mathcal{A}$ satisfies $\rho := \|\bar{\theta}' - \hat{\theta}_n^{(\mathcal{J})}\| > 0$. Then the iterates visit $\{\theta : \|\theta - \hat{\theta}_n^{(\mathcal{J})}\| \le \rho/3\}$ infinitely often and visit $\{\theta : \|\theta - \hat{\theta}_n^{(\mathcal{J})}\| \ge 2\rho/3\}$ infinitely often, so, because the increments vanish by \eqref{eq:vanishing-increments}, they cross the compact annulus
	\begin{equation*}
		\mathcal{S}_\rho := \left\{\theta \in \mathcal{B}_{\mathcal{J}} : \rho/3 \le \|\theta - \hat{\theta}_n^{(\mathcal{J})}\| \le 2\rho/3 \right\}
	\end{equation*}
	infinitely often. The continuous function $\theta \mapsto \|\nabla \bar{Q}_n(\theta)\|^2$ has no zero on $\mathcal{S}_\rho$ by the uniqueness recorded at the beginning of the proof, so $c_\rho := \inf_{\theta \in \mathcal{S}_\rho} \|\nabla \bar{Q}_n(\theta)\|^2 > 0$; the constant $c_\rho$ may depend on the data but not on $t$. Repeating the crossing argument of the previous paragraph with the distance $\|\theta_t^* - \hat{\theta}_n^{(\mathcal{J})}\|$ in place of $D_t$, the radii $\rho/3$ and $2\rho/3$ in place of $\delta$ and $2\delta$, and $c_\rho$ in place of ${c}$, each crossing eventually contributes at least $c_\rho\, \rho / (12\, C_S)$ to $\sum_t \gamma_t \|\nabla \bar{Q}_n(\theta_{t-1}^*)\|^2$, so infinitely many crossings contradict \eqref{eq:grad-summability}. Hence $\mathcal{A} = \{\hat{\theta}_n^{(\mathcal{J})}\}$, that is, $\theta_t^* \to \hat{\theta}_n^{(\mathcal{J})}$ almost surely, proving \eqref{eq:as-conv-local} for $\theta_N^*$. The averaged iterate $\bar{\theta}_N^* = N^{-1}\sum_{t=1}^N \theta_t^*$ converges to the same limit because Ces\`{a}ro averages of an almost surely convergent sequence converge to the same limit.

	Finally, we deduce the unconditional statement. Let $\varepsilon \in (0, \delta)$ be given; the case $\varepsilon \ge \delta$ follows by monotonicity. Observe the inclusion
	\begin{equation*}
		\left\{ \min_j \|\theta_N^* - \theta_\oo^{(j)}\| > \varepsilon \right\} \subseteq \left\{ \min_j \|\theta_N^* - \hat{\theta}_n^{(j)}\| > \varepsilon/2 \right\} \cup \left\{ \max_j \|\hat{\theta}_n^{(j)} - \theta_\oo^{(j)}\| > \varepsilon/2 \right\},
	\end{equation*}
	so that
	\begin{align*}
		\mathbb{P}\left( \min_j \|\theta_N^* - \theta_\oo^{(j)}\| > \varepsilon \right)
		& \le \mathbb{E}\left[ \mathbb{P}_n^*\left( \min_j \|\theta_N^* - \hat{\theta}_n^{(j)}\| > \varepsilon/2 \right) \mathbbm{1}_{E_n} \right] \\
		& \quad + \mathbb{P}(E_n^c) + \mathbb{P}\left( \max_j \|\hat{\theta}_n^{(j)} - \theta_\oo^{(j)}\| > \varepsilon/2 \right).
	\end{align*}
	By \eqref{eq:as-conv-local}, conditional on the data, $\mathbb{P}_n^*( \min_j \|\theta_N^* - \hat{\theta}_n^{(j)}\| > \varepsilon/2 ) \mathbbm{1}_{E_n} \to 0$ pointwise as $N \to \infty$ for each fixed $n$, so the first term vanishes as $N \to \infty$ by the bounded convergence theorem. Therefore,
	\begin{equation*}
		\limsup_{N \to \infty} \mathbb{P}\left( \min_j \|\theta_N^* - \theta_\oo^{(j)}\| > \varepsilon \right) \le \mathbb{P}(E_n^c) + \mathbb{P}\left( \max_j \|\hat{\theta}_n^{(j)} - \theta_\oo^{(j)}\| > \varepsilon/2 \right) \to 0
	\end{equation*}
	as $n \to \infty$, since $\mathbb{P}(E_n) \to 1$ and by Condition~\ref{asm:item:multi:consistency}. The identical argument applies to $\bar{\theta}_N^*$. This completes the proof.
\end{proof}

\subsubsection{Proof of Proposition~\ref{prop:multi:J-screening}}
{
\begin{proof}
Let $\mu_j=g(\theta_\oo^{(j)})$. Uniform convergence and consistency give $\hat g_{n,j}\pto\mu_j$ and $\hat D_{n,j}\pto D_j$. The same limits hold at $\hat\theta_n^{(j)}$. Its stationarity implies $D_j'\mu_j=0$. For $j\ne1$, criterion separation and $\mu_1=0$ imply $\|\mu_j\|^2\ge c>0$.

Since $H_jD_j=0$ and both matrices have their stated ranks, $\ker H_j=\operatorname{col}D_j$. If $H_j\mu_j=0$, then $\mu_j=D_jv$ for some $v$, giving
\[
0=v'D_j'\mu_j=\mu_j'\mu_j,
\]
a contradiction. Hence $\hat h_{n,j}\pto H_j\mu_j\ne0$. On the event of positive definiteness,
\[
J_{D,n}^{(j)}\ge
\frac{\|\hat h_{n,j}\|^2}{\lambda_{\max}(\hat V_{h,n,j})}
\pto\infty.
\]
The null limit follows directly from Lemma~\ref{lem:debiased-J:general}. Finiteness of $m$ makes rejection of the spurious candidates simultaneous.
\end{proof}
}

{\singlespacing
\renewcommand{\refname}{Appendix References}
\putbib
}
\end{bibunit}

\end{document}